# *Colloidal Systems*

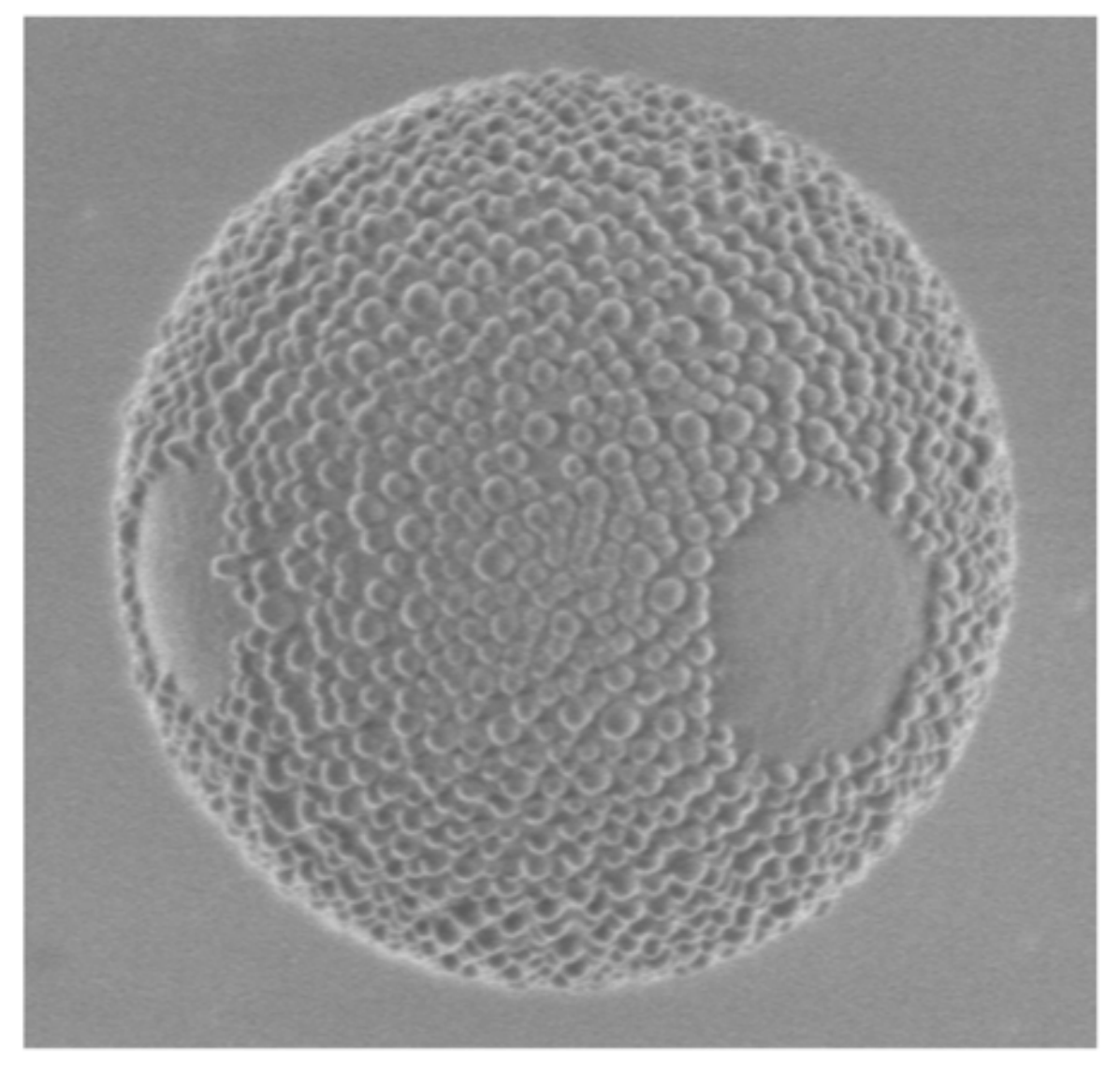

***Darrell Velegol***

***Colloidal Systems***

For more information, or to send comments or suggestions, please email velegol@psu.edu.

*Dedicated to John L. Anderson –*
*an inspiring master of colloid science, and so much more.*

*… and to my amazing wife Stephanie,*
*who brings life to my days, and who makes this work possible.*

## *Contents*

# *Preface*

Our lab group has explored colloidal systems since 1999. In the early years of fabricating assemblies or designing electrokinetic transport, we would sometimes walk into the lab, mix particles and solutions, and run the experiment to see if we got what we hoped for. Almost invariably, this approach failed. Time after time we have found the critical importance of designing experiments by examining the fundamental physics and chemistry of the system. In more recent years, as we started to explore colloidal motors and more complex colloidal systems, it has become even more important to start with the fundamentals.

On the other hand, for design purposes we might need a result only within a factor of two or even a factor of ten. Is the aggregation time 30 seconds, 30 minutes, or 30 hours? Will the particles settle in 5 minutes, 5 hours, or 5 months? Will we need a centrifuge to help the settling? Thus, navigating the design of colloidal systems can require semi-quantitative calculations, a simple recipe, or a conceptual understanding, rather than detailed calculations.

This book has one primary goal: To get you moving quickly from learning basic principles of colloid science, to designing colloidal systems for your needs. It is not intended to provide encyclopedic knowledge about all aspects of colloid and surface science; rather, I have chosen not to include many topics of genuine importance. In reaching the intended goal, I use two parallel strategies. First, I give practical results that can be used immediately. If there is a cumbersome calculation, I have worked to reframe it into a table or figure. This is the case for Hamaker constants and the hydrodynamics of spheroids, for instance. Second, I describe the physics of various colloidal phenomena, so that you as a researcher can *think* about alternative strategies.

In balancing algorithms with explanations, I provide plug-and-chug example problems to familiarize you with units, constants, and typical values. The practice problems provide extensions to the most basic theory that open new possibilities, while also giving results that are useful in practical work. Throughout the book, I provide data we commonly use for viscosities, zeta potentials, ionic phenomena, and other parameters. Rather than providing every reference and every technique, I have provided those references and techniques that we most often use in our lab. In the end, this book aims not to be an exhaustive study of colloids, but rather to be a doorway to producing desired colloidal systems as quickly as possible. And then … I hope that you will provide the magical pieces that will turn your work into something *valuable* for your research or your business.

In writing this book I have drawn on the foundations laid by hundreds of researchers, and on books and articles written by hundreds of authors. Thank you, to my friends and colleagues past and present who have contributed to this great field of ours. I can hardly express my appreciation to all the beautiful research you have published. Finally, I thank my students. My amazing students. In one song from *The King and I*, Anna sings "Getting to Know You". Part of the lyrics are

> *It's a very ancient saying,*
> *But a true and honest thought,*
> *That if you become a teacher,*
> *By your pupils you'll be taught.*

Amen.

*Darrell Velegol*
*Penn State University*
*University Park, Pennsylvania*
*www.velegol.org*
*2016 July*

# *Colloidal Systems*

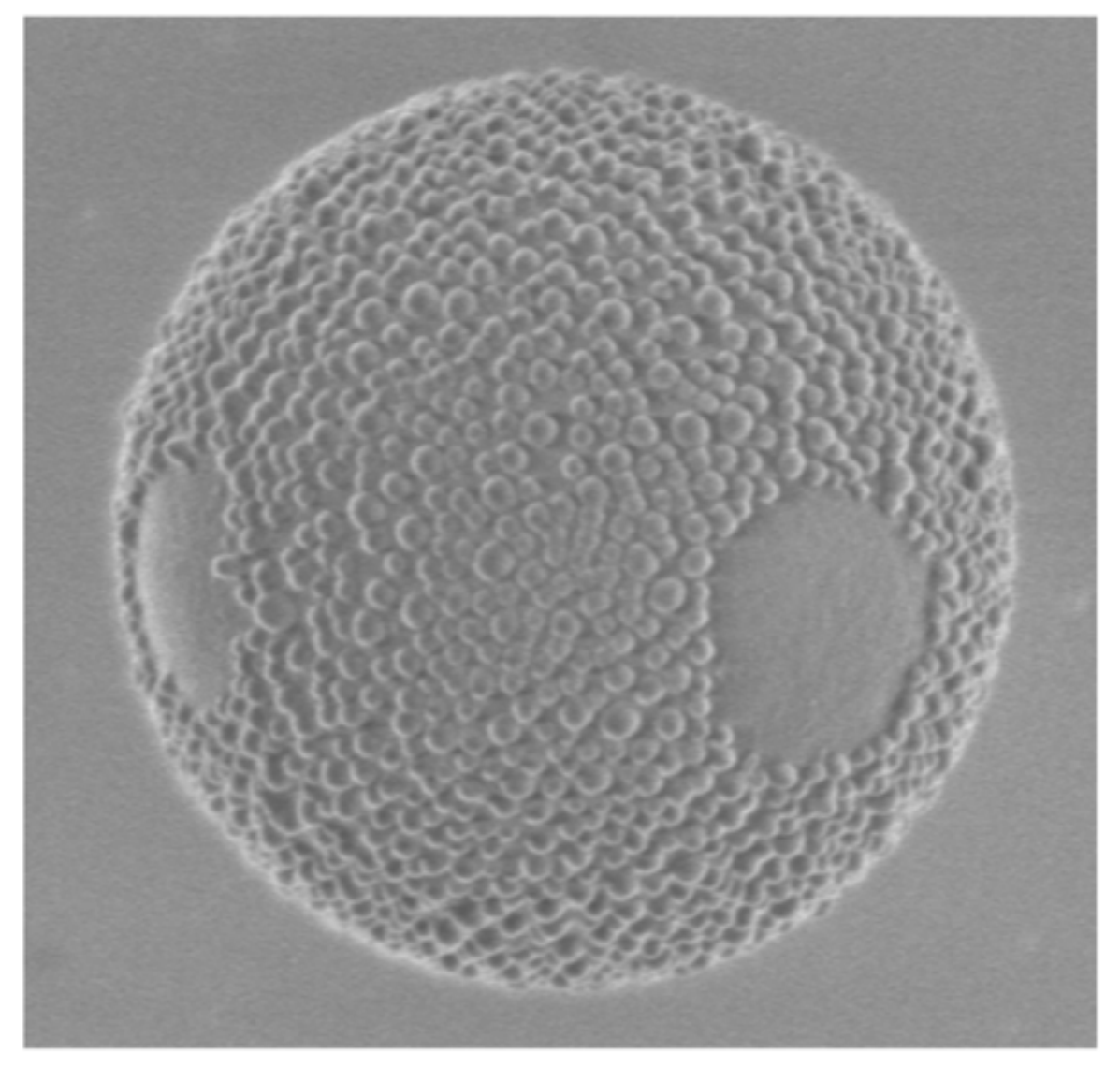

## *Darrell Velegol*

Wild Scholars Media

# 1 Overview: fabricating colloidal doublets

**References:** Berg, John C. *An Introduction to Interfaces & Colloids*. World Scientific (New Jersey) 2010. This book is among the best general colloids books I have seen – good coverage, difficult results put in graphs, insightful, and with lots of fun experiments. John Berg is a wonderful friend and colleague.

Russel, W.B.; Saville, D.A.; Schowalter, W.R. *Colloidal Dispersions*. Cambridge University Press (New York) 1989. This is the classic book covering colloidal physics in extensive mathematical depth. I suspect that experts find it more useful than beginners; it is packed with great concepts.

Hunter, Robert J. *Foundations of Colloid Science*, 2nd ed. Oxford University Press (New York) 2001.

## *Colloidal motors … a model colloidal assembly.*

In 2004 Ayusman Sen, Tom Mallouk, and their co-workers published an article in *JACS* showing a fascinating phenomenon with bi-metallic rods.[1] They made their rods in alumina membranes, according to a procedure that had been published five years earlier. The 2 μm long rods consisted of a 1 μm gold segment and a 1 μm platinum segment, and their diameter was 370 nm. When the rods were placed in a 3.7% aqueous solution of hydrogen peroxide, something unexpected happened: The rods began to move spontaneously! In fact, they moved roughly 8 μm/s – four body lengths per second. Four body lengths per second is about how fast a world class athlete runs the 1500 m race. This chapter journeys through the thought process of how a person might build and examine such self-swimming particles.

Not until more than a year later was the transport mechanism of the rods, which the researchers called “catalytic nanomotors” or “colloidal motors”, discovered. Originally, the mechanism was thought to be due to surface tension. The researchers knew that platinum decomposed $H_2O_2$ readily to produce water and oxygen, while gold did not. The hypothesis was that a gradient of surface tension was driving the movement of the rods, not altogether different from how a camphor boat moves. But then the PhD students added a thin layer of dielectric material between the platinum and gold, and they found – somewhat surprisingly – that the movement of the rods stopped. Their mindset turned to thinking that the decomposition must be happening electrochemically, and that the reaction required that electrons transport through the metallic phase. Our lab group had the privilege to be involved with this discovery, and through substantial back and forth between experiments and modeling, we together collected experimental and modeling evidence that the rods were moving by auto-electrophoresis.[2,3] The electric field that arose due to the electrochemical decomposition of $H_2O_2$ was driving the particles through the aqueous solution electrokinetically.

---

[1] Walter F. Paxton, Kevin C. Kistler, Christine C. Olmeda, Ayusman Sen, Sarah K. St. Angelo, Yanyan Cao, Thomas E. Mallouk, Paul E. Lammert and Vincent H. Crespi, “Catalytic Nanomotors: Autonomous Movement of Striped Nanorods”, *J Am Chem Soc*, **126**, 13424-13431 (2004).

By this point, probably around late 2006, our lab group had been studying the assembly of colloidal particles for a few years. We had developed a simple technique for producing doublets of particles from single spheres, a method we called Stimulate-Quench-Fuse (SQF) assembly. We wondered if we could make colloidal motors – in this case, heterodoublets of goldparticles and silver particles (Figure 1-1) – in a simple, scalable manner.

And so this is what we set out to do. In this chapter I will take you on a rapid-fire tour of the doublet fabrication process. We'll visit some of the key concepts of colloid science, highlight some of the most important equations, and use a few of the critical characterization techniques. It's OK if you don't grasp everything at this point; we'll spend the rest of the book unpacking the details.

---

[2] Kline, T. R.; Paxton, W. F.; Wang, Y.; Velegol, D.; Mallouk, T. E.; Sen, A. "Catalytic Micropumps: Microscopic Convective Fluid Flow and Pattern Formation." *Journal of the American Chemical Society*, **127**, 17150-17151 (2005).
[3] Kline, Timothy R.; Iwata, Jodi; Lammert, Paul E., Mallouk, Thomas E.; Sen, Ayusman; Velegol, Darrell. "Catalytically-driven Colloidal Patterning and Transport." *Journal of Physical Chemistry B*, **110**, 24513-24521 (2006).

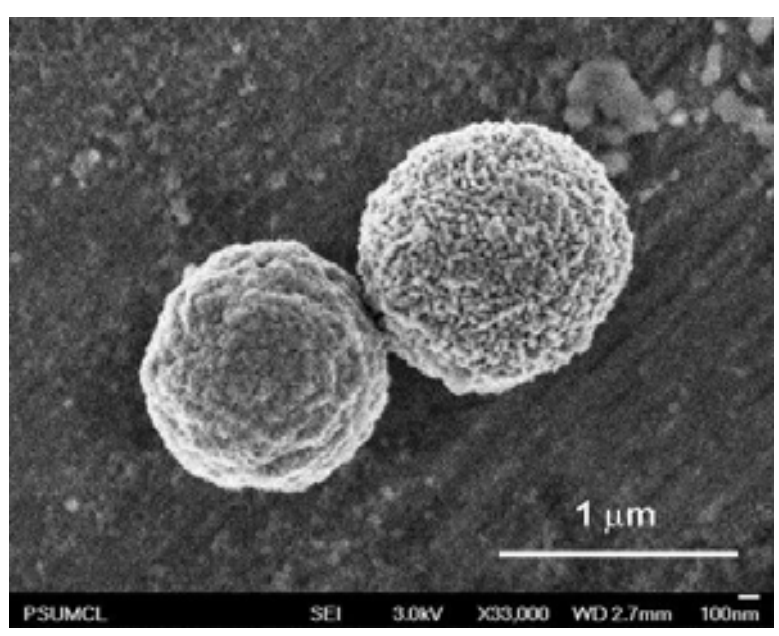

**Figure 1-1.** Gold-silver heterodoublet. This field-emission scanning electron micrograph (FESEM) shows the diameter of each particle to be about 1 μm. The gold particle (right) has a slightly different surface structure than the silver particle.

***Synthesis. Particles can be bought, or made by known recipes.***

The first step was to obtain the starting materials, which for us were gold and silver particles of diameter roughly 1 μm. Oftentimes colloidal particles can simply be purchased (Table 1-1), especially small quantities for research purposes. At first glance these particles might seem quite expensive. For a 15 mL bottle of polystyrene particles, you might pay $150. And even then, a typical bottle has perhaps 4% or 8% particles, with the rest being water. At that price, these particles cost about 10× as much as gold on a per mass basis. On the other hand, companies[4] like Dow and BASF make millions of tons of polymer particles per year, at a cost more like $1/kg, but these can be hard to obtain unless you have a contact within the company.

Some particles like silica can be readily purchased, or they can be synthesized by the well-known Stober method[5]. Particles can be made in a number of ways; for silicon we start with a wafer and use mortar and pestle crushing. If there is another type of particle we want to make, maybe calcium carbonate, we search the literature for "calcium carbonate colloidal particle synthesis". A search on Google Scholar (http://scholar.google.com) will yield hundreds or even thousands of recipes. We'll look more at particle synthesis in Chapter 7.

[4] *Polymer Dispersions and Their Industrial Applications*. Editors Dieter Urban and Koichi Takamura. Wiley-VCH (Germany) 2002. Chapter 1, "Introduction", by Dieter Urban and Dieter Distler lists a number of interesting and useful statistics about polymer lattices.
[5] Stober, Werner; Fink, Arthur; Bohn, Ernst. "Controlled Growth of Monodisperse Silica Spheres in the Micron Size Range." *J. Colloid Interface Sci.*, **26**, 62-69 (1968). This paper lists the classic recipe for silica particles.

**Table 1-1.** Purchasing colloidal particles for research use. These companies are current in 2012, and one can Google to find current listings.

| Manufacturer | Specialization |
|---|---|
| Bangs Laboratories http://www.bangslabs.com | Polystyrene and PMMA particles, silica particles |
| ThermoFisher (https://www.thermofisher.com) | monodisperse (low CV) sizes for polymer particles, including with fluorescence or dye or many other functionalizations; includes sizes from 10s of nm to 10 µm |
| Polysciences http://www.polysciences.com | polymer particles, pollen, magnetic beads, custom synthesis |

For the gold and silver particles we need for our bimetallic colloids, very standard synthesis recipes exist.[6] To get to micron size gold or silver, we use techniques that are well-established in the literature.[7-9] All of these recipes work very well, are fairly robust, and can be completed in less than one hour.

[6] Handley, Dean A. "Methods for Synthesis of Colloidal Gold," Ch 2 in *Colloidal Gold: Principles, Methods, and Applications*, Vol. 1. Ed. M.A. Hayat. Academic Press (New York) 1989. The table of contents to the chapter lists methods for making sizes from 0.82 nm to 64 nm.
[7] Goia, Dan V.; Matijevic, Egon. "Preparation of monodispersed metal particles." *New J. Chemistry*, **22**, 1203-1215.
[8] Goia, Dan V.; Matijevic, Egon. "Tailoring the particle size of monodispersed colloidal gold." Colloids and Surfaces A, **146**, 139-152 (1999).
[9] Velikov, Krassimir. P.; Zegers, Gabby E.; van Blaaderen, Alfons. " Synthesis and Characterization of Large Colloidal Silver Particles." *Langmuir*, **19**, 1384-1389 (2003).

***Colloidal forces. Particle aggregation depends on these forces.***

Say that we have synthesized our gold and silver colloidal particles. How do we go from single spherical particles of gold and silver, to *heterodoublets* of gold and silver (Figure 1-1)? One route is to use a very simple, quick method like Stimulate-Quench-Fuse assembly.[10,11] In this method we intentionally change solution conditions to encourage aggregation of the particles. This is opposite of the usual case. Usually we want particles to remain *stable* to aggregation, rather than to aggregate.

A natural question is: Why *would* particles clump together? Answer: van der Waals (VDW) forces. The mechanism of attraction is quantum mechanical in nature: The electrons within all atoms continually do their "quantum mechanical dance", and so at any instant of time, slightly more electrons exist in one region of an atom – or colloidal particle – than another. Temporary electric dipoles result on every particle, and although they might be small, they are not zero. In turn, these temporary dipoles will *induce* dipoles in neighboring atoms or particles. The temporary dipole interacts with the induced dipole to give an attractive force.

VDW forces *always* exist between atoms, and they are always attractive, at least at the atomic level. They are the force responsible for particles wanting to reduce their exposed surface area by aggregating. Studying VDW forces requires analyzing the quantum mechanics of the problem, and I save a more detailed explanation for Chapter 3. For now, we want to get some idea of the magnitude of VDW forces. A simple equation for estimating the VDW attractive energy ($\Phi_{VDW}$, measured in Joules) between two spherical particles of radius ($a$), separated by a distance ($\delta$), is

[10] Yake, Allison M.; Panella, Rocco A.; Snyder, Charles E.; Velegol, Darrell. "Fabrication of Doublets by a Salting Out – Quenching – Fusing Technique." *Langmuir*, **22**, 9135-9141 (2006). This is the original paper describing the SQF method.

[11] McDermott, Joseph J.; Velegol, Darrell. "Simple Fabrication of Metallic Colloidal Doublets Having Electrical Connectivity." *Langmuir*, **24**, 4335-4339 (2008).

$$\Phi_{VDW} = -\frac{Aa}{12\delta} \tag{1-1}$$

The Hamaker constant ($A$) hides all of the complicated quantum mechanics, and we will explore it more in Chapter 3. Typical values are $A = 1.4 \times 10^{-20}$ J for two polystyrene surfaces interacting across water, and for gold in water or vacuum $A = 20 \times 10^{-20}$ J. Almost always $A > 0$, meaning that the VDW energy is negative, which by definition is attractive. Situations for which A < 0 for particles are given in Chapter 3.

Having a "potential energy" between two particles is not altogether different from having a gravitational potential energy between earth and an object. The gravitational potential energy of an object with mass ($m$) at a height ($h$) is $\Phi_{grav} = mgh$. Just like the gravitational force is $F_g = -d\Phi_{grav}/dh = -mg$, the VDW force is given by

$$F_{VDW} = -\frac{d\Phi_{VDW}}{d\delta} = -\frac{Aa}{12\delta^2} \tag{1-2}$$

One difference between the VDW potential energy and the gravitational potential energy is that as the distance between particles or objects grows beyond about 100 nm, the VDW energy approaches zero, while the gravitational energy falls off more slowly.

If VDW forces are always attractive, then what can provide the repulsive force to keep particles from aggregating? Usually one of two types of forces is used to provide the repulsion: electrostatic or steric. Sometimes these forces are used in combination as "electrosteric respulsion". For electrostatic forces, we know that like charges repel and opposite charges attract, and we see a similar trend between particles. But electrostatic forces between particles, especially in aqueous solutions, require a lot more explanation than a simple application of Coulomb's law.

The key difference between electrostatic forces in a vacuum and electrostatic forces in aqueous solution is the *electrical double layer* (EDL). Almost any surface immersed in water becomes charged, for example by dissociated surface carboxyl groups on polymer colloids, or adsorbed ions on gold particles, or dissociated silanol groups on silica particles. The surface charges have a given surface charge density ($\rho_s$) that has a typical value of 1 μC/cm$^2$. A single charge group has the charge of a proton $e = 1.6022\times10^{-19}$ C, and so 1 μC/cm$^2$ corresponds to having one univalent charge group every 16 nm$^2$, or charge groups with an average spacing of 4 nm.

**Example 1-1**. Van der Waals energy between colloidal particles.

Estimate the VDW energy between two polystyrene particles suspended in water, if they have the same diameter $2a$ = 1.5 μm and are separated by a gap $\delta$ = 25 nm between their closest points.

*answer:* $\Phi_{VDW} = -3.5\times10^{-20}$ *J. Negative energies are attractive, by definition. To compare this with a "molecular energy scale", we use kT, since from statistical mechanics we know that every particle has an average kinetic energy of kT/2 in each direction. The Boltzmann constant* $k = 1.38065\times10^{-23}$ *J/K (the gas constant divided by Avogadro's number), and so if we choose a temperature of* $T$ = 290 K, *then* $kT = 4.00\times10^{-21}$ *J. This means that* $\Phi_{VDW} = -3.5\times10^{-20}$ J = -8.75 kT. *This is a small but significant energy that will cause aggregation of the particles, unless a repulsive energy also exists. Sometimes we list Hamaker constants in terms of kT's, as in saying that the Hamaker constant in this problem is* $A = 1.4\times10^{-20}$ J = 3.5 kT.

These charge groups give rise to an electrical potential at the surface, given in volts. Sometimes this is called the surface potential, and if the potential is measured using electrophoresis (Chapter 8), it is called the "zeta potential". In solution the surface potential decays to a bulk value of zero – electroneutrality – more quickly than Coulomb's law would suggest. Why? In solution, ions are mobile. If the charges at the surface are negative, positive ions from solution will move toward the surface to neutralize the charge. They will form a thin layer having mostly positive charges around the particle, with some negative ions mixed in. The combination of the fixed charges on the particle surface and the diffuse charges in the thin layer around the particle is called the "electrical double layer" (Figure 1-2).

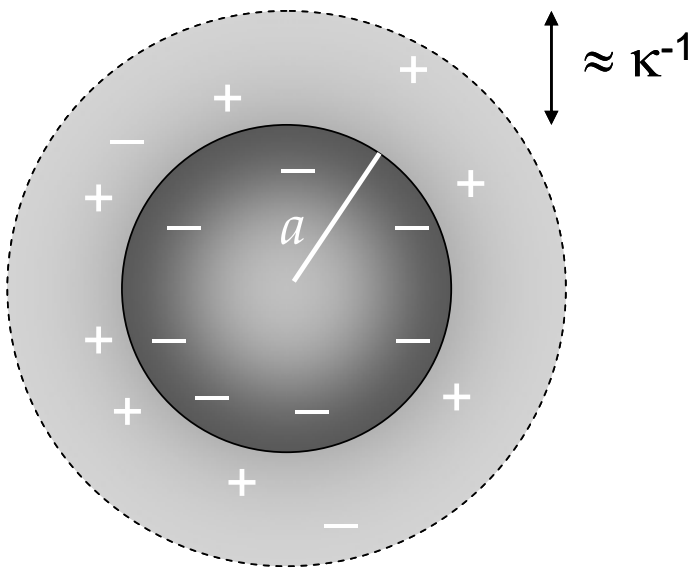

**Figure 1-2.** The electrical double layer (EDL). Particles immersed in water become charged, for example due to dissociated surface groups. Oppositely-charged counterions from solution come near the surface to neutralize the charges, and the counterions remain in a diffuse region of fluid around the particle. Together, the fixed charge layer and diffuse charge layer make up the EDL. The thickness of the EDL is characterized by the Debye length ($\kappa^{-1}$ or $\lambda_D$).

The thickness of the electrical double layer is characterized by the "Debye screening length" ($\kappa^{-1}$ or $\lambda_D$), which is often called simply the "Debye length". In aqueous solutions the Debye length is usually thin, typically a few nanometers, and depends on the ionic strength. Table 1-2 gives values of the Debye length as a function of ionic strength. Since the EDL neutralizes the charge on the particle over a thin region near the particle surface, the attractive or repulsive energy between particles is reduced from Coulomb's law by an exponential factor. The electrical potential decays away from a flat surface at a rate given roughly by the Debye-Huckel equation:

$$\psi = \psi_0 e^{-\kappa x} \qquad (1\text{-}3)$$

where $\psi_0$ is the surface potential, and $x$ is the distance from the surface. Typical magnitudes of the surface potential ($\psi$) are 10 to 150 mV, and can be negative or positive. Since these potentials arise due to the surface charges, I also list a relationship between the surface charge density (units C/m$^2$) and the surface potential:

$$\rho_s = \varepsilon \kappa \psi_0 \qquad (1\text{-}4)$$

The Debye-Huckel equation and the relationship between surface charge density and surface potential will be derived in Chapter 2.

**Table 1-2**. Debye length ($\kappa^{-1}$, in nm) at various concentrations of aqueous solution for $T$ = 293 K. Typical 1:1 salts are KCl or NaCl; a 2:1 salt is $CaCl_2$. The Debye length decreases with the square root of the ionic strength, and increases with the square root of temperature. So for instance to find the Debye length of a 23 mM KCl solution at $T$ = 300 K, we might start with the Debye length for a 1:1 salt at 1 mM, which is 9.65 nm, then multiply by two factors (300 K / 293 K)$^{0.5}$ × (1 mM / 23 mM)$^{0.5}$ = 0.211 to get 2.04 nm.

| conc (mM) | 1:1 (KCl) | 2:1 ($CaCl_2$) | 3:1 ($AlCl_3$) | 2:2 ($CaSO_4$) | 3:2 [$Al_2(SO_4)_3$] | 3:3 ($AlPO_4$) |
|---|---|---|---|---|---|---|
| 0.001 | 305 | 176 | 125 | 153 | 78.8 | 102 |
| 0.003 | 176 | 102 | 71.9 | 88.1 | 45.5 | 58.7 |
| 0.01 | 96.5 | 55.7 | 39.4 | 48.2 | 24.9 | 32.2 |
| 0.03 | 55.7 | 32.2 | 22.7 | 27.9 | 14.4 | 18.6 |
| 0.1 | 30.5 | 17.6 | 12.5 | 15.3 | 7.88 | 10.2 |
| 0.3 | 17.6 | 10.2 | 7.19 | 8.81 | 4.55 | 5.87 |
| 1 | 9.65 | 5.57 | 3.94 | 4.82 | 2.49 | 3.22 |
| 3 | 5.57 | 3.22 | 2.27 | 2.79 | 1.44 | 1.86 |
| 10 | 3.05 | 1.76 | 1.25 | 1.53 | 0.788 | 1.02 |
| 30 | 1.76 | 1.02 | 0.719 | 0.881 | 0.455 | 0.587 |
| 100 | 0.965 | 0.557 | 0.394 | 0.482 | 0.249 | 0.322 |
| 300 | 0.557 | 0.322 | 0.227 | 0.279 | 0.144 | 0.186 |
| 1000 | 0.305 | 0.176 | 0.125 | 0.153 | 0.0788 | 0.102 |

Between two particles of radius ($a$), the electrostatic energy ($\Phi_{ES}$) resulting from the electrical potentials on a particle separated by a distance of closest approach ($\delta$) is approximated by

$$\Phi_{ES} = 2\pi\varepsilon a \psi_0^2 e^{-\kappa\delta} \tag{1-5}$$

The electrical permittivity ($\varepsilon$) is given in Table 1-3 for a few common fluids. Fluids with higher dielectric constants give larger electrostatic interactions energies, not only since $\varepsilon$ appears in Eq 1-2 above, but also since particles tend to become more highly charged – and therefore have a larger magnitude of surface potential – in these fluids. And so electrostatic forces tend to be fairly large in water at short range.

**Table 1-3**. Static (zero frequency) electrical permittivities for several liquids. The permittivity of vacuum is $\varepsilon_0 = 8.8542\times10^{-12}$ C$^2$/N-m$^2$. For most liquids the electrical permittivity ($\varepsilon$) is represented by a multiple of $\varepsilon_0$ called the “dielectric constant” and a “relative permittivity” ($\varepsilon_r$). For example, at 20 C water has $\varepsilon_r = 80.1$, and so the permittivity of water at 20 C is $80.1\varepsilon_0 = 7.09\times10^{-10}$ C$^2$/N-m$^2$. The static dielectric constant depends weakly on temperature.[12]

| *fluid* | $\varepsilon_r$ at $T$ = 20 C | $\varepsilon_r$ at $T$ = 25 C | *dielectric constant ($\varepsilon_r$)* at T (C) |
|---|---|---|---|
| acetone | 21.2 | 20.7 | $\varepsilon_r = 21.2\exp[-0.00472(T-20)]$ |
| ammonia | 17.4 | 16.9 | $\varepsilon_r = 17.4 - 0.090(T-20)$ |
| benzene | 2.284 | 2.274 | $\varepsilon_r = 2.284 - 0.0020(T-20)$ |
| cyclohexane | 2.023 | 2.015 | $\varepsilon_r = 2.023 - 0.0016(T-20)$ |
| ethanol | 25.1 | 24.3 | $\varepsilon_r = 25.1\exp[-0.006217(T-20)]$ |
| methanol | 33.62 | 32.63 | $\varepsilon_r = 33.62\exp[-0.00599(T\text{-}20)]$ |
| water | 80.37 | 78.54 | $\varepsilon_r = 80.37\exp[-0.004605(T-20)]$ |

[12] *CRC Handbook of Chemistry and Physics*, 55th edition, CRC Press (Boca Raton, FL) 1974.

As a first approximation to the total energy between the particles, we can simply add the electrostatic and VDW energies to give the well-known "Derjaguin-Landau-Verwey-Overbeek" (DVLO) energy between the two particles.

$$\Phi \approx \Phi_{DLVO} = \Phi_{ES} + \Phi_{VDW} \approx 2\pi\varepsilon a \psi_0^2 e^{-\kappa\delta} - \frac{Aa}{12\delta} \qquad (1\text{-}6)$$

This equation has two "approximately equal" (≈) signs. The first of these arises since other forces between colloidal particles can be significant. In later chapters I will mention solvation and hydrophobic forces, which result from solvent ordering around particles, as well as the attractive depletion forces resulting from non-adsorbing polymer in solution. DLVO theory considers only the VDW energy, which is always present, as well as the electrostatic energy, which is usually present. The second "≈" arises since the expressions for the electrostatic and VDW energies written here are approximate. We will give more accurate expressions in Chapters 2 and 3.

**Example 1-2**. Interaction energy between two colloidal particles.

Estimate the electrostatic energy, the van der Waals attractive energy, and the total energy between two polystyrene particles suspended in water, if they have the same diameter $2a = 1.5$ μm and are separated by a gap $\delta = 25$ nm between their closest points. The surface potential of the particles is $\psi_0 = -35$ mV, and the solution is 10 mM KCl at a temperature $T = 20$ C. The Hamaker constant A for a PS-water-PS system is approximately $1.4\times10^{-20}$ J.

*answer:* $\Phi_{ES} = +1.13\times10^{-21}$ *J* $= +0.28$ *kT. Previously we found that* $\Phi_{VDW} = -35.0\times10^{-21}$ *J* $= -8.656$ *kT.* $\Phi_{DLVO} = -33.9\times10^{-21}$ *J* $= -8.38$ *kT. This energy is attractive at this distance.*

In maintaining stability, usually we consider the maximum energy barrier to aggregation ($\Phi_{max}$). In Figure 1-3 this is the barrier from the well at $\delta$ = 10 nm – this is the "secondary energy minimum", and its value is -4.9×10$^{-20}$ J – up to the top of the barrier at $\delta$ = 1.3 nm, which is +28×10$^{-20}$ J. These energies might at first seem extraordinarily small. How do we know if they will prevent or allow aggregation? After all, we can see a very deep "primary energy minimum" near zero separation, and so the thermodynamics indicate that aggregation will always prevail. Rather, stability is a question os kinetics, and in order to assess stability, we must learn about the aggregation rate.

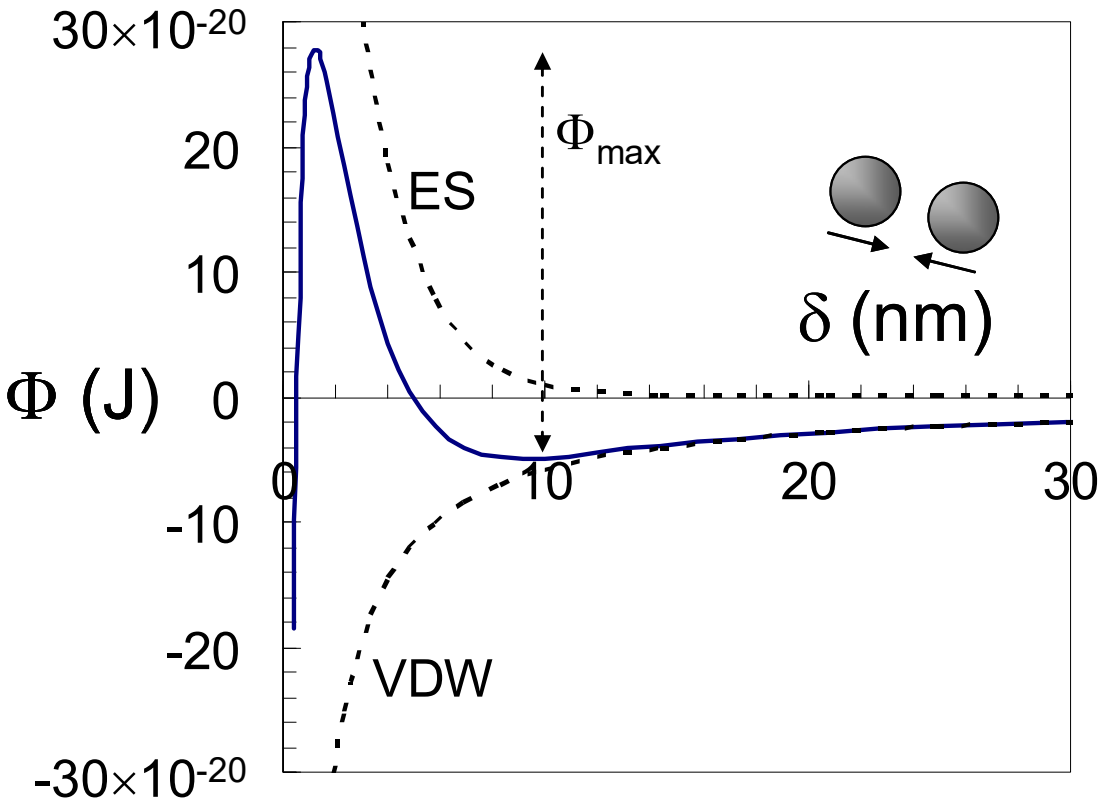

**Figure 1-3**. Energy between two particles. The plot is for two polystyrene particles with diameter 1.0 μm in 23 mM KCl solution at 293 K, so that the Debye length is $\kappa^{-1}$ = 2.0 nm. The particles each have surface potentials of -25 mV. Shown are the electrostatic energy (upper dotted line ES), the VDW energy (lower dotted line), and total energy (the "DLVO energy", Φ, solid line). The "maximum energy barrier" ($\Phi_{max}$) is shown by the vertical dashed line.

***Stability. Aggregation occurs by diffusion or shear.***

Getting back to our doublet story … in order to fabricate colloidal doublets, we are using the SQF method. The essential idea is to reduce the electrostatic repulsion between particles, enabling the VDW attractive forces to cause aggregation. Two methods that are used to reduce electrostatic repulsion are to lower the pH of the solution, which redues the number of dissociated charge groups on the negative particles, or to increase the ionic strength of the solution, which also reduces electrostatic forces. The question at this point is: If we reduce the repulsive forces and estimate that the forces between the particles are now negligible until near-contact, how long does it take the particles to "find each other" so that they can aggregate? That is, how quickly will the singlets aggregate into doublets?

Particles can aggregate only when they collide. There are two common mechanisms for collision: diffusive collisions and shear-induced collisions. We will focus here on diffusive or "Brownian" aggregation, leaving shear-induced aggregation for Chapter 6. The Smoluchowski equation gives an estimate of the aggregation time ($\tau$) for roughly half of the particles in suspension to aggregate:

$$\tau = \frac{\pi \eta a^3}{2kT\phi} W \tag{1-7}$$

As expected, the time is longer for a larger fluid viscosity ($\eta$) and for a smaller volume fraction of particles ($\phi$). The aggregation time depends also on the "stability ratio" ($W$), which identifies how many times particles must collide on average, before they adhere. One simulated prediction of $W$, for any given energy barrier ($\Phi_{max}$, see Figure 1-3), is

$$W = 1 + 0.25\left[ exp\left( \frac{\Phi_{max}}{kT} \right) - 1 \right] \tag{1-8}$$

What causes diffusive aggregation? Diffusive aggregation results first when the interparticle repulsive forces are small enough so that they cannot prevent aggregation, and simultaneously when the process of "Brownian motion" causes particles to collide. Let's take a look at the process of "Brownian motion" for our gold particles in water. It was in 1827 that Robert Brown made his observations of dancing pollen grains, which we now know as Brownian motion. A good way to visualize the process is to Google "brownian motion" and click on "images" or "video". YouTube shows several optical microscope recordings of particles undergoing Brownian motion.

**Example 1-3**. Colloidal aggregation rate.

We have a suspension of 1.1 μm diameter gold spheres in water at 298 K. The volume fraction of particles is $\phi = 0.00033$. Estimate how long it will take for half the particles to aggregate, both in a) rapid aggregation ($\Phi_{max} = 0$) and b) with electrostatic stabilization ($\Phi_{max} = +20$ kT). The viscosity of water at 298 K is 0.00089 Pa-s.

*answer: a) For rapid aggregation, $W = 1$ and so $\tau = 171$ s $\approx 3$ min. The particles will aggregate on a time scale that we could watch in the lab; they would not aggregate too fast to see, say μs, or too slow to see, say years. b) $W = 1.21\times10^8$ and $\tau = 659$ years. The particles are predicted not to aggregate on a commercial time scale of even 2 to 3 years. As a rule of thumb, particles with surface potentials $|\psi_0| > 50$ mV are stable.*

The cause of Brownian motion is the ceaseless, ever-moving action of the molecules in a system. From classical statistics mechanics, we know that every molecule in the system is moving randomly in all three dimensions, with an average kinetic energy $kT/2$ in each direction, or $3kT/2$ total. The thermal energy ($kT$) is the Boltzmann constant ($k = 1.38065\times10^{-23}$ J/K) multiplied by the absolute temperature ($T$). The average kinetic energy of a molecule is $\langle m(v_x^2 + v_y^2 + v_z^2)/2\rangle = \langle m\mathbf{v}\cdot\mathbf{v}/2\rangle = 3kT/2$, which is found either by averaging over one molecule over a long time or for many molecules at an instant. Thus, smaller particles move faster.

Our gold particles in this doublet fabrication, having a diameter of roughly 1 μm, are buffeted on all sides by the water molecules. The net effect of the thermal motion is that the gold particles also take on an average kinetic energy of $3kT/2$, and they move randomly in the fluid. But for the particles, there are in fact two processes to consider. First, the very rapid buffeting of the solvent molecules, which will in general not be exactly symmetric about the particle at any instant, produces an instantaneous force on the particle. Second, when the particle is given a force, it moves through the sea of water molecules, but it must move out all the molecules in front of it. This second process gives resistance to the particle movement, which dissipates the energy of the particle back into the solvent molecules in front of it.

**Example 1-4**. Average speed of a molecule undergoing random thermal motion.

Find the root-mean-square (RMS) speed $\langle v^2\rangle^{1/2} = \langle \mathbf{v}\cdot\mathbf{v}\rangle^{1/2}$ for a water molecule in a beaker of water, at $T$ = 293 K. The mass of a water molecule is 18 g/mol = (0.018 kg/mol) / ($6.022\times10^{23}$ / mol) = $2.989\times10^{-26}$ kg/molecule.

*answer:* $\langle v^2\rangle^{1/2}$ *= 637 m/s. Sound travels at roughly 1500 m/s in water.*

The entire process is a “fluctuation-dissipation” process, with the fluctuation being the Brownian kick the particle receives, and the dissipation being the movement of the solvent molecules in front of it. The kicks give an average kinetic energy to the particle of $3kT/2$. The fluid resistance force against the particle is given by Stokes law for a particle, which predicts the hydrodynamic resistance force ($\mathbf{F}_H$) for a particle of radius ($a$) moving through a fluid of viscosity ($\eta$) at a velocity ($\mathbf{U}$):

$$\mathbf{F}_H = -6\pi\eta a\mathbf{U} \tag{1-9}$$

The bold font indicates a vector quantity, since the particle can move in any of the 3 dimensions. The hydrodynamic force is opposite to the direction of the particle movement, as expected. Stokes law will be discussed more in Chapter 4.

The analysis to find the Brownian motion of the gold particles is somewhat involved, and I will save the details for Chapter 5. For now, I will list a couple useful results for Brownian motion that we can use to make quick and accurate estimates.

$$L^2_{RMS} = \left\langle \Delta x^2 + \Delta y^2 + \Delta z^2 \right\rangle = \left\langle \Delta\mathbf{x} \cdot \Delta\mathbf{x} \right\rangle = 6Dt \tag{1-10}$$

$$D = \frac{kT}{6\pi\eta a} \tag{1-11}$$

Equation 1-10 predicts the "root-mean-square" distance ($L_{RMS}$) that a particle will diffuse by Brownian motion in 3-dimensional space. For one dimension, the right hand side is simply one third as much, or $2Dt$, and for two dimensions, $4Dt$. The diffusion coefficient ($D$) is given by the Stokes-Einstein result in Equation 1-11. Both the $kT$ "kick factor" and the Stokes dissipation factor appear in the expression for $D$.

**Example 1-5**. Diffusion of a colloidal particle.

What is the average (root-mean-square) distance a gold particle of diameter 0.9 μm will diffuse in 3 dimensions in 1 minute? In 1 dimension? The temperature $T$ = 298 K. The viscosity of water at 298 K is 0.00089 Pa-s.

*answer:* $D = 5.45\times10^{-13}$ m²/s, $L_{RMS,3D} = 14.0$ μm, $L_{RMS,1D} = 8.09$ μm.

We are now armed with several simple and powerful results from colloidal physics, which enable us to design our process for fabricating doublets. Say we have 1.1 μm diameter gold spheres in water at 298 K, at a volume fraction of particles $\phi = 0.00033$. How do we form doublets? We first "stimulate" the aggregation by killing the electrostatic repulsive forces. We might suddenly change the pH – for our gold particles, which are for example stabilized by citric acid groups with a $pK_a$ of 3.1, we might change the pH to 1.5 or 2.0 – or we might suddenly change the ionic strength to say 500 mM KCl. Then as Example 1-3 shows, we wait roughly 3 min. In practice, we usually wait for a bit less than this, perhaps 1 minute, to avoid having too many triplets and other higher order aggregates form. The final step is to "quench" the aggregation, by restoring the electrostatic repulsion between the particles. If we were using a low pH, we would add base to bring the pH back to 6.0 or so. If we were using a high ionic strength to kill the electrostatic repulsion, we would now dilute to 10 mM to restore the repulsive force.

In the process of forming doublets, we are left with a large amount of "unreacted" singlets, as well as a small fraction of particles that went beyond doublets to form doublets, triplets, 4-lets, and so on. After all, this is a random aggregation process. We end up with a significant fraction of doublets since we quenched the aggregation early, but if we want a pure stream of say greater than 80% doublets, we have to do a separation or sorting process.

***Sorting. Colloidal assemblies must be sorted from mixtures.***

By the "separation" of colloids, I could mean separating particles from the suspending fluid, or separating one type of particle from other types. This latter type – separating one type of particles from others – I call "sorting". In order to produce a pure stream of doublets (Figure 1-4), we must sort them from the "unreacted" singlets, as well as from the higher order "side products" that occur during the aggregation process.

Several techniques exist for the sorting operation. Field flow fractionation and hydrodynamic chromatography provide very good sorting capabilities, but they both produce a fraction of a gram of colloidal material per hour. We might expect that gravitational settling would provide an easy method for sorting particles of different size or even shape. I will use the name "sedimentation" for the process of settling in a gravitational field (**g**), whether it is in the earth's gravitational field (1g) or in centrifugation (perhaps 1000s of g).

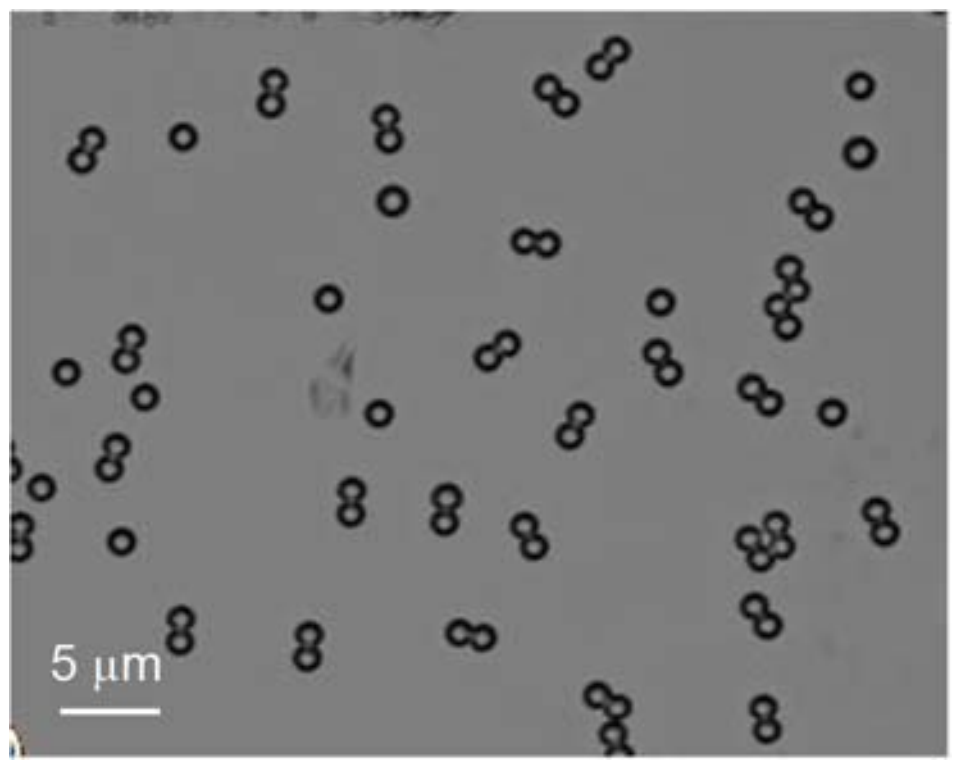

**Figure 1-4**. Colloidal doublets.10 The particles are composed of polystyrene spheres 2.4 µm in diameter, suspended in de-ionized water and settled onto a glass coverslip. These particles were sorted using a density gradient sedimentation.

An analysis of the fluid mechanics equations reveals that a sphere of density ($\rho_p$) settles in a fluid of density ($\rho_f$) and viscosity ($\eta$) during a sedimentation operation at a velocity (**U**) given by

$$\mathbf{U} = \frac{2a^2(\rho_p - \rho_f)}{9\eta}\mathbf{g} \qquad (1\text{-}12)$$

Material properties for several fluids and particles are given in Table 1-4.

**Table 1-4**. Material properties for various fluids and particles. The fluid densities are listed for pure fluids, free from air. All temperatures are in degrees C, and most equations are linearized at T = 20 C. To convert viscosity in cP to Pa-s, divide by 1000.

| material | *density, $\rho$ ( kg/m$^3$) at T (C)* | *viscosity, $\eta$ ( cP) at T (C)* |
|---|---|---|
| Fluids | | |
| acetone | 792 | $0.327 - 0.0021(T - 20)$ |
| air (dry) | $1.293\times[273/(273+T)]\times(p/1\ \text{atm})$ | $0.0181 + 0.000043(T - 20)$ |
| benzene | 899 | $0.652 - 0.0097(T - 20)$ |
| cyclohexane | 779 | 17 C, 1.02 cP |
| ethanol | $789 - 0.8455(T - 20)$ | $1.200 - 0.02315(T - 20)$ |
| glycerin (pure) | 1 261 | $1490 - 137.6(T - 20)$ |
| mercury | 13 546 – 2.5(T – 20) | $1.554 - 0.0058(T - 20)$ |
| methanol | 810 | $0.597 - 0.0076(T - 20)$ |
| water[13,14] | 10 C, 999.73 | 10 C, 1.307 cP = 0.001307 Pa-s |
| | 20 C, 998.23 | 20 C, 1.002 |
| | 25 C, 997.07 | 25 C, 0.8904 |
| | 30 C, 995.67 | 30 C, 0.7975 |
| | 40 C, 992.24 | 40 C, 0.6529 |
| Particles | | |
| glass (common) | 2 400 to 2 800 | |
| gold | 19 300 | |
| polymethylmeth acrylate | 1 190 | |
| polystyrene | 1 055 | 120 C, roughly $10^{10}$ |
| silica | 2 200 | |
| silver | 10 490 | |

[13] The density of water (kg/m$^3$) is given by $\rho_f = 998.23 - 0.192(T - 20) - 0.0051(T - 20)^2$ for 4 C < T < 50 C.

[14] The viscosity of water is given in cP by

$$\log_{10}\eta = \frac{1301}{998.333 + 8.1855(T-20) + 0.00585(T-20)^2} - 1.30233 \quad \text{for } 0\text{ C} < T < 20\text{ C}$$

and

$$log_{10}\frac{\eta}{1.002} = -\frac{1.3272(T-20) + 0.001053(T-20)^2}{T+105} \quad \text{for } 20\text{ C} < T < 100\text{ C}.$$

If the particle is a doublet settling long-axis downward, the particle settles 1.55× as fast as a singlet, whereas if the long-axis is horizontal, the particle settles 1.41× as fast. But there is a notorious difficulty – a fluid instability – in sorting by sedimentation that requires extra effort and thought.[15]

When particles are suspended in a fluid, they cannot escape their local environment quickly. As a result, the particles plus the suspending fluid in a given region behave as a single complex fluid, which takes on instabilities and in fact mixes itself. In the end sedimentation can work as a sorting technique, but requires great care. I will describe effective density sorting and density gradient sedimentation in Chapter 9.

For now, I will simply give the recipe for sorting singlets from doublets using density gradient sedimentation.[16] Using a test tube and a density gradient maker, you add a sucrose solution to the tube that has perhaps 20% sucrose at the bottom and 5% sucrose at the top. The sugar polymer Ficoll 400 (pronounced FIʹ kol) can also be used. The concentration profile of sugar is roughly linear, and the resulting solution will also have a gradient of density. On top of the sucrose solution is added the particle suspension to be separated – it is critical that the volume fraction of particles be low, perhaps $\phi <$ 0.001 or even 0.0001. With time the particles will settle down the tube in a stable manner, without swirling or mixing, and after two or more bands of particles visibly appear, they can be collected with a syringe or some similar technique. Upon sorting the doublets, we end up with a suspension that appears as shown in Figure 1-4. This optical microscope image shows greater than 80% doublets, with only a few singlets. No high order aggregates appear.

[15] Jerri, Huda A.; Sheehan, William P.; Snyder, Charles E.; Velegol, Darrell. "Prolonging Density Gradient Stability," *Langmuir*, **26**, 4725-4731 (2010).
[16] Price, C.A. *Centrifugation in Density Gradients*, Academic Press (New York) 1982. This book describes many of the instabilities that occur, and how to counteract them using density gradients.

***Electrokinetics. Electric fields drive the fluid in the EDL.***

At this point we have our colloidal motor doublets. Now we can examine if they are functional. Will they in fact move like the Sen-Mallouk bimetallic nanorods? Through careful study it was found that the nanorods move by the mechanism of auto-electrophoresis. What does that mean?

In order to understand auto-electrophoresis, I will first discuss the process of electrophoresis. Say that we have a suspension of nanorods – or now, bimetallic doublets – and that we apply an electric field ($\mathbf{E}_\infty$) to the suspension. We would see the particles move through the solution due to the applied electric field. How do they move? It might seem obvious that a charged particle would move in an electric field. Heuristically, it is often helpful to remember that if a particle has a positive surface potential ($\psi_0 > 0$ mV) it will move toward the anode (negative terminal of the electric field) and if a particle has a negative surface potential ($\psi_0 < 0$ mV) it will move toward the cathode (positive terminal of the electric field).

More specifically, electrophoresis occurs when an applied electric field ($\mathbf{E}_\infty$) acts on the charges on the particle itself, as well as on the charges within the electrical double layer (EDL, Figure 1-2). When an electric field is applied, the charges in the EDL are set into motion, since they are in a fluid. The ions come to their terminal velocity within a fraction of a picosecond, transferring their force via fluid friction to the surrounding aqueous medium. The net result is that the electric field acts upon the charged fluid of the EDL – which now behaves like a continuum of charged fluid with a volumetric charge density ($\rho_e$) – causing it to move with a rather complex flow field field. The combination of the electric field acting on the charges on the particle itself, plus the electric field causing movement of the fluid in the EDL, causes the particle to move with an electrophoretic velocity ($\mathbf{U}$). One equation that predicts $\mathbf{U}$ is the Smoluchowski equation:

$$\mathbf{U} = \frac{\varepsilon \zeta \mathbf{E}_\infty}{\eta} \tag{1-13}$$

where $\varepsilon$ is the fluid permittivity given in Table 1-3 and $\eta$ is the fluid viscosity given in Table 1-4. The zeta potential ($\zeta$) gives a value that is often used for the surface potential ($\psi_0$) of the particle. This definition of the zeta potential will be cleared up significantly, with complications pointed out, in Chapter 8. But typical values for the magnitude of $\zeta$ range from 25 to 150 mV, or 1 to 6 $kT/e$ at room temperature, and these can typically be measured using electrophoresis to within a few millivolts (mV).

**Example 1-6**. Electrophoresis of a colloidal particle.

Estimate how fast a 2.0 μm diameter polystyrene sphere will move by electrophoresis through an aqueous solution of 0.1 M KCl at T = 293 K? The applied electric field is $\mathbf{E}_\infty$ = 3 V/cm = 300 V/m. The zeta potential of the particle is $\zeta$ = +37 mV, since it is amidine-functionalized. Compare this to speed to that of a pseudo-cubic hematite ($\alpha$-$Fe_2O_3$) peanut-shaped particle with a length 1.8 μm and a diameter 0.8 μm, which at a particular pH also has $\zeta$ = +37 mV.

*answer: Both move with U = +7.9 μm/s. This is a typical electrophoretic speed, given in μm/s. The size and shape of the particle do not affect the speed, when the thickness of the double layer is thin compared to the size of the particle (large $\kappa a$). For a 1.0 μm radius particle in 100 mM KCl, $\kappa a$ = 1000.*

Electrophoresis is one type of electrokinetic phenomenon. There are others, such as diffusiophoresis, which we will explore in more detail (Chapter 8). The colloidal motor doublets also move by an electrokinetic mechanism, called auto-electrophoresis. The silver-gold doublets are charged since they are in aqueous solution, perhaps due to adsorbed ions from solution or due to added stabilizing molecules. The two metals catalyze the reaction of hydrogen peroxide in solution, and this reaction causes a spontaneous electric field to occur in the near vicinity of the particle. At its surface the gold particle produces protons (H+) that diffuse and are convected toward the silver particle, where the protons are consumed in a reaction. The moving protons are what set up the electric field in the vicinity of the particle, which – similar to the case previously described – acts on the charge groups on the particle and on the charge groups in the diffuse layer of the EDL around the particles. When the spontaneously-produced electric field acts on the charge colloidal doublets, we see movement as shown in Figure 1-5. The movement of the protons through the solution near the doublet produces an auto-electrophoresis that causes the particles to move.

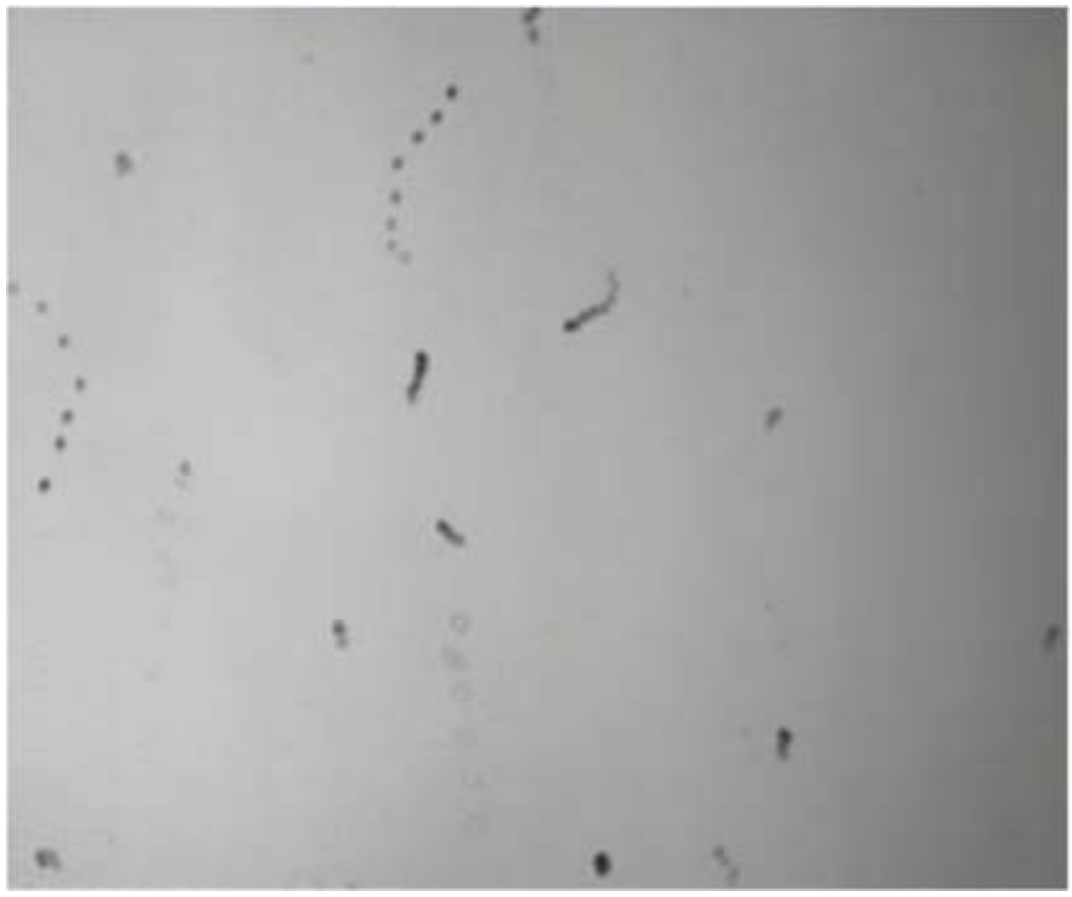

**Figure 1-5**. Auto-electrophoresis of colloidal doublet motors.[11] The trace images are taken over a time period of roughly 10 seconds, and the movement of some of the doublets is clearly visible.

Throughout this chapter, we have seen characterizations of our single particles and our doublets. We used FESEM to see the doublets close up (Figure 1-1). We used optical microscopy to see our doublets after sorting (Figure 1-4). We use zeta potential measurements to determine the surface potential on our particles. Zeta potentials can be used in predictions of electrostatic forces (Equation 1-5), in predictions of electrophoretic velocity (Equation 1-13), or even in characterizing the chemistry on the particle surface in some cases. We used video microscopy to examine the moving nanodoublets (Figure 1-5). Many physical and chemical characterizations can be performed, and a list of many additional technicals for measuring physical, chemical, electrical, and other properties is given by the Materials Characterization Lab at Penn State (http://www.mri.psu.edu/facilities/mcl). Always when one is working with safety in the lab, a paramount concern is safety.[17]

***Summary. All these processes have levels of detail.***

For every process discussed in this chapter – from synthesis to electrophoresis – there are layers of complexity. Sometimes the enhanced layers are necessary to understand the phenomena observed in the lab. My purpose in this chapter has been to take you on a rapid-fire tour of doublet production. In subsequent chapters I will take a deeper look at each of the phenomena described in this chapter. Our destination for the entire book is to reach the point where we know enough context – particle types, measurement techniques, physical processes – and enough modeling so that we can design new colloidal systems. Let’s enjoy the journey in subsequent chapters.

[17] Our Colloids Lab meeting starts weekly with the topic of “safety”. In 2010 I sat in on a Process Safety course CH E 452 taught by my colleague Bob Nedwick, which was absolutely eye-opening. Safety accidents happen unexpectedly and suddenly, and the only way to avoid most accidents is by preparation. Bob’s course emphasized this sudden nature of accidents – they can occur in even a fraction of a second, far quicker than any person can respond – as well as the importance of a systematic checklist-type approach to handling safety. I cannot recommend high enough having a systematic approach to safety in every lab.

### *Symbols*

$a$ = sphere radius [=] m. For colloidal particles, $a$ is often in nm or small μm.

$A$ = Hamaker constant [=] J. Typical value is $10^{-20}$ J.

$D$ = diffusion coefficient for particles [=] $m^2/s$.

$\delta$ = distance of closest approach between particles [=] m.

$\Delta x$, $\Delta y$, $\Delta z$ = distance traversed by diffusion [=] m

$\Delta \mathbf{x}$ = vector traversed by diffusion [=] m

$e$ = charge on a proton = $1.6022\times10^{-19}$ C.

$\varepsilon$ = permittivity [=] $C^2/N\text{-}m^2$. $\varepsilon = \varepsilon_0\varepsilon_r$.

$\varepsilon_0$ = permittivity of a vacuum = $8.8542\times10^{-12}$ $C^2/N\text{-}m^2$.

$\varepsilon_r$ = relative permittivity or dielectric constant [=] dimensionless. For water, $\varepsilon_r$ = 80 roughly, so that $\varepsilon = 7.1\times10^{-10}$ $C^2/N\text{-}m^2$.

$\mathbf{E}_\infty$ = applied electric field [=] V/m. Typical values in the lab are 1-10 V/cm.

$\mathbf{F}_H$ = hydrodynamic resistance force [=] N.

$\Phi_{DLVO}$ = DLVO energy (sum of electrostatic and VDW) [=] J.

$\Phi_{ES}$ = electrostatic energy between particles [=] J.

$\Phi_{max}$ = maximum repulsive energy barrier to aggregation between particles [=] J.

$\Phi_{VDW}$ = van der Waals (attractive) energy between particles [=] J.

$\Phi$ = total energy between particles [=] J.

$\mathbf{g}$ = gravitational vector, where $g$ = 9.81 $m/s^2$.

$\eta$ = viscosity [=] Pa-s or kg/m-s. The symbol μ is often used for viscosity.

$k$ = Boltzmann's constant = $1.38065\times10^{-23}$ J/K.

$\kappa^{-1}$ = Debye length [=] m. Typical values are 1 to 100 nm in aqueous solutions.

$\lambda_D$ = another symbol used for the Debye length. See $\kappa^{-1}$.

$\rho_e$ = volumetric charge density [=] $C/m^3$.

$\rho_f$ = fluid density [=] $kg/m^3$.

$\rho_p$ = particle density [=] $kg/m^3$.

$\rho_s$ = surface charge density [=] C/m$^2$. A typical value is 1 μC/cm$^2$ = 0.01 C/m$^2$.

$t$ = time [=] s

$T$ = temperature [=] C or K.

$\tau$ = aggregation time [=] s.

**U** = particle velocity [=] m/s.

$\zeta$ = zeta (≈surface) potential [=] V or mV. 1 V = 1 J/C. Typical values are $|\zeta|$ = 25 to 150 mV.

### *Practice Problems*

1. **Synthesis recipes**. Find a recipe to make polyvinyl acetate colloidal particles. One method is to use Google Scholar.

2. **Interparticle forces and energy**. Using Eq 1-6 plot the energy profile (i.e., distance-energy curve) shown in Figure 1-**3**. What is $\Phi_{max}$?
*answer*: roughly 33×10$^{-20}$ J.

3. **Sedimentation speed**. Calculate the sedimentation speed of a 320 nm diameter silica spheres settling at 1g in water at 25 C.
*answer: 75 nm/s.*

4. **Diffusion coefficient**. Estimate the diffusion coefficient of sucrose in water at 293 K. Sucrose molecules have a diameter of 0.68 nm.
*answer*: $D$ = 0.63×10$^{-9}$ m$^2$/s. The actual $D$ = 0.4586×10$^{-9}$ m$^2$/s at this temperature. Continuum theory gives a reasonable prediction, even though the sucrose molecule is only slightly larger than the diameter of a water molecule (0.3 nm).

5. **Aggregate rate**. Estimate how long it would take for significant aggregation to occur in the following systems:

a radius $a$ = 2000 nm, water at 293 K, $\phi$ = 0.04, $\Phi_{max}$ = 0.
*answer*: 78 s.

b radius $a$ = 50 nm, water at 293 K, $\phi$ = 0.04, $\Phi_{max}$ = 0.
*answer*: 0.0012 s.

c radius $a$ = 50 nm, water at 293 K, $\phi$ = 0.04, $\Phi_{max}$ = 15 kT.
*answer*: 16.5 minutes.

d radius $a$ = 50 nm, water at 293 K, $\phi$ = 0.04, $\Phi_{max}$ = 30 kT.
*answer*: 103 years.

6. **Stokes settling velocity**. Derive Eq 1-12 for Stokes settling in a gravitational field. Use the fact that at terminal velocity, $F_H + F_g = 0$, and put the gravitational force in terms of the buoyant density of the particle.

7. **Surface potential**. Most colloidal particles have a surface potential that is a small multiple of $kT/e$. Show that $kT/e$ = 25.7 mV at $T$ = 298 K.

8. **Colloidal size diameter**. The definition of a "colloidal particle" is subjective. One definition is by size. Derive an expression for the diameter for which the time it takes the particle to settle a distance of one diameter (2$a$) by gravity ($2a = Ut$) is the same as the average time required to diffuse that distance by Brownian motion ($2a = (6Dt)^{1/2}$). Both $U$ and $D$ depend upon $a$, as given in Chapter 1. This method gives a similar result for the "colloidal size radius" as setting the Peclet number $Pe = Ua/D = 1$.

*answer*: $2a = \left( \dfrac{36kT}{\pi(\rho_p - \rho_f)g} \right)^{1/4}$. No viscosity appears.

9. **Upper colloidal size diameter values**. Using the expression from the previous problem, estimate the upper colloidal size

diameter ($2a$) for a polystyrene sphere in water at $T$ = 293 K. Do the same for a silica sphere, and a gold sphere.
*answer*: $2a$ = 3.04 μm for PS, 1.41 μm for silica, and 0.713 μm for gold. These three materials, of greatly different buoyant density, have similar "upper colloidal size diameters".

10. **Characterization techniques**. List 20 techniques for characterization at the Penn State Materials Characterization Lab (http://www.mri.psu.edu/facilities/mcl).

# 2 Electrostatic force

**References:** Hunter, Robert J. *Zeta Potential in Colloid Science*, Academic Press (New York) 1981.
Robinson, R.A.; Stokes, R.H. *Electrolyte Solutions*, 2nd ed. Dover Publications (Mineola, New York) 1959. This book has extensive and useful lists of data for electrolyte solutions.
Russel, W.B.; Saville, D.A.; Schowalter, W.R. *Colloidal Dispersions*. Cambridge University Press (New York) 1989.

## *Electrical double layer (EDL)*

In 1993 I started as a PhD student in Professor John L. Anderson's group. To get me started into my research, he said, "Darrell, the first thing I want you to learn about is the electrical double layer. It is central to the study of colloidal forces, and also to electrokinetic phenomena and colloid science more generally." He suggested that I begin by studying Chapter V of the famous book by Arthur Adamson.[18]

What is "double" about the electrical double layer around a particle? There are two charged layers: 1) a fixed layer of charges on the particle surface, and 2) a fluid layer touching the particle surface, which contains oppositely-charged "counter-ions". Together, the net sum of the charges on the particle plus the charges in the fluid layer add to zero, meaning they are together electroneutral. The primary quantities that we want to know about the EDL are the electrical potential ($\psi$) on and around the particle, and sometimes the spatial distribution of ions in the fluid layer.

In aqueous solutions the particle surfaces almost always become charged, by a variety of mechanisms. For polymer colloids, charge groups such as carboxyl, sulfonate, or sulfate are often added covalently during the synthesis process. If the charge group is a carboxyl (-COOH), then when the pH of the fluid is greater than the pKa of the acid group – in this case, roughly 4 – the proton will dissociate, leaving a negatively-charged COO- group bonded to the surface. For silica particles, the surface groups first become silanol groups (-Si-OH) in water, and then the protons dissociates to give negatively-charged Si-O- groups at the surface. Whether the mechanism is surface group dissociation, or metal ion substitution, or adsorption of ions from solution, by some mechanism the particle surface will almost always become charged in water. We often call this the "fixed charged layer", although the "fixed" nature ignores the pico-second scale dynamics of the interface, instead describing the charges with statistical averages. Colloid scientists frequently also refer to a Stern layer, which is an additional layer of species bound directly near the fixed layer of charges. The Stern layer concept has not been especially useful in our lab, since it is complex to model and think about. We just include it as part of the fixed charges, at least until the ionic strength becomes very high.

[18] Adamson, Arthur W. *Physical Chemistry of Surfaces*, 5th ed. Wiley (New York: Wiley (1990).

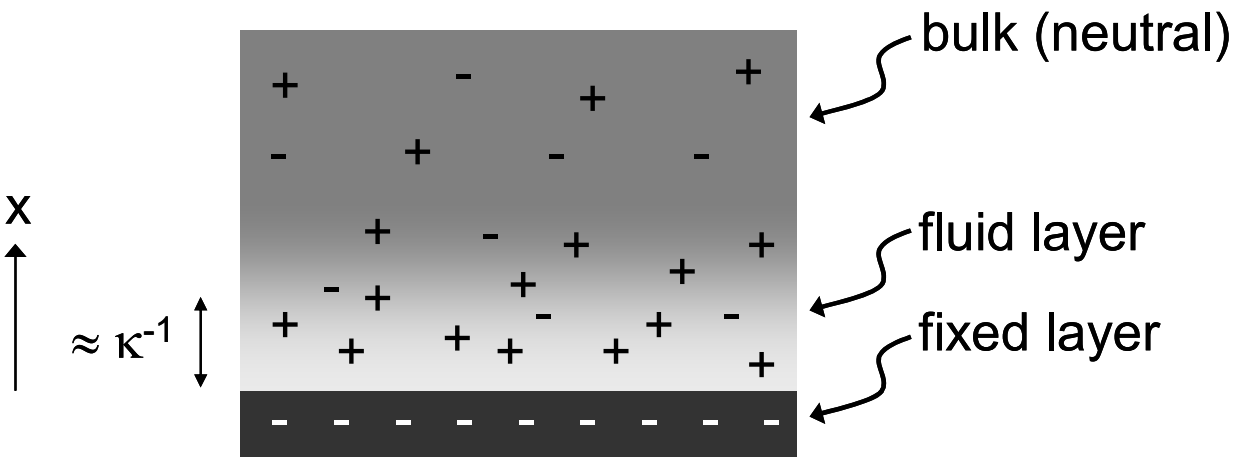

**Figure 2-1**. Electrical double layer (EDL). The "fixed charge layer" has charges bound on the surface, while the "fluid charge layer" – sometimes called the "diffuse layer" – has counter-ions, and a few co-ions, nearby in solution. This schematic does not show any Stern layer.

The charges on the particle arise due to the solvating action of the fluid on the particle. In turn the ions in the solvent re-distribute their positions in solution so that they form a layer – often only nanometers thick – that counter-balances the fixed charges on the particles. For example, if the charges on the particle surface are negative, positive ions from solution will be drawn toward the surface. The positively-charged ions are called the "counter-ions".

### *The Poisson-Boltzmann equation*

In order to obtain a quantitative description of the electrical double layer, I start with Poisson-Boltzmann (PB) physics. The PB equation is built on the Poisson equation of electrostatics and the Boltzmann equation of statistical mechanics. Let's examine each one in turn, starting with the Poisson equation.

The field of classical electromagnetism[19] has four primary equations called the Maxwell equations[20]. These are the bedrock of classical electromagnetism, and together with Newton's law and the two laws of thermodynamics, they form the guts of classical physics. The Maxwell equations resulted from decades of careful experimental work, along with a critical insight by Maxwell himself. In many more decades since, in fact in more than a century since, experiment after experiment has supported the accuracy of these equations, to the point that if the experiment were ever to disagree with the model, the experiments or the assumptions used to solve the Maxwell equations would be questioned deeply.

For the "static case" – which in colloid science effectively means when the frequency is less than $10^9$ $s^{-1}$ – I can ignore the dynamic parts of the Maxwell equation, the $\partial/\partial t$ parts, and simplify the full Maxwell equations. I will define an "electrical potential" ($\psi$, in V or mV), which is the voltage at any position in the system.[21]

---

[19] Jackson, J.D. *Classical Electrodynamics*, 3rd ed. Wiley (New York) 1999. Previous editions of this classic book on electromagnetism use "c-g-s units" (centimeter-gram-second), such as "statvolts". Many physicists are quite comfortable with these units, although throughout this book we will use SI units. In cgs units Coulomb's law is given by $F = q_1 q_2 / \varepsilon_r r_{12}^2$ .

[20] See Homework Problem 2-3 for the list of Maxwell equations for free charge and free current, as well as for a derivation of the Poisson equation.

[21] Having a static voltage also enables us to calculate the electric field (**E**), using $\mathbf{E} = -\nabla\psi$ . As a simple example, if the voltage varies in 1-dimension, from 7 V at $x = 0$ m to 3 V at $x = 0.4$ m, then the electric field $E = -$ (3 V – 7 V) / (0.4 m – 0 m) =

The voltage arises due to charge groups on the particle, and changes through the EDL. As Problem 2.8 shows, for static systems the electric potential can be very accurately-described by a single scalar equation called "the Poisson equation", or "Poisson's equation":

$$\nabla^2\psi = \frac{\partial^2\psi}{\partial x^2} + \frac{\partial^2\psi}{\partial y^2} + \frac{\partial^2\psi}{\partial z^2} = -\frac{\rho_e}{\varepsilon} \qquad (2\text{-}1)$$

This partial differential equation predicts the electrical potential ($\psi$, in V or mV) at any position (x,y,z), when the volumetric charge density ($\rho_e$, in C/m$^3$) is known everywhere. Given the Maxwell equations, the Poisson equation assumes only a) a static (time-independent) system and b) a constant and uniform material permittivity ($\varepsilon$). The permittivity provides a measure of how well a material "permits" an electric field to penetrate through it. Vacuum has a value of $\varepsilon = \varepsilon_0 = 8.854\times10^{-12}$ C$^2$/N-m$^2$, and readily permits an electric field through it; water has a value of about $80\varepsilon_0 = 7.1\times10^{-10}$ C$^2$/N-m$^2$, which counteracts an applied field. Table 1-3 gives values for a number of materials. With these two assumptions, the Poisson equation becomes the bedrock equation describing the electrical potential in the system. In principle, the permittivity can be altered, although in practice the assumption that $\varepsilon$ is constant and uniform gives results that explain experiments quite accurately.

The question at this point is, "How do we know the volumetric charge density?" This question requires more work to answer. By adding up the charge on each of the *N* ion types in the system, we know almost be definition that at any local position,

$$\rho_e = \sum_{i=1}^{N} z_i e c_i \qquad (2\text{-}2)$$

---

10 V/m. See Problem 2.8 for more details about how knowing **E** in terms of the electrical potential ($\psi$) enables us to derive the Poisson equation.

So we see that the volumetric charge density ($\rho_e$ [=] C/m$^3$) results from a concentration $c_i$ of ion type i (#/m$^3$) times the proton charge ($e = 1.6022\times10^{-19}$ C) times the valence ($z_i$ on that ion), and then we sum over all ions in solution. The units work. If we have a 10 mM concentration of NaCl, we sum over the two ions, which have equal but opposite valences, and so that sum appropriately comes to zero. We expect that near to a negative particle surface we will have a higher concentration of positive counter-ions in solution, and fewer co-ions. But how do we quantify those concentrations, and thus evaluate the charge density?

This question brings us to the second physical result used to build the PB equation: the Boltzmann equation. Boltzmann was a pioneer in believing that the material world is composed of atoms and molecules, and his failed struggle to convince other famous scientists like Mach and Ostwald eventually led to Boltzmann's suicide in 1906. The Boltzmann distribution predicts quantitatively what fraction of the time an entity – whether it is a stone, a gas molecule, or a colloidal particle – will spend in a local position of any given volume, given the energy at that local position. For example, an oxygen molecule near the ground in Boulder, Colorado has a higher potential energy than an oxygen molecule near the ground in State College, Pennsylvania, since the elevation of Boulder is higher. We know that gravitational potential energy is given by $mgh$, and since the gravitational constant ($g$) is very nearly the same in the two locations and the mass of the oxygen molecule ($m$) is identically the same, the higher value of $h$ increases the energy. We might guess, correctly, that it is more probable to find oxygen molecules in the air of State College, and so it's a bit harder to breathe in Boulder. But how do we quantify this?

In the late 1800s, Ludwig Boltzmann developed his famous "Boltzmann distribution equation". In terms of concentration of ions, the Boltzmann equation says for ion type $i$, that

$$c_i(x,y,z) = c_{i\infty}\, exp\left(\frac{-E_i(x,y,z)}{kT}\right) = c_{i\infty}\, exp\left(\frac{-z_i e\psi}{kT}\right) \tag{2-3}$$

The energy for an ion due to an electric field is $E_i = z_i e\psi$, where in this case the potential is defined to be zero far from the particles, in the bulk solution. This expression for the ion's energy is an approximation because it neglects, for instance, the VDW attraction between the ions. But especially for univalent ions, the approximation works well, and importantly, it gives us an analytical result that we can use to *think* through EDL problems. I note that the "electrical potential" has units of volts (V, where a volt is Joules/Coulomb, or 1 V = 1 J/C), and the "electric potential energy" has units of Joules (J), and these are two different things.

Equation 2-3 closes the loop. When we combine Eqs 2-1 to 2-3, inserting the Boltzmann expression for concentration into the expression for charge density, and then the expression for charge density into the Poisson equation, we obtain the full Poisson-Boltzmann equation:

$$\nabla^2\psi = -\frac{\sum_{i=1}^{N} z_i e c_{i\infty}\, exp\left(-\frac{z_i e\psi}{kT}\right)}{\varepsilon} \tag{2-4}$$

Now I'll simplify this equation so that I can get to a few useful results as quickly as possible. First, I will assume a symmetric and binary Z:Z electrolyte, like NaCl (1:1) or $Ca(SO_4)$ (2:2). In deriving the next result then, I remember that $z_+ = -z_i = Z$ and the bulk $c_{+\infty} = c_{-\infty} = c_\infty$. I will also use the definition $sinh\, x = \left(e^x - e^{-x}\right)/2$. With these slight simplifications, I obtain a fairly complete and quite useful form of the PB equation.

$$\nabla^2\left(\frac{Ze\psi}{kT}\right) = \kappa^2 \sinh\frac{Ze\psi}{kT} \tag{2-5}$$

where $\kappa^2$ is defined as[22]

$$\boxed{\kappa^2 = \frac{2Z^2e^2c_\infty}{\varepsilon kT}} \tag{2-6}$$

In the next section, I will use these equations to derive results for the electric potential (in V or mV) near a plate, between plates, and around a sphere.

The very famous parameter $\kappa^{-1}$ is called "the Debye length". It plays a key role in determining the electrostatic potential near a surface. A simple and even semi-quantitative knowledge of how the Debye length can be controlled in a solution opens surprising design abilities for manipulating colloidal particles. See Example 2-1 for an example of calculating a Debye length, or see Table 1-2 for the Debye length evaluated for various salt concentrations and types. I have always found it interesting that the Debye length does not contain a surface potential in it. It turns out that the *thickness* of the EDL depends on the bulk ionic strength. However, if we want to know the *number* or *concentration* of ions in the EDL, then the surface potential becomes important.

### *Debye-Huckel results for electric potential*

Up to this point, I have kept the PB equation in its tensor form, so that we can use it for any coordinate system. Let's look at the PB equation near to a charged plate, using only the $x$ dimension, where $x = 0$ at the surface of the plate and extends to infinity. Furthermore, I will make an approximation that $\psi$ is small, less than $kT/e$ = 25.7 mV at room temperature. For small values of $w$,

---

[22] See Problems 2-2 and 2-10 for a definition of the Debye length with mixed electrolytes.

$\sinh w = w + w^3/6 + ... \approx w$, and so the PB equation can be approximated as

$$\frac{d^2\psi}{dx^2} = \kappa^2\psi \tag{2-7}$$

In order to solve the PB equation we need two boundary conditions. Two common boundary conditions are

$$x = 0 : \psi = \psi_0 \tag{2-8}$$

$$x \to \infty : \psi \to 0 \text{ or } d\psi/dx \to 0 \tag{2-9}$$

Using a standard math text,[23] it is straightforward to find that the general solution of Eq 2-7 is $\psi = B_1 e^{\kappa x} + B_2 e^{-\kappa x}$. After using the boundary conditions to find the constants, I obtain the Debye-Huckel equation for electric potential near to a single flat plate:

$$\boxed{\psi = \psi_0 e^{-\kappa x}} \tag{2-10}$$

This simple but powerful result reveals the essential physics of the EDL. At the plate surface the electric potential is $\psi_0$, while just a few Debye lengths away, the electric potential decays to nearly zero. We used the approximation $\psi << kT/e$, but a full analysis shows reasonable accuracy even for $2kT/e \approx 51.4$ mV at room temperature. And for higher surface potentials, one often finds that electrostatic forces are so small that more precise results are in fact not needed.

[23] Kreyszig, Erwin, *Advanced Engineering Mathematics*, 7th ed. Wiley (New York) 1993. Chapter 2 discusses homogeneous, second order differential equations, under "Damped System".

**Example 2-1**. Electric potential near a flat plate.

A plate has a potential of $\psi_0$ = -39 mV at its surface. The plate is in a 5.4 mM NaCl solution at $T$ = 305 K. Find the Debye length, and then the electric potential at a distance of 7.0 nm from the plate. At this T, $\varepsilon_r$ = 76.0 (Table 1-3).

*answer: the Debye length* $\kappa^{-1}$ *= 4.13 nm.* $\psi$*(7.0 nm) = -7.16 mV.*

Let's look now at the potential around a spherical particle of radius ($a$). Variations in the angular directions are zero, due to symmetry. Using Eq 2-7 for spherical coordinates, and again assuming a small surface potential, I obtain[24]

$$\frac{1}{r^2}\frac{d}{dr}\left(r^2\frac{d\psi}{dr}\right)=\frac{d^2\psi}{dr^2}+\frac{2}{r}\frac{d\psi}{dr}=\kappa^2\psi \qquad (2\text{-}11)$$

$$r=a:\psi=\psi_0 \qquad (2\text{-}12)$$

$$r\to\infty:\psi\to 0 \text{ or } d\psi/dr\to 0 \qquad (2\text{-}13)$$

Solving the equation for $\psi(r)$ is a bit more complicated for a sphere than for a planar surface,[25] but the Debye-Huckel equation for a sphere is

$$\boxed{\psi=\psi_0\frac{a}{r}exp[-\kappa(r-a)]} \qquad (2\text{-}14)$$

Substituting this result into the differential equation and the boundary conditions reveals its veracity.

The functional form of the decay of potential is not quite purely exponential, since there is a $1/r$ in the denominator, but usually it is close. In fact, if we have a large sphere, so that if when we define $x \equiv r - a$ we see $x/a << 1$, then Eq 2-14 reduces to Eq 2-10. In practice, if $\kappa a > 10$ we can use Eq 2-10 with fairly good accuracy – the "flat earth approximation". As a result, the intuition from Eq 2-10 that the EDL potential decays exponentially is still useful.

[24] Kreyszig also discusses differential operators in various coordinate systems, in Chapter 8.
[25] Russel, W.B.; Saville, D.A.; Schowalter, W.R. *Colloidal Dispersions*. Cambridge University Press (New York) 1989. Section 4.8 describes the procedure.

Let's examine one final case, that of the electric potential between two charged plates. Once again I use Eq 2-7, but now I use the boundary conditions

$$x = -L/2 : \psi = \psi_1 , \quad x = L/2 : \psi = \psi_2 \qquad \text{(2-15)}$$

Solving the equation with these boundary conditions gives (for $-L/2 \le x \le L/2$)

$$\boxed{\psi = \left( \frac{\psi_2 e^{+\kappa L/2} - \psi_1 e^{-\kappa L/2}}{e^{+\kappa L} - e^{-\kappa L}} \right) e^{\kappa x} + \left( \frac{\psi_1 e^{+\kappa L/2} - \psi_2 e^{-\kappa L/2}}{e^{+\kappa L} - e^{-\kappa L}} \right) e^{-\kappa x}} \qquad \text{(2-16)}$$

***Surface charge density using Debye-Huckel***

We now have expressions relating the potential at any position within the fluid part of the EDL to the surface potential. Oftentimes we want to predict the surface charge density ($\rho_s$ [=] C/m$^2$), knowing the surface potential, or vice versa. There are various ways to approach this problem, but perhaps the simplest to understand is that a surface plus its EDL form an electroneutral system. Thus, the charges at the surface and the charges in the bulk fluid balance. This can be written mathematically for a flat plate as

$$\iiint_V \rho_e dV + \iint_S \rho_s dS = 0 \Rightarrow \int_0^\infty \rho_e dx + \rho_s = 0 \qquad \text{(2-17)}$$

This equation says that the surface charges must be balanced by the charges in the fluid from near the plate to infinity. A rearrangement of the Poisson equation tells us

$$\frac{d^2\psi}{dx^2} = -\frac{\rho_e}{\varepsilon} \Rightarrow \rho_e = -\varepsilon\frac{d^2\psi}{dx^2} \qquad (2\text{-}18)$$

As a result, we can integrate to find

$$\rho_s = \varepsilon\int_0^\infty \frac{d^2\psi}{dx^2}dx = \varepsilon\int_0^\infty d\left(\frac{d\psi}{dx}\right) \qquad (2\text{-}19)$$

For small potentials near a plate, $\psi = \psi_0 e^{-\kappa x}$, and so for small $\psi_0$,

$$\boxed{\rho_s = \varepsilon\psi_0\kappa} \qquad (2\text{-}20)$$

How can these variables be measured? Surface charge density ($\rho_s$) is often measured by titration experiments; surface potential ($\psi_0$) is often estimated from electrophoresis experiments; Debye length ($\kappa^{-1}$) is calculated using Eq 2-6. In particular circumstances it is often easier to measure one of these variables and calculate one, and then to use Eq 2-20 to estimate the third value.

**Example 2-2**. Surface charge density.

A plate has a surface potential of -39 mV in 5.4 mM NaCl solution at $T$ = 305 K. Find the surface charge density in $C/m^2$, and the average area per charge group, for monovalent charges. Use the result for the Debye length from Example 2-1.

*answer: -0.00635 $C/m^2$ = -0.635 $\mu C/cm^2$. Since monovalent charges are $1.6\times10^{-19}$ C, the spacing is one charge group per $25.2\times10^{-18}$ $m^2$ = $(5.0\ nm)^2$. About 1 $\mu C/cm^2$ is a typical magnitude of surface charge density.*

A similar analysis around a sphere gives

$$\boxed{\rho_s = \varepsilon\kappa\psi_0\left(\frac{1+\kappa a}{\kappa a}\right)} \qquad (2\text{-}21)$$

Between parallel plates having a surface potential ($\psi_0$) and separated by a distance ($L$), the surface charge density is given by

$$\boxed{\rho_s = \varepsilon\kappa\psi_0\left(\frac{1-e^{-\kappa L}}{1+e^{-\kappa L}}\right)} \qquad (2\text{-}22)$$

If $L \to \infty$ this equation reduces to Eq 2-20. Curiously, if Eq 2-22 is solved for $\psi_0$, an interesting result occurs:

$$\psi_0 = \frac{\rho_s}{\varepsilon\kappa}\left(\frac{1+e^{-\kappa L}}{1-e^{-\kappa L}}\right) \qquad (2\text{-}23)$$

As $L \to 0$ this equation becomes singular, meaning that a very large potential must arise from a finite $\rho_s$ for small gaps between plates.

### *Electrostatic force between plates*

Once the electrostatic potential between two plates is known, we can evaluate the force or potential energy between the two plates. There are several ways we can start this problem, and all the approaches are somewhat involved, including the derivation shown here. And so this section will require some effort to understand well.

The most general method involves using mechanical and electrostatic stress tensors,[26] but I will proceed here by a simpler

---

[26] Russel, W.B.; Saville, D.A.; Schowalter, W.R. *Colloidal Dispersions*. Cambridge University Press (New York) 1989. Chapter 4 of this book gives a derivation of the electrostatic force between plates using the mechanical and electrostatic stress tensors. For a fuller discussion on the electrostatic stress tensor see Jackson, John

route. I start with two plates, each with a surface potential ($\psi_0$) and separated by a distance ($h$). The coordinate $x$ runs perpendicular to the plates, with $x = 0$ halfway between the plates. Outside the plates the bulk pressure is $p_\infty$, while between the plates the pressure is higher due to the electrostatic force on the fluid – charged, since it is in the EDL – and also due to a higher osmotic pressure between the plates since there are more ions in the EDL than in the bulk fluid.

If the plates have the same surface potential, they will repel with an equal but opposite force. But let's imagine that we hold the plates in place by applying a force ($F_1$) onto plate 1 and a force ($F_2 = -F_1$) on plate 2. In this case the fluid between the plates will remain static, meaning that it will not accelerate, nor will it flow and dissipate energy. Our job at this point is to evaluate $F_1 = -F_2$. I will neglect end effects on the plates; mathematically this translates into having plates that are infinite in extent. I will therefore evaluate the electrostatic force per area ($f_{es}$). For plates with an area ($A$), we have $f_{es} \equiv -\lim_{A\to\infty} F_1 / A = \lim_{A\to\infty} F_2 / A$.

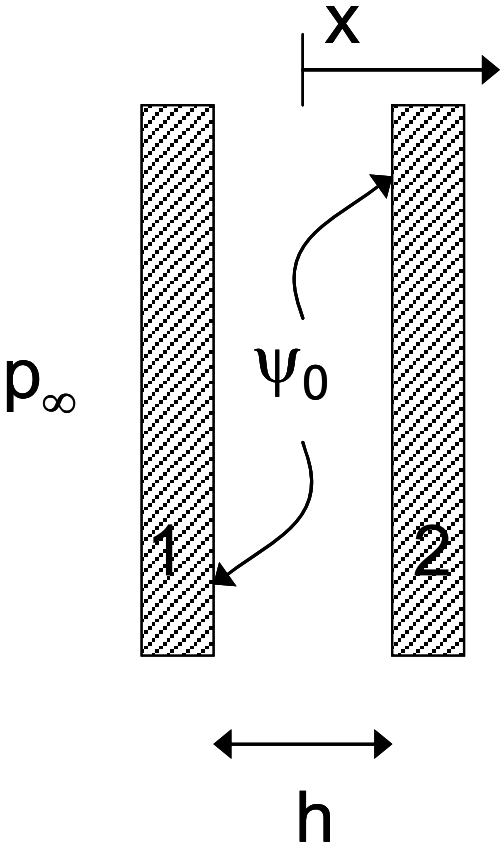

**Figure 2-2**. Two plates with electrostatic interactions. In the case shown, the electrostatic potential on the surface of both plates (1 and 2) is the same. As a result, the interaction energy is repulsive (positive). The pressure outside of the plates is $p_\infty$, while the pressure inside the plates is higher due to both electrostatic forces on the fluid, as well as osmotic pressure.

---

David. *Classical Electrodynamics*, 3rd ed, Wiley (New York) 1999 or Woodson, Herbert H.; Melcher, James R. *Electromechanical Dynamics. Part II: Fields, Forces, and Motion*. Malabar, FL: Robert E. Krieger Publishing (1968).

I will start with the equations of fluid mechanics. For a Newtonian fluid we use the Navier-Stokes equation, a vector equation for the momentum balance that I will discuss more in Chapter 4. The Navier-Stokes equation is[27]

$$\rho_f \left( \frac{\partial \mathbf{v}}{\partial \mathbf{t}} + \nabla \cdot \mathbf{vv} \right) = \eta \nabla^2 \mathbf{v} - \nabla p + \rho_e \mathbf{E} = \mathbf{0} \qquad (2\text{-}24)$$

where $\mathbf{v}$ is the fluid velocity; $\rho_f$ is the fluid density; $\eta$ is the fluid viscosity; $p$ is the isotropic pressure in the fluid; $\rho_e$ is the electrical charge density (C/m$^3$); and $\mathbf{E}$ is the local electric field. For our static case here, I set $\mathbf{v} = \mathbf{0}$. For electrostatic forces we have $\mathbf{E} = -\nabla \psi$, which has units of V/m. Charges move due to a finite electric field, whether the charges are electrons or ions. The electric field gives not just the voltage, and not just the voltage difference, but how the voltage changes with distance in any direction. As the local electric field becomes stronger – meaning that the voltage change with distance becomes steeper – the ions in that region move faster.

Since I am examining what happens between two plates that are infinite in the $y$ and $z$ directions, the $x$-direction Navier-Stokes equation with $\mathbf{v} = \mathbf{0}$ is the one that tells me what I want to know:

$$\frac{dp}{dx} = \rho_e E_x = -\rho_e \frac{d\psi}{dx} \qquad (2\text{-}25)$$

What do we have at this point? Eq 2-25 clearly shows the balance of pressure- and electrical forces on the fluid. Where the fluid is static, these two forces balance at every position. Using a re-arranged version of the Poisson equation 2-1,

[27] Here I think of a single vector equation, but I could also think of three scalar equations and write, "The Navier-Stokes equations are …"

$$\rho_e = -\varepsilon \frac{d^2\psi}{dx^2} \tag{2-26}$$

I make a substitution for the volumetric charge density to give

$$\frac{dp}{dx} = \varepsilon \left( \frac{d^2\psi}{dx^2} \right) \left( \frac{d\psi}{dx} \right) \tag{2-27}$$

The following identity, found using the chain rule from calculus, is helpful:

$$\frac{d}{dx}\left[ \left( \frac{d\psi}{dx} \right)^2 \right] = 2\left( \frac{d\psi}{dx} \right)\left( \frac{d^2\psi}{dx^2} \right) \tag{2-28}$$

As a result,

$$\frac{dp}{dx} = \frac{\varepsilon}{2} \frac{d}{dx}\left[ \left( \frac{d\psi}{dx} \right)^2 \right] \tag{2-29}$$

which can be integrated to

$$p = \frac{\varepsilon}{2}\left( \frac{d\psi}{dx} \right)^2 + B \tag{2-30}$$

Evaluating the constant ($B$) requires a bit of additional thought. The easiest place to evaluate $B$ is at the midplane, where by symmetry we know that $d\psi / dx = 0$. Therefore $B$ is the pressure at $x = 0$. What contributes to this pressure? If the plates were uncharged, the pressure at the midplane would be simply the ambient pressure ($p_\infty$). As charges are added to the plate, bringing additional ions into the electrical double layer region between the plates, there is also an osmotic pressure contribution. Because the ions are "trapped" in the EDL – there is an analogy to being trapped behind a semi-permeable membrane – the osmotic pressure at the midplane is higher in the EDL by an amount $\pi(x = 0)$. The pressure is then

$$p(x=0)-p_{\infty}=\pi(x=0)-\pi_{\infty} \tag{2-31}$$

which simply says that the pressure inside the EDL between the plates is different from the pressure outside the plates by the difference in osmotic pressure. The $\pi_{\infty}$ appears since the bulk solution will also have some finite ionic strength, even though it isn't as high as that between the plates.

Since the fluid at $x = 0$ has a slightly higher pressure, it will push outward on the neighboring fluid, which in turn will push on the fluid all the way to the wall. The electrical forces on the fluid at $x = 0$ are zero, since by symmetry $d\psi/dx=0$ there, and thus $\mathbf{E} = \mathbf{0}$. In the end the pressure force is the only force remaining:

$$f_{es}=p_{\infty}-p(x=0) \tag{2-32}$$

The job of finding the electrostatic force has thus been reduced to finding the pressure difference inside and outside the plate. But Eq 2-31 already gives us an expression for this difference.

Osmotic pressure is a well-established thermodynamic property, although its mechanical origins in terms of fluid mechanics have only recently been explored in depth.[28] The osmotic pressure of a solution with a dilute concentration of molecules or ions ($n$) is given by the van't Hoff result[29]:

---

[28] Guell, David C.; Brenner, Howard. "Physical Mechanism of Membrane Osmotic Phenomena," *Ind. Eng. Chem. Res*., **35**, 3004-3014 (1996).

[29] For a thermodynamic explanation of osmotic pressure, see Denbigh, Kenneth, The Principles of Chemical Equilibrium, 4th ed, Cambridge Press (1981), especially pages 262-264. Briefly, the chemical potential of the water decreases with the addition of solute, and increases with the addition of pressure ($\Delta p$), such that $RT\ln x_w + v\Delta p = 0$ at equilibrium, where $v$ is the molar volume of the solvent. Recognizing that the

$$\pi = nkT \tag{2-33}$$

where $n$ has units #/m$^3$ to match $kT$. Now we simplify our analysis to a Z:Z binary electrolyte. At $x = 0$, $\pi - \pi_\infty = (n_+ + n_- - 2n_\infty)kT$. The values of $n_+$ and $n_-$ are found from the Boltzmann factor for each ion or molecule, and the extra "$2n_\infty$" in the equation results since for a Z:Z electrolyte there is both a positive and negative ion in the bulk.

$$\pi - \pi_\infty = n_\infty \left[ exp\left(\frac{Ze\psi}{kT}\right) + exp\left(\frac{-Ze\psi}{kT}\right) - 2 \right] kT \tag{2-34}$$

which can be factored into

$$\pi - \pi_\infty = n_\infty kT \left[ exp\left(\frac{Ze\psi}{2kT}\right) - exp\left(\frac{-Ze\psi}{2kT}\right) \right]^2 \tag{2-35}$$

For low surface potentials, I use a Taylor series $e^w = 1 + w + w^2/2 + ...$ to obtain

$$\begin{aligned} \pi - \pi_\infty &= n_\infty kT \left[ \left(1 + \frac{Ze\psi}{2kT} + ...\right) - \left(1 - \frac{Ze\psi}{2kT} + ...\right) \right]^2 \\ &\approx \frac{Z^2 e^2 n_\infty}{kT} \psi^2 \end{aligned} \tag{2-36}$$

where $\psi$ is evaluated at the midplane ($x = 0$). Using Eq 2-16 for the potential between two plates, we find that approximately,

$$\psi(x = 0) = \frac{2\psi_0 \left(e^{\kappa h/2} - e^{-\kappa h/2}\right)}{e^{\kappa h} - e^{-\kappa h}} \approx 2\psi_0 e^{-\kappa h/2} \tag{2-37}$$

---

solute mole fraction $x_s = 1 - x_w$, defining $\pi = \Delta p$ as the additional pressure required for equilibrium, and using $x_s = n / v$ gives $\pi = nkT$.

.

Putting all the pieces together now gives

$$f_{es} = \pi(x=0) - \pi_\infty = \frac{4Z^2e^2n_\infty}{kT}\psi_0^2 e^{-\kappa h} \tag{2-38}$$

Putting this expression in terms of the Debye parameter,

$$\boxed{f_{es} = 2\varepsilon\kappa^2\psi_0^2 e^{-\kappa h}} \tag{2-39}$$

If we want to know the energy between two plates, we can integrate. Just as we know for gravity that the force $F = -mg = -dV/dh$, and thus we integrate the force to give the potential energy $V = mgh$, so the potential energy per area is

$$V_{es} = -\int_h^\infty \mathbf{f}_{es} \cdot d\mathbf{h}' = \int_\infty^h f_{es} dh' = 2\varepsilon\kappa^2\psi_0^2 \int_h^\infty e^{-\kappa h'} dh' \tag{2-40}$$

This integration leads to the final expression for the electrostatic energy per unit area between two plates:

$$\boxed{V_{es} = 2\varepsilon\kappa\psi_0^2 e^{-\kappa h} = \kappa^{-1} f_{es}} \tag{2-41}$$

**Example 2-3**. Potential energy and force due to electrostatic interaction between plates.

---

Two plates have surface potentials of -24 mV in a solution with a Debye length of 4.2 nm. The plates are separated by 15.3 nm. What is the electrostatic potential energy per area and the electrostatic force per area on the plates, for $T$ = 293 K?

*answer:* $V_{es} = +5.1\times10^{-6}$ *J/m*$^2$, $f_{es}$ = *1220 N/m*$^2$ *= 0.177 psi.*

---

Hogg, Healy, and Fuerstenau (HHF)[30] developed a more exact result between flat plates, for arbitrary but small surface potentials, and for any gap between the plates. Their result is

$$\boxed{V_{es} = \tfrac{1}{2}\varepsilon\kappa\left[\frac{2\psi_1\psi_2}{sinh\,\kappa h} + \left(\psi_1^2 + \psi_2^2\right)\left(1 - \frac{cosh\,\kappa h}{sinh\,\kappa h}\right)\right]} \qquad (2\text{-}42)$$

The first term in the brackets gives the interaction that is reducible to that in Eq 2-41. The second term includes additional physics resulting from an "image charge", in which one charged surface actually induces a charge in the other material. For plates separated by a few Debye lengths, the first term decays roughly as $exp(-\kappa h)$, while the second term decays more quickly as $exp(-2\kappa h)$.

### *Spheres and the Derjaguin approximation*

Having the electrostatic interaction energy between two flat plates enables us to proceed to the interaction between two spheres ($\Phi_{es}$, in Joules). I will employ the widely-used "Derjaguin approximation" (Figure 2-3).[31] In this approximation we estimate

$$\begin{aligned}\Phi_{es} &= V_{es}(h_0 = \delta)\pi r_0^2 + V_{es}(h_1)2\pi r_1\Delta r \\ &\quad + V_{es}(h_2)2\pi r_2\Delta r + V_{es}(h_3)2\pi r_3\Delta r + \ldots\end{aligned} \qquad (2\text{-}43)$$

If the rings are chosen as differential in size, the summation can be converted into an integral as

$$\Phi_{es} \approx \int_0^A V_{es}[h(r)]2\pi r dr \tag{2-44}$$

---

[30] Hogg, R.; Healy, T.W.; Fuerstenau, D.W. "Mutual Coagulation of Colloidal Dispersions." *Transactions of the Faraday Society*, **62**, 1638-1651 (1966). Richard Hogg was a faculty member at Penn State University for many years, including when I started in 1999.

[31] Boris Derjaguin (http://en.wikipedia.org/wiki/Boris_Derjaguin) is famous as one of the developers of the famous "DLVO theory" of colloidal interactions, described more fully in Chapter 6. He was also involved in the infamous "polywater" studies (http://en.wikipedia.org/wiki/Polywaterv) that some believe cost him a Nobel Prize.

Ordinarily this integral might be challenging analytically, but there is a helpful simplification that arises when the bottom of the spheres are *approximated as parabolas*. Normally the equations for the upper (1) and lower (2) spheres are $x^2 + y^2 + (z_1 - a - \delta)^2 = a^2$ and $x^2 + y^2 + (z_2 + a)^2 = a^2$. In this case I let $r^2 = x^2 + y^2$. The gap $h = z_1 - z_2$, and an expansion gives $z_1 = a + \delta - \sqrt{a^2 - r^2}$ $= a + \delta - a\sqrt{1 - r^2/a^2} \approx a + \delta - a(1 - r^2/2a^2)$. Then I can write the equations $z_1 \approx \delta + r^2/2a$ and $z_2 \approx -r^2/2a$, so $h = z_1 - z_2 \approx \delta + r^2/a$. Now I can write $dh = 2rdr/a$, and so the variable of integration in Eq 2-44 can in fact be changed to *h*. Thus, the parabolic approximation enables us to derive an analytical result which gives many useful insights!

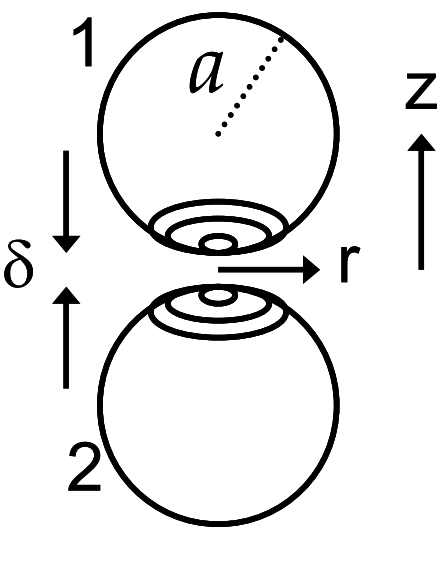

**Figure 2-3**. Derjaguin approximation. Each ring is treated as flat with a width $\Delta r$, interacting with "flat ring" opposite it. Since potential energy between flat plates is expressed as energy per area (e.g., Eq 2-41), the energy for each ring is found by multiplying by its area, and the total potential energy is summed from all the rings. In fact the size of each ring is taken as a differential element of area so that the result is integrated.

I now write

$$\Phi_{es} \approx \pi a \int_\delta^\infty V_{es}(h)dh \tag{2-45}$$

The upper limit is written as $h \to \infty$, because its exact value need not be specified, since the integrand decays rapidly with *h*. I will use Eq 2-45 for other energies in this book such as van der Waals energies, since in fact the expression is quite general. I have not yet specified anything particular to electrostatic forces.

Using the expression for $V_{es}$, integration gives the electrostatic energy between two spheres:

$$\boxed{\Phi_{es} = 2\pi\varepsilon a \psi_0^2 e^{-\kappa\delta}} \qquad (2\text{-}46)$$

This result is the one I have been aiming toward for the entire chapter. Eq 2-46 gives a prediction for the electrostatic potential energy in Joules. If a more precise result is need, but still for small surface potentials with magnitude less than about 50 mV, we can use the HHF result for spheres:

$$\boxed{\Phi_{es} = \pi\varepsilon\left(\frac{a_1 a_2}{a_1 + a_2}\right)\left[\begin{array}{l} 2\psi_1\psi_2 \ln\left(\dfrac{1+e^{-\kappa\delta}}{1-e^{-\kappa\delta}}\right) \\ +\left(\psi_1^2 + \psi_2^2\right)\ln\left(1-e^{-2\kappa\delta}\right) \end{array}\right]} \qquad (2\text{-}47)$$

The result for the force corresponding to Eq 2-46 is $F_{es} = 2\pi\varepsilon a\kappa\psi_0^2 e^{-\kappa h} = \kappa\Phi_{es}$; the HHF force can be found from differentiating $\Phi_{es}$.

Throughout this chapter I have emphasized low potential results. Why have I focused on low surface potentials, rather than the more general case? There are two reasons: The theory for large surface potentials is much more complicated, but gives little additional physics. But more importantly, when surface potentials are greater in magnitude than 50 mV, electrostatic forces become very strong – often dominant – and more precise calculations are usually not necessary.[32]

**Example 2-4**. Electrostatic potential energy between spheres.

Two 1.2 μm diameter spheres have surface potentials of -24 mV in a solution with a Debye length of 4.2 nm. The spheres are separated by 15.3 nm. What is the potential energy between the spheres, for $T = 293$ K? Compare the simple result in Eq 2-46 to that in Eq 2-47.

*answer:* $\Phi_{es} = 4.04\times10^{-20}$ *J = 10.0 kT for the simple expression. The HHF gives* $\Phi_{es} = 3.99\times10^{-20}$ *J = 9.87 kT, quite close (1.3% difference), since the gap is large relative to* $\kappa^{-1}$ *while still being small relative to a. For gaps with* $\kappa\delta$ *less than 2, the HHF result for spheres will give important differences, due to the image charge term.*

Other forces act on colloidal particles. These include solvation forces, solvophobic forces, and depletion forces. Each of these forces is important and can be used in a colloidal assembly technique. But the most ubiquitous force between particles are the van der Waals (VDW) forces. I explore these in the next chapter.

### *Dielectric media (organic fluids)*

Electrostatic interactions in dielectric media like organic fluids can be much different from that in aqueous fluids.[33] There are several reasons for this:

1) Obtaining any charges in dielectric media is hard to do. An important parameter arises from comparing the electrostatic energy

[32] Sometimes more precise results are needed. See for example Ramirez, Laura Mely; Smith, Adrian S.; Unal, Deniz B.; Colby, Ralph H.; Velegol, Darrell. "Self-assembly of Doublets from Flattened Polymer Colloids," *Langmuir*, **28**, 4086-4094 (2012).
[33] A classic article in this area is Morrison, Ian D. "Electrical charges in nonaqueous media," *Colloids and Surfaces A*, **71**, 1-37 (1993).

($V$) of two ions in solution (Coulomb's law, $V = (z_1 e)(z_2 e)/4\pi\varepsilon r$, for a separation ($r$) between the ions) with the thermal energy ($kT$) which tends to randomize ion positions throughout the solution. If we equate these energies, we find for a symmetric Z:Z electrolyte a distance ($\lambda_B$), known as the Bjerrum length, given by

$$\lambda_B = \frac{Z^2 e^2}{4\pi\varepsilon kT} \qquad (2\text{-}48)$$

In water at room temperature, $\lambda_B$ = 0.70 nm. Thus, if I have NaCl dissociated into $Na^+$ and $Cl^-$, the ions have to be really close – in fact closer than water solvation allows – in order for them to form NaCl again. In contrast, in hexane with a relative permittivity of 2.0, $\lambda_B$ = 28 nm. Thus, the ions can be spaced far apart and still attract each other back into NaCl. Interestingly, if I assume that I have one ion pair in a volume $1/\lambda_B^3$, then I can estimate a saturation concentration, which scales as $1/Z^6$. In water I find a saturation concentration for Z = 1 of 4.8 M. The actual saturation concentration of NaCl is 5.4 M, and for KCl it is 4.2 M. When $\lambda_B$ = 28 nm, the saturation concentration for Z = 1 is 0.075 mM, and actual concentrations in dielectric media are usually much lower.

2) The Debye length in the fluid can be micrometers, which is frequently larger than the particles of interest. Thus, colloidal particles can look like point charges electrostatically.

3) The time required for electrostatic operations can be much longer than usual. The free charge relaxation time ($t_{fcr}$) is given by[34]

$$t_{fcr} = \frac{\varepsilon}{\sigma} \qquad (2\text{-}49)$$

[34] Jackson, J.D. *Classical Electrodynamics*, 3rd ed. Wiley (New York) 1999.

In 10 mM aqueous KCl near room temperature, the permittivity $\varepsilon \approx 80\varepsilon_0 = 7.1\times10^{-10}\ \mathrm{C^2/N\text{-}m^2}$ and the electrical conductivity $\sigma \approx 1400\ \mu\mathrm{S/cm}$, giving $t_{\mathrm{fcr}}$ = 5 ns (nanoseconds) – fairly rapid. In organic media, one has $\varepsilon \approx 2\varepsilon_0 = 0.177\times10^{-10}\ \mathrm{C^2/N\text{-}m^2}$, but $\sigma \sim 1\ \mathrm{pS/cm}$ or even much less, giving $t_{\mathrm{fcr}}$ ~ 1 s or much more. Thus, dielectric media respond to electric fields slowly.

4) Most fixed charges on particles in aqueous suspensions are firmly bound; in organic media, the charges can detach more readily from the particle surface, meaning that the "fixed charge" is not always stable. In fact, to stabilize particles, one usually must add particular surfactants that produce charge in dielectric media. Perhaps the best known of these is Aerosol OT (AOT, with a technical name dioctyl sodium sulfosuccinate, giving a conductivity ~20 nS/cm), although there are others such as OLOA by Chevron or the nonionic Span 85 from Sigma Aldrich that can give conductivites more than 10 times higher. These surfactants can provide enough charge to stabilize many suspensions of particles in dielectric media. Note that all of these conductivites are still low.

Overall, colloidal behavior in organic or dielectric media is an important area that is being researched vigorously at present.

### *Charge nonuniformity*

Throughout this chapter, we have assumed that the surface charge density is uniform over the surface of the particle. But this might not always be the case. This is an area that my lab has spent time studying since 1999.[35] The basic picture of a randomly "patchy particle" surface is shown in Figure 2-4.

[35] There is also nice work on charge nonuniformity or heterogeneity on surfaces, and the resulting paticle capture that can occur. See for instance Santore, Maria M.; Kozlova, Natalia. "Micrometer Scale Adhesion on Nanometer-Scale Patchy Surfaces: Adhesion Rates, Adhesion Thresholds, and Curvature-Based Selectivity." *Langmuir*, **23**, 4782-4791 (2007)..

When the patch size for the charge nonuniformity is very small, indeed going all the way to the level of the placement of individual ions, then the "discrete charge effect" is very small.[36] However, if the patch size ($L$) becomes larger, say with a 10-50 nm length scale, the electrostatic forces can be significantly altered.[37]

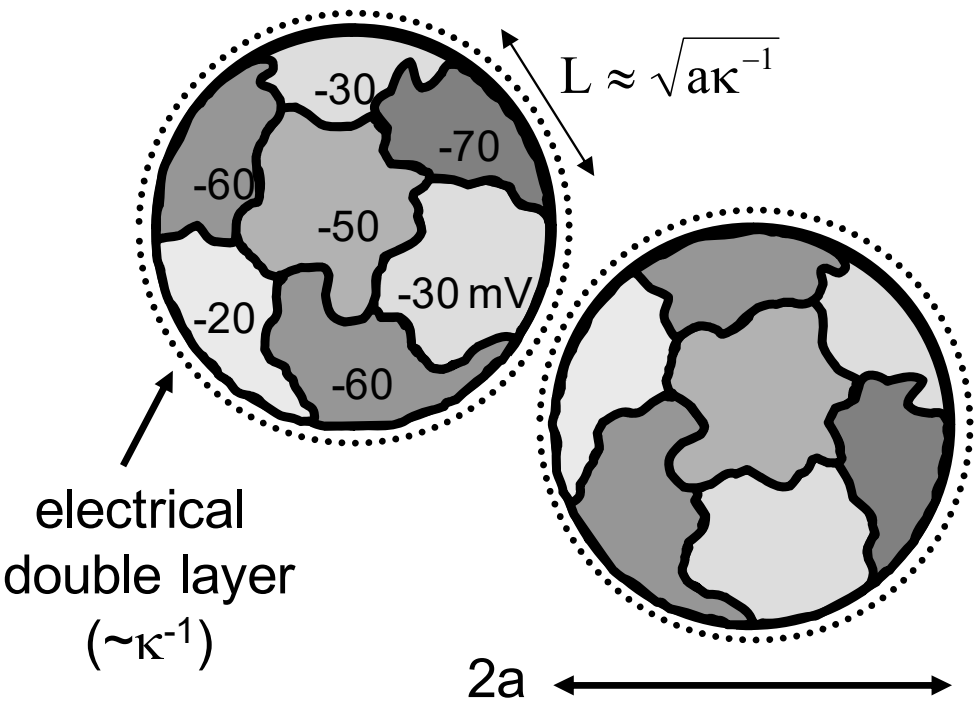

**Figure 2-4**. Charge nonuniformity on particles. The average surface potential is roughly -50 mV, which one would predict is sufficient to maintain stability in most cases. However, there might be regions on the particles where the surface potential is only -30 mV, as shown. In solution, particles undergo random Brownian rotation, and when two regions of -30 mV are adjacent, the electrostatic repulsion might not be sufficient to overcome the van der Waals attraction, and the particles could aggregate together.

The amount of the variation in electrostatic repulsion depends on the amount of variation in the surface potential. Say that we measure the surface potential for a suspension of particles. The usual electrophoresis techniques essentially measure the average zeta potential over the particles surfaces ($\langle\zeta\rangle$). For the particles in Figure 2-4, this value would be close to -50 mV. Now let's say we had a technique that could measure the standard deviation of zeta potential over the surface ($\sigma_\zeta$). One such technique, rotational electrophoresis, is discussed briefly in Ch 8.[38,39,40] For the schematic of Figure 2-4, $\sigma_\zeta \approx 20$ mV.

[36] Israelachvili, Jacob N. *Intermolecular and Surface Forces*, 2nd ed. Academic Press (New York) 1992. See Section 12.20, on p 254, for the effect of discrete charges. For typical conditions, Israelachvili provides a quick calculation showing that the effect of considering discrete charges is less than 1% different from considering those same charges as "smeared out".

[37] Velegol, Darrell; Thwar, Prasanna K. "Analytical Model for the Effect of Surface Charge Nonuniformity on Colloidal Interactions." *Langmuir*, **17**, 7687-7693 (2001). See for example Eqs 20-26.

For this suspension, we might want to know the orientation average repulsive electrostatic energy between particles ($\langle\Phi_{es}\rangle$). And we might also want to know how much the repulsion ($\sigma_\Phi$) varies, depending on which patches of the spheres are nearly touching. Remember, the particles will continue to undergo Brownian rotation, so that the regions of low potential will eventually find each other if left in suspension long enough.

Calculations show that the average electrostatic forces are close to those given by Eq 2-47. But there is a significant variation of electrostatic forces, depending on the relative orientation of the two spheres. Our quantitative estimates show that

$$\left|\frac{\sigma_\Phi}{\langle\Phi_{es}\rangle}\right| \approx \left|\frac{\sigma_\zeta}{\langle\zeta\rangle}\right| \qquad (2\text{-}50)$$

Such variations in electrostatic repulsion can cause an unstable suspension – wit the net interparticle energy being attractive at even larger separations -- when measurements of the *average* zeta potential would have predicted a stable suspension. The attractive forces are caused in large part by van der Waals forces, which are discussed in the next chapter. In colloidal assembly operations, creating patchy particles can be advantageous in fabricating particular colloidal assemblies, as described in Chapter 9.

---

[38] Feick, Jason D.; Velegol, Darrell. "The Electrophoresis of Spheroidal Particles having a Random Distribution of Zeta Potential", *Langmuir*, **16**, 10315-10321 (2000). This paper includes the mathematical physics of the rotational electrophoresis technique.

[39] Feick, Jason D.; Velegol, Darrell. "Measurements of charge nonuniformity on polystyrene latex particles." *Langmuir*, **18**, 3454-3458 (2002).

[40] Feick, Jason D.; Chukwumah, Nkiru; Noel, Alexandra E.; Velegol, Darrell. "Altering surface charge nonuniformity on individual colloidal particles." *Langmuir*, **20**, 3090-3095 (2004).

**Example 2-5** Variation of electrostatic repulsion between spheres.

As in Example 2-4, two 1.2 μm diameter spheres have surface potentials of $\langle \zeta \rangle$ = -24 mV in a solution with a Debye length of 4.2 nm. The spheres are separated by 15.3 nm. From Example 2-4, we know that the average electrostatic repulsion is $\langle \Phi_{es} \rangle = 3.99\times10^{-20}$ J = 9.87 $kT$. From the lab, we find that on a relevant length scale, $\sigma_\zeta$ = 10 mV. What is the variation in the electrostatic repulsion?

*answer: From Eq* 2-**Error! Reference source not found.**, *we see that* $\sigma_\Phi$ *= 4.1 kT. If we account for two standard deviations (8.2 kT of energy), that means the repulsion will vary roughly from 1.7 kT to 18.1 kT. The 18.1 kT would provide significant repulsion, preventing particle aggregation. However, when the low potential patches approach, giving only 1.7 kT of electrostatic repulsion, the particles will not have sufficient repulsion against attractive va der Waals interactions..*

***Symbols***

$a$ = sphere radius [=] m. For colloidal particles, $a$ is often in nm or small μm.

$c_i$ = concentration of ion type I [=] mol/m$^3$ or M.

$\delta$ = distance of closest approach between particles [=] m.

$e$ = charge on a proton = $1.6022\times10^{-19}$ C.

$\varepsilon$ = permittivity [=] C$^2$/N-m$^2$. $\varepsilon = \varepsilon_0\varepsilon_r$.

$\varepsilon_0$ = permittivity of a vacuum = $8.8542\times10^{-12}$ C$^2$/N-m$^2$.

$\varepsilon_r$ = relative permittivity or dielectric constant [=] dimensionless.

$f_{es}$ = electrostatic force per area [=] N/m$^2$. Derjaguin called this the "disjoining pressure" between surfaces.[41]

$F$ = force [=] N. For electrostatic forces I sometimes write $F_{es}$.

$\Phi_{ES}$ = electrostatic energy between particles [=] J.

$\langle \Phi_{es} \rangle$ = average $\Phi_{ES}$, for all relative orientations of two spheres.

$\eta$ = viscosity [=] Pa-s or kg/m-s.

$k$ = Boltzmann's constant = $1.38065\times10^{-23}$ J/K.

$\kappa^{-1}$ = Debye length [=] m. Typical values are 1 to 100 nm in aqueous solutions.

$n$ = number concentration of ions [=] #/m$^3$. $n_+$ is for the positive ion.

$n_\infty$ = bulk concentration of ion pairs [=] #/m$^3$.

$\rho_e$ = volumetric charge density [=] C/m$^3$.

$\rho_f$ = fluid density [=] kg/m$^3$.

$\rho_p$ = particle density [=] kg/m$^3$.

$\rho_s$ = surface charge density [=] C/m$^2$. A typical value is 1 μC/cm$^2$ = 0.01 C/m$^2$.

$\pi$ = osmotic pressure = $nkT$ [=] N/m$^2$.

$\sigma_\Phi$ = standard deviation of $\Phi_{ES}$, depending on relative orientation of two spheres [=] J.

$\sigma_\zeta$ = standard deviation of zeta potential over the surface of a particle [=] V or mV. A "patch size" must be defined.

$t$ = time [=] s

$T$ = temperature [=] C or K.

$V_{es}$ = electrostatic potential energy per area [=] J/m$^2$.

$\psi_0$ = surface potential [=] V or mV. Typical values are $|\zeta|$ = 25 to 150 mV.

$Z$ = valence of symmetric electrolyte [=] dimensionless. NaCl has $Z = 1$.

$\zeta$ = zeta potential, an estimate of the surface potential [=] V or mV.

$\langle \zeta \rangle$ = average zeta potential over the entire surface of a sphere.

---

[41] http://en.wikipedia.org/wiki/Disjoining_pressure.

***Practice Problems***

1. **Debye length for 1:1 electrolytes**. Calculate the Debye length for KCl at $T$ = 293 K and at a concentration of 0.003 M. It is often easiest to put concentration in terms of #/m$^3$ instead of mol/L; the units are often easier to make consistent in this way. Check that your result is consistent with that found in Table 1-2.
*answer: 5.57 nm*

2. **Debye length for mixed electrolytes**. Aluminum sulfate, $Al_2(SO_4)_3$, is a 3:2 electrolyte often used as a flocculating agent in water purification. It is similar to "alum". Calculate the Debye length at $T$ = 293 K and at a concentration of 0.003 M = 3 mM, using the result (derived in Problem 2-10)

$$\kappa^2 = \sum_{i=1}^{N} \frac{z_i^2 e^2 c_{i\infty}}{\varepsilon kT}$$

Check that your result is consistent with that found in Table 1-2. Note that the concentrations for each ion must be adjusted properly, since there are in fact 5 ions dissociated per molecule.
*answer: 1.44 nm*

3. **Electrostatic potential near to a surface**. A solution has a Debye length of 2.6 nm. a) Find the surface potential 8.2 nm away from a flat plate that has $\psi_0$ = -39 mV. b) Find the surface potential 8.2 nm away from a 50 nm diameter sphere that has $\psi_0$ = -39 mV.
*answer: a) -1.66 mV, b) -1.25 mV.*

4. **Surface charge density**. Two plates are separated by a distance $L$ = 44 nm. They are in an aqueous solution with $\kappa^{-1}$ = 14.1 nm. The surface potential is $\psi_0$ = +48 mV. Find the surface charge density.
*answer: 0.0022 C/m$^2$*

5. **Osmotic pressure**. A 25 mM aqueous solution of $Na_2(SO_4)$ dissolves completely in water at $T$ = 305 K. Find the osmotic pressure of the solution. Note that osmotic pressure is a property of the solution, but only manifests itself as a pressure if the ions are confined, say by a semi-permeable membrane. Also note that sodium sulfate produces 3 ions for every salt molecule.
*answer: 1.88 bar. The North Atlantic has an NaCl concentration of about 0.6 M, which due to both ions gives an osmotic pressure of roughly 30 bar.*

6. **Electrostatic potential energy between spheres**. Two 75 nm diameter spheres have surface potentials of -32 mV in a solution with a Debye length of 3.6 nm and $T$ = 293 K. Plot the electrostatic energy as a function of separation, and find the electrostatic energy in $kT$ units when the spheres are separated by 9.04 nm. Use the HHF result in Eq 2-47.
*answer:* $\Phi_{es}/kT$ = +3.3 *at 9.04 nm separation. The curve appears as the upper dotted line in the plot below. The lower doted curve is for the VDW energy, and the solid curve is for the sum (DLVO theory).*

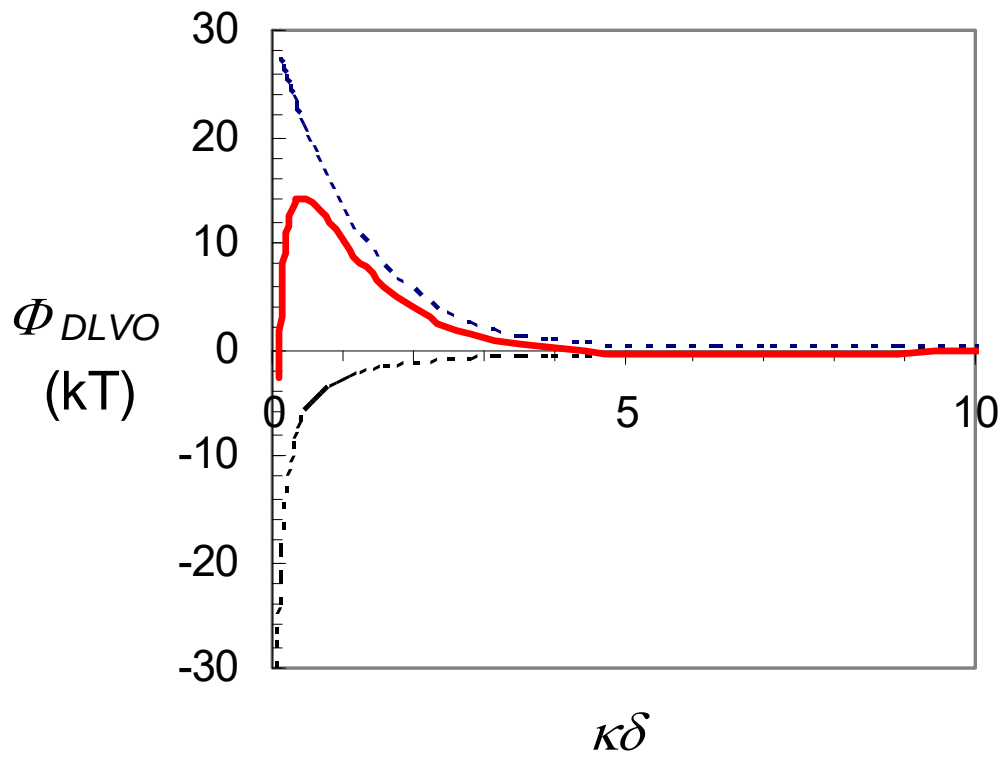

7. **Boltzmann distribution from kinetic arguments**. The Boltzmann equation quantitatively describes the equilibrium

competition between energy – in this chapter, electrical – and diffusion. The equation for the flux of the i$^{th}$ type of is often given by $\mathbf{j}_i = -D_i \nabla c_i + z_i e D_i c_i \mathbf{E} / kT + c_i \mathbf{v}$. At equilibrium $\mathbf{j}_i = \mathbf{0}$ for each ion, and the fluid velocity $\mathbf{v} = \mathbf{0}$. Starting with the flux equation, derive the Boltzmann factor distribution of ions from the ion migration equation. There is also a statistical mechanics derivation, which we do not derive here.

8. **Poisson equation**. The four Maxwell equations are $\nabla \cdot \mathbf{D} = \rho_e$ , $\nabla \times \mathbf{H} = \mathbf{J} + \partial \mathbf{D} / \partial t$ , $\nabla \cdot \mathbf{B} = 0$, $\nabla \times \mathbf{E} = -\partial \mathbf{B} / \partial t$ . For the static case, show that

a) the Maxwell equations reduce to $\nabla \cdot \mathbf{D} = \nabla \cdot (\varepsilon \mathbf{E}) = \rho_e$, $\nabla \cdot \mathbf{J} = 0$, and $\nabla \times \mathbf{E} = \mathbf{0}$ .
b) we can always define an electrical potential $\mathbf{E} = -\nabla \psi$ that satisfies $\nabla \times \mathbf{E} = \mathbf{0}$ .
c) for constant and uniform $\varepsilon$, we obtain the Poisson equation.

The Wikipedia site http://en.wikipedia.org/wiki/Poisson_equation gives insights into the derivation of the Poisson equation, as does the classic book by Jackson mentioned early in Chapter 2.

9. **Boltzmann distribution for oxygen molecules in the atmosphere**. The Boltzmann distribution is

$$p(x, y, z) = B\, exp\left[\frac{-E(x, y, z)}{kT}\right]$$
$$= exp\left[\frac{-E(x, y, z)}{kT}\right] \Big/ \iiint_V exp\left[\frac{-E(x, y, z)}{kT}\right] dv$$

where $B$ is a constant that normalizes the probability. For a gas molecule we know that $E(z) = mg(z - z_0)$, where $z_0$ is a reference height, which we can take as sea level. At sea level, where $z - z_0 =$ 0, let's estimate that the pressure $p$ = 1.00 bar and the temperature $T$ = 293 K. Estimate the pressure at the top of K2, the second tallest mountain in the world, which is in the Himalayan range. Its height is 8611 m above sea level, and a typical temperature is $T$ = 235 K. The value of $g$ is roughly constant, decreasing from 9.81 m/s$^2$ at sea level to about 9.78 m/s$^2$ at the top of K2.
*answer: 0.33 bar. Many K2 climbers now use bottled oxygen for their climb.*

10. **Debye length for mixed electrolytes**. Show that the Debye length for an arbitrary mixture of electrolytes is given by

$$\kappa^2 = \sum_{i=1}^{N} \frac{z_i^2 e^2 c_{i\infty}}{\varepsilon kT}$$

One way to do this is to start with the full PB equation 2-4, linearize $e^{-x} \approx 1 - x$, and then use the fact that in the bulk – which manifests itself in the "1" part of the expansion of the exponent – the solution is electroneutral, so that $\sum z_i e c_i = 0$.

11. **Gouy-Chapman model for large surface potentials**. For large surface potentials or large surface charge densities, the Debye-Huckel model loses accuracy. This is especially true for surface potentials greater than $2kT/e$, which is roughly 50 mV. In these cases you can use the Gouy-Chapman model of the EDL.[42] If we define $\phi \equiv Ze\psi / kT$ and $y = \kappa x$, for a binary Z:Z electrolyte, then we can find

[42] Gouy is pronounced "gwE", much like the word "we" with a hard "g" in front.

$$\boxed{\phi_{gc} = 2ln\left[\frac{1+e^{-y}\,tanh\left(\frac{\phi_0}{4}\right)}{1-e^{-y}\,tanh\left(\frac{\phi_0}{4}\right)}\right]}$$, which can be re-arranged to

$$tanh\left(\frac{\phi_{gc}}{4}\right) = tanh\left(\frac{\phi_0}{4}\right)e^{-y}$$

Notice that the modified potential has the same $exp(-\kappa x)$ dependence as the Debye-Huckel model. Derive these results. There is a useful mathematical trick for deriving this result: Multiple both sides of the full PB equation 2-5 by $d\psi / dx$ and then obtain $d(d\psi / dx)^2 = 2(d\psi / dx)(d^2\psi / dx^2)$, which is an exact differential. Now integrate both sides. You will also use the definition of $cosh\,x = (e^x + e^{-x})/2$ to obtain the identity $cosh\,x - 1 = (e^x + e^{-x} - 2)/2 = (e^{x/2} - e^{-x/2})^2/2$. Then since $sinh\,x = (e^x - e^{-x})/2$, you find $cosh\,x - 1 = 2\,sinh^2(x/2)$.

12. **Gouy-Chapman surface charge for a plate**. A careful analysis of the Maxwell equations reveals that the accurate way to find the surface charge density at an interface (*S*) between two phases A and B, given the electrical potential distribution and electrical permittivities in both phases, is by using the Gauss boundary condition

$$\rho_s = \varepsilon_B \left.\frac{\partial \psi_B}{\partial x}\right|_S - \varepsilon_A \left.\frac{\partial \psi_A}{\partial x}\right|_S$$

where *x* points into the "A" phase, and the operation at *S* indicates that the boundary condition is evaluated at the interface. Often it happens that the permittivity of the suspending fluid – say, water, which has a dielectric constant of roughly 80 – has a much higher permittivity than the particle. If the permittivity of "B" is much less than that of "A", we can ignore the "B" phase. Use the Gauss boundary condition, along with the Gouy-Chapman potential, to derive the following expression for the surface charge density for large surface potentials.

$$\rho_s = 2(2\varepsilon kTn_\infty)^{1/2}\, sinh\left(\frac{Ze\psi_0}{2kT}\right)$$

13. **Gouy-Chapman surface charge for a plate**. A plate has a surface potential of -114 mV in 5.4 mM NaCl solution at $T = 305$ K. Find the surface charge density in C/m$^2$, comparing results from the simplified (for low surface potentials) Eq 2-20 and the result from the previous problem.
*answer: approximation gives -0.019 C/m$^2$; the Gouy-Chapman gives -0.037 C/m$^2$. The approximation is off by roughly a factor of 2.*

14. **Gouy-Chapman plot**. Plot the dimensionless potential ($\phi$) versus the dimensionless gap ( $y = \kappa x$ ) for $\phi_0 = 1$, 2, and 5 from $y =$ 0 to 5. Plot the results for the Debye-Huckel approximation, the Gouy-Chapman model, and a numerical solution (e.g., with Mathematica, using the NDSolve expression).

15. **Gouy-Chapman-type results for a sphere**. When the surface potential on a sphere is large, Eqs 2-14 and 2-21 are no longer valid. In fact, no analytical solution is available. But good approximations for large $\kappa a$ >10, are (Hunter, *Zeta Potential*, 1981, p 49)

$$\frac{Ze\psi}{kT} \approx \frac{4a}{r} tanh^{-1}\left[e^{-\kappa(r-a)}\, tanh\left(\frac{Ze\psi_0}{4kT}\right)\right],$$

$$\rho_s \approx \varepsilon\kappa\left(\frac{kT}{Ze}\right)\left(2\, sinh\frac{Ze\psi_0}{2kT} + \frac{4}{\kappa a} tanh\frac{Ze\psi_0}{4kT}\right)$$

The first of these equations is simply the result from Problem 2-12 times $a/r$. Say you have 350 nm diameter ($2a$) particles with $\psi_0$ = -90 mV, suspended in a KCl solution with $\kappa^{-1}$ = 4.2 nm and $T$ = 295 K. Find the surface charge density using the equation above, and compare the result with that obtained from the approximation in Eq 2-21 for low potentials. Also, find the electric potential ($\psi$) a distance 5.7 nm away from the surface, and compare the value with the approximation in Eq 2-14. Note that in finding the value 5.7 nm away from the surface, $r - a$ = 5.7 nm.
*answers: $\kappa a$ = 41.67, approximate $\rho_s$ = -0.0156 C/m$^2$, approximate $\psi$(5.7 nm) = -22.4 mV; more precise $\rho_s$ = -0.0248 C/m$^2$, more precise $\psi$(5.7 nm) = -18.2 mV.*

16. **Constant potential compared with constant charge**. When two plates approach closely, one must choose between a "constant surface potential" model or a "constant surface charge density" model, often described as "constant potential" or "constant charge".[43] The reason is highlighted in Eqs Eq 2-22 and Eq 2-23, which show that for the charge to remain constant as two plates are brought closer together (smaller $L$), the potential must increase dramatically. The potential energy between the two plates is still given by the HHF model for flat plates, but for a constant potential model, the potential simply becomes smaller for a constant charge system. Plot the potential energy between two plates with a constant $\rho_s$ = 0.010 C/m$^2$ and $\kappa^{-1}$ = 3.2 nm as a function of the separation ($L$), for constant charge.

[43] It turns out that there is yet a third model, the "charge regulation model", that more properly accounts for the changing surface potential as a changing dissociation constant on surface charge groups. See Chan, D.; Perram, J.W.; White, L.R.; and Healy, T.W. "Regulation of Surface Potential at Amphoteric Surfaces during Particle-Particle Interaction," *Faraday Trans. I*, **5**, 1046-1057 (1975).

17. **Surface charge density on a sphere**. For the case when the permittivity of the particle is very small compared with that of the fluid, use the Gauss boundary condition (Problem 2-12) to derive Eq 2-21 starting from Eq 2-14.

18. **Electrostatic forces with large surface potentials**. For surface potentials greater than 50 mV in magnitude, there are several routes for predicting electrostatic forces between spheres. One could use the results found in Ohshima et al, JCIS, 89, 484 (1982). One could also use superposition, as given in Sader et al, *JCIS* **171**, 46-54 (1995). Use these two results to plot the potential energy as a function of separation between two polystyrene spheres of diameter 120 nm having surface potentials -90 mV, in a solution with Debye length 2.4 nm.

19. **Capacitance**. The electrical double layer can be thought of as a type of capacitor. We know that capacitance is given by $C = \varepsilon A / d$ . Say we have an EDL with $\kappa^{-1}$ = 1.0 nm, over an area of 3 $m^2$, in water with $\varepsilon_r$ = 78.0. If $d = \kappa^{-1}$, what is the capacitance?
*answer: 2.0* $C^2/N\text{-}m$ *= 2.0 F. Ordinary capacitors often have capacitance measured in milliFarads. Electrical double layer capacitors (EDLCs), which are being researched for use in energy storage as "supercapacitors", can have C > 1000 F (http://en.wikipedia.org/wiki/Electric_double-layer_capacitor).*

# 3 van der Waals force

## *Van der Waals interactions between two atoms*

In the previous chapter we examined electrostatic interactions, which occur for many colloidal systems. But there is one force that is even more ubiquitous: the van der Waals (VDW) interactions[44]. Of all the forces that must be examined for colloidal particles – depletion forces, solvation forces, solvophobic forces, hydrogen bonding (Figure 3-1) – the most ubiquitous is the VDW force. This force was hypothesized in 1873 by Johannes Diderik van der Waals, who integrated the concept of a molecular attraction into describing vapor-liquid equilibrium.[45] His insight earned him the Nobel Prize in Physics in 1910, even though van der Waals did not know about the mechanism behind this force.

---

[44] The word "interaction" in the literature usually corresponds to "energy", with units of Joules or similar.

[45] Van der Waals took us from $PV = RT$, where here $V$ is the molar volume, to $\left(P + a/V^2\right)\left(V - b\right) = RT$. The "*a*" represented an attractive force between the gas molecules, which we now assess as the VDW forces. The "*b*" represented the finite volume of the gas molecules; this introduces a repulsive force in the equation, which is required for this gas model to describe the liquid phase.

It is commonly thought that the VDW interaction is weak. So why do we colloid scientists spend so much time with it? First, VDW interactions are weak for atomic or molecular systems – often giving energies less than 0.1 *kT*. On the other hand, for colloidal system, where the particles typically contain $10^3$ to $10^{12}$ atoms, VDW interactions can readily amount to several 100 *kT* of attraction, due to the larger size of the particles. Second, VDW interactions are ubiquitous. Except in very particular cases,[46] colloidal particles always feel VDW forces.

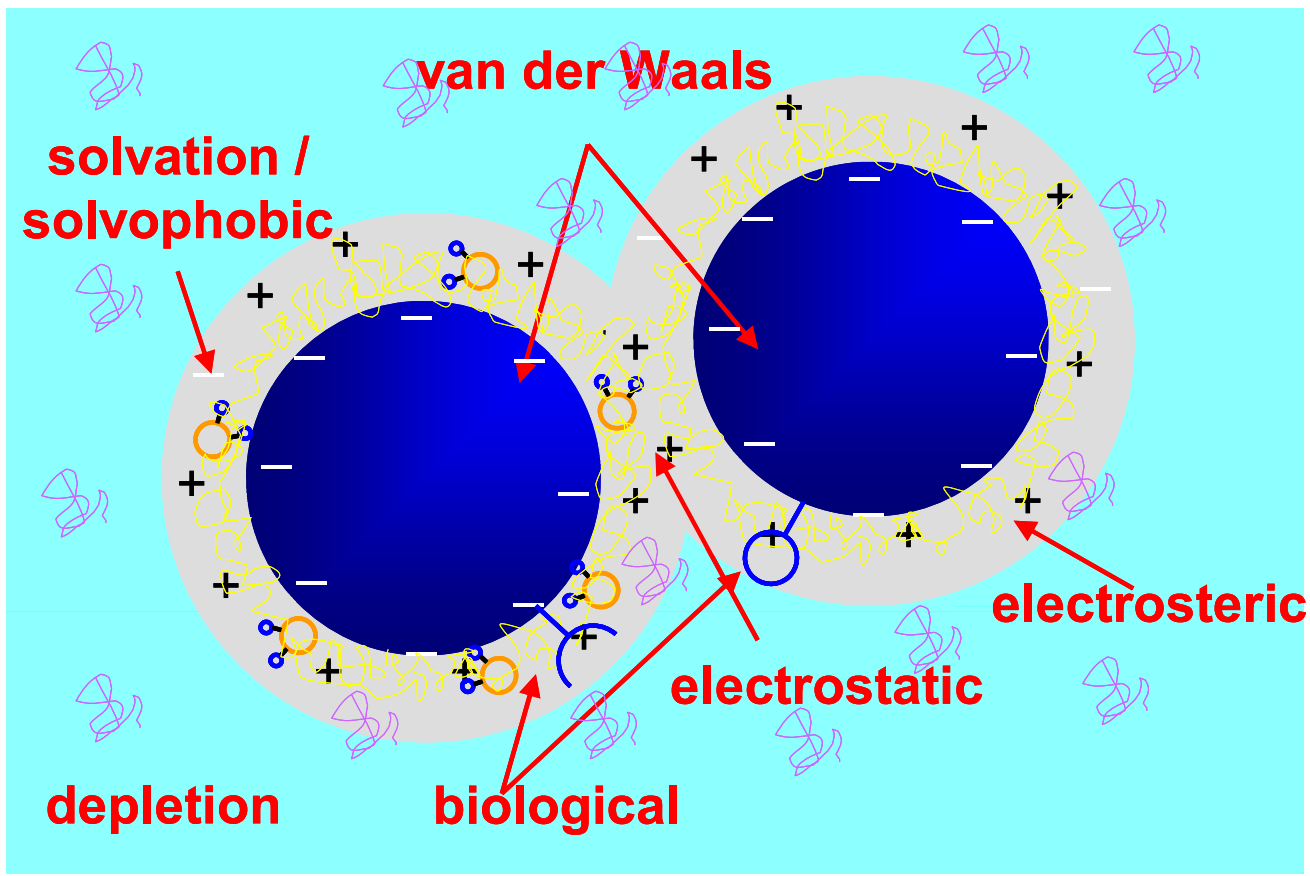

**Figure 3-1**. Forces between colloidal particles.[47]

The physics of the VDW interaction is quantum mechanical. Within the atoms of an electron, the electrons are continually performing their quantum mechanical dance, as Heisenberg would dictate. As a result of the electron movement, small and instantaneous dipoles form within the atoms (Figure 3-2). These temporary dipoles produce an electric field, which can in fact cause slight dipoles in the surrounding atoms. It is the interaction of the temporary dipoles and their induced dipoles that causes the VDW attraction. Does it seem strange that the molecules spontaneously produce dipoles, and then are acted upon by other dipoles? Isn't that perpetual motion, violating the 2$^{nd}$ law of thermodynamics? No it isn't. VDW attractions cause particles to attract each other together, but there is no continuous production of work. After the particles are aggregated, no further work is done by the VDW forces. Thus, having dispersed particles is a high energy state.

[46] The interaction can become exceedingly small in the artificial case where the fluid is "refractive index-matched" to the particle.

[47] This figure is adapted from Nel, Andre E.; Mädler, Lutz; Velegol, Darrell; Xia, Tian; Hoek, Eric M.; Somasundaran, Ponisseril; Klaessig, Fred; Castranova, Vince; Thompson, Mike. "Bio-physicochemical interactions at the nano-bio interface: implications for biological use and safety assessment of engineered nanoparticles." *Nature Materials*, **8**, 543-557 123 (2009).

Classical electrostatics predicts the attractive energy between two dipoles. However, at any instant of time, we know neither the direction or magnitude of the dipoles. In order to explain and predict VDW forces, we must delve into quantum physical considerations.

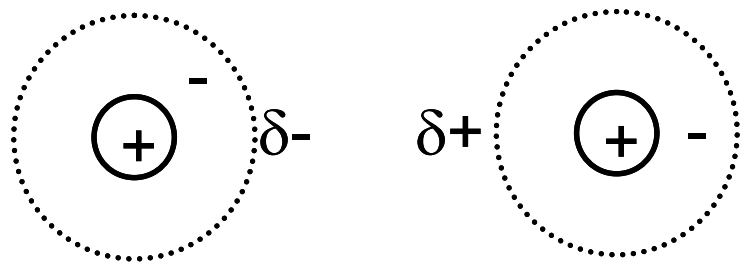

**Figure 3-2**. Instantaneous temporary dipole and induced dipole on two hydrogen atoms. The electron in the atom on the left has spontaneously moved to one side of the atom, and this in turn causes the electron in the atom on the right to move to the far right side, since like charges repel. The resulting dipoles are arranged with the negative tail of one dipole near to the positive end of the other. Attraction results.

### *London's approach*

The detailed physics for VDW attraction between two hydrogen atoms, starting with second order perturbations to the orbital solutions of quantum mechanics for the hydrogen atom, have been known for three-fourths of a century.[48] London found that the change in energy ($\Delta V_{VDW,2}$) of two atoms when they are brought from infinite separation to a finite center-to-center separation ($r$) is

$$\Delta V_{VDW,2} = -\frac{3\hbar\omega_0\alpha_0^2}{4(4\pi\varepsilon_0)^2 r^6} \tag{3-1}$$

---

[48] Pauling, Linus; Wilson, E. Bright. Introduction to Quantum Mechanics, McGraw-Hill (New York) 1935. This wonderful book lays out the solution clearly.

Several constants and parameters are used: $\hbar = h / 2\pi$, where $h = 6.626068\times10^{-34}$ J-s is Planck's constant. In classical mechanics the value of $h$ is set identically to zero. The atom is characterized by a static (zero frequency) polarizability ($\alpha_0$) and a characteristic frequency ($\omega_0$). The polarizability provides a measure of how easily the electrons are moved to one side of the atom in the presence of an electric field, whereas the characteristic frequency indicates how quickly the electrons change polarization within the atom. Finally, the vacuum permittivity ($\varepsilon_0 = 8.854\times10^{-12}$ C$^2$/N-m$^2$) tells us how readily an electric field moves through vacuum.

**Example 3-1**. VDW attraction between two H atoms.

A hydrogen atom has a radius $a_0 = e^2 / 8\pi\varepsilon_0\hbar\omega_0$ = 52.9 pm, a zero-frequency polarizability $\alpha_0 / 4\pi\varepsilon_0 = 0.65\times10^{-30}$ m$^3$, and a characteristic frequency $\omega_0 = 2.073\times10^{16}$ s$^{-1}$ (same units as radians/s). Find the VDW energy between two H atoms whose nuclei are separated by 1.0 nm.

*answer:* $\Delta V_{VDW} = -6.9\times10^{-25}$ *J.*

### *The coupled dipole method (CDM)*

Extending London's approach to multiple atoms is very difficult. There is, however, an alternative approach that captures the essential parts of the quantum mechanics. The approach is called the "coupled dipole method" (CDM),[49] and the CDM accounts for the "electrical conversation" among all atoms in the system. Let's first take a sub-atomic look at how each atom responds to an electric field.

The CDM treats all atoms or groups of atoms as being polarizable when an electric field (**E**) is applied. That is, the electrons respond to **E** to give a net dipole. The applied **E** field can in fact change in time with some frequency ($\omega$). At zero or low frequency, the electrons within a given atom have plenty of time to respond to the electric field, and so they polarize – move to one side of the atom on average – to the maximum extent. As a result, the atom takes on a dipole (**p**), with units of C-m, which is charge times distance, according to $\mathbf{p} = \alpha\mathbf{E}$. In essence, we have stuffed the complex quantum mechanics of the electrons' motion into the single polarizability ($\alpha$).

As the frequency of **E** increases, the electrons have to move back and forth within the atom. At a frequency of say 1000 $s^{-1}$, the electrons are easily able to keep up. It is almost as if a static **E** field were being applied. As the frequency gets higher, perhaps $10^{14}$ $s^{-1}$, the electrons no longer have time to accelerate completely, so they are not able to polarize the atom to the maximum extent. At very high frequencies, say $10^{18}$ $s^{-1}$, the **E** field is reversing so quickly that before an electron can move in one direction, due to mass inertia,, the **E** field has already changed direction. In this case, on average, the electrons do not polarize at all. As we see, the polarization depends on the frequency of how fast **E** is being reversed. Thus, the polarizability ($\alpha$) is not a single number, but a function that depends on frequency.

---

[49] Kim, Hye-Young; Sofo, Jorge O.; Velegol, Darrell; Cole, Milton W.; Lucas, Amand A. "Van der Waals forces between dielectric nanoclusters." *Langmuir*, **23**, 1735-1740 (2007). Seeds of the coupled dipole method date back to London. My friend Milton Cole is a wonderful teacher who taught me a lot about quantum mechanics during our many years or working on this problem.

One common equation for expressing the frequency-dependent polarizability $\alpha(\omega)$ is the Drude model:

$$\alpha(\omega) = \frac{\alpha_0}{1 - (\omega / \omega_0)^2} \qquad (3\text{-}2)$$

where, as before, $\alpha_0$ is the static (zero frequency) polarizability and $\omega_0$ is the characteristic frequency, or resonant frequency. When the **E** field is applied at a frequency $\omega = \omega_0$, it is like pushing a child on a swing at the resonant frequency of the pendulum: A little push causes the child to swinger ever higher. Mathematically, $\alpha$ becomes infinite when $\omega = \omega_0$. I show a plot of the polarizability function in Figure 3-3. I have also plotted $\alpha(i\omega)$, which has no physical meaning, but which is sometimes used as a mathematical artifice to simplify calculations.

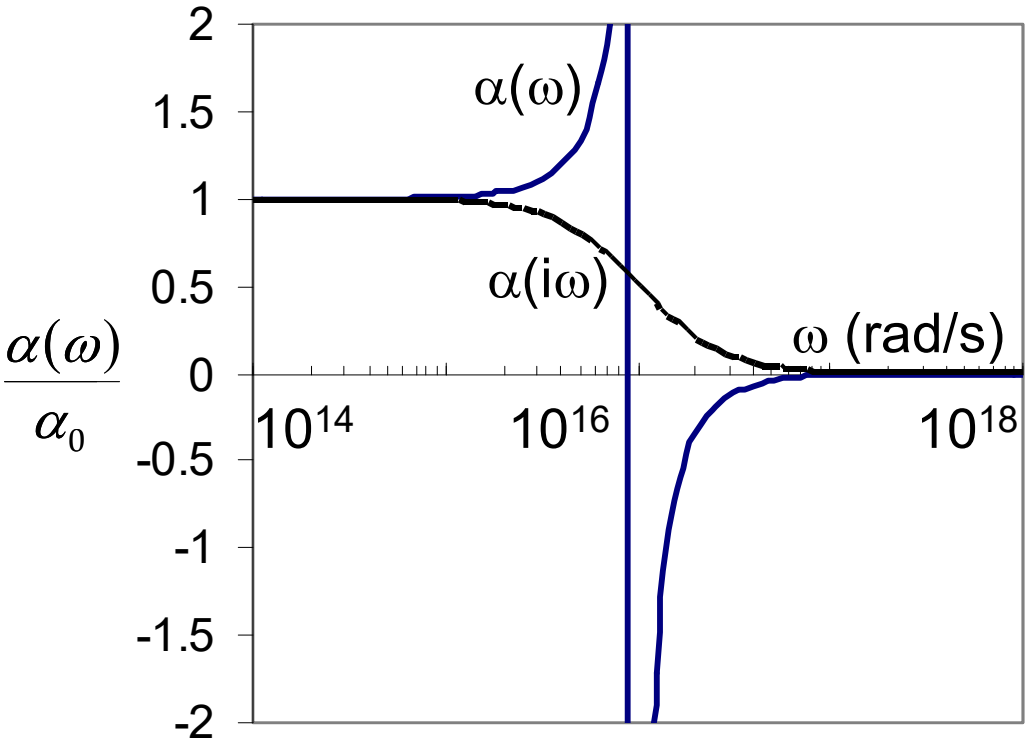

**Figure 3-3**. Polarizability function. This atom has $\omega_0 = 10^{16}$ rad/s. When expressing $\alpha$ as a function of frequency $\omega$ (rad/s = Hz), the function diverges at $\omega_0$. Sometimes the polarizability is expressed as a function of "complex frequency" ($i\omega$), which has no physical meaning but is useful mathematically, since this function does not diverge and is monotonic.

Now imagine that we have two identical atoms in the universe. Every energy must have a reference point, so that I can define where "zero energy" exists. For the VDW energy, the reference state is set where the atoms are separated by an infinite distance. The VDW energy ($\Delta V_{\mathrm{VDW}}$) is thus defined as

$$\Delta V_{VDW}(r) = V(r) - V(r \to \infty) \tag{3-3}$$

where $V$ is the work required to position the atoms in given locations. By subtracting the work required to assemble the system at infinite separation, we remove energies such as assembling the nucleus and similar physics, enabling us to focus on the VDW interaction alone.

Let's define the position of one atom as being at the origin ($\mathbf{x}_1 = \mathbf{0}$), and the other atom we will place along the $x$-axis at $x_2 = r$. For these atoms the polarizations are

$$\mathbf{p}_1 = \alpha \mathbf{E}(\mathbf{x}_1),\ \mathbf{p}_2 = \alpha \mathbf{E}(\mathbf{x}_2) \tag{3-4}$$

where $\alpha$ is the polarizability of each atom, which is the first place where quantum mechanics enters this problem. The electric field at $\mathbf{x}_1$ and $\mathbf{x}_2$ could be due to an applied field ($\mathbf{E}_0$), but it could also be due to a dipole field resulting from the other atom. A standard text like JD Jackson's gives the dipole electric field at $\mathbf{x}_1$ resulting from a dipole ($\mathbf{p}_2$) at $\mathbf{x}_2$ as

$$\begin{aligned} \mathbf{E}(\mathbf{x}_1) &= \mathbf{E}_0 + \frac{(3\mathbf{nn} - \mathbf{I}) \cdot \mathbf{p}_2}{4\pi\varepsilon_0 r^3} \\ \mathbf{E}(\mathbf{x}_2) &= \mathbf{E}_0 + \frac{(3\mathbf{nn} - \mathbf{I}) \cdot \mathbf{p}_1}{4\pi\varepsilon_0 r^3} \end{aligned} \tag{3-5}$$

In this case the normal vector (**n**) from dipole 1 to 2 is simply $\mathbf{n} = \mathbf{i}_x$, the unit vector along the *x*-axis, since the atoms are along the x-axis..[50]

Now I am in a position to combine the previous Eqs 3-2 to 3-5. Since I know that VDW forces occur even in the absence of an applied field $\mathbf{E}_0$, I set $\mathbf{E}_0 = \mathbf{0}$, and combine and re-arrange the equations to give

$$\begin{aligned} \omega_0^2 \mathbf{p}_1 - \alpha_0 \omega_0^2 \left( \frac{3\mathbf{nn} - \mathbf{I}}{4\pi\varepsilon_0 r^3} \right) \cdot \mathbf{p}_2 &= \omega^2 \mathbf{p}_1 \\ -\alpha_0 \omega_0^2 \left( \frac{3\mathbf{nn} - \mathbf{I}}{4\pi\varepsilon_0 r^3} \right) \cdot \mathbf{p}_1 + \omega_0^2 \mathbf{p}_2 &= \omega^2 \mathbf{p}_2 \end{aligned} \tag{3-6}$$

These are two vector equations for two vector quantities ($\mathbf{p}_1$ and $\mathbf{p}_2$). I have left them in this form because they reveal something very important: We have an eigenvalue problem.[51] After dividing both sides by $\omega_0^2$ and defining $\gamma = \alpha_0 / 4\pi\varepsilon_0 r^3$, I obtain the 6 scalar equations as

$$3\begin{pmatrix} n_x n_x & n_x n_x & n_x n_x \\ n_x n_x & n_x n_x & n_x n_x \\ n_x n_x & n_x n_x & n_x n_x \end{pmatrix} - \begin{pmatrix} 1 & 0 & 0 \\ 0 & 1 & 0 \\ 0 & 0 & 1 \end{pmatrix}$$

---

[50] A dyad like **nn** is a "tensor" quantity. The normal vector can be written as $\mathbf{n} = n_x \mathbf{i}_x + n_y \mathbf{i}_y + n_z \mathbf{i}_z$, and then the expression 3**nn** – **I** could be written

[51] Here is another type of eigenvalue problem: $4x + 3y = \lambda x,\ \ x + 6y = \lambda y$. We know that if a proper set of linear equations has a solution, that this is the one and only solution. However, in an eigenvalue problem, $\lambda$ is not given at first. In fact, the goal of an eigenvalue problem is not to solve for *x* and *y*, which would both be 0 for this problem. After all, in quantum mechanics, often a solution of 0 is not permitted physically. Rather, in this problem, what we are really solving for is $\lambda$, that is, values of $\lambda$ that allow other solutions for *x* and *y*. For example, if $\lambda = 3$ (an "eigenvalue"), then the two equations become *redundant*, and the system of equations becomes *underdetermined*. Both equations say $x + 3y = 0$, and so we have an infinite number of solutions $x = -3y$. So $\lambda = 3$ is the eigenvalue for this system of equations, and the eigenvector $(x,y) = (-3, 1)$ is a solution, as is (-6, 2), as is (-9, 3), and so on. We often list just the *normalized eigenvector*, which is $(-3,1)/\sqrt{(-3)^2 + 1^2}$ = (-0.9487, 0.3162).

$$\begin{pmatrix} 1 & 0 & 0 & -2\gamma & 0 & 0 \\ 0 & 1 & 0 & 0 & \gamma & 0 \\ 0 & 0 & 1 & 0 & 0 & \gamma \\ -2\gamma & 0 & 0 & 1 & 0 & 0 \\ 0 & \gamma & 0 & 0 & 1 & 0 \\ 0 & 0 & \gamma & 0 & 0 & 1 \end{pmatrix} \cdot \begin{pmatrix} p_1^x \\ p_1^y \\ p_1^z \\ p_2^x \\ p_2^y \\ p_x^z \end{pmatrix} = \left(\frac{\omega}{\omega_0}\right)^2 \begin{pmatrix} p_1^x \\ p_1^y \\ p_1^z \\ p_2^x \\ p_2^y \\ p_x^z \end{pmatrix} \tag{3-7}$$

Solving Eq 3-7 will give 6 eigenvalues for $(\omega/\omega_0)^2$. Eq 3-7 describes the coupling between the two atoms – it describes their "dipole conversation". The eigenvalues gives the frequencies "at which the dipoles talk". Upon typing these equations into Mathematica and symbolically solving for the eigenvalues, I obtain

$$\left(\frac{\omega}{\omega_0}\right)^2 = \{1-2\gamma, 1-\gamma, 1-\gamma, 1+2\gamma, 1+\gamma, 1+\gamma\} \tag{3-8}$$

The frequencies $(\omega/\omega_0)$ are always positive, and so taking the square root of each of these 6 values gives the characteristic frequencies of the system:

$$\frac{\omega}{\omega_0} = \left\{\sqrt{1-2\gamma}, \sqrt{1-\gamma}, \sqrt{1-\gamma}, \sqrt{1+2\gamma}, \sqrt{1+\gamma}, \sqrt{1+\gamma}\right\} \tag{3-9}$$

Earlier I indicated the first place where quantum mechanics enters, with the polarizability ($\alpha$). Now I introduce the second place where quantum mechanics enters the problem: I make the approximation that each atom behaves as a quantum harmonic oscillator. Quantum mechanics tells us the energy of a harmonic oscillator. Each of the two atoms can have oscillations in each of 3 directions – 6 modes of oscillation. When the atoms have an infinite separation they behave independently, and since the energy of each mode of oscillation is $\hbar\omega_0/2$ from quantum mechanics,[52] this results in $6\hbar\omega_0/2$. The overall energy changes when two atoms are brought from an infinite separation to a finite separation ($r$) is the VDW energy, and it is written as

$$\Delta V_{VDW} = \sum_{i=1}^{6} \frac{\hbar\omega_i}{2} - 6\frac{\hbar\omega_0}{2} \tag{3-10}$$

where the $\omega_i$ are from the eigenvalues listed in Eq 3-9.

How does this expression compare to the London expression listed in Eq 3-1? We cannot add the square root sign expressions directly, but for small values of $\gamma$, I take a Taylor series for the values in Eq 3-9, obtaining

$$\frac{\omega}{\omega_0} \approx \{1-\gamma-\gamma^2/2,\ 1-\gamma/2-\gamma^2/8,\ 1-\gamma/2-\gamma^2/8,\ 1+\gamma/2-\gamma^2/8,\ 1+\gamma/2-\gamma^2/8,\ 1+\gamma-\gamma^2/2\} \tag{3-11}$$

Approximating the sum in Eq 3-10 using this expansion, I find

$$\Delta V_{VDW} = \frac{\hbar}{2}\omega_0\left(6-\frac{3\gamma^2}{2}\right) - 6\frac{\hbar\omega_0}{2} = -\frac{\hbar}{2}\omega_0\left(\frac{3\gamma^2}{2}\right) \tag{3-12}$$

Inserting $\gamma = \alpha_0/4\pi\varepsilon_0 r^3$ gives

$$\Delta V_{VDW} = -\frac{3\hbar\omega_0}{4}\left(\frac{\alpha_0}{4\pi\varepsilon_0 r^3}\right)^2 \tag{3-13}$$

---

[52] The energy for a harmonic oscillator is more fully written as $coth(\hbar\omega/2kT)\hbar\omega/2$ for an oscillator with a frequency ($\omega$), but this simplifies to the expression given for

which is identical with London's result from Eq 3-1. In the sum, the order ($\gamma$) terms sum to zero. The leading term is of order $\gamma^2$. If the value of $\gamma$ were not small, I would have kept more terms in the Taylor series, or even the exact eigenvalues.

In the end, with only two pieces of quantum information – 1) the expression for the energy of a harmonic oscillator and 2) the polarizability of each atom – the dipole conversation has been re-framed as an eigenvalue problem that gives the van der Waals attractive energy between two atoms.[53] The VDW energy results from the slight change in frequency of the modes of the *system*, as compared with the frequencies of the original atoms. Once the atoms are brought to a finite separation ($r$), the entire system must be considered.

What assumptions were used in reproducing Eq 3-1 with the CDM? There are several:

- point dipoles. The atoms have no size.
- electrostatics (neglects finite speed of light, "retardation")
- description of atom by a simple polarizability (e.g., Drude)
- neutral atoms (not ions)
- no permanent dipole (like water)

In Eq 3-1 the VDW energy is seen to vary as $1/r^6$. This same dependence on $r$ is the origin of the second term in the Lennard-Jones parameterization of the interaction energy between atoms, which is given by

$$\Delta V_{L-J}(r) = 4\varepsilon\left[\left(\frac{\sigma}{r}\right)^{12} - \left(\frac{\sigma}{r}\right)^{6}\right] \qquad (3\text{-}14)$$

[53] Amand Lucas and Milton Cole were the real drivers who led us in the development of this technique, which is a type of "nanoscale Lifshitz theory". I was delighted to be part of the effort.

where $\varepsilon$ gives the depth of the potential well, and $r = \sigma$ gives the distance where $\Delta V_{L\text{-}J} = 0$. The twelfth power on the first term simply provides a simple way to express the hard-sphere repulsion between the atoms.

In the calculation in this section, I have used 2 atoms. If there were more atoms, I would simply expand Eq 3-7 and solve for additional eigenvalues. In principle, the CDM described in this section can be used for any number of atoms, subject to limitations of the computer to find eigenvalues. I can readily use 5000 atoms on a personal computer with Mathematica. On the other hand, 5000 atoms represents a particle that is roughly 21 atoms in diameter, or just a few nanometers. A particle that is even 100 nm in diameter has more than ten million atoms. Since the calculation for eigenvalues scales as the number of atoms cubed, this becomes a challenging calculation. And so while the CDM technique reveals the basic physics of VDW interactions, and is itself useful for studying nanoparticles, we might want to know whether for larger particles, is there an easier way?

### *Hamaker approach*

Doing eigenvalue calculations for $10^6$ or $10^{10}$ atoms is not a trivial task. Another approach is to assume that every atom of particle 1 interacts with every atom of particle 2, and that "pairwise addition" is valid. That is, when two atoms interact, we approximate that surrounding atoms do not affect the interaction. One method for calculating all these interactions is to take a summation, but doing a sum with that many atoms turns out not to be trivial either. Researchers in the 1930s tried some integration techniques, in which they converted the sums of all the particles to volume integrals over two spheres, giving an integral over *x*-*y*-*z* for each particle, or 6 integrations. This method also turned out not to be trivial.

In 1937 a key breakthrough came. Hamaker published a clever integration technique that enabled an exact integral over every part of both particles.[54] Here I give only the final result. Say that I start with two spheres (1 and 2), of radii ($a_1$ and $a_2$). The spheres have a separation ($\delta$) between them at the point of their closest approach. The geometry is shown in Fig 3-4.

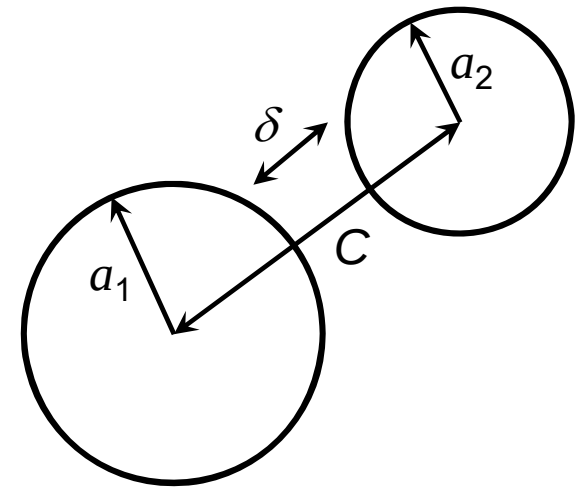

**Figure 3-4**. Geometry for two spheres interacting through VDW forces. Either the larger or the smaller sphere can be "sphere 1".

Hamaker assumed that the interaction energy ($\Delta e_{1\text{-}2}$) between an atom in particle 1 and an atom in particle 2 could be written as $\Delta e_{1-2} = -\lambda_{12} / r_{12}^6$, where $r_{12}$ is the separation between atoms 1 and 2, and $\lambda_{12}$ is the coefficient describing the interaction strength. This $1/r^6$ form comes from London, given earlier in this chapter. Next the summation is converted into an integral:

$$\Delta\Phi_{VDW} = -\sum_{i=1}^{N_1}\sum_{j=1}^{N_2}\frac{\lambda_{ij}}{r_{ij}^6} \approx -\int_{V_1}\int_{V_2}\frac{\lambda_{12}}{r^6}\,d(n_1V_1)\,d(n_2V_2) \tag{3-15}$$

where $\Delta\Phi_{\mathrm{VDW}}$ is the VDW energy between two spherical particles, $n_1$ describes the density (#/m$^3$) of atoms in particle 1, and $n_2$ describes the density (#/m$^3$) of atoms in particle 2.

Hamaker let the material in particle 1 be the same as that in particle 2, so the density is described by just $n$ and the London-VDW coefficient is described by just $\lambda$. He then defined a constant ($A$) – now known as the "Hamaker constant" – as $A = \pi^2 \lambda n^2$. He also defined geometric parameters: $x = \delta / 2a_1$ and $y = a_2 / a_1$.

---

[54] Hamaker, H.C. "The London-van der Waals Attraction between Spherical Particles." *Physica*, **IV**, 1058-1072 (1937).

After doing some calculus and algebra, Hamaker derived the following result for the VDW energy:

$$\Phi_{vdw} = -\frac{A}{12}\left(\frac{y}{x^2+xy+x}+\frac{y}{x^2+xy+x+y}+2\,ln\frac{x^2+xy+x}{x^2+xy+x+y}\right) \tag{3-16}$$

The equation is symmetric: Either sphere can be “1”.

Two interesting results occur upon taking limits. First let us examine the case when $\delta << a$ and $a = a_1 = a_2$, which gives

$$\Phi_{vdw} = -\frac{Aa}{12\delta} \tag{3-17}$$

When Hamaker published this result in 1937, it revealed quite a surprise about colloidal particles. Prior to his publication, colloid scientists thought that since VDW energies decay so rapidly – the $1/r^6$ dependence found for two atoms is very steep – that VDW interactions must play almost no role in colloidal interations.[55] As Eq 3-17 shows, in fact the VDW energy decays very slowly for small gaps, and can be of huge significance. It turns out that it is the VDW forces that cause particles to aggregate together in most cases.[56] Similar to Eq 3-17 is taking the limit for $\delta << a$ for the case $a = a_1$ and $a_2 \rightarrow \infty$. In this case $\Phi_{vdw} = -Aa/6\delta$.[57]

---

[55] For two point atoms separated by a distance $r$, the VDW energy decreases as $1/r^6$. For a point atom near a semi-infinite body, the VDW energy decreases as $1/r^3$. For two semi-infinite bodies separated by a distance ($h$), the energy per area decreases as $1/h^2$. For two spheres separated by a gap ($\delta$), the effective area of interaction, which can be estimated by using the Derjaguin approximation found later in this chapter, is $A \sim a\delta$. Thus, if two spheres interact, the energy scales as $1/\delta^2$ for the energy per area times the effective area, giving the $1/\delta$ dependence.

[56] The VDW interactions also cause substances like hexane or benzene to be liquids at room temperature, instead of gases. And while hydrogen bonding is important in keeping water as a liquid, in fact the VDW interations are also important for water.

[57] Eq 3-17 makes it appear that the VDW energy goes to negative infinity. But if we say that the closest that two spheres can approach is $\delta = 0.3$ nm, roughly the size of

In the other limit, when $\delta >> a$ for the case $a = a_1 = a_2$, I might hypothesize that I should recover a $1/r^6$ dependence, since the particles are so far apart that they simply act as large "atoms". The limit turns out to be

$$\Phi_{vdw} = -\frac{16A}{9}\left(\frac{a}{\delta}\right)^6 \qquad (3\text{-}18)$$

The distance dependence is $1/\delta^6$ as expected, and the interaction also depends upon the products of the particle volume (i.e., $a^3 \times a^3$ gives $a^6$), meaning that particles with more mass interact more strongly by VDW forces. For the limit $\delta >> a$ and $a = a_1$ and $a_2 \rightarrow \infty$, I find $\Phi_{vdw} = -2Aa^3/9\delta^3$, consistent with the interaction of an atom and a semi-infinite body.

The dependence on $\delta$ is thus seen to depend on distance. If I define $\Phi_{VDW} = -c\delta^{-n}$ for the case $a = a_1 = a_2$, I find that $n$ varies as shown in Figure 3-5 for $a = a_1 = a_2$. Homework Problem 3-9 provides more detail about how to calculate $n$.

**Example 3-2**. VDW attraction between two particles.

Two polystyrene particles are separated by $\delta = 20$ nm in water. Their diameter is 750 nm. The Hamaker constant for the system is $A = 1.4\times10^{-20}$ J. Calculate the VDW energy of interaction using Eq 3-16 and Eq 3-17.

*answer: full equation gives* $\Phi_{VDW} = -1.58\times10^{-20}$ *J, approximation gives* $\Phi_{VDW} = -2.2\times10^{-20}$ *J.*

---

an atom, then we find a finite attractive energy. For example, if $A = 1.4\times10^{-20}$ J, $a = 500$ nm, $\delta = 0.3$ nm, and $T = 298$ K, then when two spheres are touching, their VDW energy is -473 kT. This is a large attractive energy, but not infinite.

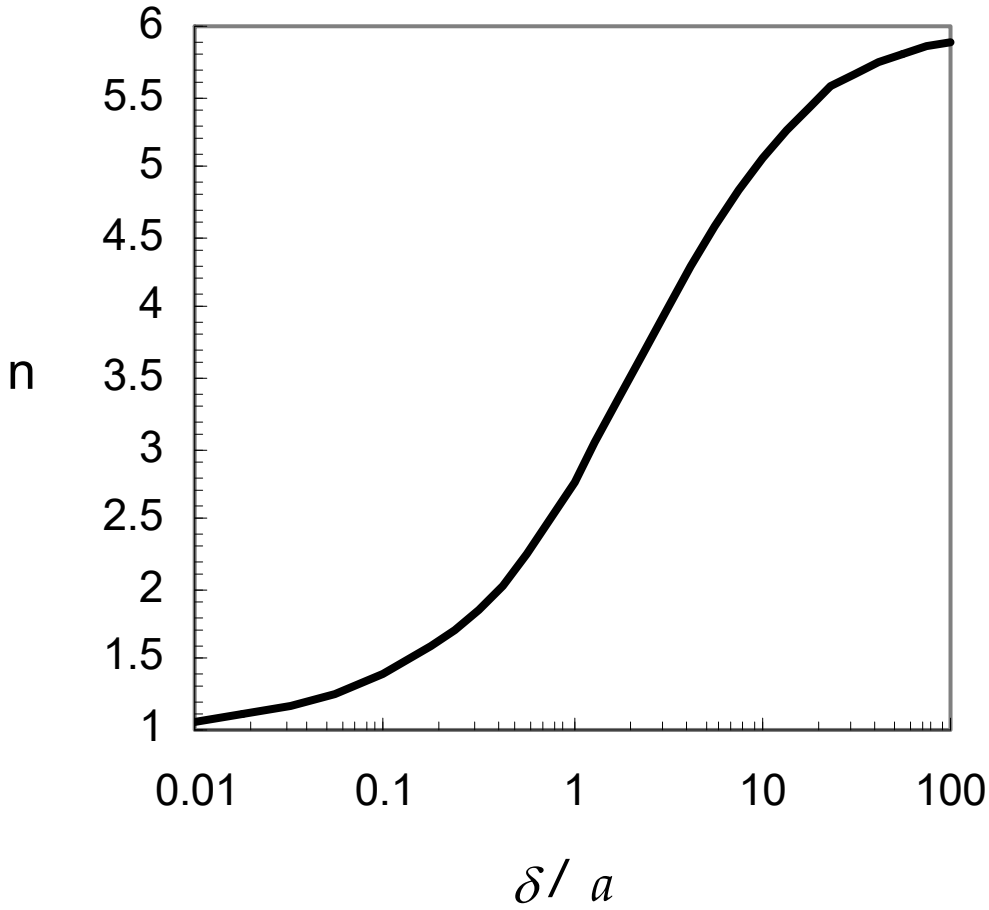

**Figure 3-5**. Value of *n* for $\Phi_{VDW} = -c\delta^{-n}$ , based on Eq 3-16. See Homework Problem 3-10 for more detail about calculating *n*, and the note about the coefficient *c*.

What assumptions does Hamaker use in his analysis? Several. 1) Hamaker uses "geometric spheres", instead of particles having real atoms. As the particles come close enough that the discrete atomic nature matters, Hamaker's analysis begins to fail. 2) He assumes 2-body additivity. That is, he assumes that one differential region from particle 1 interacts with one differential region of particle 2, without being influenced by other regions of the particles or fluid. This approximation fails for highly polarizable materials, especially metals. 3) He neglects "retardation"[58]. Implicit in his calculation is the fact that the speed of light is infinite, which of course it is not. When a temporary dipole gives an electric field, that electric field is not felt at a finite distance away in an instantaneous manner. It takes time. On the one hand, we might think, "Well, we're talking about nanometers, and the electric field is moving at the speed of light. The time scale is say 100 nm divided by $3\times10^{8}$ m/s, or $3\times10^{-16}$ s. Does the speed of light really matter here?" Yes, because the electrons within the atoms change their distribution on a comparable time scale. The most capable theory for addressing the latter two issues is "Lifshitz theory".[59]

[58] Casimir, H.B.G. and Polder, D. "The Influence of Retardation on the London-van der Waals Forces." *Physical Review*, **73**, 360-372 (1948).

### *Lifshitz theory*

Lifshitz theory – sometimes called "DLP" theory after the authors of an early paper[59] – accounts fully for retardation and for multi-body effects. The geometry of the problem is shown in Figure 3-6 for two semi-infinite bodies A and B separated by a distance ($H$) in a medium (m). The theory is involved, but in principle, Lifshitz theory is a continuum version of the coupled dipole method that also accounts for retaradation. Here I simply list the final *algorithm*, with little explanation, for calculating the Hamaker constant ($A$) for the system A-m-B.

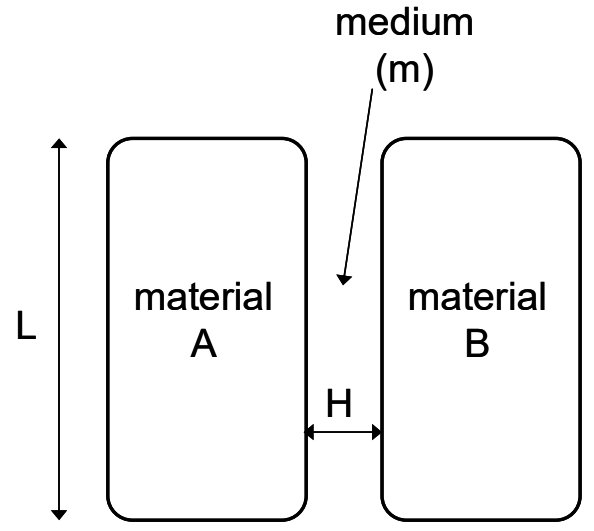

**Figure 3-6**. Geometry for two semi-infinite bodies (A and B) interacting across a medium (m).

Lifshitz theory requires that the real permittivities at complex frequencies $\varepsilon(i\xi)$ be known.[60] These are $\varepsilon_A(i\xi)$ for material A, $\varepsilon_m(i\xi)$ for the medium m, and $\varepsilon_B(i\xi)$ for material B. The letter $i=\sqrt{-1}$, and for mathematical convenience eigenfrequencies

[59] Dzyaloshinskii, I.E.; Lifshitz, E.M.; Pitaevskii, L.P. "The General Theory of Van der Waals Forces," *Adv. Phys.* **10**, 165-209 (1961). This paper is not light reading.
[60] The "real permittivity at complex frequencies" is a mathematical construct, with no physical meaning that I have ever seen described.

$$\xi_n = \frac{4\pi^2 kT}{h} n, \quad n = 0, 1, 2, \ldots \tag{3-19}$$

are introduced. In this equation, $k$ is the Boltzmann constant; $T$ is the absolute temperature; and $h$ is Planck's constant.

Despite the presence of resonant electromagnetic absorption peaks in a material, $\varepsilon(i\xi)$ is a monotonically decaying, well-behaved function, similar to $\alpha(i\omega)$ in Fig 3-3.[61] Data for $\varepsilon(i\xi)$ are usually presented in the following form:

$$\frac{\varepsilon(i\xi)}{\varepsilon_0} = 1 + \frac{d}{1+\xi\tau} + \sum_j \frac{f_j}{1 + g_j \cdot \left(\frac{\xi}{\omega_j}\right) + \left(\frac{\xi}{\omega_j}\right)^2} \tag{3-20}$$

The physical meanings of the various contributions are described in Hunter (1986) and Russel *et al.* (1989). The parameters $d$, $f_j$, $g_j$, and $\omega_j$ are known for several materials.

**Example 3-3**. Calculating a permittivity at complex frequency.

Polystyrene has four peaks in its dielectric data, and $d = 0$. The peaks are represented below. Find $\varepsilon(i\omega) / \varepsilon_0$ for polystyrene at $\omega = 0.50\times10^{16}$.

| $j$ | $\omega_j$ (rad/s) | $f_j$ (no units) | $g_j$ (no units) |
|---|---|---|---|
| 1 | $1.03\times10^{16}$ | 0.362 | 0.102 |
| 2 | $1.78\times10^{16}$ | 0.367 | 0.318 |
| 3 | $2.26\times10^{16}$ | 0.494 | 0.357 |
| 4 | $3.25\times10^{16}$ | 0.339 | 0.572 |

*answer: 2.34.*

[61] See p 162 of Jackson's book on *Classical Electrodynamics*, 3rd ed. For simple materials, the molecular polarizability and bulk material permittivity are related by the Clausius-Mossotti equation $(\varepsilon - \varepsilon_0)/(\varepsilon + 2\varepsilon_0) = n\alpha / 3$ .

For materials with more limited spectroscopic data, the dielectric function is approximated by the following equation:

$$\frac{\varepsilon(i\xi)}{\varepsilon_0} = 1 + \frac{C_{IR}}{1+(\xi/\omega_{IR})^2} + \frac{C_{UV}}{1+(\xi/\omega_{UV})^2} \qquad (3\text{-}21)$$

**Example 3-4**. Calculating a permittivity at complex frequency.

Using data from Table 3-1, calculate $\varepsilon(i\omega)$ / $\varepsilon_0$ for fused quartz at $\omega = 0.50\times10^{16}$.

*answer: 1.94*

The expression for the potential of mean force per area ($\Delta V_{VDW}$) between the plates A and B is given in terms of an integral over a dummy variable $x$:

$$\Delta V_{VDW} = \frac{kT}{8\pi H^2}\left[\tfrac{1}{2}\int_{\kappa H}^{\infty} x \ln D(x, n=0)dx + \sum_{n=1}^{\infty}\int_{r_n}^{\infty} x \ln D(x,n)dx\right] \qquad (3\text{-}22)$$

Note that in including the Debye length in the limit of integration, that we account for changes in the VDW forces due to changes in ionic strength. These affect only the zero frequency contribution.

**Table 3-1.** Spectroscopic data for several materials. Data from p. 221 Hunter (Ch 04).

| Material | $C_{IR}$ | $\omega_{IR}$ (rad/s) | $C_{UV}$ | $\omega_{UV}$ (rad/s) |
|---|---|---|---|---|
| Crystalline quartz (avg) | 1.93 | $2.093\text{x}10^{14}$ | 1.359 | $2.032\text{x}10^{16}$ |
| Fused quartz | 1.70 | $1.880\text{x}10^{14}$ | 1.098 | $2.024\text{x}10^{16}$ |
| Fused silica | 1.71 | $1.880\text{x}10^{14}$ | 1.098 | $2.033\text{x}10^{16}$ |
| Poly(vinylchloride) | 0.9 | $5.540\text{x}10^{14}$ | 1.333 | $1.815\text{x}10^{16}$ |
| Poly(styrene) | 0.2 | $5.540\text{x}10^{14}$ | 1.424 | $1.432\text{x}10^{16}$ |

In the expression for VDW energies, we define the following:

$$D = \left(1 - \bar{\Delta}_{Bm}\bar{\Delta}_{Am}e^{-x}\right)\left(1 - \Delta_{Bm}\Delta_{Am}e^{-x}\right) \tag{3-23}$$

$$\bar{\Delta}_{Am} = \frac{\varepsilon_A x - \varepsilon_m x_A}{\varepsilon_A x + \varepsilon_m x_A} \tag{3-24}$$

$$\bar{\Delta}_{Bm} = \frac{\varepsilon_B x - \varepsilon_m x_B}{\varepsilon_B x + \varepsilon_m x_B} \tag{3-25}$$

$$\Delta_{Am} = \frac{x - x_A}{x + x_A} \tag{3-26}$$

$$\Delta_{Bm} = \frac{x - x_B}{x + x_B} \tag{3-27}$$

$$x_A = \sqrt{x^2 - r_n^{\ 2}\left(1 - \varepsilon_A / \varepsilon_m\right)} \tag{3-28}$$

$$x_B = \sqrt{x^2 - r_n^{\ 2}\left(1 - \varepsilon_B / \varepsilon_m\right)} \tag{3-29}$$

Only the ratios of permittivities appear in these expressions, so the permittivities can be relative permittivities or absolute permittivities. The limit of integration $r_n$ is given by

$$r_n = 2H\varepsilon_m^{\ 1/2}\xi_n / c \tag{3-30}$$

where $c$ is the speed of light in vacuum.

The potential of mean force between two plates separated by a distance ($H$) is defined in terms of a flat plate Hamaker constant $A(H)$ as

$$\Delta V_{VDW} = -\frac{A(H)}{12\pi H^2} \tag{3-31}$$

Thus, we obtain

$$\boxed{\frac{A(H)}{kT} = -\frac{3}{2}\left[\tfrac{1}{2}\int_{\kappa H}^{\infty} x \ln D(x, n=0)dx + \sum_{n=1}^{\infty}\int_{r_n}^{\infty} x \ln D(x,n)dx\right]} \tag{3-32}$$

Like the Hamaker approach, Lifshitz theory has several assumptions and limitations. 1) The material must be continuum. For materials having discrete atom effects or nanoscale features, the CDM would have to be used. 2) Lifshitz theory has been worked out only for certain geometries, namely semi-infinite bodies, spheres, cylinders, and points.[62]

***Spheres and the Derjaguin approximation***

In order to use Lifshitz theory to calculate the VDW energy between two spheres, I appeal to the Derjaguin approximation, which I discussed in the section "Spheres and the Derjaguin approximation" in Chapter 2. I use an equation very similar to Eq 2-45, with $H = \delta + r^2/a_1 + r^2/a_2$ and Eq 3-31, to obtain

$$\Phi_{VDW} = \int_0^{2\pi}\int_0^{a_1} \Delta V_{VDW}(H) r\, dr\, d\theta \approx -\frac{1}{6}\left(\frac{a_1 a_2}{a_1 + a_2}\right)\int_{\delta}^{a_1} \frac{A(H)}{H^2} dH \tag{3-33}$$

Doing the integration gives

$$\boxed{\Phi_{VDW} = -\frac{A_{eff}(\delta)}{6\delta}\left(\frac{a_1 a_2}{a_1 + a_2}\right)} \tag{3-34}$$

[62] Langbein, Dieter. *Theory of Van der Waals Attraction*. (Springer Tracts in Modern Physics, Volume 72). Springer-Verlag: New York (1974).

This equation gives the same result that Hamaker obtained for small gaps, except that is uses an effective Hamaker constant

$$A_{eff}(\delta) = \delta \int_{\delta}^{\infty} \frac{A(H)}{H^2} dH \qquad (3\text{-}35)$$

that accounts for "retardation", which accounts for the finite speed of light in VDW energies. Usually the $A_{\mathrm{eff}}$ must be calculated numerically, since $A(H)$ varies in a complicated way.

***Hamaker constant tables for various A-m-B systems***

In this section I provide expressions for the Hamaker constants of various material systems. The calculations include the effects of retardation, and they are done using Eq 3-32 for $A(H)$ for two semi-infinite bodies, and Eq 3-35 for $A_{\mathrm{eff}}(\delta)$ for two spheres. Figure 3-6 gives the geometry. The values for the dielectric functions are taken from Table 4.1 (p 221) of Hunter[63] for most materials, with water and polystyrene taken from Parsegian[64]. Material systems A-m-B have a contribution due to high frequencies ($A_{\mathrm{h}}$), and aqueous systems also have a zero-frequency contribution ($A_{\mathrm{z}}$) that tends to fade quickly with distance for most materials.

The calculated values for the Hamaker constants, including the effect of retardation, are fit in Tables 3-2 (semi-infinite bodies) and 3-3 (two spheres). For semi-infinite bodies I fit to

$$A(w) = \exp\left[k_1 + k_2 \ln w + k_3 (\ln w)^2\right] \qquad (3\text{-}36)$$

---

[63] Hunter, Robert J. *Foundations of Colloid Science*, Vol I. Clarendon Press / Oxford University Press (New York) 1986. The Preface to the book gives significant credit to Lee White for Chapter 4.
[64] Parsegian, V.A. "Long Range van der Waals Forces." in Physical Chemistry: Enriching Topics from Colloid and Surface Science. eds H. van Olphen and Karol J. Mysels. La Jolla, CA: Theorex (1975).

where the Hamaker constant $A$ is given in electron volts (1 eV = $1.602 \times 10^{-19}$ J), $w = H$ / 1 nm, and $1 < w < 100$. Thus, for a gap of 37 nm, $w = 37$. For larger gaps not only does retardation diminish the Hamker constant, but as given in Eq 3-31, since $H^2$ is in the denominator, the contribution to VDW attraction becomes small.

For spherical particles I fit an effective Hamaker constant ($A_{eff}$, Eq 3-35 approximation) in Tables 3-3 to

$$A_{eff}(w) = \exp\left[k_4 + k_5 \ln s + k_6 (\ln s)^2\right] \qquad (3\text{-}37)$$

where the separation $s = \delta$ / 1 nm, and $1 < s < 100$:

The materials listed are

fused silica = fs
hexane = h
gold = Au
poly(methylmethacrylate) = pm
polystyrene = ps
poly(vinyl chloride) = pvc
sapphire = sap ($\alpha$-$Al_2O_3$) = sp
vacuum = v (approximately the same as air)
water with no salt, or infinite Debye length = w
water with Debye length 100 nm = w100

In the tables I give the Hamaker constant near zero separation ($A_0$, at about 0.1 nm separation). I also give the maximum percent error for semi-infinite bodies and spheres from 1 to 100 nm. I emphasize that the fits should not be extrapolated outside this range, because sometimes they fit badly. That said, since the $k_3$ and $k_6$ are always negative, it means that at large gaps that the prediction is that the Hamaker constant approaches zero, which is correct.

For some systems, especially those where material B = vacuum, the VDW interactions are repulsive, giving a negative value of $A$.

For these systems I fit to $A(w) = -\exp\left[k_1 + k_2 \ln w + k_3 (\ln w)^2\right]$ and $A_{eff}(w) = -\exp\left[k_4 + k_5 \ln s + k_6 (\ln s)^2\right]$

For mixed systems A-m-B, where A ≠ B, one sometimes uses combining rules. One example is[65]

$$A_{AmB} \approx \sqrt{A_{AmA} A_{BmB}} \tag{3-38}$$

Sometimes this rule works well (e.g., ps-w-fs), while in other cases it does not (e.g., ps-w-Au).

**Example 3-5**. Evaluating a Hamaker constant at a given separation..

Find the Hamaker constant for a Au-h-Au (gold-hexane-gold) system at a separation of 1 and 10 nm, for two semi-infinite bodies and 2 spheres.

*answer: Note that the results below are all less than the values at a separation of 0.1 nm.*

*semi-infinite, h = 1 nm, A = 0.872 eV*

*semi-infinite, h = 10 nm, A = 0.773 eV*

*sphere-sphere, h = 1 nm,* $A_{eff}$ *= 0.834 eV*

*sphere-sphere, h = 10 nm,* $A_{eff}$ *= 0.638 eV*

---

[65] See Israelachvili, Jacob N. *Intermolecular and Surface Forces*, 2nd ed. Academic Press (New York) 1992. Section 11.9 provides several "combining relations", and it also mentions a few limitations, especially in dielectric media like water.

**Table 3-2.** Coefficients for Hamaker constants (Eq 3-36) for semi-infinite systems A / medium / B. For systems marked with a *, the range extends to only 30 nm, probably because of the low value. For systems with negative Hamaker constants, such as the Au-w1-v system, the modified equation mentioned in the text was used, which includes the extra negative sign.

| A-m-B | $A_0$ (eV) | $k_1$ | $k_2$ | $k_3$ | max % |
|---|---|---|---|---|---|
| Au-h-Au | 0.878 | -0.136 | 0.0732 | -0.0546 | 8.4 |
| Au-v-Au | 1.327 | 0.273 | 0.0596 | -0.056 | 3.2 |
| Au-w-Au | 0.913 | -0.059 | 0.037 | -0.049 | 5.5 |
| Au-w100-Au | 0.913 | -0.098 | 0.0681 | -0.0531 | 4.1 |
| Au-w100-ps | 0.160 | -1.708 | -0.0078 | -0.0664 | 17.6 |
| Au-w100-v | -0.312 | -1.106 | -0.0038 | -0.0632 | 15.6 |
| fs-h-fs | 0.0226 | -3.838 | -0.1366 | -0.0435 | 15.5 |
| fs-v-fs | 0.408 | -0.917 | 0.0269 | -0.0997 | 3.0 |
| fs-w-fs | 0.0429 | -3.173 | 0.0065 | -0.0408 | 11.1 |
| fs-w100-fs* | 0.0365 | -3.646 | -0.0081 | -0.1071 | 4.9 |
| fs-w100-ps* | 0.0539 | -3.157 | -0.0076 | -0.1099 | 6.6 |
| fs-w100-v | -0.060 | -2.616 | -0.0038 | -0.1037 | 8.3 |
| h-w100-h* | 0.019 | -4.848 | -0.0077 | -0.1566 | 16.9 |
| h-v-h | 0.254 | -1.343 | 0.009 | -0.1001 | 7.5 |
| h-w100-v | 0.0106 | | | | |
| pm-w100-pm* | 0.0496 | -3.267 | 0.0149 | -0.0837 | 6 |
| ps-h-ps | 0.0643 | -2.778 | -0.0632 | -0.0983 | 2.7 |
| ps-v-ps | 0.5433 | -0.645 | 0.0395 | -0.1051 | 3.3 |
| ps-w-ps | 0.0903 | -2.481 | 0.0437 | -0.072 | 13.2 |
| ps-w100-ps | 0.0852 | -2.626 | 0.025 | -0.1171 | 4.7 |
| ps-w100-pm | 0.0624 | -2.973 | -0.0077 | -0.0967 | 5.8 |
| ps-w100-v | -0.111 | -2.11 | 0.015 | -0.1123 | 5.3 |
| pvc-w100-pvc | 0.0646 | -2.918 | -0.0043 | -0.0905 | 6.3 |
| sp-w100-sp | 0.2883 | -1.291 | -0.0043 | -0.0994 | 5 |
| v-w100-v | 0.2557 | -1.4 | -0.0042 | -0.1054 | 4.1 |

**Table 3-3.** Coefficients for Hamaker constants (Eq 3-37) for two sphere systems A / medium / B. For systems marked with a *, the range extends to only 30 nm, probably because of the low value. For systems with negative Hamaker constants, such as the Au-w1-v system, the modified equation mentioned in the text was used, which includes the extra negative sign.

| A-m-B | $A_0$ (eV) | $k_4$ | $k_5$ | $k_6$ | max % |
|---|---|---|---|---|---|
| Au-h-Au | 0.873 | -0.181 | -0.032 | -0.0367 | 1.9 |
| Au-v-Au | 1.317 | 0.217 | -0.0444 | -0.041 | 3.8 |
| Au-w-Au | 0.908 | -0.145 | -0.033 | -0.037 | 2.2 |
| Au-w100-Au | 0.908 | -0.1437 | -0.0347 | -0.0347 | 3.0 |
| Au-w100-ps | 0.160 | -1.841 | -0.0231 | -0.0626 | 6.3 |
| Au-w100-v | -0.310 | -1.232 | 0.0153 | -0.0811 | 1.6 |
| fs-h-fs | 0.0220 | -3.942 | -0.2809 | -0.0283 | 14.6 |
| fs-v-fs | 0.400 | -1.056 | -0.1065 | -0.0942 | 11.1 |
| fs-w-fs | 0.0425 | -3.177 | -0.199 | 0.0052 | 6.1 |
| fs-w100-fs* | 0.0318 | -3.977 | -0.0262 | -0.0668 | 18.4 |
| fs-w100-ps* | 0.0488 | -3.322 | -0.1652 | -0.0783 | 1.9 |
| fs-w100-v | -0.060 | -2.616 | -0.0038 | -0.1037 | 8.3 |
| h-w100-h* | 0.0151 | -5.101 | -0.1472 | -0.0459 | 29.7 |
| h-v-h | 0.250 | -1.519 | -0.0926 | -0.0952 | 4.6 |
| h-w100-v | 0.0069 | | | | |
| pm-w100-pm* | 0.0448 | -3.503 | 0.0142 | -0.0936 | 17.2 |
| ps-h-ps | 0.0625 | -2.967 | -0.2214 | -0.0856 | 1.8 |
| ps-v-ps | 0.5327 | -0.78 | -0.114 | -0.0949 | 1.5 |
| ps-w-ps | 0.0887 | -2.55 | -0.1076 | -0.056 | 6.7 |
| ps-w100-ps | 0.0795 | -2.791 | -0.1686 | -0.0812 | 5.6 |
| ps-w100-pm | 0.0573 | -3.114 | -0.1767 | -0.0579 | 7.4 |
| ps-w100-v | -0.111 | -2.383 | 0.0142 | -0.1418 | 9.0 |
| pvc-w100-pvc | 0.0596 | -3.049 | -0.1554 | -0.0575 | 6.6 |
| sp-w100-sp | 0.2802 | -1.414 | -0.1965 | -0.0785 | 3.2 |
| v-w100-v | 0.248 | -1.574 | -0.1491 | -0.0866 | 4.9 |

## *Symbols*

$A$ = Hamaker constant [=] J. I express $A$ as a sum of high frequency (h) and zero frequency (z) term, with $A = A_h + A_z$ .

$c_1, c_2, c_3, c_4$ = fitted constants to express $A(\delta)$ [=] for particular units.

$\Delta\Phi_{VDW}$ = VDW attractive energy between 2 spheres [=] J. Sometimes it is listed simply as $\Phi_{VDW}$.
$\Delta V_{VDW}$ = VDW attractive energy between 2 atoms [=] J.
$\xi_n$ = eigenfrequencies for summing in Lifshitz theory.
$f_j$ = strength of a particular frequency mode for a material [=] none.
$g_j$ = dampening of a particular frequency for a material [=] none.
$h$ = Planck's constant = $6.626068\times10^{-34}$ J-s.
$\hbar = h/2\pi = 1.054571\times10^{-34}$ J-s.
$H$ = separation distance between two surfaces [=] m
$n_i$ = the density of atoms in particle i [=] #/m$^3$
$r$ = separation distance [=] m.
$\omega$ = oscillation rate [=] rad/s. For atoms, $\omega \approx 10^{16}$ rad/s.
$\omega_i$ = resonant frequency for a material [=] rad/s.
$v$ = energy/area [=] J/m$^2$.

***Practice Problems***

1. **VDW energy**. Two silica particles are separated by $\delta$ = 30 nm in water. They have diameters 400 nm and 600 nm. The Hamaker constant for the system is $A = 0.83\times10^{-20}$ J. Calculate the VDW energy of interaction using Eq 3-16 and Eq 3-17.
*answer: full equation gives $\Phi_{VDW} = -0.30\times10^{-20}$ J, approximation gives $\Phi_{VDW} = -0.55\times10^{-20}$ J. The approximate equation is useful for thinking about the physics, but the Hamaker equation is better in most cases.*

2. **Dielectric function (water)**. Plot the function $\varepsilon(\omega) / \varepsilon_0$ for water, for $\omega$ from 1.0 to $10^{17}$ Hz (which is rad/s). 1.0 Hz will give almost the same result 0 Hz. Use Eq 3-20. The data for water is taken from Russel *et al*. (1989). They give

$d$ = 74.8
$1/\tau = 1.05\text{x}10^{11}$ rad/s

| j | $\omega_j$ (rad/s) | $f_j$ (no units) | $g_j$ (no units) |
|---|---|---|---|
| 1 | $3.34\text{x}10^{13}$ | 1.46 | 0.725 |
| 2 | $1.11\text{x}10^{14}$ | 0.73 | 0.55 |
| 3 | $1.49\text{x}10^{14}$ | 0.151 | 0.30 |
| 4 | $3.23\text{x}10^{14}$ | 0.142 | 0.13 |
| 5 | $6.78\text{x}10^{14}$ | 0.077 | 0.13 |
| 6 | $1.33\text{x}10^{16}$ | 0.0393 | 0.062 |
| 7 | $1.62\text{x}10^{16}$ | 0.0567 | 0.088 |
| 8 | $1.84\text{x}10^{16}$ | 0.0923* | 0.135 |
| 9 | $2.10\text{x}10^{16}$ | 0.156 | 0.158 |
| 10 | $2.41\text{x}10^{16}$ | 0.152 | 0.200 |
| 11 | $3.00\text{x}10^{16}$ | 0.271 | 0.340 |

3. **Hamaker constant**. Calculate the Hamaker constant for polystyrene-water-polystyrene at $T$ = 298 K and $h$ = 0.1 nm, using Eq 3-32 and the data in Figure 3-3. Use a Debye length of 10 nm. *answer:* $1.4\times10^{-20}$ *J.*

4. **Hamaker constant**. Plot the Hamaker constant for polystyrene-water-polystyrene at $T$ = 298 K from $h$ = 0.1 to 100 nm, using Eq 3-32. Do this for Debye lengths of 200 nm (i.e., very large), 10 nm, and 1 nm.

5. **Eigenvalues**. Use Mathematica to verify the eigenvalues in Eq 3-8.

6. **Hamaker result for spheres**. Take the limits to show that in fact Eq 3-16 becomes either Eq 3-17 or Eq 3-18.

7 **London equation for VDW forces between atoms**. Calculate the VDW attraction between 2 argon atoms separated by 1.0 nm, at $T$ = 0. The precise temperature is not important, since the polarizability changes little with $T$, at least up to room temperature. Is the energy large or small compared with two H atoms (from Example 3-1 in the book)? The $\alpha / 4\pi\varepsilon_0 = 1.59\ \text{A}^3$ and the characteristic frequency is $\omega_0 = I_1 / \hbar$, where the first ionization energy $I_1$ = 1521 kJ/mol for Ar.

*answer: -2.88 J/mol or -4.79×10^-24 J. This value is almost 7× larger in magnitude than that for two H atoms.*

8. **Size of an atom**. Use the following result from quantum mechanics to calculate the "size" of a H atom:

$$a_0 = \frac{4\pi\varepsilon_0\hbar^2}{m_e e^2} = \frac{e^2}{8\pi\varepsilon_0\hbar\omega_0}$$

$$\omega_0 = \frac{m_e e^4}{32\pi^2\varepsilon_0^2\hbar^3}$$

answer: $a_0$ = 0.0529177 nm (Bohr radius)

9. **VDW energy between spheres**. For a non-retarded Hamaker const $A$ = 3.2 kT and $T$ = 295 K, plot $\Phi_{vdw}$ for gaps from 5 nm to 1000 nm (use log scale on $x$-axis) for two spheres with radius a = 100 nm. Use the full Hamaker expression.

10. **Power law from Hamaker constant**. The VDW energy can be expressed as $\Phi_{VDW} = -c / r^n = -cr^{-n}$, so that if I compare the non-retarded VDW energy at two distances $\delta_1$ and $\delta_2$, I find

$$\frac{\Phi_{VDW,2}}{\Phi_{VDW,1}} = \frac{-c\delta_2^{-n}}{-c\delta_1^{-n}} = \left(\frac{\delta_2}{\delta_1}\right)^{-n}$$

or

$$n = \frac{\ln\left(\Phi_{VDW,2} / \Phi_{VDW,1}\right)}{\ln\left(\delta_1 / \delta_2\right)}$$

If I take the limit as $\delta_1$ approaches $\delta_2$, at any given value of $\delta_1$, then I can evaluate the non-retarded VDW potential energy from Eq 3-16 at each distance and evaluate *n*. Use this technique to evaluate *n* for $\delta / a = 0.281$. While I could say $c = -Aa/12$ or $c = -16Aa^6/9$ based on the limits at small or large separations, the inconsistency of the coefficient for the near and far cases makes the present approach easier to follow, while giving some feel for the exponent *n*.
*answer: n = 1.80. Even at this somewhat small gap, the power of n is fairly different from 1.0.*

10. **Evaluating Hamaker constants**. Find the Hamaker constant for a ps-w100-fs (i.e., polystyrene, water with 100 nm Debye length, fused silica) at a separation of 1 and 10 nm, for two semi-infinite bodies and 2 spheres.
*answer: Note that the results below are all less than the values at a separation of 0.1 nm.*
*semi-infinite, h = 1 nm, A = 0.0426 eV*
*semi-infinite, h = 10 nm, A = 0.0233 eV*
*sphere-sphere, h = 1 nm,* $A_{eff}$ *= 0.0361 eV*
*sphere-sphere, h = 10 nm,* $A_{eff}$ *= 0.0163 eV*

# 4 Hydrodynamics

**References:** Kim & Karrila. *Microhydrodynamics*. Boston: Butterworth-Heinemann (1991)
Happel & Brenner. *Low Reynolds Number Hydrodynamics*. Boston: Kluwer (1991)
Van de Ven. *Colloidal Hydrodynamics*. New York: Academic Press (1989).

## *Stokes law*

Let's say that I'm at a blackboard in a classroom, and I pick up the chalk eraser and pound it against my hand. The chalk dust flies from the eraser and settles, but it does so slowly, maybe centimeters per second. On the other hand, when I pick up a piece of chalk that I would write with, and I drop the chalk, it falls very quickly to the ground. It is easy to see the difference, but how can I *quantify* the difference? Why does a piece of chalk dust that is 10 μm in size – that is, colloidal in size – behave differently from a piece of writing chalk that is 1 cm in size?

When colloidal particles move through a fluid, they encounter a resistance from the fluid, whether the fluid is air, water, or benzene. If a spherical colloidal particle is translating through a fluid at a velocity (**U**), the particle encounters a hydrodynamic drag force given by Stokes law[66]:

$$\boxed{\mathbf{F}_h = -6\pi\eta a\mathbf{U}} \qquad (4\text{-}1)$$

where $\eta$ is the viscosity of the fluid (units centipoise cP, Pa-s, or kg/m-s) and $a$ is the radius of the particle. If the particle is also rotating at an angular velocity ($\mathbf{\Omega}$, units rad/s), the particle encounters a hydrodynamic drag torque

$$\boxed{\mathbf{T}_h = -8\pi\eta a^3 \Omega} \tag{4-2}$$

---

[66] The full derivation of these hydrodynamic results is rather involved. The velocity fields can be derived using a stream function approach, nicely-described in the book by Happel & Brenner, and the derivation of Stokes law is explained there also. Another approach for deriving Stokes law, through the use of singularity solutions, is given in the book by Kim & Karrila.

Oftentimes in fluid mechanics, we find that the hydrodynamic force depends on $U^2$. What is different for a colloidal particle? Why is the resistance *linear* in the translational and angular velocity? In order to answer that question, and to identify other limitations to Eqs 4-1 and 4-2, I take us back to look at the starting equation of fluid mechanics, which is the Navier-Stokes equation.

**Example 4-1**. Hydrodynamic drag on a sphere.

A spherical particle with diameter 130 nm is moving in a fluid with a viscosity of 0.85 cP = 0.00085 Pa-s. a) Find the hydrodynamic drag force when the translation velocity is $\mathbf{U} = 2.1\times10^{-6}$ m/s in the $x$-direction, or $U$ = 2.1 μm/s. b) Find the hydrodynamic torque when the angular velocity is $\mathbf{\Omega}$ = 0.2 rad/s around the $x$-axis, or $\Omega$ = 0.2 rad/s.

*answers: a) -2.19×10⁻¹⁵ N, b) -1.17×10⁻²⁴ N-m. Both values are negative, meaning opposite in direction to the motion.*

The Navier-Stokes equation for fluid mechanics applies to Newtonian fluids, such as water, oils, or air. They consist of a vector equation that describes the momentum balance for the *fluid* (terms are labeled *A*-*F*):

$$\rho\left(\frac{\partial \mathbf{v}}{\partial t} + \mathbf{v}\cdot\nabla\mathbf{v}\right) = \eta\nabla^2\mathbf{v} - \nabla p + \rho\mathbf{g} + \rho_e\mathbf{E} \tag{4-3}$$

*A* *B* *C* *D* *E* *F*

and a scalar equation for continuity, which gives conservation of mass:

$$\frac{\partial \rho}{\partial t} + \nabla \cdot (\rho \mathbf{v}) = 0 \tag{4-4}$$

In these equations $\rho$ is the fluid density; **v** is the fluid velocity; $t$ is time; $p$ is pressure; **g** is the gravitational vector (9.81 m/s$^2$ in the vertical direction); $\rho_e$ is the volumetric charge density (C/m$^3$); **E** is the electric field (V/m); and $\nabla$ is the gradient operator (1/m).

Let's take a typical system to see which terms of the Navier-Stokes equation are important. I use "scaling concepts" to get very rough approximations of terms. Experience[67] reveals that these estimates are valid usually to within at least a factor of 10. The result of the following example will show that the fluid inertia terms, which are labeled $A$ and $B$ in Eq 4-3, are very small.

Consider this example: a 1.0 μm diameter particle accelerates from a speed $U = 0$ to $U = 1$ μm/s, and we examine the particle after $\tau = 1.0$ s. The movement occurs in water with $\eta = 1.0$ cP = 0.0010 kg/m-s and $\rho = 1000$ kg/m$^3$. Below are my estimates for each term $A$ through $C$ from Eq 4-3. Time derivatives will scale with $\tau$, and spatial derivatives will scale with the particle radius ($a$). First I'll consider the Navier-Stokes equations without gravity or an applied electric field, so that I ignore those terms for now.

$A = \rho \partial \mathbf{v} / \partial t \approx \rho U / \tau =$

(1000 kg/m$^3$) × (10$^{-6}$ m/s) / (1.0 s) = 0.001 N/m$^3$.

$B = \rho \mathbf{v} \cdot \nabla \mathbf{v} \approx \rho U \cdot U / a =$

(1000 kg/m$^3$) × (10$^{-6}$ m/s)$^2$ / (0.5×10$^{-6}$ m) = 0.001 N/m$^3$.

$C = \eta \nabla^2 \mathbf{v} \approx \eta U / a^2 = 4000$ N/m$^3$.

---

[67] I remember being in my first graduate fluid mechanics course, seeing scaling arguments. And I wondered, "How in the world do they know these work?" The answer: experience. Scaling arguments turn out to be surprisingly effective.

$D = \nabla p$ . This term is harder to estimate.

However, since we see that the value of the viscous force (*C*) is 6 orders of magnitude greater than the inertial terms (*A* and *B*), the pressure force (*D*) must be the one balancing the viscous force. These magnitudes are typical: Inertial forces are inconsequential for colloidal systems, unless the speed (*U*) becomes very large, or the time scale of observation is very short, comparable to the acceleration time of the colloidal particle. Typically, a colloidal particle accelerates to a steady speed according to a time scale

$$\tau_0 = \rho a^2 / \eta \tag{4-5}$$

For the example system I've given here, that time corresponds to less than 1 μs. As a result, when a force is applied to a colloidal particle, it reaches *terminal velocity* almost instantaneously. Inertia plays almost no role in colloidal systems.

How might I estimate the force on a sphere, to check Eq 4-1? I recognize that force (N) = stress (N/m$^2$) × area (m$^2$). By definition for a Newtonian fluid, the hydrodynamic shear stress ($\sigma_h$) is

$$\sigma_h = \eta \frac{\partial v_\theta}{\partial r} \tag{4-6}$$

As the fluid moves faster around the particle (higher $v_\theta$), the stress rises. As the viscosity of the fluid ($\eta$) is larger, the stress is larger. To estimate the magnitude of the stress, I use the same scaling arguments as before, and so $\sigma_h \approx \eta U / a$ . The area of a sphere is $4\pi a^2$, and so the hydrodynamic drag on a sphere moving at a speed (*U*), due to the shearing of the fluid over its surface, will be roughly

$$F_h \approx -4\pi a^2 \cdot \frac{\eta U}{a} = -4\pi \eta a U \tag{4-7}$$

I include the negative sign, since the hydrodynamic force acts in the opposite direction to the movement of the sphere. The pressure term also contributes to the force, with the same scaling. While this analysis is far from an exact solution, the scaling arguments have revealed the origins of Stokes law, given by Eq 4-1.

For the torque, I can check Eq 4-2 in a similar way. Since torque = force × lever arm, I must include the distance from the center of the sphere to where the drag occurs. This distance is the radius (*a*). The speed of the fluid at the surface of the sphere is roughly $v \approx a\Omega$, and Combining these estimates gives

$$T_h = -\left(\eta \frac{a\Omega}{a} \times 4\pi a^2\right)a = -4\pi\eta a^3 \Omega \qquad (4\text{-}8)$$

One measure often used to evaluate the importance of inertia for a particle in a fluid is the particle Reynolds number:

$$Re = \frac{\rho U a}{\eta} \qquad (4\text{-}9)$$

The Reynolds number gives the ratio of inertial forces divided by viscous forces. When $Re < 0.1$, inertia is relatively unimportant, and Stokes law can be used with great accuracy.

**Example 4-2**. Reynolds number for a sphere.

Consider Example 4-1, where we found a 130 nm diameter sphere had a speed of $2.1\times10^{-6}$ m/s, moving through a fluid with a viscosity of 0.85 cP = 0.00085 Pa-s and a density 995 kg/m$^3$. Evaluate the Reynolds number.

*answer: Re = 1.6×10*$^{-7}$*. Inertia would have a negligible importance in this system.*

Knowing Stokes law enables us to derive the settling speed of a particle. The gravitational force on a sphere is *mg*, but we use the buoyant mass of the particle, since if the particle has the same density as the fluid, it will not settle. The force on a sphere is then

$$F_g = \frac{4}{3}\pi a^3 \left(\rho_p - \rho\right) g \tag{4-10}$$

where $\rho_p$ is the density of the particle, and $\rho$ is the density of the fluid. On earth the value of $g$ = 9.81 m/s$^2$.

As we have already seen, when a sphere is settling at low Reynolds number, the particle almost immediately comes to terminal velocity. This means that the downward force of gravity is balanced by the upward drag force, such that the particle is no longer accelerating. Setting the sum of the forces equal to zero gives

$$F = 0 = F_h + F_g = -6\pi\eta a U + \frac{4}{3}\pi a^3 \left(\rho_p - \rho\right) g \tag{4-11}$$

Solving for $U$ yields Stokes law for settling:

$$\boxed{U = \frac{2a^2\left(\rho_p - \rho\right) g}{9\eta}} \tag{4-12}$$

**Example 4-3**. Settling speed.

In dusting a chalkboard eraser, small particles settle in the air. How fast will a 10 μm diameter sphere of calcite (specific gravity SG = 2.71, so density is 2710 kg/m$^3$) settle by gravity through air, which has a viscosity $1.8\times10^{-5}$ kg/m-s and a density of 1.2 kg/m$^3$? Does Stokes law still hold?

*answers: U = 8.2 mm/s. Since the Re = 0.0027, yes Stokes law will give an accurate value for the settling rate.*

Sometimes it is helpful to know the fluid velocity field around a spherical particle as it moves. The velocity is expressed with a radial component ($v_r$), in which the fluid moves toward or away from the center of the sphere (unit vector $\mathbf{i}_r$), and a tangential component ($v_\theta$) having a unit vector ($\mathbf{i}_\theta$). The flow field ($\mathbf{v}$) and pressure field ($p$) around an isolated sphere of radius ($a$), moving with velocity ($\mathbf{U} = U\mathbf{i}_z$) in a stagnant fluid, is given by[68]

$$\frac{\mathbf{v}}{U} = -\frac{\cos\theta}{2}\left[\left(\frac{a}{r}\right)^3 - 3\left(\frac{a}{r}\right)\right]\mathbf{i}_r - \frac{\sin\theta}{4}\left[\left(\frac{a}{r}\right)^3 + 3\left(\frac{a}{r}\right)\right]\mathbf{i}_\theta \qquad (4\text{-}13)$$

The pressure field is given by

$$p - p_\infty = \frac{3\eta U\cos\theta}{2a}\left(\frac{a}{r}\right)^2 \qquad (4\text{-}14)$$

Fig 4-1 gives a plot of the fluid velocity field.

In this section I estimated that $F_h = -\beta\pi\eta aU$, starting from the equations of motion. For spherical particles $\beta = -6$. What about other shapes? What is their value for $\beta$? The basic formulation is similar, but there are a few differences for other shapes.

### *Spheroids and doublets*

A spherical particle has three axes, all the same, so that $a_1 = a_2 = a_3 = a$. A spheroid also has three axes, but only two of them are the same; spheroids can be more rod-shaped (prolate) or more flattened (oblate), as shown in Fig 4-2. An ellipsoid can have all

---

[68] If the sphere is stationary and the far-field fluid has a velocity $\mathbf{u} = -v_\infty\mathbf{i}_x$, then

$$\frac{\mathbf{u}}{v_\infty} = \mathbf{i}_z - \frac{\mathbf{v}}{U} = \cos\theta\left[1 - \frac{3}{2}\left(\frac{a}{r}\right) + \frac{1}{2}\left(\frac{a}{r}\right)^3\right]\mathbf{i}_r - \sin\theta\left[1 - \frac{3}{4}\left(\frac{a}{r}\right) - \frac{1}{4}\left(\frac{a}{r}\right)^3\right]\mathbf{i}_\theta$$

three axes be different, such that a spheroid or a sphere is a simplified ellipsoid.

I will express forces and torques for spheroids as follows:[69]

$$\boxed{\mathbf{F}_H = -6\pi\eta(a_1 a_2 a_3)^{1/3}(\beta_1 U_1 \mathbf{i}_1 + \beta_2 U_2 \mathbf{i}_2 + \beta_3 U_3 \mathbf{i}_3)} \qquad (4\text{-}15)$$

$$\boxed{\mathbf{T}_H = -8\pi\eta(a_1 a_2 a_3)(\gamma_1 \Omega_1 \mathbf{i}_1 + \gamma_2 \Omega_2 \mathbf{i}_2 + \gamma_3 \Omega_3 \mathbf{i}_3)} \qquad (4\text{-}16)$$

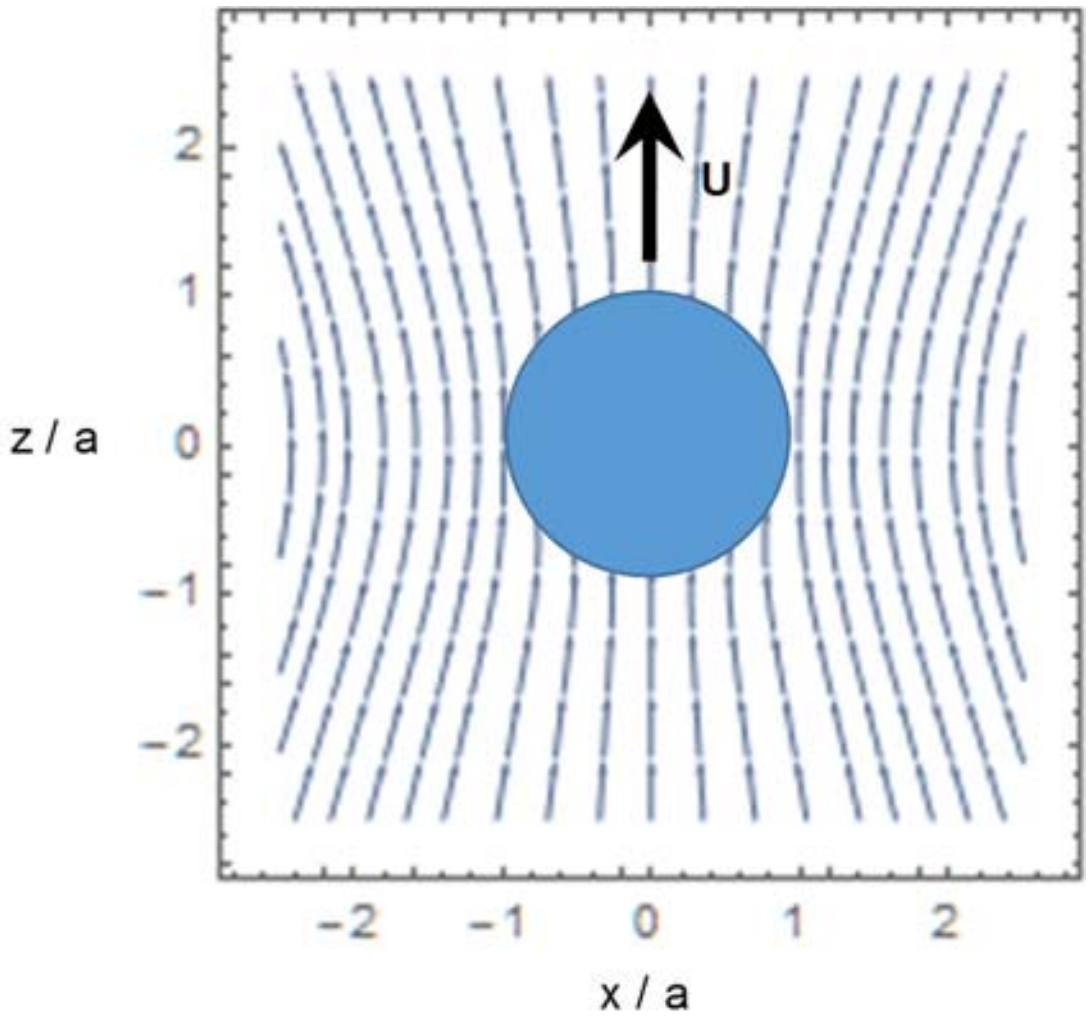

**Figure 4-1**. Streamlines of flow for a moving sphere. As the sphere translates at a speed ($U$), it must push fluid out from in front of it, and draw fluid in to the space where it vacated.

---

[69] Sometimes Lamb's formula for an infinite rod is used to estimate the hydrodynamic drag on a cylinder or spheroid moving parallel to its axis: $F = 4\pi\eta UL/(0.5 - \gamma - \ln \mathrm{Re}/4)$, where $\mathrm{Re} = \rho Ua/\eta$ and the Euler constant $\gamma \approx$ 0.57722. However, this equation works only when the boundaries are >500 diameters away! Furthermore, this equation includes *Re*, and thus density and inertia, which we might not expect to be important. Happel & Brenner show in their section 7-7 that for a cylinder of length ($L$) and radius ($a$) translating at a speed ($U$) through a stationary cylinder of radius ($b$), where $b >> a$, that $F = 4\pi\eta UL/(\ln b/a + 1)$. If the cylinder is translating perpendicular to its axis, $F = 4\pi\eta UL/(\ln b/a - 1)$, which has a larger drag.

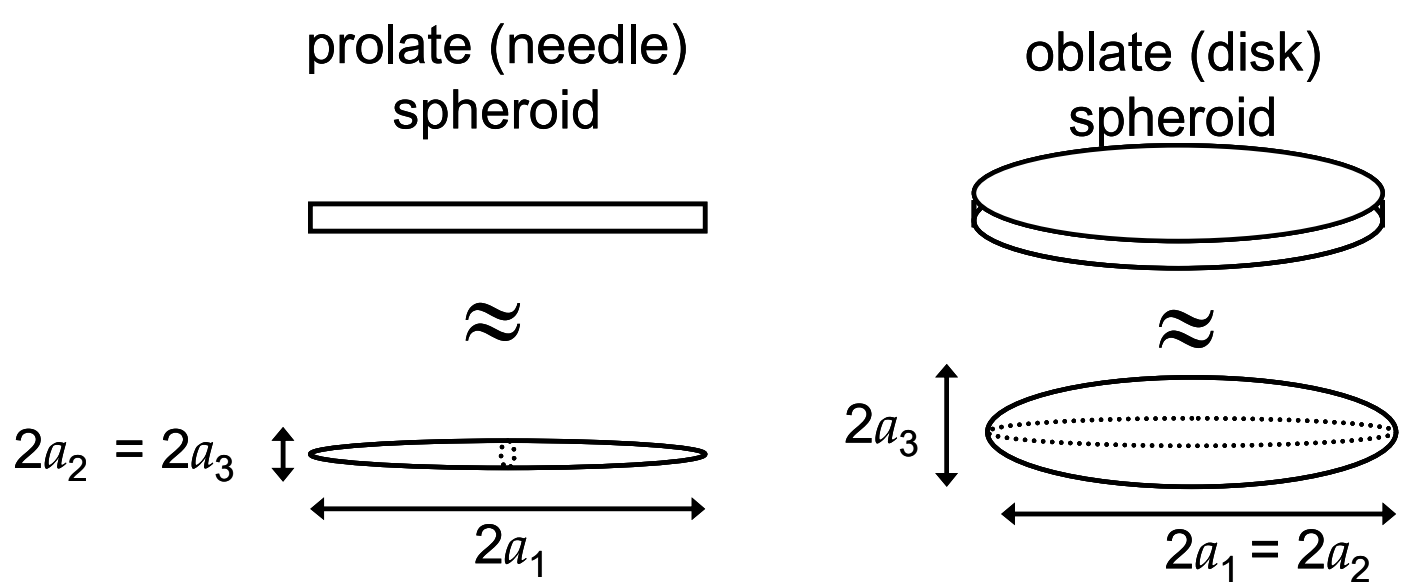

**Figure 4-2**. Oblate and prolate spheroids. This figure defines the axes, as used in this chapter.

**Table 4-1.** Single particle hydrodynamics. As an example, if a prolate spheroid is moving along the "1 axis" (i.e., $U_1$ finite), then it is moving along its long axis. Note that according to the table, when the axis ratio is 10 or less, that Stokes law for a sphere works to within a factor or 2, if $a$ is replaced by $(a_1 a_2 a_3)^{1/3}$. This is shown because the values of the $\beta$s are less than 2.

| Shape | $\beta_1$ | $\beta_2$ | $\gamma_1$ | $\gamma_2$ |
|---|---|---|---|---|
| prolate spheroid (rod-like)<br>1 = long semi-axis<br>2 = 3 = short semi-axis | | | | |
| $a_1 / a_2 = 1$ (sphere) | 1.000 | 1.000 | 1.000 | 1.000 |
| $a_1 / a_2 = 2$ | 0.956 | 1.094 | 0.807 | 1.505 |
| $a_1 / a_2 = 5$ | 1.044 | 1.387 | 0.706 | 4.641 |
| $a_1 / a_2 = 10$ | 1.229 | 1.769 | 0.680 | 13.37 |
| $a_1 / a_2 = 100$ | 2.993 | 4.954 | 0.667 | 694.7 |
| oblate spheroid (plate-like)<br>2 = 3 = long semi-axis<br>1 = short semi-axis | | | | |
| $a_1 / a_2 = 1$ (sphere) | 1.000 | 1.000 | 1.000 | 1.000 |
| $a_1 / a_2 = 0.5$ | 0.999 | 1.141 | 1.132 | 1.410 |
| $a_1 / a_2 = 0.2$ | 1.128 | 1.473 | 2.240 | 2.672 |
| $a_1 / a_2 = 0.1$ | 1.321 | 1.837 | 4.305 | 4.789 |
| $a_1 / a_2 = 0.01$ | 2.649 | 3.940 | 42.45 | 42.98 |

Doublets of two spheres, each sphere having a radius ($a$), have a higher resistance than does a sphere translating through a fluid. However, the resistance is not double. If a doublet is translating parallel to its long axis, the equivalent Stokes law becomes

$$F_{h,||} = -6\pi\eta a U_{||}(1.29) \tag{4-17}$$

If the doublet is translating perpendicular to its long axis, the equivalent Stokes law becomes

$$F_{h,\perp} = -6\pi\eta a U_{\perp}(1.41) \tag{4-18}$$

The torque required to spin a doublet around "like a propeller", that is, on an axis perpendicular to the line of centers, is

$$T_h = -29.92\pi\eta a^3 \Omega \tag{4-19}$$

For larger clusters of spheres, there are results available.[70]

**Example 4-3**. Force and torque on a prolate spheroid.

A prolate spheroid with $2a_2 = 2a_3 = 140$ nm and "length" $2a_1 = 1400$ nm is moving in a liquid with viscosity 0.80 cP. Find the force required to move the spheroid along its axis at $U_1 = 3.2$ μm/s, and perpendicular to its axis at $U_2 = 3.2$ μm/s. Find the torque required to spin along its long axis at $\Omega_1 = 2.8$ rad/s, and to spin "like a propeller" at $\Omega_2 = 2.8$ rad/s.

*answer:* $F_1 = +8.94\times10^{-15}$ *N,* $F_2 = +12.9\times10^{-15}$ *N,* $T_1 = +1.31\times10^{-22}$ *N-m,* $T_2 = +25.8\times10^{-22}$ *N-m. Note that since the hydrodynamic force is resistive, that a positive force must be applied to cause the particles to move.*

---

[70] Filippov, A.V. "Drag and Torque on Clusters of N Arbitrary Spheres at Low Reynolds Number," *J. Colloid Interface Sci.*, **229**, 184-195 (2000).

***Sphere-sphere interactions***

When two spheres are translating at a finite distance apart, they have sphere-sphere interactions, imparting a hydrodynamic force on each other. Hydrodynamics is a long-range force, generally decaying slowly with distance to the first power. The situation is shown in Figure 4-3 for when two spheres move perpendicular ( $\perp$) to their line of centers, or parallel (||) to their line of centers.

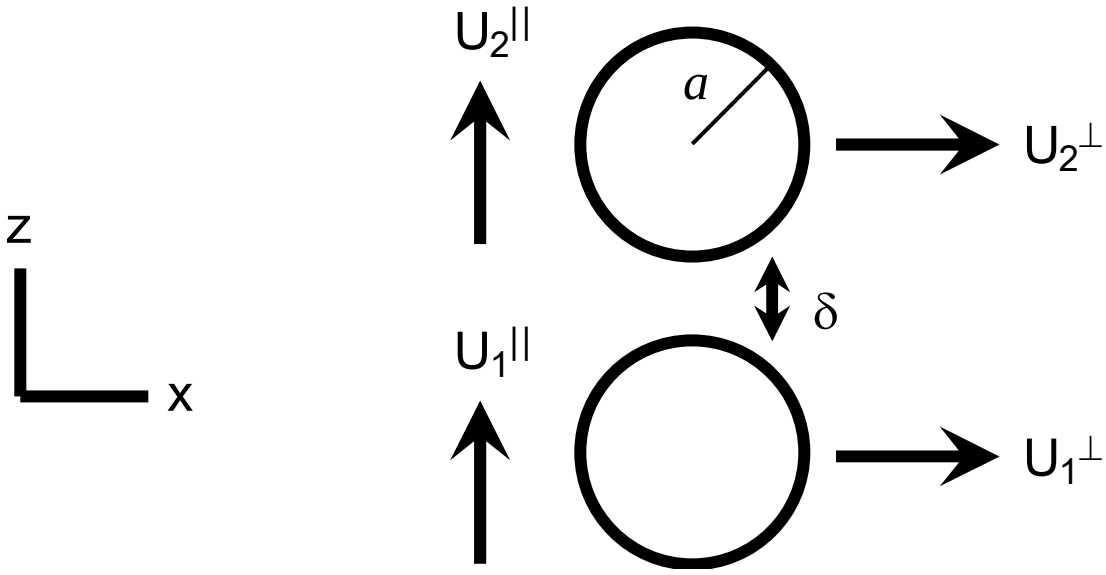

**Figure 4-3**. Hydrodynamic interactions for spherical particles of radius ($a$), at a closest distance ($\delta$) apart. When the spheres move in the *x* direction, they move perpendicular ( $\perp$) to their line of centers. When they move in the z direction, they move parallel (||) to their line of centers. The dimensionless separation is $\varepsilon = \delta / a$, and the dimensionless center-to-center separation is $(\delta + 2a)/a = \varepsilon + 2$.

The hydrodynamics of widely separated spheres (roughly, $\delta / a > 2$) is described well by the method of reflections, as given in Happel & Brenner and Kim & Karrila. Likewise, when two spheres are very near (roughly $\delta / a < 0.2$), they interact much more strongly, as described well by lubrication hydrodynamics, as given in Kim & Karrila. We can express the forces as given in Eq 4-20a-d. Notice that these expressions look similar to Stokes law Eq 4-1, except that they include the effects of both spheres. Also note the negative sign in each equation, since the hydrodynamic force is always opposite (resistive) to the movement of the sphere.

$$\frac{F_1^{\parallel}}{-6\pi\eta a} = X_{11}U_1^{\parallel} + X_{12}U_2^{\parallel}$$
$$\frac{F_2^{\parallel}}{-6\pi\eta a} = X_{12}U_1^{\parallel} + X_{11}U_2^{\parallel}$$
$$\frac{F_1^{\perp}}{-6\pi\eta a} = Y_{11}U_1^{\perp} + Y_{12}U_2^{\perp}$$
$$\frac{F_2^{\perp}}{-6\pi\eta a} = Y_{12}U_1^{\perp} + Y_{11}U_2^{\perp} \qquad \text{(4-20a-d)}$$

The functions[71] $X_{11}$ and $Y_{11}$ can be expressed as given in Eq 4-21-a-d. Note the negative signs in Eq 4-21-b and d. The first term in the numerators are for the lubrication results, while the second terms are for the reflection results. Figure 4-4 plots the results.

$$X_{11} = \frac{\frac{1}{\varepsilon^2}\left(\frac{1}{4\varepsilon} - \frac{9}{40}\ln\varepsilon + 0.9954\right) + \varepsilon^2\left(1 + \frac{9}{4}\left(\frac{1}{2+\varepsilon}\right)^2 + \frac{93}{16}\left(\frac{1}{2+\varepsilon}\right)^4\right)}{\frac{1}{\varepsilon^2} + \varepsilon^2} \qquad \text{(4-21)}$$

[71] These $X_{11}$ and $Y_{11}$ functions are taken from Kim & Karrila Ch 11. They define the functions slightly differently, and use a superscript "A". Their functions are given in the near field (lubrication theory for small separations) and far field (reflections calculations for large separations). I blend these into a single function here by taking

$$X_{11} = \frac{\left(\frac{1}{\varepsilon^n}\right)X_{11}^{near} + \left(\varepsilon^n\right)X_{11}^{far}}{\frac{1}{\varepsilon^n} + \varepsilon^n}$$

I use $n = 2$. The maximum error in the function is <3%, and it occurs near $\varepsilon = \delta / a \approx 0.60$ .

$$X_{12} = -\frac{\frac{1}{\varepsilon^2}\left(\frac{1}{4\varepsilon} - \frac{9}{40}\ln\varepsilon + 0.3502\right) + \varepsilon^2\left(\frac{3}{2}\left(\frac{1}{2+\varepsilon}\right) + \frac{19}{8}\left(\frac{1}{2+\varepsilon}\right)^3\right)}{\frac{1}{\varepsilon^2} + \varepsilon^2} \qquad (4\text{-}21b)$$

$$Y_{11} = \frac{\frac{1}{\varepsilon^2}\left(-\frac{1}{6}\ln\varepsilon + 0.9983\right) + \varepsilon^2\left(1 + \frac{9}{16}\left(\frac{1}{2+\varepsilon}\right)^2 + \frac{465}{256}\left(\frac{1}{2+\varepsilon}\right)^4\right)}{\frac{1}{\varepsilon^2} + \varepsilon^2} \qquad (4\text{-}21c)$$

$$Y_{12} = -\frac{\frac{1}{\varepsilon^2}\left(-\frac{1}{6}\ln\varepsilon + 0.2737\right) + \varepsilon^2\left(\frac{3}{4}\left(\frac{1}{2+\varepsilon}\right) + \frac{59}{64}\left(\frac{1}{2+\varepsilon}\right)^3\right)}{\frac{1}{\varepsilon^2} + \varepsilon^2} \qquad (4\text{-}21d)$$

For results on torques on two spheres, the same sections of Kim & Karrila – especially Chs 7 and 11 – have the solutions. When many spheres are interesting and one needs more precise hydrodynamic results, the state of the art is the Stokesian dynamics methods from John Brady.[72]

[72] There are many publications on this technique. One of the early reviews is Brady, John F. and Bossis, Georges, "Stokesian Dynamics." *Annual Review of Fluid Mechanics*, **20** , 111-157 (1988).

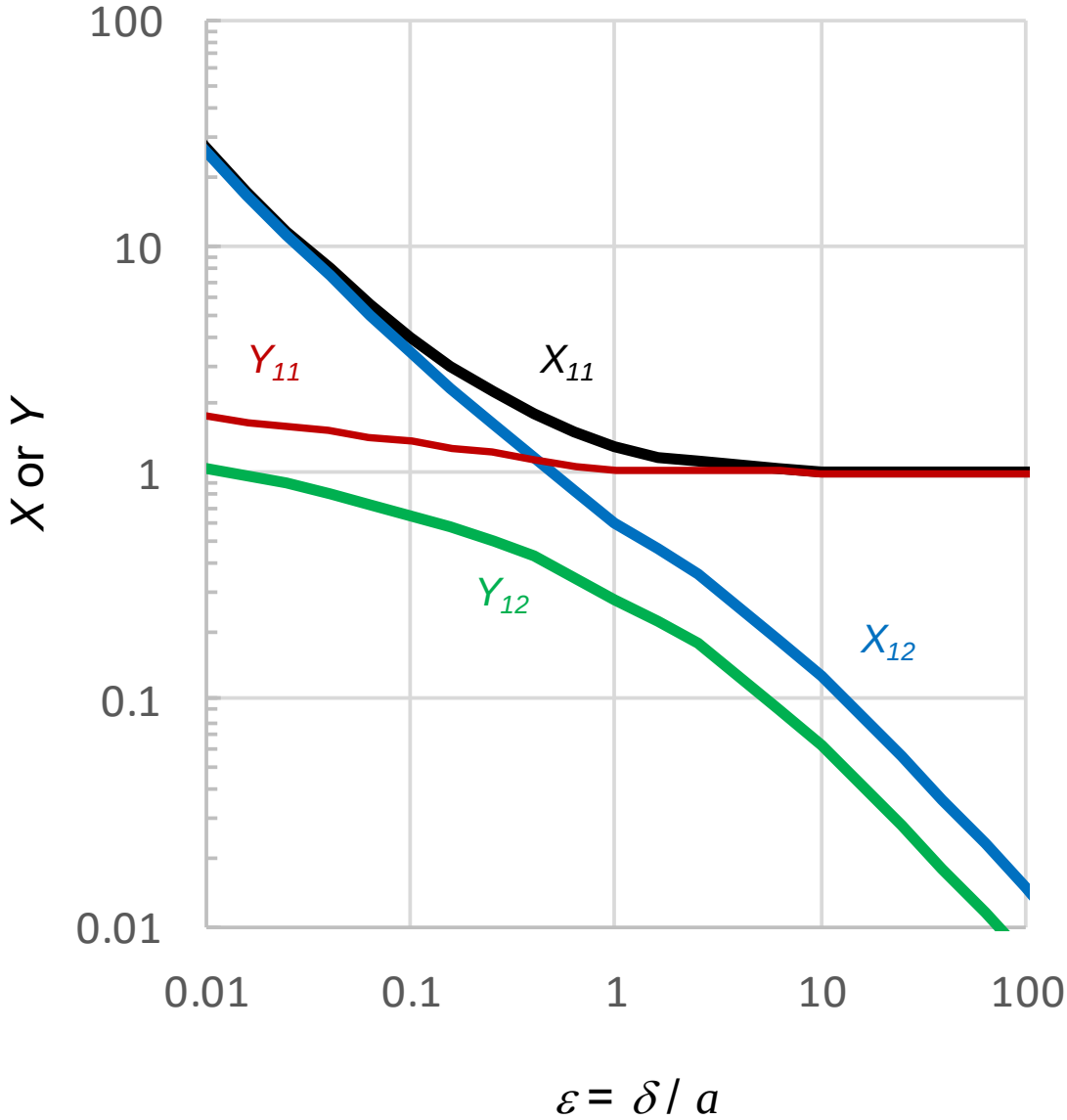

**Figure 4-4**. Hydrodynamic interactions for spherical particles of radius ($a$), at a closest distance ($\delta$) apart. When the spheres move in the *x* direction, they move perpendicular ($\perp$) to their line of centers. When they move in the z direction, they move parallel (||) to their line of centers. The dimensionless separation is $\varepsilon = \delta / a$, and the dimensionless center-to-center separation is $\left(\delta + 2a\right)/ a = \varepsilon + 2$ .

**Example 4-4**. Hydrodynamic calculations for a moving sphere.

Sphere 1 and sphere 2 have radius $a$ = 1.7 µm and are separated by 527 nm. Both are moving along their line of centers. Sphere 1 is moving at a speed +2.5 µm/s, while sphere 2 is moving at a speed +0.8 µm/s. They are moving through water which has a viscosity 0.85 cP. What is the force on sphere 1 and sphere 2? What if they were instead moving perpendicular to their line of centers at the same speed?

*answers: In this case* $\varepsilon = 0.31$, *so* $X_{11} = 2.061$, $X_{12} = -1.415$, $Y_{11} = 1.193$, $Y_{12} = -0.468$. *For the case along their line of centers,* $F_1^{\parallel} = -0.110$ *pN,* $F_2^{\parallel} = +0.051$ *pN,* $F_1^{\perp} = -0.071$ *pN,* $F_2^{\perp} = +0.0059$ *pN.*

### *Sphere-plate interactions*

Similar to sphere-sphere interactions, hydrodynamic results exist for sphere-plate interactions (Figure 4-5). The far field reflection results for a sphere of radius ($a$) separated by a distance ($\delta$) from a plate are[73]

$$
\begin{aligned}
6\pi\eta a U^{\parallel} &= \left[1 - \frac{9}{8}\left(\frac{a}{\delta}\right) + \frac{1}{2}\left(\frac{a}{\delta}\right)^3\right] F^{\parallel} \\
6\pi\eta a U^{\perp} &= \left[1 - \frac{9}{16}\left(\frac{a}{\delta}\right) + \frac{1}{8}\left(\frac{a}{\delta}\right)^3\right] F^{\perp}
\end{aligned}
\qquad (4\text{-}22)
$$

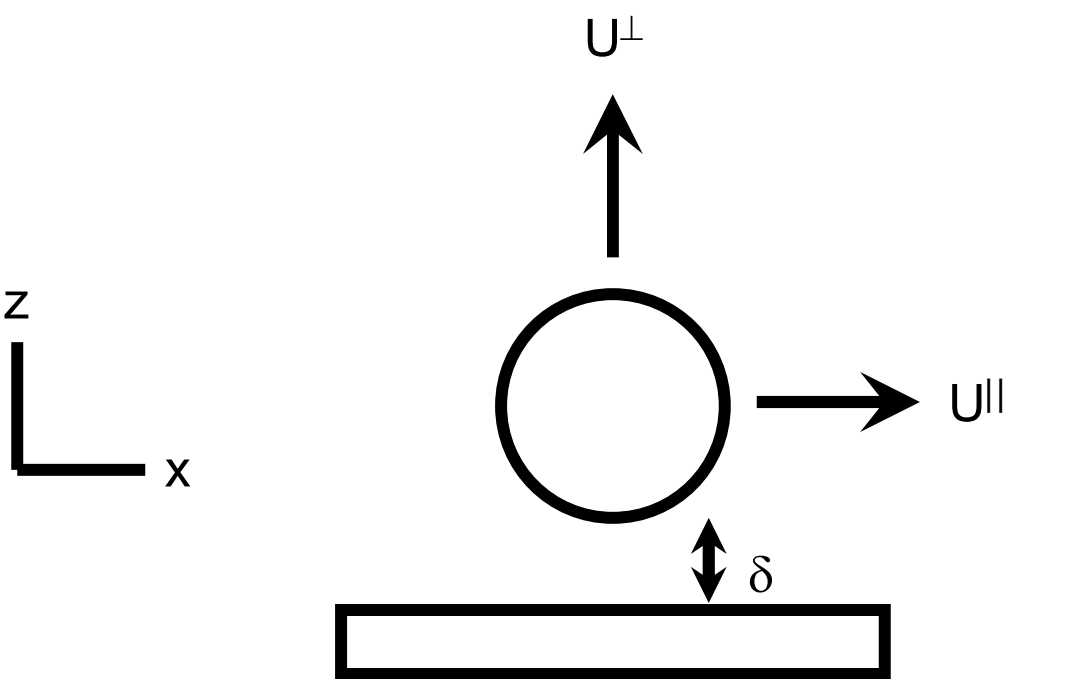

**Figure 4-5**. Hydrodynamic interactions for a spherical particle of radius ($a$), at a closest distance (δ) from a plate.

### *Hard-sphere rheology*

When a suspension of particles flows, the combination of fluid plus particles – often called a “complex fluid” – can be considered to have its own macroscopic viscosity. Einstein was the first to calculate the modified viscosity of a complex fluid ($\eta_{cf}$) of particles plus suspending medium, finding that the modified viscosity is higher than the viscosity of the pure fluid ($\eta$). He gave us the

[73] Kim & Karrila Ch 12 gives this result on p 316. They list results for angular velocities there as well.

Einstein viscosity correction, which provides that at a low volume fraction ($\phi$) of spheres,

$$\frac{\eta_{cf}}{\eta} = 1 + 2.5\phi \tag{4-23}$$

For more concentrated suspensions, a Krieger-Dougherty equation is used for hard spheres (i.e., spheres used without considering interparticle forces or electrical double layers). For small shear rates,

$$\frac{\eta_{cf}}{\eta} = \left(1 - \frac{\phi}{0.63}\right)^{-2} \tag{4-24}$$

At high shear rates this changes slightly to

$$\frac{\eta_{cf}}{\eta} = \left(1 - \frac{\phi}{0.71}\right)^{-2} \tag{4-25}$$

**Example 4-5**. Viscosity of a concentrated suspension.

A suspension of particles suspended in water is being pumped through a tube. The water has a viscosity of 1.0 cP. Find the low shear and high shear viscosities of the complex fluid, when a) $\phi = 0.01$, b) $\phi = 0.20$, c) $\phi = 0.50$.

*answers: a) low shear $\eta_{cf}$ = 1.033 (Einstein eq gives 1.025), high shear $\eta_{cf}$ = 1.029, b) low shear $\eta_{cf}$ = 2.15, high shear $\eta_{cf}$ = 1.94, c) b) low shear $\eta_{cf}$ = 23.5, high shear $\eta_{cf}$ = 11.4. Not until the volume fraction exceeds 10% or 20% does the change in viscosity become large.*

***Summary. Stokes law is widely applicable.***

Hydrodynamic results are used for more than just evaluating the force and torque on particles as they move. The results can also be used to evaluate the importance of Brownian motion, which is the random movement of particles in a fluid due to the impact of solvent molecules on the particles. I evaluate Brownian motion quantitatively in the next chapter.

### *Symbols*

$a$ = sphere radius [=] m. For ellipsoids, semi-axes $a_1$, $a_2$, $a_3$.
$\beta$ = coefficient for Stokes law; for spheres, $\beta = 6$.
$\mathbf{E}$ = electric field [=] V/m = J/C-m
$\mathbf{F}_h$ = hydrodynamic drag force [=] N
$\phi$ = volume fraction of particles [=] dimensionless
$\mathbf{g}$ = gravitational acceleration vector. $g = 9.81$ m/s$^2$ on Earth.
$\eta$ = viscosity of the fluid [=] kg/m-s = Pa-s; sometimes cP
$\eta_{cf}$ = viscosity of the complex fluid [=] kg/m-s = Pa-s
$\mathbf{i}_\theta$, $\mathbf{i}_r$ = unit vectors [=] dimensionless
$\mathbf{\Omega}$ = particle angular velocity [=] 1/s.
$p$ = pressure [=] N/m$^2$ = Pa
$p_\infty$ = ambient fluid pressure [=] N/m$^2$
$\theta$ = angle [=] rad
$\rho$ = fluid density [=] kg/m$^3$
$\rho_e$ = volumetric charge density [=] C/m$^3$
$\rho_p$ = particle density [=] kg/m$^3$
$r$ = radial distance [=] m
$Re$ = Reynolds number [=] dimensionless.
$\sigma_h$ = hydrodynamic stress [=] N/m$^2$
$t$ = time
$\tau_0$ = time scale for particle acceleration
$\mathbf{T}_h$ = hydrodynamic drag torque [=] N
$\mathbf{U}$ = particle velocity [=] m/s
$\mathbf{v}$ = fluid velocity [=] m/s; used as $v_r$ or $v_\theta$ or other.

***Practice Problems***

1. **Stokes law**. A spherical particle with diameter $2a$ = 13 μm is moving in a fluid with a viscosity of 0.65 cP = 0.00065 Pa-s. a) Find the hydrodynamic resistive force when the translation velocity is **U** = $4.7\times10^{-6}$ m/s in the *x*-direction, or $U$ = 4.7 μm/s. b) Find the hydrodynamic resistive torque when the angular velocity is **Ω** = 0.82 rad/s around the *x*-axis, or $\Omega$ = 0.82 rad/s.
*answers: a) $-3.74\times10^{-13}$ N = -0.374 pN, b) $-3.68\times10^{-18}$ N-m. Both values are negative, meaning opposite in direction to the motion.*

2. **Reynolds number**. Consider Problem 4-1. The fluid has a density of 990 kg/m$^3$. Calculate the particle Reynolds number, and evaluate whether Stokes law should still apply.
*answers: Re = $47\times10^{-6}$. Inertia would have a negligible importance in this system, and Stokes law will still apply.*

3. **Stokes law**. In dusting a chalkboard eraser, small particles settle in the air. How fast will a 20 μm diameter sphere of calcite (specific gravity SG = 2.71, so density is 2710 kg/m$^3$) settle by gravity through air, which has a viscosity $1.8\times10^{-5}$ kg/m-s and a density of 1.2 kg/m$^3$? Does Stokes law still hold? Evaluate the Reynolds number.
*answers: U = 32.8 mm/s. Since the Re = 0.022, yes Stokes law will give an accurate value for the settling rate.*

4. **Limit of Stokes law**. Chalk is usually cylindrical, but say that we have a piece of chalk that is spherical, with a diameter of 1.0 cm. How fast does Stokes law predict that the chalk will settle in air, under the same conditions as the previous problem? Does Stokes law still hold? Evaluate the Reynolds number.
*answers: U = 8200 m/s from Stokes law. Since the Re = $2.7\times10^{6}$, Stokes law does not even come close to being applicable.*

*When the particle Reynolds number $Re = \rho Ua / \eta$ is larger than 0.1, the first correction is known for Stokes law for a sphere. Proudman and Pearson found that*[74]

$$\mathbf{F} = -6\pi\eta a\mathbf{U}\left(1 + \frac{3}{8}Re + \frac{9}{40}Re^2 \ln Re + O\left(Re^2\right)\right)$$

*However, due to the large value of Re, this equation is not helpful either. Instead, when the particle Re has 1000 < Re < 350,000, the following empirical equation is within 13%:*[75]

$$U = 1.73\sqrt{\frac{2ga\left(\rho_p - \rho\right)}{\rho}}$$

*The terminal speed will by 25.7 m/s – which will take time to reach – resulting in Re = 8600, so that our equation holds. On a similar note, a settling leaf has a Re of roughly 1000 to 10,000, and so Stokes law is far from being applicable.*

5. **Ionic conductivity**. The "limiting conductivities" at $T = 0$ C are 26.5 $cm^2/\Omega$-mol for Na+ and 41.0 $cm^2/\Omega$-mol for Cl-. Thus, at 0 C, the infinite dilution conductivity for NaCl is (26.5 + 41.0) $cm^2/\Omega$-mol = 66.5 $cm^2/\Omega$-mol. As the temperature is increased, the viscosity goes down, and Stokes law predicts that the ions will have a higher speed for the same electrical force on them. Use Stokes law to predict the conductivity at 25 C. The viscosity of water at 0 C is 1.787 cP, while at 25 C it is 0.893 cP.[76]

---

[74] Proudman, Ian and Pearson, J.R.A. "Expansion at Small Reynolds numbers for the flow past a sphere and a circular cylinder." *J. Fluid Mech.*, **2**, 237-268 (1957).
[75] *Perry's Chemical Engineers' Handbook*, 6th ed, Ed. Don W. Green. McGraw-Hill (New York) 1984. Information about settling particles is on p 5-63 to 5-68. See also http://en.wikipedia.org/wiki/Settling.
[76] See http://www.engineeringtoolbox.com/water-dynamic-kinematic-viscosity-d_596.html.

*answer: 133.0 cm²/Ω-mol. The experimental value is 126.45 cm²/Ω-mol. The change in viscosity gives most of the change in conductivity.*

6. **Time to reach steady-state**. A spherical colloidal particles comes to terminal velocity very quickly in a fluid. This time can be estimated from $\tau = 2a^2 \rho_p / 9\eta$. A gold sphere ($\rho_p$ = 19,000 kg/m³) with $a$ = 230 nm is being accelerated by gravity through an oil with $\eta$ = 32 cP and $\rho$ = 850 kg/m³. Estimate the time to reach terminal velocity.
*answer: 7.0 nsec.*

7. **Spheroid settling**. A cylindrical gold rod (specific gravity SG = 19) with length 8.6 μm and diameter 86 nm (100:1 axis ratio) is settling through water (SG = 1.00) by gravity at $T$ = 293 K (so viscosity = 1.0 cP). a) Derive an expression for the settling rate. b) Calculate the settling rate when the rod settles with its axis parallel to gravity (vertical orientation) or c) perpendicular to gravity (horizontal orientation). Approximate the rod as a prolate spheroid in this problem, with length $2a_1$ and diameter $2a_2$.
*answers: settling rate is*

$$U = \frac{\left(a_1 a_2^2\right)^{2/3}\left(\rho_p - \rho\right)g}{3\eta\beta}$$

*b) For the parallel case, Table 4-1 gives $\beta$ = 2.993, so U = 0.78 μm/s. c) For the perpendicular case, Table 4-1 gives $\beta$ = 4.954, so U = 0.47 μm/s.*

8. **Suspension viscosity**. A suspension of particles suspended in water is being pumped through a tube. The water has a viscosity of 1.0 cP. Find the low shear and high shear viscosities of the complex fluid, when a) $\phi$ = 0.01, b) $\phi$ = 0.20, c) $\phi$ = 0.50.

*answers: a) low shear* $\eta_{cf}$ *= 1.033 cP, high shear* $\eta_{cf}$ *= 1.029 cP, b) low shear* $\eta_{cf}$ *= 2.15 cP, high shear* $\eta_{cf}$ *= 1.94 cP, c) b) low shear* $\eta_{cf}$ *= 23.5 cP, high shear* $\eta_{cf}$ *= 11.4 cP. Not until the volume fraction exceeds 10% or 20% does the change in viscosity become large. This small increase in viscosity with volume fraction is one reason why it's beneficial to sell polymer as particles, rather than as solution.*

9. **Stokes law for droplets**. A 2.4 μm diameter droplet having a viscosity of $\eta_{\text{droplet}} = 4.5$ cP is settling in a fluid with a viscosity $\eta =$ 2.0 cP at 0.28 μm/s. Find the force on the droplet. The result could be estimated with Stokes law, but a more exact result for when the sphere is not rigid is

$$F_H = -6\pi\eta a U\left(\frac{1+\dfrac{2\eta}{3\eta_{dropplet}}}{1+\dfrac{\eta}{\eta_{dropplet}}}\right)$$

*answer:* $F_h = -1.14\times10^{-14}$ *N. This value is about 10% less that for a rigid sphere, which occurs because part of the stress is relieved by internal circulation within the droplet. With droplets, experiments often show that they behave like rigid particles. This can occur if the surface has surfactants on it, which make free circulation at the surface difficult.*

10. **Stokes law for doublets**. A doublet particle is composed of two spheres each having a diameter $2a = 800$ nm. The doublet is moving along its long axis through a fluid with a viscosity of 0.65 cP = 0.00065 Pa-s at $U = 0.78\times10^{-6}$ m/s. Calculate the drag on the doublet using a) Eq 4-17 and b) Eq 4-15, with axes $2a_1 = 4a$ and $2a_2 = 2a_3 = 2a$.

*answers: a) -4.93×10^-15^ N, b) -4.60×10^-15^ N. The spheroid estimate is only 7% smaller than the precise doublet result. One reason for this is that the equivalent spheroid we have chosen has a different volume and surface area from the doublet.*

11. **Two sphere hydrodynamics**. Sphere 1 and sphere 2 have radius $a$ = 930 nm and are separated by 1142 nm. Both are moving along their line of centers. Sphere 1 is moving at a speed +1.27 μm/s, while sphere 2 is moving at a speed -0.48 μm/s. They are moving through water which has a viscosity 0.94 cP. What is the force on sphere 1 and sphere 2? What if they were instead moving perpendicular to their line of centers at the same speed?

*answers: In this case* $\varepsilon = 1.228$, *so* $X_{11} = 1.234$, $X_{12} = -0.527$, $Y_{11} = 1.038$, $Y_{12} = -0.2536$. *For the case along their line of centers,* $F_1^{\parallel} = -0.0300$ *pN,* $F_2^{\parallel} = +0.0208$ *pN,* $F_1^{\perp} = -0.0234$ *pN,* $F_2^{\perp} = +0.0135$ *pN.*

# 5 Brownian motion

**References:** Russel, W.B.; Saville, D.A.; Schowalter, *Colloidal Dispersions*. Cambridge University Press (New York) 1989. Ch 3.

Ninham, B.W. and Michael N. Barber, *Random and Restricted Walks : Theory and Applications*, vol. 10 in Mathematics and Its Applications Series. Gordon and Breach 1970.

Van de Ven, Theo G.M. *Colloidal Hydrodynamics*, Academic Press (New York) 1989. p 64.

Kittel, Charles. *Elementary Statistical Physics*, Dover (New York) 2004 (original 1958). See Section 31 for a nice solution to the Langevin equation.

## *Stokes-Einstein equation for translational Brownian motion*

In 1812 Robert Brown observed dancing pollen grains. When looking at many pollen grains in his microscope, Brown saw what appeared to be random movements of the pollen grains, and the movement never stopped. His observation might have been due to vibrations in his microscope, or a "life force" within the pollen grains, or poor eye-sight. But in this case, his observation was real. At the time, the explanation for the dancing pollen grains was missing.

Then in 1905 a clerk in the Swiss patent office proposed a model to explain the random motion.[77] As it turned out, 1905 was a fairly good year for this scientist, Dr. Einstein, and subsequent experiments by Jean Baptiste Perrin later revealed that Einstein's "theory of the Brownian movement" was correct. Perrin was even able to use the experiments to predict Avogadro's number to within 5%.[78]

Einstein examined the following problem. Say we have a set of particles settling by gravity toward a plate. As the particles accumulate near the plate, their concentration rises. If they also

---

[77] Albert Einstein, *Investigations on the Theory of the Brownian Movement*, Dover (New York) 1956.

[78] Perrin, Jean Baptise. *Les Atomes*, Nostrand 1916.

*diffuse* back into the solution, an equilibrium is eventually reached in which the settling due to gravity ($J_{grav}$) balances the back-diffusion due to the concentration gradient of particles ($J_{diff}$). For particles in a fluid, the number of particles per area settling due to gravity is

$$J_{grav} = Uc \tag{5-1}$$

where $U$ is the particle settling speed, $c$ is the local particle concentration, and the flux $J_{grav}$ has units of #/m$^2$-s. The diffusion is given by

$$J_{diff} = -D\frac{dc}{dx} \tag{5-2}$$

where $D$ is the diffusion coefficient of the particles – unknown here as yet – and $x$ is the distance upward from the plate.

At equilibrium these two effects balance, so that

$$J_{grav} + J_{diff} = 0 = Uc - D\frac{dc}{dx} \tag{5-3}$$

From the previous chapter, we know that $U = F/f$, where $f$ is the frictional drag on the particle, given by $f = 6\pi\eta a$ for a sphere, and $F$ is the external force on the sphere, in this case due to gravity. From physics we know that the gravitational force can be written as $F = -dV/dx$, where the potential energy ($V$) is given by

$$V = \frac{4}{3}\pi a^3\left(\rho_p - \rho\right)gx \tag{5-4}$$

This equation has accounted for the buoyant mass of the particle. Combining these results gives

$$fU = -\frac{dV}{dx} \tag{5-5}$$

$$-\frac{1}{f}\frac{dV}{dx}c - D\frac{dc}{dx} = 0 \tag{5-6}$$

From calculus we know that $d\ln c = dc/c$, and so we can rewrite the equation as

$$\frac{1}{f}\frac{dV}{dx} + D\frac{d\,ln\,c}{dx} = 0 \tag{5-7}$$

The general solution, for uniform *f* and *D*, is

$$c = B\exp\left(\frac{-V}{Df}\right) \tag{5-8}$$

In order to find the constant (*B*), we cannot use a boundary condition in the ordinary way. For example, one boundary condition might be that as $x \to \infty$, that $c \to 0$. But this condition is satisfied regardless of the value of *B*. Instead, we use a "normalization condition", meaning that what we know in fact is that the particle must be *somewhere*. If we start with a number (*N*) of particles, distributed over a surface area (*S*), then

$$S\int_0^\infty c\,dx = N \tag{5-9}$$

A substitution gives the final result for the constant *B*.

But here is where the critical insight from Einstein entered. Einstein knew Boltzmann's result, that

$$c = B\exp\left(\frac{-V}{kT}\right) \tag{5-10}$$

The combination of results from statistical mechanics and fluid mechanics is still a cutting edge technique in many fluid mechanics problems. The Stokes-Einstein equation is obtained independently from statistical mechanics, and it has a broad applicability. I used this equation in Ch 2 concerning the electrical double layer, where the energy was electrical rather than gravitational. Einstein recognized that the two equations must match, and so $Df = kT$. Equating these two pieces gives the Stokes-Einstein equation:

$$D = \frac{kT}{f} \tag{5-11}$$

For a sphere, we obtain the specific case of the Stokes-Einstein equation:

$$\boxed{D = \frac{kT}{6\pi\eta a}} \tag{5-12}$$

**Example 5-1**. Diffusion coefficient of a colloidal particle.

A spherical colloidal particle diameter 170 nm is in a fluid with a viscosity of 0.55 cP at $T = 285$ K. Calculate the diffusion coefficient for the sphere.

*answer:* $D = 4.46\times10^{-12}$ $m^2/s$. *This result holds when the sphere is isolated, not near to other particles.*

What is happening at the molecular scale? Solvent molecules are ever moving, with $kT/2$ of kinetic energy in each direction ($x$-$y$-

*z*). As they move, they collide into the colloidal particle on a time scale of roughly $10^{-13}$ s.[79] The particle accelerates with a time scale of $\tau = 9a^2\rho_p / 2\eta$. For a radius of $a$ = 100 nm, a silica particle (SG = 2.2) moving in water at 293 K (so viscosity = 1.0 cP = 0.0010 Pa-s) has a time scale closer to $10^{-7}$ s. Averaged over all the collisions, one might think that the collisions on one side of the particle would balance the collisions on the other side. Mostly they do, but not quite, and the slight asymmetry is where Brownian motion arises. That is, the slight random asymmetry in collisions is the *fluctuation* that causes movement. When a few more collisions happen, say, on the right side of the particle than on the left, the particle moves to the left. As the particle does so, it must push other solvent molecules out of the way, *dissipating* the momentum it has gained from previous solvent molecules to the solvent molecules that must be displaced. Such *fluctuation-dissipation* is ever happening in solution.

### *Brownian motion and diffusion*

The random motion of particles leads to a random walk of the individual particles. For an individual particle, it is not possible to predict how far the particle would move in a given time step. If we had $10^6$ particles, or if we wanted to know the average displacement the particle would traverse, we can estimate this with diffusion theory since we know we can calculate the diffusion coefficient ($D$) from the Stokes-Einstein relation.

Fick's first law of diffusion, giving the vector flux (**J**, in units of #/m$^2$-s) of a species, is

$$\mathbf{J} = -D\nabla n \tag{5-13}$$

---

[79] From a continuum perspective, I can also attain this time scale if I use $\tau = 9a^2\rho_f / 2\eta$, using a diameter of 0.3 nm for a water molecule.

In this equation $n$ is the concentration of the species, say in #/m$^3$. Putting this equation into the microscopic mass balance $\partial n / \partial t + \nabla \cdot \mathbf{J} = 0$ gives Fick's second law of diffusion:

$$\frac{\partial n}{\partial t} = D\nabla^2 n \tag{5-14}$$

This equation can be solved in 1-dimension to give the displacement ($\Delta x$) in a given time interval ($\Delta t$) as

$$\left\langle \Delta x^2 \right\rangle = 2D\Delta t \tag{5-15}$$

and in 3-D as

$$\left\langle \Delta x^2 + \Delta y^2 + \Delta z^2 \right\rangle = \left\langle \Delta \mathbf{x} \cdot \Delta \mathbf{x} \right\rangle = 6D\Delta t \tag{5-16}$$

**Example 5-2**. Average diffusional distance of a colloidal particle.

Spherical colloidal particles with diameter 170 nm are suspended in a fluid with a viscosity of 0.55 cP at $T$ = 285 K. We calculated the diffusion coefficient for the spheres in the previous example. Calculate the root-mean-square (rms) distance traveled for the spheres in 3-dimensions, after 47 s.

*answer:* $\sqrt{\left\langle \Delta x^2 + \Delta y^2 + \Delta z^2 \right\rangle} = \; = 35\ \mu m.$

### *Rotational Brownian motion*

Rotational Brownian motion works similarly to translational motion, except that if the solvent molecules glance off the larger particle, they impart a torque onto the sphere. Just as with translational Brownian motion, where when a slight asymmetry in the impulses from the solvent molecules cause the particle to translate, here when a slight asymmetry in torques arises, the particle rotates. The rotational diffusion coefficient ($D_r$) provides a quantitative measure of the rotation of the particles:

$$D_r = \frac{kT}{g} \tag{5-17}$$

For a sphere, we obtain the specific case of a rotational Stokes-Einstein relationship:

$$\boxed{D_r = \frac{kT}{8\pi\eta a^3}} \tag{5-18}$$

The units of $D_r$ are $s^{-1}$, which are different from the units for the translational diffusion coefficient ($m^2/s$).

**Example 5-3**. Rotational diffusion coefficient of a colloidal particle.

A spherical colloidal particle diameter 170 nm is in a fluid with a viscosity of 0.55 cP at $T = 285$ K. Calculate the diffusion coefficient for the sphere.

*answer:* $D_r = 463\ s^{-1}$.

For small times the angular movement of the particle is given by a similar result to that for translation:

$$\langle \Delta\theta^2 \rangle = 2D_r \Delta t \tag{5-19}$$

For spheres we know that $g = 8\pi\eta a^3$. Other particle shapes have different values for $g$, as we learned in Ch 4 on hydrodynamics. For doublets rotating along their “propeller” direction, $g = 29.92\pi\eta a^3$, where $a$ is the radius of each sphere in the doublet. For a spheroid, we can use the results from Table 4-1. A comparison of the rotational diffusion for a doublet and a spheroid with $a_2/a_1$ = 2 is given in homework 5-3.

**Example 5-4**. rms rotation of a colloidal particle.

Spherical colloidal particles with diameter 170 nm are suspended in a fluid with a viscosity of 0.55 cP at $T$ = 285 K. We calculated the rotational diffusion coefficient for the spheres in the previous example. Calculate the average root-mean-square (rms) rotation traversed for the spheres, after 0.0017 s.

*answer:* $\sqrt{\langle \Delta\theta^2 \rangle}$ = *1.25 rad.*

---

Sometimes it is important to know how much time is required for the orientation of a particle to "randomize". For example, a spheroid has a different translational diffusion coefficient along its two axes. Over time, however, the direction of the 1-axis of the spheroid changes. As a result, if we examine the motion along the *x*-axis in the laboratory, we can find an "average" diffusion coefficient. But how long should we wait before we know we have a good average? Milliseconds? Hours? A good estimate of the time to randomize the direction is given by a few multiples of the rotational diffusion time. That is, the time to randomize the orientation can be estimated by

$$\tau_{random} \approx 5 / D_r \tag{5-20}$$

For example, the sphere in Example 5-3 randomizes in roughly 5/463 $s^{-1}$ = 0.011 s.

***Brownian dynamics simulations***

Situations often arise where it is difficult to find an analytical solution to a diffusion problem. One approach is to conduct a Brownian dynamics simulation (BDS). You allow the species – say, the colloidal particles – to undergo Brownian motion and convective motion (e.g., due to gravity) step by step, in small time intervals (Figure 5-1). Time marches by, and the translational and rotational state of the particles are followed in a computer. By running the simulations hundreds or thousands of times, one can find averages of where the species diffuse. BDS account for diffusion and convective motion. At each time step, the change is accounted for incrementally.

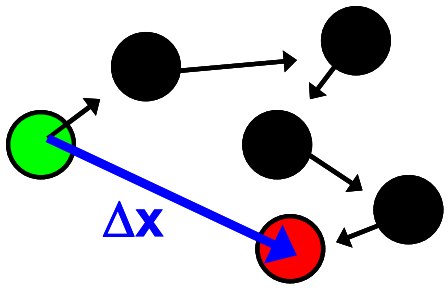

**Figure 5-1**. Brownian dynamics simulation. The particle starts at some position (green), marches randomly in time (black particles), and after some amount of time exists at some location (red).

In each step, the convective motion is

$$\Delta\mathbf{x}_{conv} = \mathbf{U}\Delta t \tag{5-21}$$

In order to find the diffusive or Brownian motion at each step, we draw from the "point source solution". The solution to Eq 5-14 is

$$\frac{n(\Delta x, \Delta t)}{N} = \frac{1}{\sqrt{4\pi D\Delta t}} \exp\left(\frac{-\Delta x^2}{4D\Delta t}\right) \tag{5-22}$$

Starting with $N$ particles concentrated at some single position at some time, if we look at any $\Delta x$ (in 1-D) and $\Delta t$ later, we can evaluate the concentration of particles. This equation has a Gaussian distribution with a variance $2D\Delta t$, and so we can find the probability of a concentration $n/N$ of a particle having a displacement at any $\Delta t$. The average "step" is $s = \sqrt{2D\Delta t}$ .

In order to use this result, I start with a time step $\Delta t$, choose a Gaussian random number,[80] which usually has a value from -3 to +3, and then evaluate the "step" by the random number times $s$. In the end, if I start from a position $\mathbf{x}(t)$, then after some time step $\Delta t$, a given particle will have moved to $\mathbf{x}(t + \Delta t)$ due to convection (velocity $\mathbf{U}$, in units m/s) and a diffusion ($\Delta\mathbf{x}$, chosen from a Gaussian random number). Table 5-1 gives the algorithm of a BDS.

**Table 5-1**. Algorithm for a Brownian dynamics simulation (BDS).

1. Choose the number ($N$) of particles you will simulate.
2. Choose the boundaries of the container the particles will move in, whether it is a box, or a sphere, or some other more complex shape.
3. Choose a starting position ($\mathbf{x}_0$) and angles ($\theta$ and $\phi$) for each particle.
4. Choose a time step ($\Delta t$).
5. Evaluate the step size for convection = $s = \sqrt{2D\Delta t}$ and angular motion in the $\theta$ direction $\sigma_\theta^2 = 2D_r\Delta t$ .
6. *Begin the march in time for each particle.* First evaluate the step in the $\phi$ direction $\sigma_\phi^2 = 2D_r\Delta t / \sin^2\theta$ . Note that this step depends on θ, and therefore cannot be chosen once for always at the beginning of the simulation.

[80] Some computer programs give a "uniform" random number, from 0 to 1, in which any number in that range can be chosen with equal probability. Let's call these $u(0,1)$ numbers. We want numbers distributed according to a bell curve, which are Gaussian distributed. That is, there is a higher probability that the number will be chosen near to the average ($\mu$) than to say a number one standard deviation ($\sigma$) away. Let's call $g(0,1)$ a Gaussian random number having the average = 0 and the standard deviation = 1. It turns out that you can convert two uniformly-distributed numbers ($u_1$ and $u_2$) to two Gaussian random numbers ($g_1$ and $g_2$) using the transformation

$$g_1 = \sqrt{-2\ln u_1}\cos 2\pi u_2, \quad g_2 = \sqrt{-2\ln u_1}\sin 2\pi u_2$$

7. Choose 5 Gaussian random numbers, $g_i(0,1)$, with a mean of 0 and a standard deviation of 1. Some particles lacking any symmetry might require a 6th random number to specify their movement.
8. Evaluate the translational movement

$$\mathbf{x}(t + \Delta t) = \mathbf{x}(t) + \mathbf{U}\Delta t + \Delta\mathbf{x} + \nabla_r \cdot \mathbf{D}\Delta t \qquad (5\text{-}23)$$

The value of $\Delta x$ for the time step is $sg_1$; the value of $\Delta y$ for the time step is $sg_2$; $\Delta z$ for the time step is $sg_3$. The convective velocity (**U**) is calculated using the known forces, and later in the book we will examine electrophoretic and other contributions to the velocity.

9. Evaluate the angular motions

$$\theta(t + \Delta t) = \theta(t) + \Omega \cdot \mathbf{i}_\theta \Delta t + \Delta\theta \qquad (5\text{-}24)$$

$$\phi(t + \Delta t) = \phi(t) + \Omega \cdot \mathbf{i}_\phi \Delta t + \Delta\phi \qquad (5\text{-}25)$$

in which the Brownian rotations $\Delta\theta = \sigma_\theta g_4$ and $\Delta\phi = \sigma_\phi g_5$.

10. As the simulation proceeds, account for constraints such as wall boundaries or the presence of other particles. If the simulation uses a very dilute or non-interacting sample of particles, these can sometimes be ignored.
11. Repeat starting with step 6, until the simulation reaches the desired time of action, or some other final condition is met.

---

In Eq 5-23 there is a term involving the gradient of the diffusion coefficient. Throughout this entire chapter I have considered that the diffusion coefficient is uniform. Sometimes it will not be, due to hydrodynamic resistances near boundaries, or due to different solution conditions. In these instances, there is a net drift of the particle, as given in the equation. Details are found elsewhere.[81]

### *Brownian space sampling*

As a particle undergoes translational Brownian motion in an unbounded space, the particle samples the space (Figure 5-2). Here, instead of measuring the average squared distance traversed as in Eqs 5-15 and 5-16, I give expressions for the total number ($n$) of equispaced regions (of volume $L^N$) sampled after a particle moves with a diffusion coefficient ($D$) for a time ($t$).[82] The results for the average number $<n>$ of regions sampled after a time ($t$) can be correlated as

$$\langle n \rangle = a\left(\frac{2Dt}{L^2}\right)^b \tag{5-26}$$

and the average time $<t>$ required to sample $n$ regions as

$$\left\langle \frac{2Dt}{L^2} \right\rangle = cn^d \tag{5-27}$$

**Table 5-2**. Brownian space sampling parameters for Eqs 5-26 and 5-27.

| **dimensions (*N*)** | ***a*** | ***b*** | ***c*** | ***d*** |
|---|---|---|---|---|
| 1 | 1.67 | 0.497 | 0.412 | 2.035 |
| 2 | 2.40 | 0.851 | 0.222 | 1.29 |
| 3 | 4.63 | 0.970 | 0.188 | 1.047 |
| 4 | 9.01 | 0.979 | 0.0914 | 1.043 |

[81] Ermak, Donald L.; McCammon, J.A. "Brownian dynamics with hydrodynamic interactions." *J. Chem. Phys.*, **69** (1978). The book by Russel, Saville, Schowalter has an explanation in their Ch 3 that many readers will find helpful.

[82] Velegol, Darrell. "Brownian sampling in an unbounded space." *Journal of Colloid and Interface Science*, **274**, 334-336 (2004).

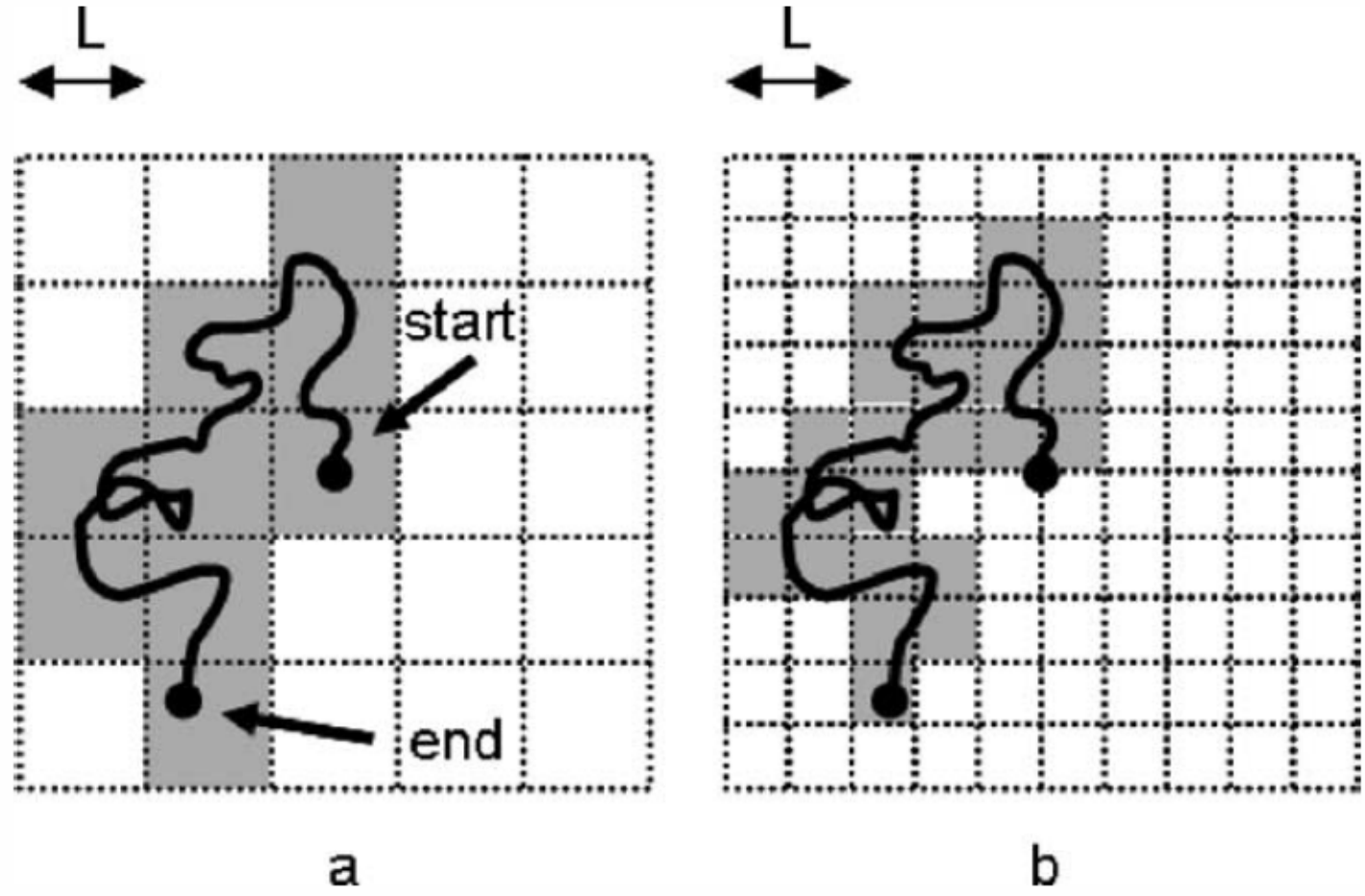

**Figure 5-2**. Particle undergoing translational Brownian motion over an unbounded 2-D space. The dotted grid lines guide the eye in defining regions, although the space itself is unbounded. As the particle traverses through a square, that square "turns gray". With time, we count the regions that turn gray. a) Each region has an area $L \times L$. Nine regions (total area $9L^2$) are sampled. b) Each region now has an area $(L / 2)^2$. Both (a) and (b) have the same trajectory, but in (b), $n = 25$ regions are sampled.

***Summary. The Stokes-Einstein equation works well.***

The Stokes-Einstein equation gives remarkably good results for the diffusion coefficient of "particles" from microns in size, and even down to the size of an ion. The rotational diffusion coefficient is also given by a Stokes-Einstein result.

Brownian dynamics simulations (BDS) give a flexible, simple, and powerful method of solving convective-diffusion problems. Their disadvantage is that for a large number of particles, the computational expense can become high, but for cases where the number of particles examined can be limited, BDS represent a powerful method of solution.

### *Symbols*

$a$ = sphere radius [=] m
$D$ = translational diffusion coefficient [=] $m^2/s$
$D_r$ = rotational diffusion coefficient [=] $s^{-1}$
$f$ = fluid friction coefficient for a particle = $6\pi\eta a$ for a sphere
$g$ = rotational friction coefficient = $8\pi\eta a^3$ for a sphere
$g_1$, $g_2$ = Gaussian random numbers
$\eta$ = viscosity [=] Pa-s
$k$ = Boltzmann constant ($1.38\times10^{-23}$ J/K)
$L$ = diffusion height [=] m
$N$ = dimensions (e.g., 3 for 3 dimensions)
$\Omega$ = particle angular velocity [=] 1/s.
$\phi$ = volume fraction of particles [=] dimensionless
$\rho_p$ = particle density [=] $kg/m^3$
$\sqrt{\langle \Delta\theta^2 \rangle}$ = root-mean-square angle traversed [=] rad
$T$ = absolute temperature [=] K
$t$ = time [=] s
$\tau_0$ = time scale for particle acceleration
$u_1$, $u_2$ = uniform random numbers between 0 and 1.
$U$ = particle velocity [=] m/s
$\langle z \rangle$ = ensemble average of all $z$

### *Practice Problems*

1. **Diffusion coefficient**. A spherical colloidal particle with diameter 1330 nm is in water at $T$ = 293 K. Calculate the translational and rotational diffusion coefficients for the sphere.
*answers:* $D = 3.23\times10^{-13}$ $m^2/s$, $D_r = 0.547$ $s^{-1}$.

2. **Diffusional distance traversed**. For the previous problem, we calculated the diffusion coefficients for the spheres. Calculate the average root-mean-square (rms) distance traveled for the spheres in 2-dimensions, after 0.72 s. In 1-D we used $2D\Delta t$, while in 3-D we used $6D\Delta t$. For 2-D we use $4D\Delta t$. Also calculate the rotational angle traversed.

*answers:* $\sqrt{\langle \Delta x^2 + \Delta y^2 \rangle} =$ *= 0.96* $\mu m$, $\sqrt{\langle \Delta\theta^2 \rangle}$ *= 0.89 rad.*

3. **Diffusion of a doublet compared with a spheroid**. We have a doublet (two spheres connected together) in which each sphere has a diameter 700 nm. Evaluate the following seven diffusion coefficients: translation along the long axis and along the short axis, rotation about the long axis and the short axis, and evaluate these four diffusion coefficients by using results for a doublet, and approximating with results for a spheroid with $2a_1 = 1400$ nm (i.e., two spheres, each 700 nm) and $2a_2 = 2a_3 = 700$ nm. The particles reside in water at 293 K.

*answers*: $D_{1,doublet} = 4.75\times10^{-13}$ $m^2/s$, $D_{2,doublet} = 4.35\times10^{-13}$ $m^2/s$, $D_{1,spheroid} = 5.09\times10^{-13}$ $m^2/s$, $D_{2,spheroid} = 4.47\times10^{-13}$ $m^2/s$, $D_{r1,spheroid} = 2.32$ $s^{-1}$, $D_{r2,doublet} = 1.00$ $s^{-1}$, $D_{r2,spheroid} = 1.25$ $s^{-1}$.

4. **Orientation randomization time**. Estimate how long does it take the sphere in Problem 5-1 to randomize.

*answer: 9.14 s.*

5. **Diffusion height**. When particles exist in a suspension, they settle downward due to gravity. But as they concentrate at the lower plate, there arises a net upward movement by diffusion, since there is a concentration gradient of particles. The final distribution of heights is given by a Boltzmann distribution, as given in Eq 5-10. In this problem we want to estimate that diffusion height.

The energy of a particle is given by

$$V(x) = mgx = \frac{4}{3}\pi a^3 (\rho_p - \rho_f) gx$$

Since $p(x) = c_1 \exp(-V(x)/kT)$ represents the probability distribution of particle concentration, we have

$$p(x) = c_1 \exp\left( -\frac{\frac{4}{3}\pi a^3 (\rho_p - \rho_f) gx}{kT} \right) = c_1 \exp\left( -\frac{x}{L} \right)$$

As a result we have a "diffusion height"

$$\boxed{L = \frac{3kT}{4\pi a^3 (\rho_p - \rho_f) g}}$$

What is the diffusion height for 150 nm diameter polystyrene (SG = 1.055) particles in water at 1g gravity at $T$ = 300 K?

*answer: 4.3 mm. That is, you will find particles at a height 4.3 mm with a probability $e^{-1}$ = 36.8% as much as at 0.0 mm. Small particles have a large diffusion height, especially when they are almost neutrally buoyant like polystyrene.*

6. **Brownian dynamics simulation**. Solve problem 5-5 using a Brownian dynamics simulation. You might not get exactly 4.3 mm, but you should get a result that is fairly close.

7. **Point source solution**. Show that Eq 5-22 (point source solution) satisfies Eq 5-14 (diffusion equation). One way to do this is to put the left hand side (LHS) and right hand side (RHS) into Mathematica, and to subtract them symbolically. When the Simplify command is used, the difference will be 0. Note: Eq 5-22 also satisfies a boundary condition and initial condition, that at $\Delta t = 0$, the concentration of particles at $\Delta x = 0$ is infinite, but such that the total number of particles there is $N$.

8. **Diffusional coverage**. A 35 nm diameter spherical particle is undergoing Brownian motion in an unbounded 3-D space. The fluid has a viscosity of 0.80 cP and T = 30 C. How many cubic regions of volume $(10\ \mu m)^3$ will the particle samples in 1 hour on average? *answer:* $D = 15.85\times10^{-12}\ m^2/s$, *and* $L = 10\times10^{-6}\ m$. *Since* $a = 4.63$ *and* $b = 0.970$ *for this situation, after* $t = 3600\ s$, $\langle n \rangle = 4300$.

# 6 Stability

**References:** Russel, W.B.; Saville, D.A.; Schowalter, *Colloidal Dispersions*. Cambridge University Press (New York) 1989. See Ch 08.
Hiemenz, Paul C. & Rajagopalan, Raj. *Principles of Colloid and Surface Chemistry*, 3rd ed, CRC Press (Boca Raton FL) 1997. See Ch 13.
Israelachvili, Jacob N. *Intermolecular and Surface Forces*, 2nd ed. Academic Press (New York) 1992. Ch 12.

## *Diffusion-limited aggregation*

As particles move through solution, either by diffusion or convection, they collide. If the particles have strong interparticle repulsive forces, they are stable. If the interparticle forces are not repulsive, the collisions result in aggregates or clusters.

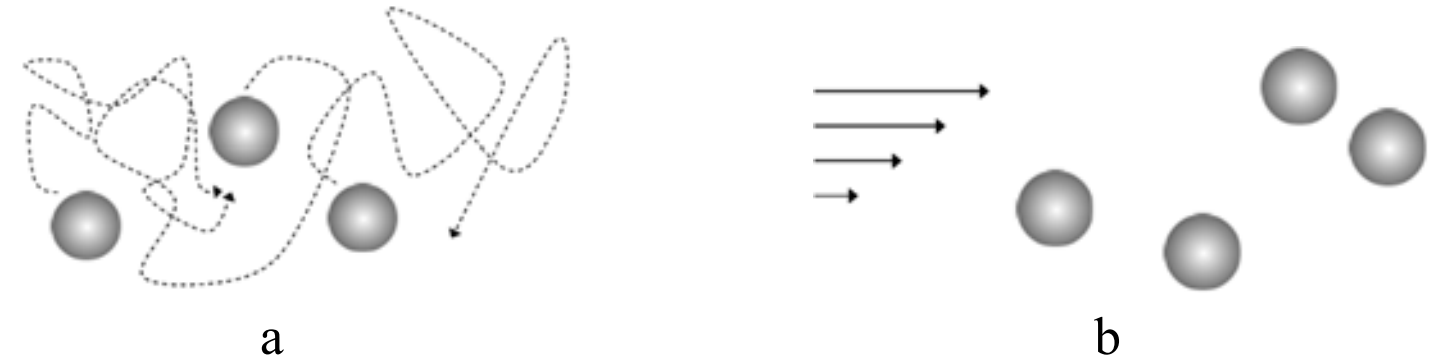

**Figure 6-1**. Aggregation of colloidal particles. a) Diffusion-limited aggregation, in which particles undergo random Brownian motion, and adhere when they collide. b) Shear-induced aggregation, in which particles traveling on "faster flowlines" collide with particles traveling on slower flow lines. As shown, the two particles in the lower left will collide, as will the two particles in the upper right of (b).

In this chapter I examine how quickly the particles will collide. Let's first examine rapid Brownian aggregation, sometimes called diffusion-limited aggregation. This case makes the assumption that every time particles collide, they adhere to each other. The system starts with a concentration of single spherical particles ($n_0$, in #/m$^3$), and with time, the number of singlets diminishes as the singlets aggregate into doublets, triplets, and high order aggregates.

First I consider collisions with a fixed test particle (black particle in Figure 6-2). As a gray particle undergoes random Brownian motion, it might collide with the black particle, meaning that the center of the gray particle is a distance $2a$ from the center of the black test particle. When the collision occurs, we model that the gray particle no longer exists – it is consumed. As a result, the concentration of singlets at $r = 2a$ is given by

$$n(r = 2a) = 0 \tag{6-1}$$

This is our first boundary condition.

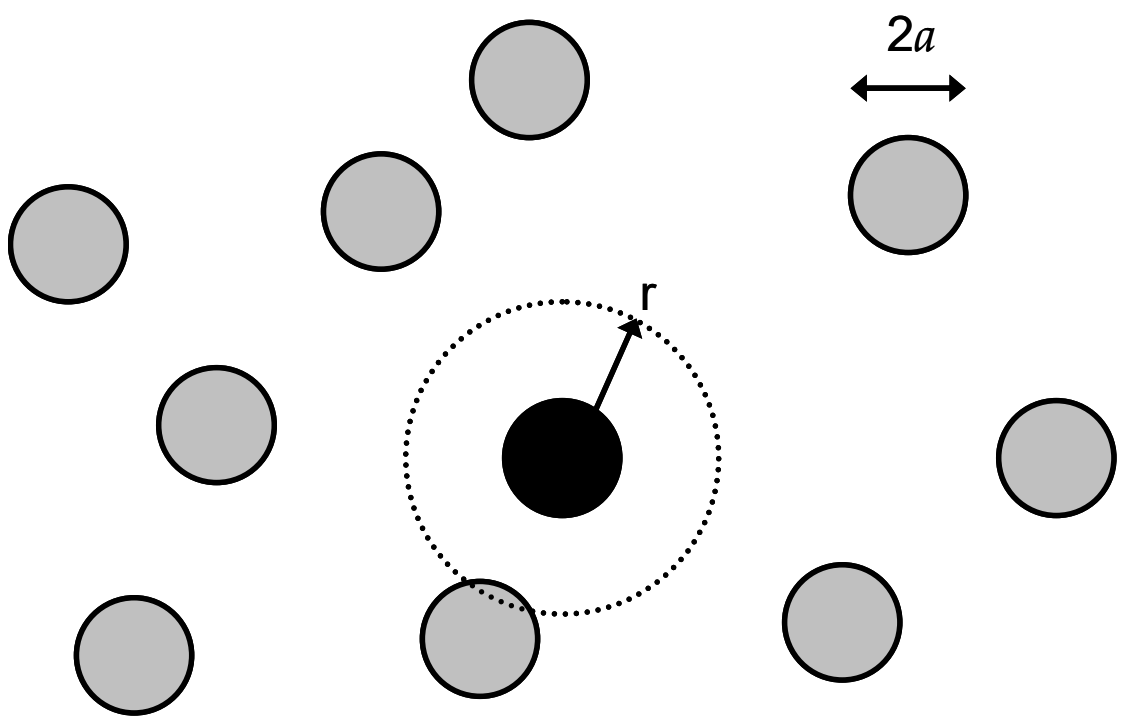

**Figure 6-2**. Collisions with a test particle (black). The spherical particles have a radius ($a$). Around the test particle, we draw an imaginary boundary with radius (r).

Over a small time interval, the concentration of gray singlets far away from the black test particle remains unchanged. This is a pseudo-steady approximation, since in fact the concentration of particles diminishes from $n_0$ with time. Let's make this pseudo-steady approximation, so that we'll say that the concentration of particles in the bulk at any time gives our second boundary condition:

$$n(r \to \infty) \to n_\infty(t) \tag{6-2}$$

where we have an initial condition

$$n_\infty(t = 0) = n_0 \tag{6-3}$$

Because the gray singlets are disappearing as they collide with the fixed black test particle, there is a continuous concentration gradient of gray particles – low near the black particle, and higher in the bulk. As a result, the gray particles diffuse toward the black particle. The flux ($j$, in #/m$^2$-s) of this diffusion is given by Fick's first law:

$$j = -D\frac{dn}{dr} \tag{6-4}$$

Since gray particles are not being created nor destroyed except when they collide with the black particle, the number of particles crossing an imaginary boundary of radius ($r$) is the same as the number crossing a boundary at 1.3$r$ or 1.8$r$ or 7.2$r$ or any other value. This rate of particle transfer ($J_1$, in #/s, the number of particles moving toward the fixed test particle) is then simply the area times the flux:

$$J_1 = jA = \left(-D\frac{dn}{dr}\right)\left(4\pi r^2\right) \tag{6-5}$$

Treating $J_1$ as constant, in a pseudo-steady manner, and solving Eq 6-5 with the boundary conditions Eqs 6-1 and 6-2 gives

$$J_1 = -8\pi a D n_\infty \tag{6-6}$$

Until this point I have considered the black particle as fixed in space. Of course, the black particle is diffusing also. The net effect of the black particle moving, which I'll show next, is that $D$ is replaced with $2D$.

We want to find changes in the distance between the black particle and a given gray particle. I'll call these "1" and "2" for now, so that we want to know $\Delta(\mathbf{x}_2 - \mathbf{x}_1)$. We know how each particle diffuses separately:

$$\langle \Delta\mathbf{x}_1 \cdot \Delta\mathbf{x}_1 \rangle = \langle \Delta\mathbf{x}_2 \cdot \Delta\mathbf{x}_2 \rangle = 6D\Delta t \tag{6-7}$$

Because the motions of particle 1 and particle 2 are random,

$$\langle \Delta\mathbf{x}_1 \cdot \Delta\mathbf{x}_2 \rangle = 0 \tag{6-8}$$

The average goes to zero since when particle 1 moves right, particle 2 is as likely to go to the right as to the left. The multiplied movements thus cancel out.

Now I examine the following difference:

$$\begin{aligned} &\langle \Delta(\mathbf{x}_2 - \Delta\mathbf{x}_1) \cdot (\mathbf{x}_2 - \Delta\mathbf{x}_1) \rangle = \\ &\langle \Delta\mathbf{x}_2 \cdot \Delta\mathbf{x}_2 \rangle - 2\langle \Delta\mathbf{x}_2 \cdot \Delta\mathbf{x}_2 \rangle + \langle \Delta\mathbf{x}_1 \cdot \Delta\mathbf{x}_1 \rangle = \\ &6D\Delta t - 0 + 6D\Delta t = 12D\Delta t \end{aligned} \tag{6-9}$$

When both particles can move by diffusion, the difference in their positions changes twice as fast as when one of the particles is fixed, and so I replace $D$ with $2D$ as I mentioned above.

There is yet one more element of the process that we must include. I have evaluated how many particles will collide with the black test particle, labeled "1" here. But in fact, any of the particles can hit with any of the particles, and really we are interested in the total number of collisions with time ($J$), in a given volume. As a result, I multiple the number of total collisions by the concentration ($n_\infty$) of particles. At this point I have $J = -8\pi a D n_\infty \cdot 2 \cdot n_\infty$, or

$$J = -16\pi a D n_\infty^2 \quad \text{(6-10)}$$

The units are #/m$^3$-s, or collision rate per volume.

By writing the diffusion coefficient in terms of the Stokes-Einstein equation, $D = kT/6\pi\eta a$, I obtain

$$J = -\frac{8kT}{3\eta} n_\infty^2 \quad \text{(6-11)}$$

This final expression for $J$ (units #/m$^3$-s) gives the total rate of collisions per volume. Interestingly, it turns out that $J$ does not depend on $a$. The omission of particle radius occurs because although smaller particles have more rapid Brownian motion, which increases $J$, smaller particles also have less surface area to accept a collision, which decreases $J$.

An assumption up to this point is that every time particles collide, they adhere and are no longer singlets. This is expressed as

$$\frac{dn_\infty}{dt} = J = -\frac{8kT}{3\eta} n_\infty^2 \quad \text{(6-12)}$$

Upon solving Eq 6-12 and applying the initial condition Eq 6-3, I obtain

$$\frac{1}{n_\infty} - \frac{1}{n_0} = \frac{8kT}{3\eta} t \quad \text{(6-13)}$$

I can rewrite this equation as

$$\frac{n_0}{n_\infty} - 1 = \frac{8kTn_0}{3\eta} t = \frac{t}{\tau_R} \quad \text{(6-14)}$$

Eq 6-14 introduces the idea that there is a time scale ($\tau_R$) to rapid aggregation. By defining a volume fraction of particles

$$\phi = \frac{4}{3}\pi a^3 n_0 \qquad (6\text{-}15)$$

which is the volume of one particle times the number of particles per volume of suspension, and inserting this expression into Eq 6-14, I obtain the final time scale ($\tau_R$, in seconds) for rapid particle aggregation in the diffusion-limit:

$$\boxed{\tau_R = \frac{\pi \eta a^3}{2kT\phi}} \qquad (6\text{-}16)$$

According to Eq 6-14, when $t = \tau_R$, then $n_\infty / n_0 = ½$.[83]

**Example 6-1**. Rapid aggregation.

Estimate the time for ½ the particles to aggregate, in an aqueous suspension with $a$ = 100 nm, $T$ = 295 K, and $\phi$ = 0.01, when the ionic strength is 500 mM KCl.

*answer: $\tau_R$ = 0.037 s. The 500 mM KCl kills the electrostatic repulsion, enabling the particles to have diffusion-limiting, rapid aggregation.*

### *Stability ratio (W)*

A shelf life of 0.037 s, as given in Example 6-1, is far too short for most applications. Typically we would want a shelf life of a year or more. We cannot stop the particles from engaging in Brownian motion and colliding, but we can include repulsive colloidal forces

[83] As aggregation proceeds beyond doublets to larger flocs, a distribution of sizes appears. See Section 8.7 (p 279) in the book by Russel, Saville, and Schowalter, *Colloidal Dispersions* (1989). Eq 8.7.4 gives one result for floc size distribution.

– especially electrostatic, steric, and electrosteric forces – that make it such that only 1 out of 10 times when the particles collide, they adhere. Or 1 out of 100, or 1 out of $10^9$. The ratio of the number of collisions to the number of adherences is called the stability ratio ($W$), defined as

$$W = \frac{J_{collisions}}{J_{adherences}} \tag{6-17}$$

Eq 6-10 gives $J_{\text{collisions}}$ (called simply "$J$" in that equation). A high value of $W$ indicates stability, because it means that few particles are aggregating when they collide; they simply bounce off each other.

Based on our use of the Boltzmann expression, we might expect that aggregation probability is Boltzmann-distributed, or at least close to it. That is, the probability that two particles can overcome a repulsive energy barrier ($\Phi_{\text{max}}$) decreases exponentially. Indeed this is the case, and in fact, an approximate expression has been found in the literature as[84]

$$\boxed{W = 1 + 0.25\left[exp\left(\frac{\Phi_{max}}{kT}\right) - 1\right]} \tag{6-18}$$

This equation gives $W$ for a limited range of conditions. However, for most conditions, we would expect the character of $W$ to remain exponential in the repulsive energy barrier ($\Phi_{\text{max}}$) between the particles. In a later section of Chapter 6 we will learn how to calculate $\Phi_{\text{max}}$ using "DLVO theory".

Knowing how many collisions are required in order for two particles to adhere enables us to estimate not only the rapid flocculation time ($\tau_R$), but also the actual flocculation time ($\tau$). I simply multiply the diffusion-limited time by $W$ to obtain

$$\boxed{\tau = \frac{\pi\eta a^3}{2kT\phi}W} \qquad (6\text{-}19)$$

[84] Prieve, D.C. and Ruckenstein, E. "Role of surface chemistry in primary and secondary coagulation and heterocoagulation." *J. Colloid Interface Sci*., **73**, 539-555 (1980).

**Example 6-2**. Slow aggregation.

Estimate the time for ½ the particles to aggregate, in an aqueous suspension with $a$ = 100 nm, $T$ = 295 K, and $\phi$ = 0.01, when the ionic strength is low, such that $\Phi_{max} = 9.0\times10^{-20}$ J.

*answer:* $\tau = 3.7\times10^7$ *s = 1.2 years. We already know from Example* 6-*1 that* $\tau_R$ *= 0.037 s. For the given energy barrier,* $\Phi_{max}/kT = 22.1$*, so that* $W = 9.98\times10^8$*, giving the final estimate for time of stability.*

***Derjaguin-Landau-Verwey-Overbeek (DLVO) theory***

In previous chapters I have explored both electrostatic and van der Waals interactions between particles. In assessing the value of $\Phi_{max}$, we use a very simple method, first used by two Soviet scientists Derjaguin and Landau, and then by two Dutch scientists Verwey and Overbeek. They simple added the interactions together:

$$\Phi \approx \Phi_{DLVO} = \Phi_{ES} + \Phi_{VDW} \approx 2\pi\varepsilon a\psi_0^2 e^{-\kappa\delta} - \frac{Aa}{12\delta} \qquad (6\text{-}20)$$

This is the same as Eq 1-6. At this stage we know how to replace the simplified expressions after the approximate equals sign, with better expressions like the full HHF expression for electrostatic stability of the full Hamaker expression for the VDW energies.

It is common in the literature to disparage the accuracy of the DLVO theory. However, the model has valuable predictive ability. Indeed when you add salt, particles adhere in solution. And indeed gold, which has a high Hamaker constant and therefore strong VDW interactions, is more attractive than polystyrene. Usually when DLVO theory is employed, the following assumptions are made: 1) The particles are assumed to be geometrically ideal, for example geometric spheres. This of course is seldom the case, since the surface often has a hairy polymer layer or surface roughness. 2) The VDW and electrostatic forces are assumed to arise from the same geometric surface. Many particles have an interface that is some nanometers thick, and therefore the charges exist not at a single plane, but over a region of interface. Furthermore, the placement of charges can be somewhat dynamic if the polymer chains move. As a result, the simple geometric approximation is not quite correct. Other assumptions sometimes used in the particular electrostatic and VDW models include low surface potentials, the Derjaguin approximation, and sometimes the neglect of certain forces. But again, overall, the model provides valuable predictive ability.

Let me start with Eq 6-20 and find the force by taking $F = -d\Phi / d\delta$. Now I set $F = 0$ and solve for the positions where the force between the particles is 0. From Figure 1-3, I expect two locations: one in the secondary energy minimum, and one at the peak of the barrier between the particles. Setting $F = 0$ from Eq 6-20 gives me the following equation:

$$e^{-\kappa\delta}\left(\kappa\delta\right)^2 = \frac{A}{24\pi\varepsilon\kappa^{-1}\psi_0^2} \qquad (6\text{-}21)$$

If I treat the right-hand-side as a single parameter, which includes the Hamaker constant (VDW forces), as well as the Debye length and surface potentials (electrostatic forces), I can solve for the equilibrium positions of the particles. These solutions are given in Figure 6-3.

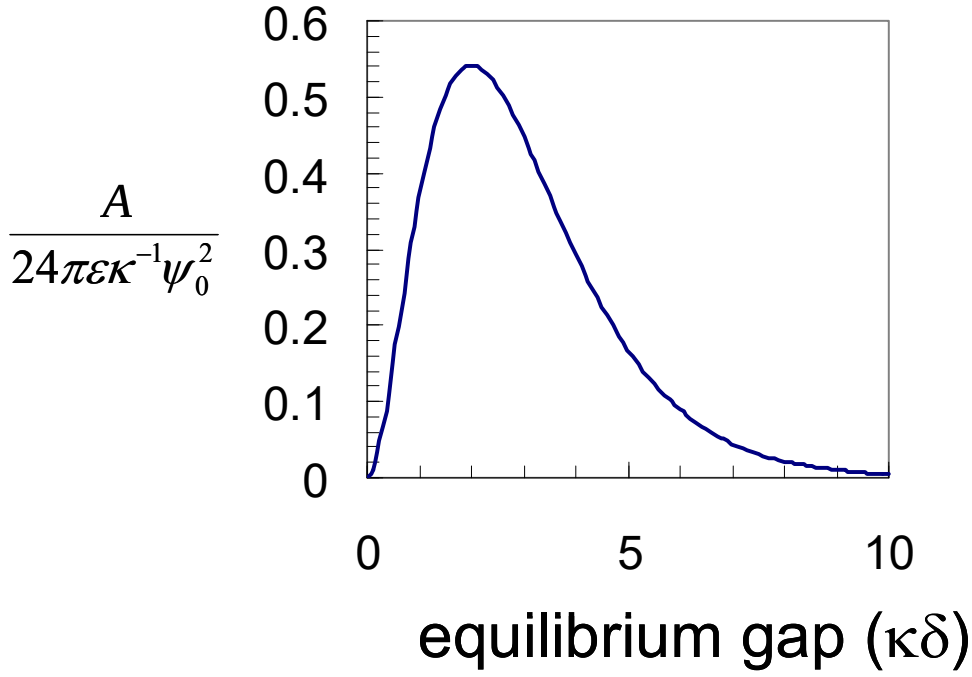

**Figure 6-3**. Equilibrium gaps between particles, from simplified DLVO theory. There are two equilibrium gaps, until the parameter on the y-axis exceeds 0.55.

In Figure 6-3 we see that there are two equilibrium gaps predicted for a given value of $A/24\pi\varepsilon\kappa^{-1}\psi_0^2$, until that parameter exceeds 0.55. At this point the particles have no equilibrium gap between them and undergo rapid aggregation. There are various ways of defining a "critical coagulation concentration", but this is one of the most straightforward ways.[85] When $A/24\pi\varepsilon\kappa^{-1}\psi_0^2 = 0.55$, we can solve for the Debye length ($\kappa^{-1}$), or the ionic strength that gives that Debye length. When the ionic strength is raised above this value, the system is no longer stable.[86]

### *Shear-induced aggregation*

In the preceding sections, I have discussed Brownian aggregation, which results from the random motion of the particles. Figure 6-1 also shows "shear-induced aggregation". Modeling

[85] The The Schulze-Hardy rule, derivable from DLVO theory, describes how the critical coagulation concentration depends on $1/Z^6$.

[86] The stablity literature indicates that at high ionic strengths, even above 10 mM, DLVO theory can start to fail. Behrens, Sven Holger; Christl, Daniel Iso; Emmerzael, Rudi; Schurtenberger, Peter; Borkovec, Michal. "Charging and Aggregation Properties of Carboxyl Latex Particles: Experiments versus DLVO Theory." *Langmuir*, **16**, 2566-2575 (2000).

shear-induced aggregation is more involved, but Russel et al. give the shear-induced time as

$$\boxed{\tau^{sh} = \frac{3\pi}{4Pe}\tau_R W^{sh}} \tag{6-22}$$

The Peclet number (Pe) gives the strength of the shear motion relative to the diffusive motion, and is given by

$$\boxed{Pe = \frac{3\pi\eta a^3}{kT}\gamma} \tag{6-23}$$

where $\gamma$ is the shear rate (units 1/s) in the system. For laminar flow in a pipe of radius ($R$), $\gamma = dv_z / dr$, whose magnitude can be expressed as $\gamma \approx U / R$, where $U$ is the average fluid speed. Typically values of $W^{sh}$ are between 0.5 and 20, which are at least within an order of magnitude of unity.

***Steric stabilty***

When greater stability is needed than can be provided by electrostatic repulsion, a steric or electrosteric stabilizer is often used. These are often polymeric molecules that prevent the close approach of particles, so that thermodynamically, they cannot aggregate. A common stabilizer is PVA (polyvinyl alcohol). More information is given in the book by Napper.[87]

***Depletion forces***

If two particles exist in solution, with radius ($a$), and in the same solution are non-adsorbing polymers or nanoparticles of radius ($R$), then "depletion attraction" can occur. Let's assume in this discussion that the depletion interaction is caused by the small nanoparticles. Just as sugar or salt or proteins can cause an osmotic pressure $\Pi = nkT$, where *n* is the concentration of all species, such as the proteins, or both types of ions, or other, so particles also can cause an osmotic pressure.

[87] Napper, Donald H. "Polymeric Stabilization of Colloidal Dispersions" (1984).

Significantly, if the gap ($\delta$) between the larger colloidal particles becomes smaller than the diameter (2*R*) of the nanoparticles, the nanoparticles can be excluded from the gap. The area of this exclusion is

$$A = \pi\left(2aR - a\delta + R^2 - \delta^2 / 4\right) \qquad (6\text{-}24)$$

There is thus an imbalance, such that the osmotic pressure outside the gap is greater than the osmotic pressure within the gap. This imbalance pushes the spheres together, causing a depletion attraction.

The depletion force between the particles is simply the difference in osmotic pressure outside the gap minus inside the gap, times the area of exclusion:

$$F = \begin{cases} +\infty & \textit{for} \quad \delta < 0 \\ -n_\infty kT\pi a\left(2R - \delta + R^2/a - \delta^2/4a\right) & \textit{for} \quad 0 \le \delta \le 2R \\ 0 & \textit{for} \quad \delta > 2R \end{cases} \qquad (6\text{-}25)$$

This force can then be integrated, since $F = -d\Phi / d\delta$, to give a depletion potential energy

$$-\frac{\Phi_{dep}}{n_\infty kT} = \begin{cases} +\infty & for \quad \delta < 0 \\ \pi\left( 2aR^2 + \frac{a\delta^2}{2} - 2aR\delta - R^2\delta + \frac{4}{3}R^3 + \frac{1}{12}\delta^3 \right) & for \quad 0 \le \delta \le 2R \\ 0 & for \quad \delta > 2R \end{cases} \tag{6-26}$$

Depletion energies are typically only a few $kT$ in magnitude, and they become zero if the gap exceeds $2R$. And yet, depletion energies can often cause substantial changes in the behavior of a colloidal suspension. The depletion interaction is added onto the DLVO interaction, to give the total interaction energy. Indeed, if other interactions are present, they also are usually added on in a linear manner.

Of recent interest is that fact that we can also have depletion *repulsion*, if the concentration of depletant particles is very high. The reason is that there become so many depletant molecules or nanoparticles, that they are difficult to remove from between the two larger colloidal particles. It is analogous to having plywood on the floor, but with round marbles between the plywood and floor. The plywood will never touch the floor, because the marbles prevent it. Similarly, when the concentration of depletant species becomes very high, they act like stabilizing marbles between two larger surfaces.[88]

### ***Summary. Flocculation is rate-based.***

I have examined flocculation from the two perspectives of diffusion and of reaction kinetics, and in juxtaposing these two perspectives, we have the diffusion-limited flocculation time. The actual time of stability is given by multiplying by a "stability ratio" ($W$), which arises due to repulsive interaction forces between the particles. These interaction forces are given by a sum of the various interactions, including electrostatic plus VDW interactions – which together give classic "DLVO theory" – as well as depletion forces and sometimes yet other forces.

---

[88] Sharma, Amber; Tan, Su Nee; Walz, John Y. "Measurement of colloidal stability in solutions of simple, nonadsorbing polyelectrolytes." *J. Colloid Interface Sci.*, **190**, 392 (1997). Figure 7 of this paper has a beautiful image of 11 test tubes (a-k) showing the effect of depletion attraction and depletion repulsion. This is one of my favorite colloids images in the literature, for its simplicity and beauty in showing the result.

***Symbols***

$a$ = sphere radius [=] m
$D$ = translational diffusion coefficient [=] $m^2/s$
$\phi$ = volume fraction of particles [=] dimensionless
$\Phi$ = potential of mean force [=] J
$\Phi_{max}$ = max peak – lowest valley
$G$ = hydrodynamic function
$\eta$ = viscosity [=] Pa s
$J$ = collision rate [=] #/$m^3$-s
$kT$ = thermal energy [=] J
$T$ = absolute temperature [=] K
$t$ = time [=] s
$\tau_R$ = estimate of rapid flocculation time [=] s
$U$ = particle velocity or fluid speed in a pipe [=] m/s
$W$ = stability ratio [=] none

***Practice Problems***

1. **Depletion energy**. Plot the depletion energy for the following system: the radius $a$ = 700 nm for the particle, $R$ = 9.0 nm for the depletant molecule, the depletant concentration $n_\infty$ = 120 mM, and the temperature $T$ = 302 K. List the depletion energy at $\delta$ = 7.0 nm, 14.0 nm, and 25.0 nm.
*answers:* $\Phi_{dep}(7.0\ nm) = -4.05\times10^{-20}$ *J,* $\Phi_{dep}(14.0\ nm) = -0.54\times10^{-20}$ *J,* $\Phi_{dep}(25.0\ nm) = 0$ *J.*

2. **Excluded area and depletion force**. Derive the excluded area between two spheres with radius ($a$) when the depletant has a radius ($R$). It is easiest to draw a couple right triangles and use the Pythagorean theorem. You will find:

$$A = \pi\left(2aR - a\delta + R^2 - \delta^2 / 4\right)$$

Then use this area to derive the depletant force and energy in Eqs 6-25 and 6-26.

3. **Flocculation time**. The system from Problem 6-1 has a volume fraction of particles $\phi = 0.13$ in water with [KCl] = 11.9 mM. The particles are silica with surface potentials -18 mV. The non-retarded Hamaker constant is $0.83\times10^{-20}$ J. Estimate 1) the diffusion-limited flocculation time, 2) the value of the maximum energy barrier ($\Phi_{max}$), using the HHF equation, the full Hamaker equation, and the full depletion result, 3) the value of W, and 4) the actual flocculation time.
*answers: 1)* $\tau_0 = 0.81$ *s, 2)* $\Phi_{max} = 25.5$ *kT, 3)* $W = 3.04\times10^{10}$*, 4)* $\tau = 2.47\times10^{10}$ *= 783 years.*

4. **Sensitivity of stability**. Typically it is very challenging to measure a zeta potential, which is often used as a measure of the surface potential, to within better than 2 or 3 mV. If the system in Problem 6-3 really had a surface potential of -16 mV, what would be the flocculation time?
*answers:* $\tau = 5800$ *s. The stability is very sensitive to the surface potential. A sensitivity analysis is usually warranted.*

5. **Depletion repulsion literature**. Read the following article carefully:

Wasan, Darsh; Nikolov, Alex; Henderson, Douglas, "New Vistas in Dispersion Science and Engineering," *AIChE J.*, **49**, 550-556 (2003).

6. **Steric stabilizer (PVA)**. List three journal articles from the literature that use poly(vinyl alcohol) as a steric stabilizer. For the format of the citations, use the same as I used in Problem 6-5, and give a 2 sentence summary of each article.

7. **Shear-induced aggregation**. We have a long tube with radius $R$ = 3.0 mm. Water at $T$ = 293 K is flowing at an average speed of $U$ = 5.0 cm/s, carrying particles of radius $a$ = 400 nm. The shear rate is defined as $\gamma = dv_z / dr$ for this system, and thus depends upon the radial position we choose. But to find the importance of shear-induced aggregation, we can estimate $\gamma \approx U / R$. Evaluate 1) the Reynolds number $\mathrm{Re} = \rho UD / \eta$ of the system, to make sure we're in the laminar regime, 2) an estimate of the shear rate ($\gamma$) in sec$^{-1}$, 3) the Peclet number, as defined by Russel et al, and 4) the factor by which shear causes aggregation, over diffusion-limited aggregation. *answers: 1) Re = 300 (still laminar), 2) $\gamma$ = 16.7, 3) Pe = 2.49, 4) enhancement over diffusion-limited = roughly 3, from Russel et al Fig 8.18. I have interpolated between the low shear and high shear curves.*

8. **Stability scenarios**. We have polymer colloids, stabilized with carboxyl groups (pKa = 4 or 4.5). But we need to have a suspension at pH = 2.0 and and ionic strength 20 mM. Evaluate the three following scenarios for stabilizing the system: 1) change the surface charge groups to sulfonate, which has pKa = 2.0. 2) Use a steric stabilizer, such as PVA. 3) Use depletion repulsion, with a high concentration of carboxylated nanoparticles.

# *7 Synthesis*

## *General considerations of particle synthesis*

Synthesizing colloidal particles is part science, part art. In this chapter I describe some principles that our lab uses in distinguishing these two parts. The aim is to provide the tools and confidence you need to start synthesizing your own particles easily, out of materials including polymers, oxides, or metals. In colloidal assembly operations, these elementary particles form the "atoms" used later for assembly.

Table 7-1 gives the primary factors we examine when considering a particle synthesis. As discussed in Chapter 1, many recipes already exist for particle synthesis. In my lab we find these using various search engines; one of my first-to-use is www.scholar.google.com. I type in keywords for various materials, sizes, shapes, or any other parameters I want, and proceed from there. However, some recipes – for example, emulsion polymerization and Stober silica – are so common, that I list some of these in Table 7-2 for convenience. Oftentimes a good way to find a recipe is to look at an existing recipe in a literature article, and then to look at the reference list at the end of that article or chapter.

**Table 7-1**. Factors to consider in particle synthesis.

| Factor | Considerations |
|---|---|
| material | Compatible (pH, T, chemistry, etc)? Biodegradable? |
| synthesis method | Recipe type (emulsion polymerization, sol-gel, etc)? |
| colloidal stability | Time required? Charge, steric, electrosteric, depletion repulsion, nanoparticles at surface[89] (Pickering emulsion) other OK to use? pH or salt conditions required? |
| functionalization | Ligands? Fluorescent groups? Effect on subsequent assembly operations? Competes with colloidal stability? Janus particles?[90][91] |
| post treatment | Heat annealing? Cooling? Coagulum removal? Competes with stability or functionalization? |
| cost effectiveness | Good properties? Fast to make, or to learn how to make? Cheap to make, in terms of both materials and equipment? Better to purchase or to synthesize? |
| safety | Chemical toxicity, flammability, corrosiveness, and similar? Carcinogenic? High T and P? Harsh pH? |
| sustainability | Sustainable (i.e., easily replaced by Nature)? |

[89] *Colloidal Particles at Liquid Interfaces*, 1st Ed. Eds. Bernard P. Binks, Tommy S. Horozov. Cambridge (2008).

[90] Jiang, Shan; Chen, Qian; Tripathy, Mukta; Luijten, Erik; Schweizer, Kenneth S.; Granick, Steve. "Janus Particle Synthesis and Assembly." Advanced Materials, **22**, 1060–1071 (2010). A Google Scholar search of "Steve Granick Janus" shows many highly-cited articles, going back to 2006.

[91] Park, Bum Jun; Brugarolasa, Teresa; Lee, Daeyeon. "Janus particles at an oil–water interface." *Soft Matter*, **7**, 6413-6417 (2011).

**Table 7-2**. The synthesis of representative colloidal particles. These particles act as the building blocks – or "colloidal atoms" – for colloidal structures and devices. The references serve as starting points for synthesizing seven common classes of particles.

| material class | material | references for size, shape, class |
|---|---|---|
| metal | silver, gold | 5-50 nm,[92] 500-5000 nm,[93] rods,[94,95] striped,[96] hollow[97] |
| | cobalt | magnetic nanoparticles[98] |
| | carbon nanotubes | various info[99,100] |
| | chromium | pyramids about 100 nm in size[101] |
| oxides | silica ($SiO_2$) | classic Stober silica,[102,103] silver core/silica shell[104] |
| | titania ($TiO_2$) | |
| | zinc oxide (ZnO) | |
| | hematite | |

---

[92] D.A. Handley, "Methods for synthesis of colloidal gold," p 13-32 Ch 2 in *Colloidal Gold: Principles, Methods, and Applications*, vol 1, ed. M.A. Hayat. Academic Press, New York, New York (1989) ISBN: 0-12-333927-8. This chapter lists about a dozen methods for producing gold particles of all sizes, with many functionalities.
[93] K.P. Velikov, G.E. Zegers, and A. van Blaaderen, "Synthesis and characterization of large colloidal silver particles," *Langmuir* **19**, 1384-1389 (2003).
[94] B.M.I. van der Zande, M.R. Bohmer, L.G.J. Fokkink, and C. Schonenberger, "Aqueous gold sols of rod-shaped particles," *J. Phys. Chem. B* **101**, 852-854 (1997).
[95] B.M.I. van der Zande, M.R. Bohmer, L.G.J. Fokkink, and C. Schonenberger, "Colloidal dispersions of gold rods: synthesis and optical properties," *Langmuir* **16**, 451-458 (2000).
[96] B.R. Martin, D.J. Dermody, B.D. Reiss, M. Fang, L.A. Lyon, M.J. Natan, and T.E. Mallouk, "Orthogonal self-assembly on colloidal gold-platinum nanorods," *Adv. Mater.* **11**, 1021-1025 (1999).
[97] Y. Sun, B. Mayers, and Y. Xia, "Metal nanostructures with hollow interiors," *Adv. Mater.* **15**, 641-646 (2003).
[98] C. Petit, A. Taleb, and M.-P. Pileni, "Self-organization of magnetic nanosized cobalt particles," *Adv. Mater.* **10**, 259-261 (1998).
[99] S. Iijima and T. Ichihashi, "Single-shell carbon nanotubes of 1-nm diameter," *Nature* **363**, 603-605 (1993). This is a follow up to Iijima's initial paper in *Nature* **354**, 56-58 (1991).
[100] T. Guo, P. Nikolaev, A. Thess, D.T. Colbert, and R.E. Smalley, "Catalytic growth of single-walled nanotubes by laser vaporization," *Chem. Phys. Lett.* **243**, 49-54 (1995).
[101] J. Henzie, E.-S. Kwak, and T.W. Odom, "Mesoscale metallic pyramids with nanoscale tips," *Nano Lett.* **5**, 1199-1202 (2005).
[102] W. Stober, A. Fink, and E. Bohn, "Controlled growth of monodisperse silica spheres in the micron size range", *J. Colloid Interface Sci.* **26**, 62-69 (1968).
[103] Bangs Laboratories (www.bangslabs.com).
[104] J. Wang, W.B. White, and J.H. Adair, "Evaluation of dispersion methods for silica-based composite nanoparticles," *J. Am. Ceram. Soc.* **89**, 2359–2363 (2006).

| | | |
|---|---|---|
| | Au core, $SiO_2$ shell | rutile[105,106] and anatase[106,107]<br>nanoparticles[108]<br>pseudo-cubes, spheroids, peanuts[109]<br>core-shell[110] |
| semiconductors | silicon<br>GaAs<br>CdTe<br>carbon nanotubes | silicon particles[111,112]<br>semiconductor[113,114]<br>semiconductor[115]<br>can be metal or semiconductor[99,100] |

[105] E. Matijevic, M. Budnik, and L. Meites, "Preparation and mechanism of formation of titanium dioxide hydrosols of narrow size distribution," *J. Colloid Interface Sci.* **61**, 302-311 (1977).

[106] M. Visca and E. Matijevic, "Preparation of uniform colloidal dispersions by chemical reactions in aerosols: I. Spherical particles of titanium dioxide," *J. Colloid Interface Sci.* **68**, 308-319 (1979).

[107] N. Spanos, I. Georgiadou, and A. Lycourghiotis, "Investigation of rutile, anatase, and industrial titania/water solution interfaces using potentiometric titration and microelectrophoresis," *J. Colloid Interface Sci.* **172**, 374-382 (1995).

[108] M. Jitianu and D.V. Goia, "Zinc oxide colloids with controlled size, shape, and structure," *J. Colloid Interface Sci*. **309**, 78-85 (2007).

[109] O. Liu and K. Osseo-Asare, "Synthesis of monodisperse Al-substituted hematite particles from highly condensed metal hydroxide gels," *J. Colloid Interface Sci.* **231**, 401–403 (2000).

[110] K. Kamata, Y. Lu, and Y. Xia, "Synthesis and characterization of monodispersed core-shell spherical colloids with movable cores," *J. Am. Chem. Soc.* **125**, 2384-2385 (2003).

[111] L.E. Pell, A.D. Schricker, F.V. Mikulec, and B.A. Korgel, "Synthesis of amorphous silicon colloids by trisilane thermolysis in high temperature supercritical solvents," *Langmuir* **20**, 6546-6548 (2004).

[112] N. Aoki, "Process for producing silicon colloid" U.S. Patent 5,811,030 (filed 13 October 1994, issued 22 September 1998) [http://patft.uspto.gov/netacgi/nph-Parser?u=%2Fnetahtml%2Fsrchnum.htm&Sect1=PTO1&Sect2=HITOFF&p=1&r=1&l=50&f=G&d=PALL&s1=5811030.PN.&OS=PN/5811030&RS=PN/5811030].

[113] M. A. Olshavsky, A. N. Goldstein, and A. P. Alivisatos, "Organometallic synthesis of gallium-arsenide crystallites, exhibiting quantum confinement," *J. Am. Chem. Soc.* **112**, 9438-9439 (1990).

[114] M.A. Malik, P. O'Brien, S. Norager, and J. Smith, "Gallium arsenide nanoparticles: synthesis and characterisation," *J. Mater. Chem.* **13**, 2591-2595 (2003).

[115] N. Gaponik, D.V. Talapin, A.L. Rogach, K. Hoppe, E.V. Shevchenko, A. Kornowski, A.E. Iler, and H. Weller, "Thiol-capping of CdTe nanocrystals: an alternative to organometallic synthetic routes," *J. Phys. Chem. B* **106**, 7177-7185 (2002)

| polymers | general<br>polystyrene<br>PMMA<br>PVAc<br>spheroids<br>SU-8 epoxy | emulsion polymerization[116,117,118,119]<br>spheres,[120,121] spheroids,[122] >3 μm[123]<br>fluorescent,[124] particles and rods[125]<br>common polymer colloid[126]<br>stretched or rods[127,125]<br>rods[128] |
|---|---|---|

[116] R.G. Gilbert, *Emulsion Polymerization*. Academic Press, New York (1995).
[117] Robert M. Fitch, *Polymer Colloids*, Academic Press, New York (1997).
[118] D. Urban and K. Takamura (eds.), *Polymer Dispersions and Their Industrial Applications*, Wiley-VCH, New York, New York (2002).
[119] Interfacial Dynamics Corporation (www.idclatex.com), Polysciences (www.polysciences.com), Duke Scientific (www.dukescientific.com), Bangs Laboratories (www.bangslabs.com).
[120] J.W. Goodwin, J. Hearn, C.C. Ho, and R.H. Ottewill, "The preparation and characterization of polymer lattices formed in the absence of surface active agents," *Br. Polym. J.* **5**, 347-362 (1973).
[121] J.W. Goodwin, J. Hearn, C.C. Ho, and R.H. Ottewill, "Studies on the preparation and characterization of monodisperse polystyrene lattices," *Colloid & Polym. Sci.*, **252**, 464-471 (1974).
[122] C.C Ho, A. Keller, J.A. Odell, and R.H. Ottewill, "Preparation of monodisperse ellipsoidal polystyrene particles," *Colloid Polymer Sci.* **271**, 469-479 (1993).
[123] J.-S. Song, F. Tronc, and M.A. Winnik, "Two-stage dispersion polymerization toward monodisperse, controlled micrometer-sized copolymer particles," *J. Am. Chem. Soc.* **126**, 6562-6563 (2004).
[124] G. Bosma, C. Pathmamanoharan, E.H.A. de Hoog, W.K. Kegel, A. van Blaaderen, and H.N.W. Lekkerkerker, "Preparation of monodisperse, fluorescent PMMA–latex colloids by dispersion polymerization," *J. Colloid Interface Sci.* **245**, 292–300 (2002) [doi:10.1006]. This paper uses "dispersion polymerization", as opposed to "emulsion polymerization", and thus not only provides a method for producing fluorescent particles, but provides an alternative method for making polymer colloids in general.
[125] A. Mohraz and M.J. Solomon, "Direct visualization of colloidal rod assembly by confocal microscopy," *Langmuir* **21**, 5298-5306 (2005) [doi:10.1021/la046908a] They stretched their PMMA rods in a manner similar to Ho et al.
[126] N. Sosa, R.D. Peralta, R.G. Lopez, L.F. Ramos, I. Katime, C. Cesteros, E. Mendizabal, and J.E. Puig, "A comparison of the characteristics of poly(vinyl acetate) latex with high solid content made by emulsion and semi-continuous microemulsion polymerization," *Polymer* **42**, 6923-6928 (2001).
[127] C.C. Ho, A. Keller, A., J.A. Odell, and R.H. Ottewill, "Monodisperse ellipsoidal polystyrene latex particles: preparation and characterisation," *Polymer Int.* **30**, 207-211 (1993).
[128] R.G. Alargova, K.H. Bhatt, V.N. Paunov, and O.D. Velev, "Scalable synthesis of a new class of polymer microrods by a liquid-liquid dispersion technique," *Adv. Mater.* **16**, 1653-1657 (2004).

| | | |
|---|---|---|
| waxes / oils | ferrofluid droplets<br>wax particles | magnetic droplets[129,130]<br>wax[131] |
| complex particles | liposomes<br>colloidosomes<br>Janus particles<br>patterned particles | stabilized by nanoparticles[132]<br>particles outside[133]<br>many types[134,135,136,137,138,139,140,141,142] |

---

[129] R.E. Rosenweig, *Ferrohydrodynamics*, Cambridge University Press, New York, New York (1985) ISBN: 0521256240. Ferrofluid droplets have various magnetic properties.

[130] P. Enzel, N.B. Adelman, K.J. Beckman, D.J. Campbell, A.B. Ellis, and G.C. Lisenky, "Preparation and properties of an aqueous ferrofluid," *J. Chemical Education* **76**, 943-948 (1999) [http://jchemed.chem.wisc.edu/Journal/Issues/1999/Jul/abs943.html].

[131] A.M. Belfort, "Method of preparing a stable wax dispersion using β-1,4 glucan," U.S. Patent 3,442,676 (filed 29 Dec. 1965, issued 6 May 1969).

[132] L. Zhang and S. Granick, "How to stabilize phospholipid liposomes (using nanoparticles)," *Nano Lett.* **6**, 694-698 (2006) [doi:10.1021/nl052455y]. Stability of liposomes is a common challenge.

[133] A.D. Dinsmore, M.F. Hsu, M.G. Nikolaides, M. Marquez, A.R. Bausch, and D.A. Weitz, "Colloidosomes: selectively permeable capsules composed of colloidal particles," *Science* **298**, 1006-1009 (2002).

[134] C. Casagrande, M. Veyssié, *C.R. Acad. Sci.* (Paris), **306** II, 1423 (1988).

[135] de Gennes, Pierre-Gilles, "Soft matter," Nobel Lecture, 9 December 1991 [http://nobelprize.org/nobel_prizes/physics/laureates/1991/gennes-lecture.pdf].

[136] A. Perro, S. Reculusa, F. Pereira, M.-H. Delville, C. Mingotaud, E. Duguet, E. Bourgeat-Lami, and S. Ravaine, "Towards large amounts of *Janus* nanoparticles through a protection-deprotection route," *Chem. Commun.*, 5542-5543 (2005).

[137] B. Gruning, U. Holtschmidt, and G. Koerner, "Particles, modified at their surface by hydrophilic and hydrophobic groups," U.S. Patent number 4,715,986 (filed 26 Mar 1985, issued 29 Dec 1987).

[138] A. Perro, S. Reculusa, S. Ravaine, E. Bourgeat-Lamic, and Etienne Duguet, "Design and synthesis of Janus micro- and nanoparticles," *J. Mater. Chem.* **15**, 3745–3760 (2005) [doi:10.1039/b505099e]. This is a very nice review of Janus particle research up to its 2005 date.

[139] O.D. Velev, A.M. Lenhoff, and E.W. Kaler, "A class of microstructured particles through colloidal crystallization," *Science* **287**, 2240-2243 (2000).

[140] L. Hong, A. Cacciuto, E. Luijten, and S. Granick, "Clusters of charged Janus spheres," *Nano Lett.* **6**, 2510-2514 (2006).

[141] L. Hong, S. Jiang, and S. Granick, "Simple method to produce Janus colloidal particles in large quantity," *Langmuir* **22**, 9495-9499 (2006).

[142] O. Cayre, V.N. Paunov, and O.D. Velev, "Fabrication of asymmetrically coated colloid particles by microcontact printing techniques," *J. Mater. Chem.* **13**, 2445–2450 (2003).

| | | |
|---|---|---|
| | | coated,[143] metal,[144,145] gold nanodots,[146,147] polymer particles,[148,149] Fibonacci coating,[150] raspberry coating,[151] modeling,[152] various complex coatings[153] |
| biocolloids | bacteria | 100 nm (*Mycoplasma*) to 600 μm (*Epulopiscium fishelsoni*), includes many *E. coli* for instance[154] |
| | viruses | tobacco mosaic virus PV-135.[154] |
| | yeast | Saccharomyces cerevisiae.[154] |

[143] B. Vincent, C.A. Young, and T. Tadros, "Adsorption of small, positive particles onto large, negative particles in the presence of polymer," *J. Chem. Soc. Faraday I* **76**, 665-673 (1980).

[144] J.C. Love, B.D. Gates, D.B. Wolfe, K.E. Paul, and G.M. Whitesides, "Fabrication and wetting properties of metallic half-shells with submicron diameters," *Nano Lett.* **2**, 891-894 (2002).

[145] C. Charnay, A. Lee, S.-Q. Man, C.E. Moran, C. Radloff, R.K. Bradley, and N.J. Halas, "Reduced symmetry metallodielectric nanoparticles: chemical synthesis and plasmonic properties," *J. Phys. Chem. B* **107**, 7327-7333 (2003).

[146] G. Zhang, D. Wang, and H. Mohwald, "Decoration of microspheres with gold nanodots—giving colloidal spheres valences," *Angew. Chem. Int. Ed*. **44**, 7767 –7770 (2005).

[147] G. Zhang, D. Wang, and H. Mohwald, "Patterning microsphere surfaces by templating colloidal crystals," *Nano Lett.*, **5** 143-146 (2005).

[148] C.E. Snyder, A.M. Yake, J.D. Feick, and D. Velegol, "Nanoscale functionalization and site-specific assembly of colloids by particle lithography," *Langmuir* **21**, 4813-4815 (2005).

[149] J.-Q. Cui and I. Kretzschmar, "Surface-anisotropic polystyrene spheres by electroless deposition," *Langmuir* **22**, 8281-8284 (2006).

[150] C. Li, X. Zhang, and Z. Cao, "Triangular and Fibonacci number patterns driven by stress on core/shell microstructures," *Science* **309**, 909-911 (2005).

[151] G. Li, X. Yang, F. Bai, and W. Huang, "Raspberry-like composite polymer particles by self-assemble heterocoagulation based on a charge compensation process," *J. Colloid Interface Sci.* **297**, 705–710 (2006).

[152] Y. Duda and F. Vazquez, ""Modeling of composite latex particle morphology by off-lattice Monte Carlo simulation," *Langmuir* **21**, 1096-1102 (2005) [doi:10.1039/b505099e]. This paper provides modeling for obtaining various patterns on particles using morphological control.

[153] Y.-S. Cho, G.-R. Yi, J.-M. Lim, S.-H. Kim, V.N. Manoharan, D.J. Pine, and S.-M. Yang, "Self-organization of bidisperse colloids in water droplets," *J. Am. Chem. Soc.* **127**, 15968-15975 (2005).

[154] http://www.atcc.org/ sells many types of biological cells and cell lines.

***Clever breakthrough: Multiple unit operations***

Most new synthesis operations contain some clever development that enables the process. In the next two sections I discuss only two of these, concerning emulsion polymerization and Stober silica. These are meant as examples to show the kind of thinking required to synthesize new particles.

When van der Hoff developed his emulsion polymerization process in the 1950s,[155] it was revolutionary in that he put all the typical unit operations of the process – reactor, storage tank, heat exchanger – into a single pot. Figure 7-1 shows various aspects of the process that emphasize the combined "unit operations". Figure 7-2 shows a photo from our lab of an emulsion polymerization apparatus. The recipe for polystyrene formed by emulsion polymerization is then given.

Emulsion polymerization starts with an organic monomer having a double-bonded carbon-carbon bond (C=C). A free-radical reaction is initiated, oftentimes by a persulfate initiator such as $K_2S_2O_8$ or $(NH_3)_2S_2O_8$. A polymer chain develops, typically reaching a molecular weight of 100,000. These polymer chains then collect to form small polymeric particles, which continue to aggregate to form yet larger particles. At some critical size – dictated by the electrostatic repulsion between the spheres, in many cases – the particles become stable and no longer aggregate. Some monomer might yet enter the particles and enlarge the particles.

[155] E. B. Bradford & J. W. Vanderhoff, "Electron Microscopy of Monodisperse Latexes," *J. Appl. Phys*, . **26**, 864-871 (1955). Vanderhoff was at Dow Chemical at the time, and his work was perhaps the first that used emulsion polymerization.

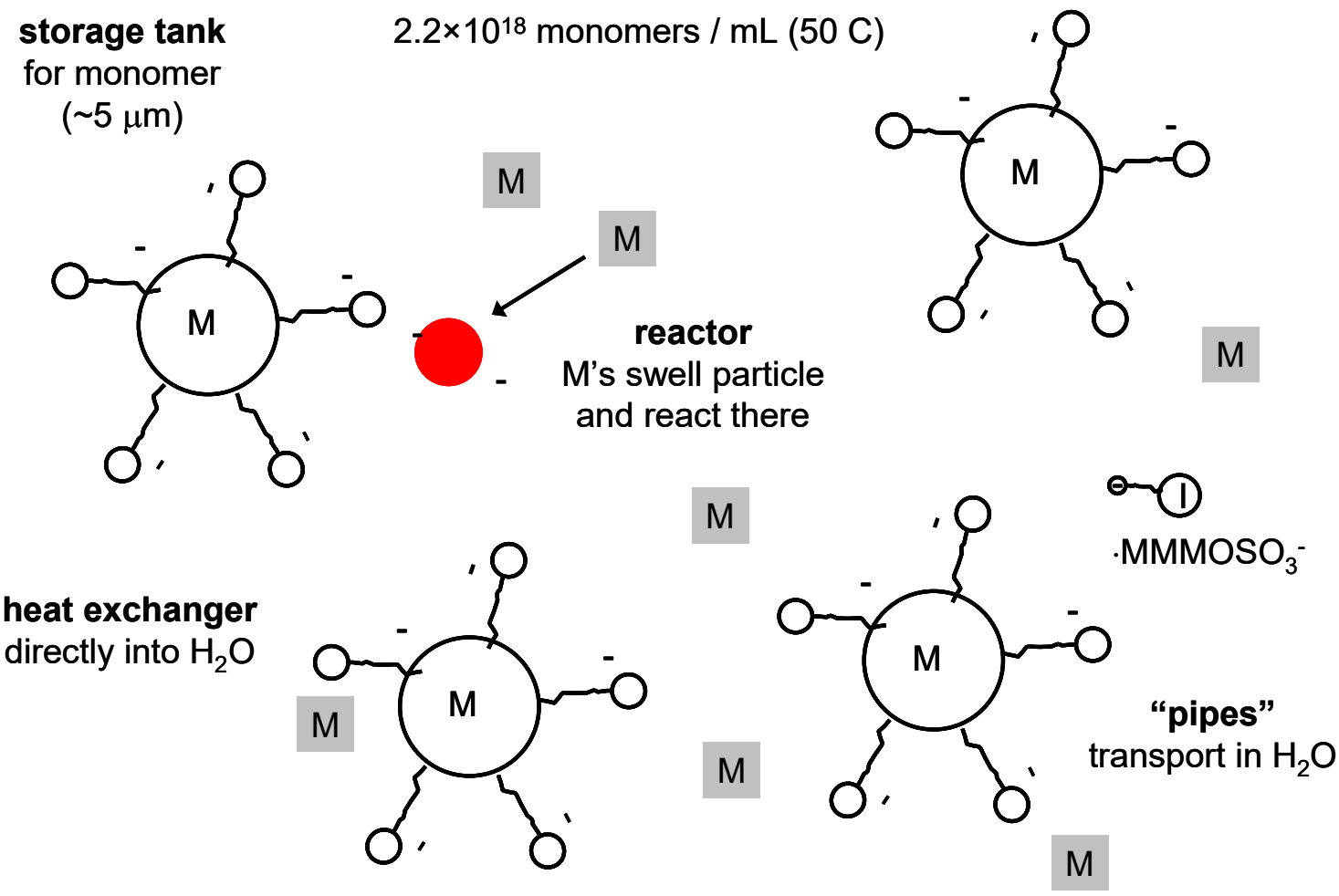

**Figure 7-1**. "Unit operations" in an emulsion polymerization. The figure shows the "unit operations" like "reactor" and "storage tank". The heat exchange is directly into the surrounding solution.

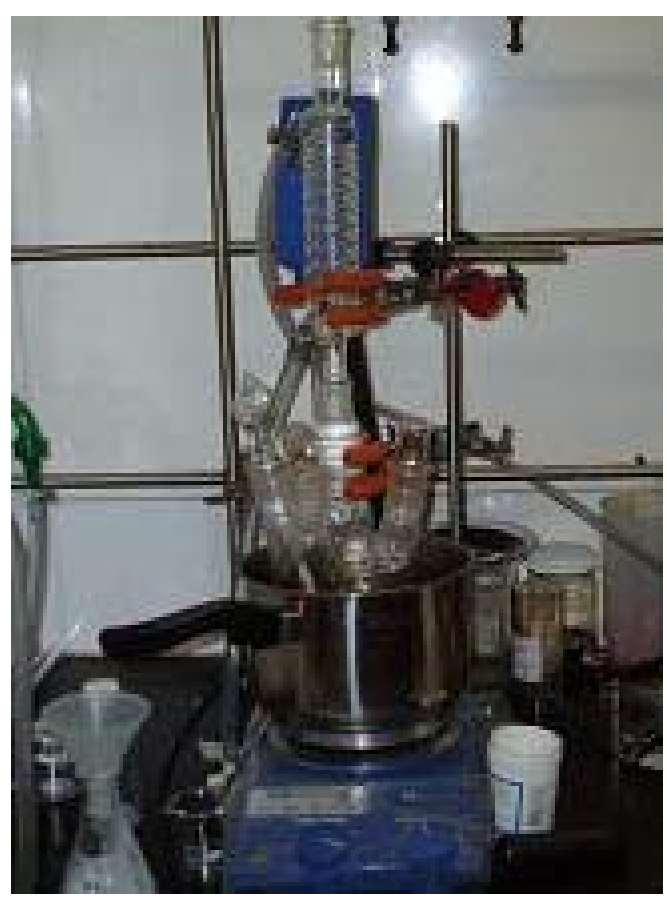

**Figure 7-2**. An apparatus for conducting emulsion polymerization. This photo is from the Velegol lab at Penn State.

**Table 7-3**. Emulsion polymerization recipe for surfactant-free polystyrene latex.[156] This is sometimes called the Goodwin process.

**PREP:** 60 minutes; SYNTH: 24 hours; POST: 15 minutes

**MAKES:** polystyrene particles, size 100-1000 nm, at roughly 90% yield. Particles have sulfate (negative) groups on them.

**SAFETY:** carcinogen (styrene). High T.

73 g styrene
670 g water (Millipore or doubly-distilled)
NaCl (0-40 mM)
1.0 g $NaHCO_3$ buffer (pH 7.0 $\pm$ 0.5)
1.0 g $K_2S_2O_8$ initiator (2.5 mM)

**1.** Purify key ingredients. If necessary, re-crystallize the initiator and distill the styrene to purify them prior to use.

**2.** Add the water and the styrene to a 1 L round-bottom Pyrex flask. Bring the system to temperature, usually between 70-80 C. This can be done in a water bath.

**3.** Add the initiator to about 20 g water, and add to the round-bottom flask. Use a magnetic stir-bar and mix for 24 hours at about 350 rev/min.

**4.** Allow the vessel to cool for a few minutes.

**5.** Clean the particles. Decant through glass wool to remove coagulum skum, then dialyze for several days to remove excess monomer, NaCl, and initiator. When the conductivity of the solution is roughly constant, the dialysis is complete (usually about 5 changes).

[156] Goodwin, J.W.; Hearn, J.; Ho, C.C.; Ottewill, R.H. "The Preparation and Characterisation of Polymer Latices Formed in the Absence of Surface Active Agents." *Br. Polym. J.*, **5**, 347-362 (1973).

### *Clever breakthough: Forming reactant in situ*[157]

In 1968 Stober and co-workers developed a condensation reaction process for producing monodisperse silica particles in the micron size range. Producing particles larger than those in the small nanometer range enabled the particles to be stable, preventing aggregation into larger sizes, which tends to form irregularly-shaped particles. At the conclusion of the reaction, the particles form a turbid white suspension. Often the reaction proceeds in less than 30 minutes, although it can take considerably longer, especially for longer alkane chains with the tetra-alkyl silicate.

A key intermediate is silicic acid, but this reactant is chemically unstable. A key advance that Stober and co-workers developed was to produce the silicic acid in situ, from the tetra-alkyl silicates.

**Table 7-3**. Stober recipe for producing monodisperse silica particles.[157,158]

**PREP:** 60 minutes; SYNTH: 2 hours; POST: 15 minutes

**MAKES:** $SiO_2$ particles, size 100-1000 nm. Typical batch is 100 mL. Particles have silanol stabilizing groups.

**SAFETY:** flammable materials, high pH, high T.

Methanol, ethanol, n-propanol, or n-butanol
Tetra-alkyl silicate (e.g., methyl or ethyl)
Anhydrous ammonia (99.99%)
14.2 M $NH_4OH$
NaOH pellets

**1.** Prepare key ingredients. Prepare alcohol solutions saturated with ammonia (the catalyst), either by bubbling through 99.99% pure anhydrous ammonia gas through the alcohol, or by using a saturated aqueous solution of $NH_4OH$, with the volume of water accounted for.

---

[157] Stober., Werner; Fink, Arthur; Bohn, Ernst. "Controlled Growth of Monodisperse Silica Spheres in the Micron Size Range." *Journal of Colloid and Interface Science*, **26**, 62-69 (1968). More detail is given in this classic paper.

[158] The classic book on silica is *The Chemistry of Silica: Solubility, Polymerization, Colloid and Surface Properties and Biochemistry of Silica*, by Ralph K. Iler (1979).

**2.** Mix the reagents. Mix the pure alcohol, saturated alcoholic ammonia solution, ammonium hydroxide, and water in an Erlenmeyer flask with ground-glass stoppers.

**3.** Add the tetra-methyl silicate (or other alkyl). Continuously mix with a magnetic stir bar. The total amount of solution is 50-110 mL at this point.

**4.** The reaction should start within 10 minutes, and the opalescence of the solution will increase. Most of the reation occurs within 15 minutes, and the result will be a turbid white solution.

**5.** Clean the particles. Carefully clean the particles, ridding the excess solution properly and safely. For pure particles, dialysis will help. As a rule of thumb, using methanol and tetra-methyl silicate produces smaller particles (~100 nm) and more quickly, while using butanol and tetra-butyl silicate produces larger particles (~2000 nm) and more slowly.

***Summary. 1000s of recipes exist in the literature.***

The synthesis or modification of a new type of particle usually involves both some science and some art. Because so many recipes already exist in the literature – including modifications to previously-published recipes – a quick way to get started is to find the recipe you are looking for. This chapter provides guidance on some of the factors we consider when synthesizing particles.

***Practice Problems***

1. **pKa's**. Acid or base groups have a value at which half their charges are dissociated.[159] List the pKa for the following surfaces

[159] http://en.wikipedia.org/wiki/Acid_dissociation_constant. A fairly large list at

in water, based on literature references: silica, sulfate, sulfonate, amidine, carboxyl. Assume that the suspending fluid is water.

2. **Keeping particles charged**. In order for a colloidal particle to have a significant charge, the pH must be a couple units away from the pKa of the charge groups on the particle. Derive the Henderson-Hasselbach equation[160]

$$pH = pK_a + \log\frac{[A-]}{[HA]}$$

Then use the H-H equation to find the pH at which only 3% of the charge groups are dissociated on a silica surface in water ($pK_a$ of silica here is 2.0).
*answer: pH = 0.49. Note that some acidic groups have pKa values much less than 0. In general, to have 50% dissociation pH = pKa, to have 10% dissociation pH = pKa - 1, to have 1% dissociation pH = pKa - 2, to have 90% dissociation pH = pKa + 1, and to have 99% dissociation pH = pKa + 2.*

3. **Gold recipe**. Write a recipe for producing 5.2 nm gold particles. You might find the book chapter from Handley to be useful. Use the "Betty Crocker" format given in the chapter for the polystyrene recipe.

4. **Your favorite particle**. Choose a type of particle to synthesize, and summarize the recipe in a "Betty Crocker" format, with ingredients, time of preparation, ease of preparation, safety, and steps. Please include a PDF of the reference you used.

---

http://chemweb.unp.ac.za/chemistry/Physical_Data/pKa_compilation.pdf gives many pKa values. More are at http://evans.harvard.edu/pdf/evans_pKa_table.pdf

[160] http://en.wikipedia.org/wiki/Henderson%E2%80%93Hasselbalch_equation.

5. **Functionalization**. Let's say that you want to graft surface fluorescent groups to silica ($SiO_2$) particles. Briefly describe a method from the literature, and include the PDF of any literature that you use.

6. **GLAD patterning**. Let's say that you want to form a patterned surface on a silica particle. Use one or two paragraphs to describe how you might do this using "glancing angle deposition".[161]

7. **Janus particles**. Let's say that you want to synthesize Janus particles, in which one half of the particle is made of one type of material, and the other half of the particle has a surface with another material. Write a recipe for how you might make a Janus particle with gold on one half and any other material on the other half.

---

[161] Pawar, Amar B.; Kretzschmar, Ilona. "Patchy Particles by Glancing Angle Deposition." *Langmuir*, **24**, 355-358 (2008).

# 8 Electrokinetics

## *Huckel equation of electrophoresis*

My first experiments as a graduate student were to examine the electrophoresis of small particles. It seems natural that when an electric field is applied to charged particles, that they should move. But when electrophoresis was first observed by Reuss,[162] it was probably unappreciated that colloidal particles are charged when they are suspended in solution.

The simplest way to think about electrophoresis is that you have a particle with a total charge ($q$) being moved in an applied electric field ($\mathbf{E}_\infty$). Just as in the case where we found Stokes law of settling by balancing the gravitational force with the hydrodynamic drag, here we balance the electrical force with the hydrodynamic drag. That is,

$$\mathbf{F} = m\mathbf{a} = \mathbf{0} = \mathbf{F}_h + \mathbf{F}_E \tag{8-1}$$

For a point charge, the electrical force $\mathbf{F}_E = q\mathbf{E}$. We remember from the chapter on hydrodynamics that $\mathbf{F}_h = -6\pi\eta a\mathbf{U}$. I can therefore solve for the velocity ($\mathbf{U}$), obtaining

$$\mathbf{U} = \frac{q\mathbf{E}}{6\pi\eta a} \tag{8-2}$$

[162] Reuss, F.F. *Mem.Soc.Imperiale Naturalistes de Moscow*, **2**, 327 (1809).

where as before, $a$ is the particle radius and $\eta$ is the solution viscosity. Now say the charge is distributed over the surface of a small sphere with a charge density ($\rho_s$), such that $q = 4\pi a^2 \rho_s$. "Small" in this case is relative to the Debye length, so that we take the limit of $\kappa a \to 0$. From Eq 2-21, which says $\rho_s = \varepsilon\kappa\psi_0(1+\kappa a)/\kappa a$, we see that for $\kappa a \to 0$,

$$\psi_0 = \frac{\rho_s a}{\varepsilon} \tag{8-3}$$

Inserting these latter expressions into Eq 8-2 gives a second version of the Huckel equation, for a sphere.

$$\boxed{\mathbf{U} = \frac{2\varepsilon\psi_0\mathbf{E}}{3\eta}} \tag{8-4}$$

**Example 8-1**. Huckel equation for electrophoresis.

A ZetaPals device finds that the "zeta potential" – similar to $\psi_0$ when measured with electrophoresis – is -42 mV. The measurement was taken in water at T = 298 K. Calculate the electric field ($E$) required to produce a speed of $U$ = +4.6 µm/s.

*answer: E = -2.06 V/cm.*

For the Huckel equation, there is an inherent assumption that the electric field acting on every charge on the particle is the same. For small particles this is very nearly true. However, when the size of the particle becomes larger, especially as $\kappa a \to 0.5$ or 1, this assumption starts to fail. We will need to examine in greater detail the coupling of the electric field and the hydrodynamics within the electrical double layer (EDL). As a first step, let's examine "electro-osmosis".

### *Electroosmosis*

When exposed to water, most surfaces will become charged. An electrical double layer (EDL) results, with permanent charges on the solid surface, and mobile charges in the fluid. When an electric field ($\mathbf{E}_\infty$) is applied, the field acts on the charges. Just like any other small "particle" in a fluid, the ions immediately come to terminal velocity, thus transmitting all of their force to the fluid. As a result, the fluid itself acts as if it is charged, with a given volumetric charge density as analyzed in Chapter 2. The fluid therefore moves at a velocity ($\mathbf{v}_{eo}$), as seen in Figure 8-1.

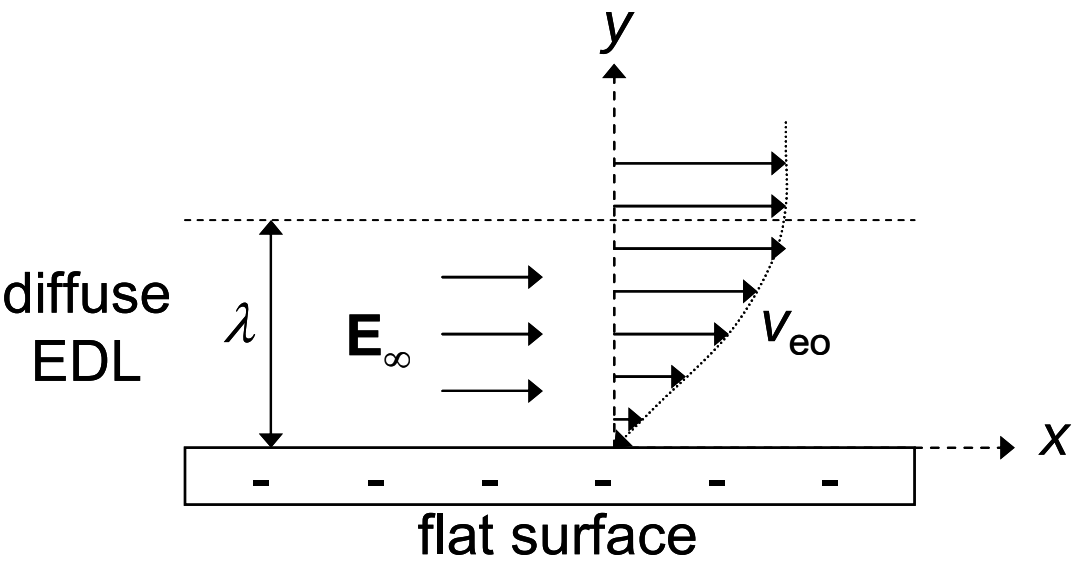

**Figure 8-1**. Schematic of electroosmosis. A flat surface is immersed in a fluid – oftentimes water – and becomes charged. Here the surface is shown to be negative. An electric field ($\mathbf{E}_\infty$) is applied parallel to the surface. The electric field acts on the diffuse part of the electrical double layer (EDL), which has a net charge, and causes the fluid to move. The thickness ($\lambda$) of the EDL is typically 4 or 5 times $\kappa^{-1}$. The resulting electroosmotic velocity ($\mathbf{v}_{eo}$) changes with distance away from the plate.

How can we analyze the fluid flow? We need to use the "electrokinetic equations". Let's first present a somewhat simplified version of the electrokinetic equations. The electric field is parallel to the surface, and so we might expect that the fluid velocity will be only in the $x$-direction. The $x$-component of the Stokes hydrodynamic equations is[163]

$$\eta\left(\frac{\partial^2 v_x}{\partial x^2}+\frac{\partial^2 v_x}{\partial y^2}+\frac{\partial^2 v_x}{\partial z^2}\right)-\frac{\partial p}{\partial x}+\rho_e E_s=0 \qquad (8\text{-}5)$$

and the continuity equation is

$$\frac{\partial v_x}{\partial x}+\frac{\partial v_y}{\partial y}+\frac{\partial v_z}{\partial z}=0 \qquad (8\text{-}6)$$

We assume that the plate is infinite in width, so $\partial v_x/\partial z=0$, and we have already assumed that $v_y=v_z=0$. As a result, the continuity equation tells us that $\partial v_x/\partial x=0$, and since this is always zero, so is $\partial^2 v_x/\partial x^2=0$. The hydrodynamics equations, which are the momentum balance, thus become

---

[163] The full set of electrokinetic equations are listed here in their vector form. The Navier-Stokes equation describes the fluid dynamics; usually for electrokinetic problems, we are at low Re number, and so the creeping flow equations apply for the momentum balance and continuity:

$$\eta\nabla^2\mathbf{v}-\nabla p+\rho_e\mathbf{E}=\mathbf{0}, \qquad \nabla\cdot\mathbf{v}=0$$

If the applied field is static (zero frequency) or low frequency (usually less than $10^9$ $s^{-1}$, then the full Maxwell equations for electrodynamics reduce to the Poisson equation of electrostatics:

$$\nabla^2\psi=-\frac{\rho_e}{\varepsilon}, \qquad \rho_e=\sum_i z_i e n_i$$

In some cases the applied electric field causes the EDL to distort in shape, and it is no longer sufficient to assume an EDL described by the Poisson-Boltzmann equation. In these cases we must use the ion migration equation, with ion mobility $\mu_i$ , and conservation of mass equation, for each ion

$$\mathbf{j}_i=-D_i\nabla n_i-z_i e\mu_i n_i\nabla\psi+n_i\mathbf{v}, \qquad \frac{\partial n_i}{\partial t}+\nabla\cdot\mathbf{j}_i=0$$

$$\eta \frac{\partial^2 v_x}{\partial y^2} - \frac{\partial p}{\partial x} + \rho_e E_s = 0 \tag{8-7}$$

There are two more useful simplifications. We will apply no pressure gradient for now. If a pressure gradient were applied, we could treat the problem separately and superimpose the solutions, since the Stokes equations are linear in velocity.

Furthermore, we will rearrange the Poisson equation to be

$$\rho_e = -\varepsilon \left( \frac{\partial^2 \psi}{\partial x^2} + \frac{\partial^2 \psi}{\partial y^2} + \frac{\partial^2 \psi}{\partial z^2} \right) \tag{8-8}$$

Prior to using this form of the Poisson equation, we make a scaling analysis. We are applying a uniform electric field in the *x*-direction, so $\partial E_x / \partial x = 0$. Since for static fields we know $E_x = -\partial \psi / \partial x$, we see that $\partial^2 \psi / \partial x^2 = 0$. No electric field is applied in the *y*-direction, but there is an electric field and variation of the electrical potential, due to the EDL. In fact, the electric field in the *y* direction scales as $E_y \sim \psi_0 / \kappa^{-1}$, and typically has a magnitude of greater than $10^6$ V/m. There is no electric field in the *z*-direction, either due to the EDL or an applied field. Thus, $\rho_e = -\varepsilon \partial^2 \psi / \partial y^2$.

Substituting these equations into the Stokes equations gives

$$\eta \frac{\partial^2 v_x}{\partial y^2} - \varepsilon \frac{\partial^2 \psi}{\partial y^2} E_s = 0 \tag{8-9}$$

Re-arranging this equation begins to reveal the final form of the electroosmotic velocity:

$$\frac{\partial^2}{\partial y^2} \left( v_x - \frac{\varepsilon E_s}{\eta} \psi \right) = 0 \tag{8-10}$$

Boundary conditions exist at the plate surface, so that

$$v_x(y=0)=0, \qquad \frac{\partial v_x(y\to\infty)}{\partial y}\to 0 \tag{8-11}$$

and

$$\psi(y=0)=\zeta, \qquad \frac{\partial \psi(y\to\infty)}{\partial y}\to 0 \tag{8-12}$$

Here I have set $y = 0$ as the "slip plane", and so the zeta potential ($\zeta$) is defined as the potential at the slip plane. Integrating Eq 8-10 once and applying the boundary conditions at infinity reveals that the first constant of integration is zero. Integrating again gives

$$v_x = -\frac{\varepsilon E_s}{\eta}(\zeta-\psi) \tag{8-13}$$

From any model of the EDL, whether the Debye-Huckel or the Gouy-Chapman models, we know that as $y$ grows beyond a few Debye lengths, the electrical potential $\psi \to 0$. As a result, within a few Debye lengths of the surface, the fluid velocity parallel to the surface is given by what is often called the electroosmotic slip velocity:

$$\boxed{v_x = -\frac{\varepsilon\zeta E_s}{\eta}} \tag{8-14}$$

### ***Smoluchowski equation of electrophoresis***

Several approaches exist for evaluating the electrophoresis of a particle with infinitesimally thin EDLs, where $\kappa a \to \infty$. I choose to examine the case when the particle is spherical, and leave the generalized result to the literature.[164] First, I use the technique of "matched asymptotic expansion": In the region very close to the sphere, within the EDL, I treat the sphere as a flat plate. At every region over the surface of the sphere, at a small distance away from the surface, I can then treat the local velocity a few Debye lengths away from the surface as the electroosmotic slip velocity from Eq 8-14. I then use an important result, that the particle velocity (**U**) is the negative of the area-averaged fluid velocity over a surface around the particle.[165] That is,

$$\mathbf{U} = -\langle \mathbf{v} \rangle = -\frac{\int_0^{2\pi}\int_0^{\pi} \mathbf{v} \sin\theta d\theta d\phi}{4\pi} \qquad (8\text{-}15)$$

where the 4π is the area of a unit sphere. Following John L. Anderson,[166] I examine a surface $S^+$ that is a slight distance away from the actual sphere surface ($S$), as shown in Figure 8-2.

When the particle is insulating, and also when its dielectric constant is much less than the surrounding fluid – the dielectric constant of many polymers is 2, while that of water is 80 – then the electric field penetrates the spheres to a very small extent.[167] The **E**

---

[164] Morrison, F.A. Jr. "Electrophoresis of a Particle of Arbitrary Shape." *J. Colloid Int. Sci.*, **34**, 210-214 (1970). During my PhD qualifying exam, I confidently told my committee that "Van Morrison showed in 1970 that the Smoluchowski equation applied to particles of any shape, if they had thin EDLs." Of course, it turns out that the famous singer of "Gloria" and "Brown Eyed Girl" did not actually derive the result; rather, it was Frank Morrison, who was at the University of Illinois.

[165] Anderson, John L.; Prieve, Dennis C. "Diffusiophoresis Caused by Gradients of Strongly Adsorbing Solutes," *Langmuir*, **7**, 403-406 (1991). See Eq 18 and Appendix B of that paper.

[166] Anderson, John L. "Colloid Transport by Interfacial Forces." *Ann. Rev. Fluid Mech.*, **21**, 61-99 (1989). This article is an absolute classic. John L.Anderson was my PhD advisor, and remains a close friend. He later became President of the Illinois Institute of Technology in Chicago. He is a master of colloid science and a world changer, and this book is dedicated to him.

[167] O'Brien, Richard W.; White, Lee R. "Electrophoretic Mobility of a Spherical Colloidal Particle." *J. Chem. Soc., Faraday Trans. II*, **74**, 1607-1626 (1978). They

field bends around the particle. The exact electric field around an insulating sphere in a uniform electric field is well known,[168] and at the edge of the sphere in the solution is

$$\mathbf{E} = -\frac{3}{2}E_{\infty}\sin\theta\mathbf{i}_{\theta} \tag{8-16}$$

where $\theta = 0$ points along the direction of the electric field, and $\mathbf{i}_{\theta}$ is the unit vector tangential to the sphere surface.

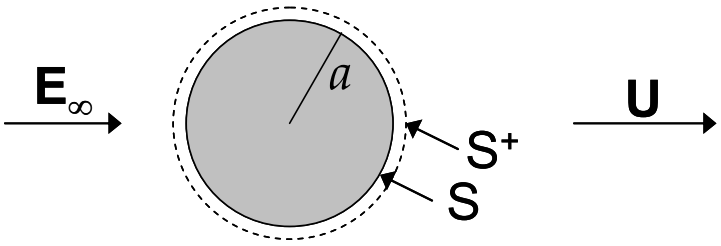

**Figure 8-2**. Sphere undergoing electrophoresis. When the EDL is thin, the edge of the EDL, often called $S^+$, is very nearly the same as the actual sphere surface (S), certainly so on the length scale of the particle radius ($a$). In this problem we take the z-direction to be along the electric field.

Since the slip velocity is given by $\mathbf{v} = -\varepsilon\zeta\mathbf{E}/\eta$, I can make the substitutions to obtain

$$\mathbf{U} = -\int_0^{2\pi}\int_0^{\pi}\left(\frac{-\varepsilon\zeta}{4\pi\eta}\right)\left(\frac{-3E_{\infty}\sin\theta}{2}\mathbf{i}_{\theta}\right)\sin\theta d\theta d\phi \tag{8-17}$$

We assume that $\zeta$ is uniform, and use the unit vector identity

$$\mathbf{i}_{\theta} = \cos\theta\cos\phi\mathbf{i}_x + \cos\theta\sin\phi\mathbf{i}_y - \sin\theta\mathbf{i}_z \tag{8-18}$$

---

showed in Section 5 of this very famous paper that it turns out that this same result is true, regardless of the boundary condition at the surface. That is, whether the particle is metallic or insulating, the same result holds.

[168] Jackson, John David. *Classical Electrodynamics*, 3rd ed. Wiley (New York) 1999. P 158 has equations 4.54, which give the electric potential inside and outside the sphere. The electric field can be calculated from $\mathbf{E} = -\nabla\Phi$, where Jackson calls $\Phi$ the electric potential. This book is the classic text for electrodynamics.

to obtain

$$\mathbf{U} = -\left(\frac{3\varepsilon E_{\infty}\zeta}{2\cdot 4\pi\eta}\right)\int_{0}^{2\pi}\int_{0}^{\pi}\left(\cos\theta\cos\phi\mathbf{i}_{x} + \cos\theta\sin\phi\mathbf{i}_{y} - \sin\theta\mathbf{i}_{z}\right)\sin^{2}\theta d\theta d\phi \tag{8-19}$$

The terms for the *x*-direction and *y*-direction integrate to zero, but the *z*-direction remains. Integrating the $\phi$ term can be done to give

$$\mathbf{U} = \frac{3\pi\varepsilon\zeta\mathbf{E}_{\infty}}{4\pi\eta}\mathbf{i}_{z}\int_{0}^{\pi}\sin^{3}\theta d\theta = \frac{3\varepsilon\zeta\mathbf{E}_{\infty}}{4\eta}\int_{0}^{\pi}\sin^{3}\theta d\theta \tag{8-20}$$

The integral can be analytically evaluated to be 4/3, so that we finally obtain the Smoluchowski equation for the electrophoretic velocity of a spherical particle with a uniform zeta potential, in a uniform applied electric field:

$$\boxed{\mathbf{U} = \frac{\varepsilon\zeta\mathbf{E}_{\infty}}{\eta}} \tag{8-21}$$

Morrison showed that this result holds for particles of arbitrary shape, so long as the zeta potential is uniform over the surface and the applied field is uniform.

**Example 8-1**. Smoluchowski equation for electrophoresis.

A ZetaPals device finds that the zeta potential when measured with electrophoresis – is -42 mV. The measurement was taken in water at T = 298 K. Calculate the electric field ($E$) required to produce a speed of $U$ = +4.6 μm/s.

*answer: E = -1.37 V/cm.*

***Henry equation for intermediate $\kappa a$ and low $\zeta$***

The Huckel equation applies to cases where $\kappa a = 0$ and $Ze\zeta / kT < 1$, while the Smoluchowksi equation applies to any value of $\zeta$ as long as $\kappa a \to \infty$. What happens when $\kappa a$ is *finite*, for a small $\zeta$ potential? Henry solved this problem,[169] showing for a sphere that

$$\boxed{\mathbf{U} = \frac{2\varepsilon\zeta\mathbf{E}_\infty}{3\eta} f(\kappa a)} \tag{8-22}$$

$$f(\kappa a) = 1 + \frac{(\kappa a)^2}{16} - \frac{5(\kappa a)^3}{48} - \frac{(\kappa a)^4}{96} + \frac{(\kappa a)^5}{96} - \left[\frac{(\kappa a)^4}{8} - \frac{(\kappa a)^6}{96}\right] e^{\kappa a} \int_\infty^{\kappa a} \frac{e^{-t}}{t} dt \tag{8-23}$$

Table 8-1 gives some values for $f(\kappa a)$.

**Table 8-1.** Table of $f(\kappa a)$ for the Henry equation.

| $\kappa a$ | $f$ | $\kappa a$ | $f$ |
|---|---|---|---|
| 0 | 1.000 | 30.0 | 1.382 |
| 0.5 | 1.009 | 40.0 | 1.407 |
| 1.0 | 1.027 | 50.0 | 1.423 |
| 2.0 | 1.065 | 75.0 | 1.446 |
| 3.0 | 1.101 | 100.0 | 1.458 |
| 4.0 | 1.132 | 200.0 | 1.478 |
| 5.0 | 1.160 | 500.0 | 1.488 |
| 10.0 | 1.253 | 1000.0 | 1.496 |
| 20.0 | 1.341 | $\infty$ | 1.500 |

[169] See Hunter's book.

### *O'Brien & White retardation*

In 1978 Richard O'Brien and Lee White published a famous paper on electrophoresis that showed that "double layer polarization" could actually give a maximum value of the electrophoresis speed. This polarization occurs as the shape of the double layer around a sphere is no longer spherical in an electric field, but rather becomes oblong and exerts an electric field in the opposite direction, slowing down the particle (Figure 8-3).

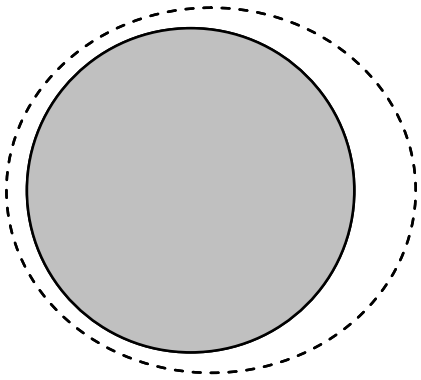

**Figure 8-3**. Double layer polarization. As the EDL distorts, it applies a backward pointing electric field that slows down the particle.

O'Brien and White produced several very famous figures that show this maximum electrophoretic velocity. Their results can be summarized by a simpler equation that reveals some important points about their solution:[170]

[170] p. 109 of Hunter's *Zeta Potential*. This is a simplication due to O'Brien, of a more full expression given in the book by S.S. Dukhin & B.V. Derjaguin, Vol 7., ch 3 in *Surface and Colloid Science* (ed. E. Matijevic), 1974. When the Dukhin number $Du \approx \exp\left(Ze\left|\zeta\right|/2kT\right)/\kappa a << 1$, the thin EDL approximation holds fairly well. I had the opportunity to visit the business (Dispersion Technology) and the home of Andrei Dukhin and his family in Summer 2014. They were such gracious hosts. Andrei's father Stanislov (the S.S. in the book above) was there, sharp-minded and pulling out colloids literature as if he had been reading it that morning.

$$f = \frac{3}{2} - \frac{6\left\{\frac{1}{2} - \frac{kT\ln 2}{Ze\zeta}\left[1 - \exp\left(-\frac{Ze\zeta}{kT}\right)\right]\right\}}{2 + \frac{\kappa a}{1 + \frac{3m}{Z^2}}\exp\left(-\frac{Ze\zeta}{2kT}\right)} \tag{8-24}$$

When $\kappa a \to \infty$, this equation says that $f$ becomes 3/2, as expected. The equation is a valid approximation for a Z:Z electrolyte. The error in the equation is of order $1/\kappa a$, and so it is especially useful for larger $\kappa a$. For KCl at room temperature, $m = 0.184$. For NaCl at room temperature, $m = 0.213$. Figure 8-4 shows a comparison of different models for electrophoresis. The O'Brien-White model is the most accurate of these since it applies for all $\kappa a$.

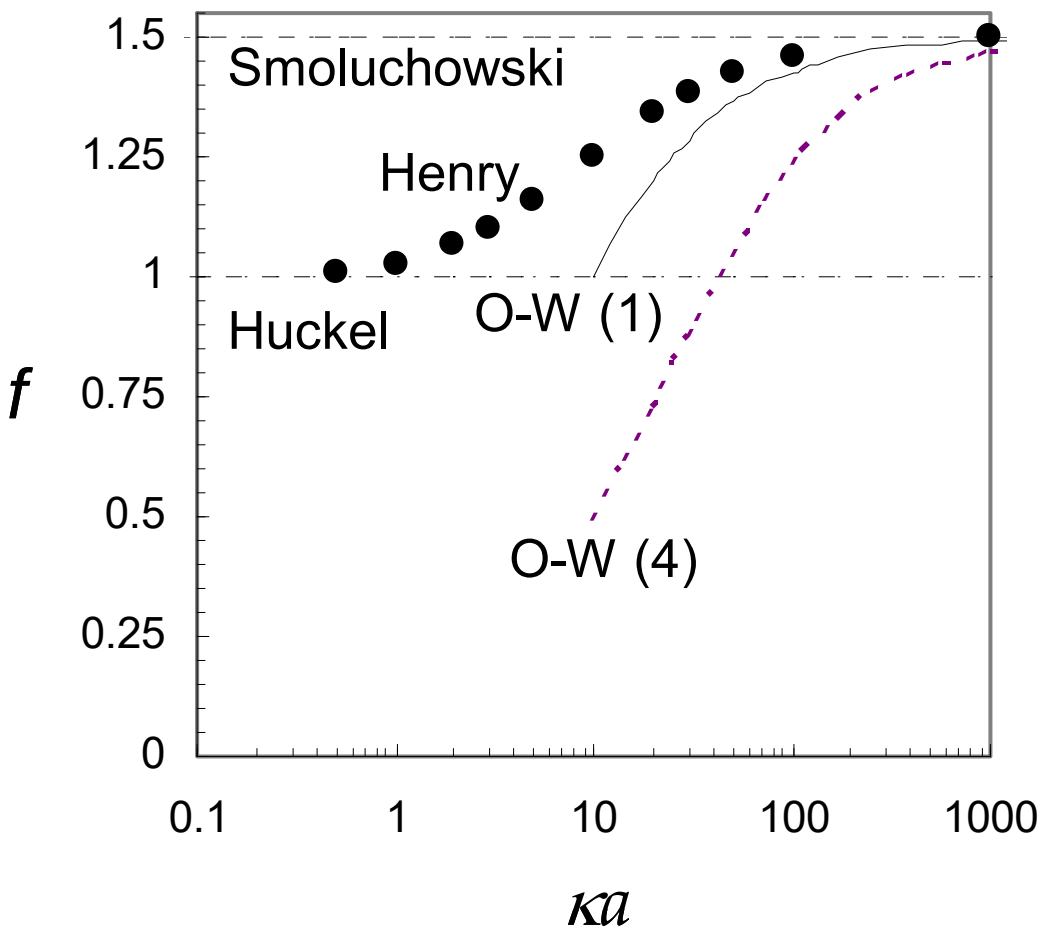

**Figure 8-4**. Electrophoresis of spheres by various models. $U = (2\varepsilon\zeta E_\infty / 3\eta)f$. The O'Brien-White (O-W) model depends on the surface potential; shown are $Ze\zeta / kT = 1$ and $Ze\zeta / kT = 4$.

### *Diffusiophoresis*

When particles are placed in a salt gradient, they can move by the process of diffusiophoresis. How does diffusiophoresis work, and what is it? First, imagine that we have a gradient of NaCl, perhaps 50 mM at a position $x = 0$ and 1 mM at a position $x = 2$ mm. There is thus a gradient of NaCl of 24.5 M/m. The Na+ and Cl- ions will start to diffuse immediately from the high to the low concentration; however, the Na+ has $D = 1.33\times10^{-9}$ m$^2$/s, while Cl- has $D = 2.03\times10^{-9}$ m$^2$/s. As a result, the Cl- tries to race ahead of the Na+. But nature doesn't want all the Cl- to rush ahead, because that would give a large charge separation, which requires an enormous energy. On average then, a backward acting electric field spontaneously arises that speeds up Na+ and slows down Cl-. This spontaneous electric field is known for any mixture of ions of type i, in any concentration profile ($n_i$), when the ions each have diffusion coefficients ($D_i$):[171]

$$\mathbf{E} = \frac{kT}{e}\frac{\sum_i z_i D_i \nabla n_i}{\sum_i z_i^2 D_i n_i} \tag{8-25}$$

For a binary, symmetric electric (i.e., $Z{:}Z$, like NaCl which has $Z$ = 1, with $z_+ = +1$ and $z_- = -1$):

$$\mathbf{E} = \frac{kT}{Ze}\frac{\left(D_+ - D_1\right)}{\left(D_+ + D_1\right)}\frac{\nabla n}{n} \tag{8-26}$$

[171] Chiang, Tso-Yi; Velegol, Darrell. "Multi-ion diffusiophoresis." *J. Colloid Interface Sci*., **424**, 120-123 (2014). This paper also shows that the usual assumptions in diffusiophoresis theory, of electroneutrality and zero current, while not exactly true, are very close. This follows the book by John Newman on *Electrochemical Systems*, 2nd edition..

This electric field acts on not only the ions in solution, but also the particles and any charged surface. In the end, having a gradient of ionic strength gives rise to a spontaneous electric field, which causes electrophoresis of particles in the suspension. If the ions have the same diffusion coefficient – for example K+ has a diffusion coefficient of $D = 1.96\times10^{-9}$ $m^2/s$, quite close to that of Cl- – then this phenomenon is reduced in importance.

There is another mechanism by which particles move in a gradient of ionic strength: chemiphoresis. When an EDL occurs near a surface, the pulling of the ions toward the surface causes an increase in pressure in the EDL, similar to how gravity causes a pressure in the atmosphere. When the salt concentration is increased, the pressure within the EDL is increased, and the precise amount can be found from the Stokes equations, setting the velocity equal to zero. As a result, if there is a gradient of ionic strength, then the fluid pressure in EDL is higher where the ionic strength is higher. This causes the fluid to move from high pressure to low pressure along the surface within the EDL. There is no flow from high pressure to low pressure normal to the surface, since the electrical forces counter-act the pressure forces; but there is a flow horizontal to the surface, since there is no counter-balancing force other than viscous drag.

Including both of these effects gives the diffusiophoretic velocity of a particle in a gradient of ionic strength ($\nabla n_0$, units $\#/m^4$, which I usually converted from molar units). The results are:

$$\boxed{\mathbf{U} = \frac{2Ze}{\kappa^2\eta}\left[\left(\frac{D_+ - D_-}{D_+ + D_-}\right)\zeta - \frac{2kT}{Ze}\ln\left(1-\gamma^2\right)\right]\nabla n_0} \qquad (8\text{-}27)$$

$$\gamma = \tanh\left(\frac{Ze\zeta}{4kT}\right) \qquad (8\text{-}28)$$

When the Debye length is small, so is the diffusiophoretic velocity – although by plugging in numbers one sees that the speed does not go to zero, even for ionic strengths >1 M. When the difference in diffusion coefficients is small, the electrophoretic term involving the diffusion coefficients becomes small.

Diffusiophoresis opens up interesting possibilities for moving particles with local chemical gradients, rather than externally-applied electric fields. It can be used to cause small particles to undergo chemotaxis and phototaxis motion, similar to that of bacteria. In addition, diffusiophoresis works in both electrolyte and non-electrolyte systems.[172,173,174] Furthermore, chemically-driven transport – far from being a rare and esoteric laboratory phenomenon – is quite common in almost every physico-chemical operation, as given in the Table 8-2,[175] although it is usually not recognized!

---

[172] Staffeld, Peter O.; Quinn, John A. "Diffusion-induced banding of colloid particles via diffusiophoresis: 1. Electrolytes." *J. Colloid Interface Science*, **130**, 69-87 (1989).

[173] Staffeld, Peter O.; Quinn, John A. "Diffusion-induced banding of colloid particles via diffusiophoresis: 2. Non-electrolytes." *J. Colloid Interface Science*, **130**, 88-100 (1989).

[174] John A. Quinn (see https://en.wikipedia.org/wiki/John_A._Quinn and http://www.cbe.seas.upenn.edu/about-people/faculty/profile-quinn.php) was my "academic grandfather" – that is, the PhD advisor of my PhD advisor John L. Anderson. John Quinn earned his PhD at Princeton in 1958 (with Joe Elgin), and he recognized Leon Lapidus and Richard Wilhelm as important mentors. John Quinn had a brilliant career in Chemical Engineering, and was a member of the National Academy of Engineering. Several of his students are also now NAE members, including John Anderson. Starting in 2004 the Department of Chemical Engineering at the University of Pennsylvania honored John Quinn at the annual "Quinn Fest" (http://www.cbe.seas.upenn.edu/about-cbe/events/quinn.php), which I first attended in 2011. When he died in February 2016, I wrote this email to my own PhD students: *John Quinn was amazing in so many ways, as you can read from the Wiki link above. But the Wiki link doesn't convey his sense of warmth, affection, and family, which made him enormously influential to everyone who knew him … like John Anderson. You have strong academic roots, which you can be proud of as you continue to imagine great dreams and win great victories in this world* ☺

[175] Velegol, Darrell; Garg, Astha; Guha, Rajarshi; Kar, Abhishek; Kumar, Manish. "Origins of concentration gradients for diffusiophoresis," *Soft Matter*, **12**, 4686-4703 (2016). This paper gives a much more detailed account of Table 8-2.

**Table 8-2.** Categories of concentration gradient origins from common physico-chemical operations.

| Category | Examples |
|---|---|
| Molecular exclusion of one or more ions | Dialysis, membranes |
| Chemical reaction with asymmetry of reactivity | Colloidal motors (e.g., Ch 01) |
| Diffusive mixing of two ionic strengths | Dead-end pores. |
| Externally-imposed | Across a track-etch membrane |
| Salt or crystal dissolution | Dissolving $CaCO_3$ or NaCl. |
| Sedimentation | Centrifugation operations |
| Molecular crystallization fro super-saturated state | Geological mineralization, kidney stones |
| Evaporation and condensation | Hygroscopic salts, coffee rings |
| Solubility or activity variation (e.g., due to T gradient) | Heated regions of a microfluidic network |

### *Dielectrophoresis*

Dielectrophoresis, sometimes abbreviated DEP in the literature, occurs when the applied electric field polarizes a particle.[176,177] Usually a large electric field is required, say >100 V/cm, and the phenomena is most important for particles large enough that Brownian motion does not dominate (i.e., > 1 μm in diameter), but small enough that gravity does not dominate (usually, < 1 mm in diameter).

---

[176] Pohl, Herbert A. *Dielectrophoresis*, Cambridge University Press (New York) 1978.

[177] Jones, Thomas B. *Electromechanics of Particles*, Cambridge University Press (New York) 1995.

A gradient of electric field is essential. The gradient causes one side of the particle to polarize more than the other side. Then as the electric field operates on the very gradient that it just created, the field acts on one side of the particle more than the other. This provides the net force that causes the particle to move. One power of the electric field, as a gradient, polarizes the particle differentially, while the other power of electric field causes the particle to move. The equation describing this force is

$$\mathbf{F}_{DEP} = 2\pi\varepsilon_f a^3 \left( \frac{\varepsilon_p - \varepsilon_f}{\varepsilon_p + 2\varepsilon_f} \right) \nabla \mathbf{E}_0^2 \qquad (8\text{-}29)$$

In order to find the velocity of the particle, $\mathbf{F}_{dep}$ is balanced against the hydrodynamic force from Stokes law.

For aqueous systems, the dielectrophoretic velocity can be more complicated. At low frequencies, below a cutoff frequency, the EDL polarizes to give the particle no net dipole moment, and therefore no dielectrophoresis. At high frequencies of the AC electric field, typically above 1 MHz, the EDL cannot reform so easily, and the particle and its double layer behave more like a purely dielectric particle.

There is a related phenomena to dielectrophoresis, called magnetophoresis. The physics has many similarities, except that the particle is magnetic – perhaps a polymer particle with some magnetite nanoparticles in it – and there is a gradient of magnetic field:

$$\mathbf{F}_{MAP} = 2\pi\mu_f a^3 \left( \frac{\mu_p - \mu_f}{\mu_p + 2\mu_f} \right) \nabla \mathbf{H}_0^2 \qquad (8\text{-}30)$$

Both dielectrophoresis and magnetophoresis scale as the sphere radius cubed.

### *Summary*

Electrokinetic phenomena provide a large array of possibilities of moving particles in solution. We have seen standard electrophoresis, but also diffusiophoresis and effects that are high order in electric field (dielectrophoresis). There is another interesting phenomena, related to the camphor boat toys often used by children, called the Marongoni effect. The Marongoni velocity of a particle is given by[178]

$$\mathbf{U} = \frac{a}{3\eta_{\text{int}} + 2\eta_{ext}} \left( -\frac{\partial \gamma}{\partial n} \right) \nabla n_0 \qquad (8\text{-}31)$$

Note that if the particle is rigid, such that the internal viscosity acts as infinite, there is no Marongoni movement. That is, both the surrounding fluid and the droplet must be fluid.

### *Symbols*

$a$ = sphere radius [=] m
$\varepsilon$ = fluid permittivity [=] $C^2/N\text{-}m^2$
$\mathbf{E}_\infty$ = applied electric field [=] V/m
$\eta$ = fluid viscosity [=] Pa-s
$q$ = charge [=] C
$\mathbf{U}$ = particle velocity [=] m/s
$\mathbf{v}_{eo}$ = "slip velocity" [=] m/s
$\psi$ = electric potential [=] V
$\zeta$ = zeta potential [=] V

---

[178] Young, Goldstein, Block, *JFM*, **6**, 350-356 (1959).

***Practice Problems***

1 We have a 900 nm diameter particle undergoing electrophoresis in an aqueous 1.0 mM KCl solution at $T$ = 293 K. It is found that the particle moves at a rate $U$ = -11.9 μm/s when $E_\infty$ = +3.0 V/cm. Find the zeta potential of the particle according to the Huckel model, the Smoluchowski model, the Henry model, and the O'Brien-White model. Comment on which model you would use.
*answer: $\kappa a$ = 46.9, Huckel gives -84 mV, Smoluchowski gives -56 mV, Henry gives -60 mV (similar to Smolochowski), O'Brien-White gives (after iterations) -70 mV. The O'Brien-White model accounts best for the EDL polarization, and would be the most accurate of these models.*

2 Say we have 900 nm diameter spherical particles in aqueous 1.0 mM KCl solution at $T$ = 293 K. Using the simplified O'Brien-White model given in Eq 8-24, plot $U$ versus $\zeta$ for the particle, for $0 \le \zeta \le 200$ mV.
*answer: The maximum speed occurs at about 125 mV. Beyond that, the speed starts to fall due to EDL polarization.*

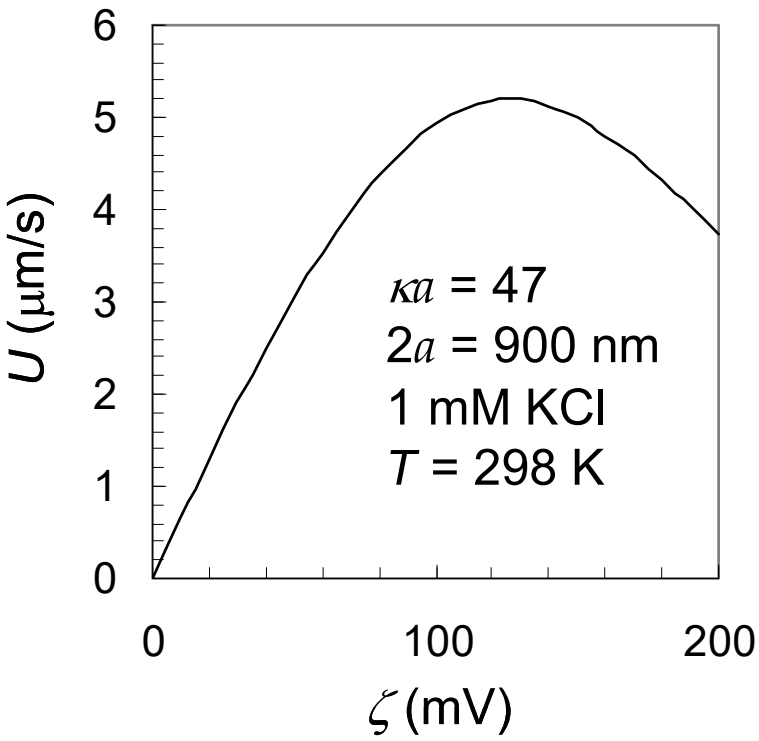

3 The definition of conductivity ($K$) follows from Ohm's law for a solution, where $i = KE_\infty$. In this equation, $i$ has units $C/m^2$-s and $E_\infty$ has units V/m (same as J/C-m). Use the Huckel equation of electrophoresis to derive a general expression for conductivity,

$$K = \frac{Z^2 e^2 n_\infty}{6\pi\eta}\left(\frac{1}{a_+} + \frac{1}{a_-}\right)$$

where as previously used, $Z$ is the valence of a symmetric $Z$:$Z$ electrolyte, $e$ is the proton charge ($1.602\times10^{-19}$ C), $n_\infty$ is the bulk salt concentration (usually I convert M to #/m$^3$ to keep consistent units), $\eta$ is the solution viscosity, and the $a$'s are the radii of the ions.

4 Estimate the conductivity ($K$, units $\Omega^{-1}$m$^{-1}$) of 1.0 mM CsI in water at $T$ = 298 K. The ionic radii are $a_{Cs+}$ = 167 pm and $a_{I-}$ = 220 pm.
*answer: The estimate based on Problem 3 is 0.0097 $\Omega^{-1}$m$^{-1}$. The actual experimental value is 0.0154 $\Omega^{-1}$m$^{-1}$. Even for ions smaller than the water molecules, the Huckel equation (continuum physics) gives a value within roughly 50%. It is interesting that the experimental conductivity due to the Cs+ and I- ions are within 0.5% of each other.*

5 An irregularly-shaped particle has a very thin EDL, suspended in an aqueous solution. The Debye length is 3.0 nm, much smaller than the particle size. When $T$ = 293 K, the particle undergoes electrophoresis at a rate of $U$ = -8.6 μm/s when $E_\infty$ = +3.0 V/cm. How fast would we expect the particle to move in a gradient of NaCl of 1000 M/m (molar per meter), which can be achieved by having a difference of 1 M salt across 1 mm of solution? At 293 K, the diffusion coefficients of the ions are $D_{Na+} = 1.18\times10^{-9}$ m$^2$/s and $D_{Cl-} = 1.81\times10^{-9}$ m$^2$/s.
*answer: $\zeta$ = -40.4 mV from the electrophoresis experiment, and the resulting diffusiophoretic speed is +15.6 μm/s.*

6 A 10.0 μm diameter silica particle is placed in an AC electric field at $10^7$ $s^{-1}$, well above the cutoff frequency of the aqueous solution. The gradient of electric field is $10^{14}$ $V/m^2$ and the electric field at the position of the particle is 100 V/cm. A useful identity is that $\nabla \mathbf{E}_\infty^2 = 2\mathbf{E}_\infty \cdot \nabla \mathbf{E}_\infty$ What is the dielectrophoretic velocity of the particle?

*answer: -0.53 μm/s.*

# 9 Assembly

**References:** Velegol, Darrell. "Assembling colloidal devices by controlling interparticle forces," *Journal of Nanophotonics*, **1**, 012502 (p 1-25) (2007). In this chapter I cite much of my own work, although the field has become very large in recent years.

Velegol, Darrell; Jerri, Huda A.; McDermott, Joseph J.; Chaturvedi, Neetu. "Micro-factories for Colloidal Assemblies," *AIChE Journal*, **56**, 564-569 (2010).

## *Possibilities and Challenges in Colloidal Assembly*

How do we fabricate "colloidal devices"? What types of colloidal devices or machines might we want to build? This chapter focuses on the first of these questions, although the second is of major importance .[179] While a few colloidal devices were considered since the 1970s, the concept of a "colloidal device" is mostly new (Table 9-1).[180] At this point there are several challenges to expanding the use of colloidal devices:

- *ideas*. I see applications as a critical challenge to the field of "colloidal devices". What will we produce that has value to customers?
- *general materials*. Many assembly techniques are limited for example to particular polymers or metals. But real

[179] Over the past decades, we have engaged research because "it might lead to something important, as it has before". We can cite numerous examples. But currently I think this diffusion-limited approach is too expensive, and wastes a huge amount of effort. Furthermore, this path encourages the world of professional science to separate from technology. There are an enormous number of *valuable* problems in this world – needed by clients and communities all around the world – to more than supply very interesting fundamental problems. My vote is to profit – yes, gain real profits – from the valuable problems, in order to provide ample resources for the fundamental research. This will encourage science to pay more attention to applications and the details required to make them work right. I also think it will allow researchers to do more of both valuable problems and fundamental research.

[180] Velegol, Darrell. "Assembling colloidal devices by controlling interparticle forces," *Journal of Nanophotonics*, **1**, 012502 (p 1-25) (2007).

devices might require polymers, metals, oxides, semiconductors, hydrogels, or biological components.

- *scalability*. Real applications often require many grams or even kilograms of materials, not micrograms. We must overcome current limitations.
- *sustainability*. Colloidal devices are mostly new, making it easier to include sustainability from the start.

**Table 9-1**. Some examples of colloidal devices (from Ref 180).

| device | use or potential use |
|---|---|
| wavelength filtering device[181] | Filtering particular wavelengths of radiation. |
| electrophoretic ink[182] | Flexible display with very fine resolution; currently commercialized by Sony and others. |
| colloidal separator[183] | Separation or sorting of microscale particles (e.g., colloids, bacteria) with a microfluidic system. |
| pumps and valves[184,185] | Pumps and valves made of particles, used in microfluidic devices. |
| display[186] | Flexible display self-assembled from small electronic components (>100 μm parts) |
| barcode identification tags | Small identification tags for information or sensing.[187,188] |
| sensor[189] | Detection of particles with over a million unique identifications. |

[181] S.A. Asher, "Crystalline colloidal narrow band radiation filter," U.S. Patent number 4,627,689 (filed 8 Dec 1983, issued 9 Dec 1986). This is the first artificial device for using colloidal crystals as a photonic manipulator.

[182] B. Comiskey, J.D. Albert, H. Yoshizawa, and J. Jacobson, "An electrophoretic ink for all-printed reflective electronic displays," *Nature* **394**, 253-255 (1998).

[183] J. Oakey, J. Allely, and D.W. M. Marr, "Laminar-flow-based separations at the microscale," *Biotechnol. Prog*. **18**, 1439-1442 (2002).

[184] A. Terray, J. Oakey, and D.W.M. Marr, "Microfluidic control using colloidal devices," *Science* **296**, 1841-1844 (2002).

[185] A. Terray, J. Oakey, and D. W. M. Marr, "Fabrication of linear colloidal structures for microfluidic applications," *Applied Phys. Lett.* **81**, 1555-1557 (2002).

[186] H.O. Jacobs, A.R. Tao, A. Schwartz, and D.H. Gracias, and G.M. Whitesides, "Fabrication of a cylindrical display by patterned assembly," *Science* **296**, 323-325 (2002).

[187] S.R. Nicewarner-Pena, R.G. Freeman, B.D. Reiss, L He, D.J. Pena, I.D. Walton, R. Cromer, C.D. Keating, and M.J. Natan, "Submicrometer metallic barcodes," *Science* **294**, 137-141 (2001).

[188] J.-M. Nam, C.S. Thaxton, and C.A. Mirkin, "Nanoparticle-based bio–bar codes for the ultrasensitive detection of proteins," *Science* **301**, 1884-1886 (2003).

[189] D. C. Pregibon, M. Toner, and P.S. Doyle, "Multifunctional encoded particles for high-throughput biomolecule analysis," *Science* **315**, 1393-1396 (2007).

Several classes of fabrication exist for building colloidal assemblies. One class is "top down assembly", in which a machine "reaches down its hands" and builds the assembly. Examples include groups of optical traps,[184] groups of atomic force microscope heads,[190] or nanofabrication techniques. Another class is "bottom up assembly",[191] in which chemical or physical information is encoded onto or into the particles themselves, and they are subsequently allowed to self-assemble into small devices, usually based on settling into a thermodynamic minimum energy state. A third class combines the two, having aspects that are top down, and aspects that are bottom up. In the last few years, a fourth class of assembly has been developed, in which ensembles of "active matter" particles[192] can be collected together in an evolutionary fashion; in this technique, the particles do not find a thermodynamic minimum, since energy is continually supplied to the system. Our groups research on active matter has been ongoing for many years.[193,194,195]

Out of these techniques, only two will be discussed in this chapter, in order to give a flavor for assembly operations. One technique is the "Stimulate-Quench-Fuse" (SQF) technique, and another is the particle lithography technique. Both techniques are from my lab, and I discuss these not only because I think they are powerful methods, not only because they involve less-expensive bottom-up assembly techniques, but primarily because I am most able to discuss their operation.

---

[190] Ginger, David S.; Zhang, Hua; Mirkin, Chad A. "The Evolution of Dip-Pen Nanolithography." *Angew. Chem. Int. Ed*., **43**, 30-45 (2004).

[191] Grzelczak, Marek; Vermant, Jan; Furst, Eric M.; Liz-Marzán, Luis M. "Directed Self-Assembly of Nanoparticles." *ACS Nano*, **4**, 3591-3605 (2010).

[192] The colloidal motors discussed in Ch 1 open up this field of active matter, and in my mind played an essential role in re-invigorating the field of colloid science.

[193] Kline, T. R.; Paxton, W. F.; Wang, Y.; Velegol, D.; Mallouk, T. E.; Sen, A. "Catalytic Micropumps: Microscopic Convective Fluid Flow and Pattern Formation." *Journal of the American Chemical Society*, **127**, 17150-17151 (2005).

[194] Hong, Yiying; Blackman, Nicole M.K.; Kopp, Nathaniel D.; Velegol, Darrell; Sen, Ayusman. "Chemotaxis of Non-Biological Nanorods." *Physical Review Letters*, **99**, 178103 (p 1-4) (2007). See Highlight "Nano Steams Ahead" in *Nature*, **450**, 5 (2007).

[195] Das, Sambeeta; Garg, Astha, Campbell, Andrew I.; Howse, Jonathan; Sen, Ayusman; Velegol, Darrell; Golestanian, Ramin; Ebbens, Stephen J. "Boundaries can Steer Active Janus Spheres," *Nature Communications*, **6**, 8999 (2015).

However, it is important to recognize that all techniques have niche capabilities and provide useful opportunities; therefore, other techniques are listed in Table 9-2. *Controlled aggregation* is similar to second order chemical reactions. Aggregation is controlled through particle functionalization or manipulation of system thermodynamics.[196,197,198,199,200] *Template-driven assembly* is similar to catalysis. Specific assemblies can be made by varying template geometry and order of particle addition.[201,202,203,204,205] The wells of the template can be made by physical wells that use gravitational force, but also by force wells created through optical or acoustic trapping. *Coalescence and dewetting* requires the addition of a functionalizing solvent or monomer. Solution conditions are then changed to induce phase separation into two distinct regions, similar to a decomposition reaction.[206,207,208,209] *Field-driven assembly* works as particles begin to assemble under the influence of an externally applied field, forming chains or cross-linked structures.[210,211,212,213,214] Magnetic, electric, shear, and other fields may be used to produce chains or sheets of particles, with similarities to polymerization.

---

[196] Yake AM, Panella RA, Snyder CE, Velegol D. Fabrication of Colloidal Doublets by a Salting Out-Quenching-Fusing Technique. *Langmuir*. 2006; **22**(22):9135-9141.
[197] Snyder CE, Ong M, Velegol D. In-Solution Assembly of Colloidal Water. *Soft Matter*. 2009; **5**:1263-1268.
[198] Hiddessen AL, Rodgers SD, Weitz DA, Hammer DA. Assembly of Binary Colloidal Structures via Specific Biological Adhesion. *Langmuir*. 2000; **16**(25):9744-9753.
[199] Johnson PM, van Kats CM, van Blaaderen A. Synthesis of Colloidal Silica Dumbbells. *Langmuir*. 2005; **21**(24):11510-11517.
[200] Zhao K, Mason TG. Directing Colloidal Self-Assembly Through Roughness-Controlled Depletion Attractions. *Phys. Rev. Lett.* 2007; **99**:268301-268304.
[201] Manoharan VN, Elsesser MT, Pine DJ. Dense Packing and Symmetry in Small Clusters of Microspheres. *Science*. 2003; **301**(5632):483-487.
[202] Xia Y, Yin Y, Lu Y, McLellan J. Template-Assisted Self-Assembly of Spherical Colloids into Complex and Controllable Structures. *Adv. Funct. Mater.* 2003; **13**(12):907-918.
[203] Pantina JP, Furst EM. Directed Assembly and Rupture Mechanics of Colloidal Aggregates. *Langmuir*. 2004; **20**(10):3940-3946.
[204] Hernandez CJ, Mason TG. Colloidal Alphabet Soup: Monodisperse Dispersions of Shape-Designed LithoParticles. *J. Phys.Chem. C.* 2007; **111**(12):4477-4480.
[205] Shi J, Mao X, Ahmed D, Colletti A, Huang TJ. Focussing Microparticles in a Microfluidic Channel with Standing Surface Acoustic Waves (SSAW). *Lab Chip*. 2008; **8**(2):221-223.

### *Stimulate-Quench-Fuse (SQF) Technique*

The SQF method was described briefly in Chapter 1. The basis of the technique is given in Figure 9-1. It has been known for milennia – even if not appreciated – that particles aggregate when they are immersed in salt water. This is the basis for the formation of river deltas near the ocean: Particles that are happy to be suspended in 2 mM ionic strength in river water, rapidly aggregate when exposed to the 600 mM NaCl of the ocean, and the large aggregates settle out of suspension quickly. The high ionic strength kills the electrostatic repulsions, enabling the ubiquitous VDW interactions to cause aggregation among the particles.

---

[206] Fujimoto K, Nakahama K, Shidara M, Kawaguchi H. Preparation of Unsymmetrical Microspheres at the Interfaces. *Langmuir*. 1999; **15**(13):4630-4635.
[207] Lu Y, Xiong H, Jiang X, Xia Y. Asymmetric Dimers Can Be Formed by Dewetting Half-Shells of Gold Deposited on the Surfaces of Spherical Oxide Colloids. *J. Am. Chem Soc.* 2003; **125**:12724-12725.
[208] Kim JW, Larsen RJ, Weitz DA. Synthesis of Nonspherical Colloidal Particles with Anisotropic Properties. *J. Am. Chem Soc.* 2006; **128**(44):14374-14377.
[209] Kim JJ, Shin K, Suh KD. Preparation of Organic-Inorganic Doublet Particles Using Seeded Polymerization. *Macromol. Res.* 2007; **15**(7):601-604.
[210] Yeh SR, Seul M, Shraiman BI. Assembly of Ordered Colloidal Aggregates by Electric-Field-Induced Fluid Flow. *Nature*. 1997; **386**:57-59.
[211] Furst EM, Suzuki C, Fermigier M, Gast AP. Permanently Linked Monodisperse Paramagnetic Chains. *Langmuir*. 1998; **14**(26):7334-7336.
[212] Hermanson KD, Lumsdon SO, Williams JP, Kaler EW, Velev OD. Dielectrophoretic Assembly of Electrically Functional Microwires from Nanoparticle Suspensions. *Science*. 2001; **294**(5544):1082-1086.
[213] Doyle PS, Bibette J, Bancaud A, Viovy JL. Self-Assembled Magnetic Matrices for DNA Separation Chips. *Science*. 2002; **295**(5563):2237.
[214] Roh KH, Martin DC, Lahann J. Biphasic Janus Particles with Nanoscale Anisotropy. *Nat. Mater.* 2005; **4**:759-763.

**Table 9-2**. Various types of assembly techniques from the literature. (Figure taken from Ref 215).

|  |  |
|---|---|
| Controlled Aggregation | A + B → AB |
| Template-Driven Assembly | A + B Template → AB |
| Coalescence and Dewetting | A + B → C → AB |
| Field- Driven Assembly | nA Field → $A_n$ |

The SQF technique was conceived in a lab meeting, because our lab needed some doublets for an experiments we were running. We realized that if we could add salt and start the formation of doublets, that only small steps would be needed to turn this random particle aggregation into a useful and powerful assembly technique. After a set amount of time – readily predicted by the rapid flocculation time – the aggregation process is "quenched" by adding a large amount of de-ionized or pure water. This step halts the aggregation, leaving the suspension with many small aggregates, usually just doublets, although triplets will form if left for longer. In order to make the doublet assemblies more permanent, especially if a polymer colloid is involved, a "fusing" step is done, in which the particles are heated to above the glass transition temperature ($T_g$) of the polymer (i.e., $T > T_g$). This can be accomplished either by raising T or lowering $T_g$.[216]

[215] Velegol, Darrell; Jerri, Huda A.; McDermott, Joseph J.; Chaturvedi, Neetu. "Micro-factories for Colloidal Assemblies," *AIChE Journal*, **56**, 564-569 (2010).

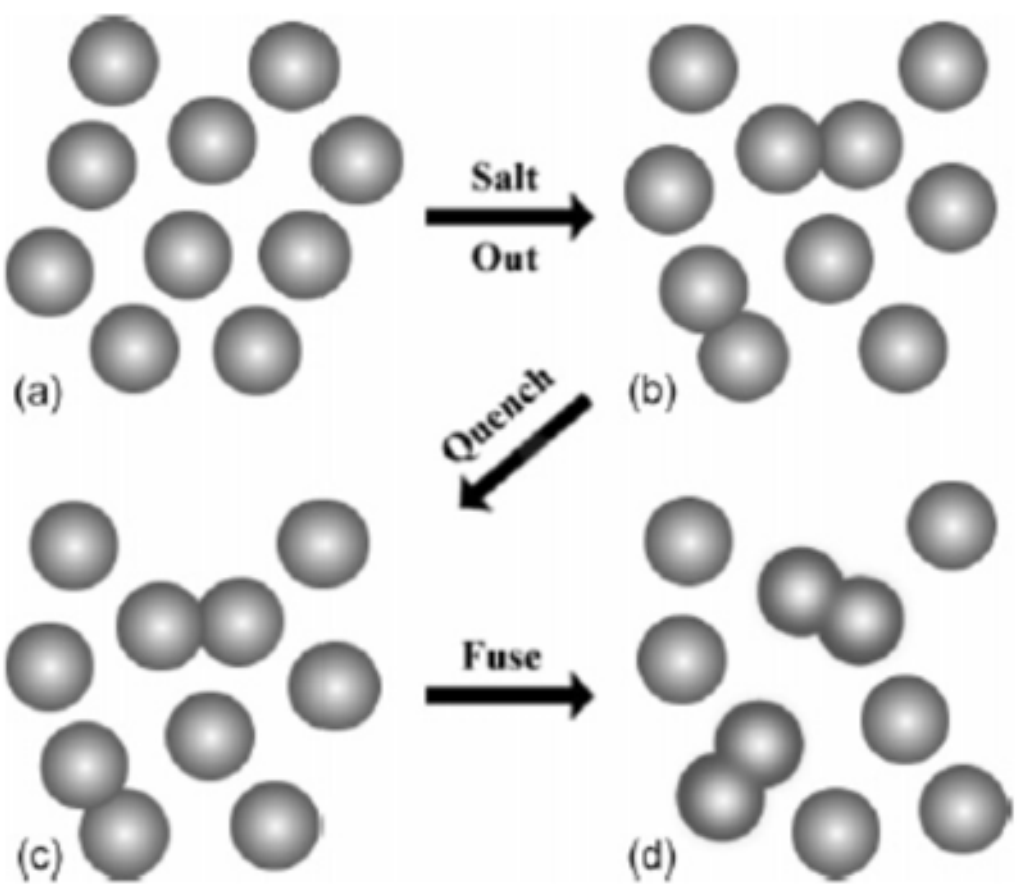

**Figure 9-1**. Schematic of the stimulate-quenching-fusing technique. (a) Starting with a suspension of singlet particles, (b) we introduce a high ionic strength to stimulate diffusion-limited aggregation due to van der Waals forces. (c) The ionic strength is quickly diluted after roughly the rapid flocculation time, which effectively quenches the aggregation reaction due to large electrostatic repulsive forces. (d) To fuse the doublets together permanently, we heat them to above their glass transition temperature or chemically fuse them. (Figure taken from Ref 196).

Other types of particles can also be formed. For instance, we have modified the SQF technique to produce trimers of gold-silicon-silver (Figure 9-2), giving a colloidal motor as in Chapter 1, but with a semiconductor particle in between. The assembly has three different materials, has the particular sequence Au-Si-Ag, is strongly held together, and is readily scaled up in production. Because the assembly has a random aspect to it, the "bond angle" of the three pieces – gold, silicon, and silver – is not controlled.

[216] Ramirez, Laura Mely; Smith, Adrian S.; Unal, Deniz B.; Colby, Ralph H.; Velegol, Darrell. "Self-assembly of Doublets from Flattened Polymer Colloids," *Langmuir*, **28**, 4086-4094 (2012). In this paper we added a bit of toluene to the top of the water, which enabled us to lower $T_g$ so that particle flattening could be done at room temperature.

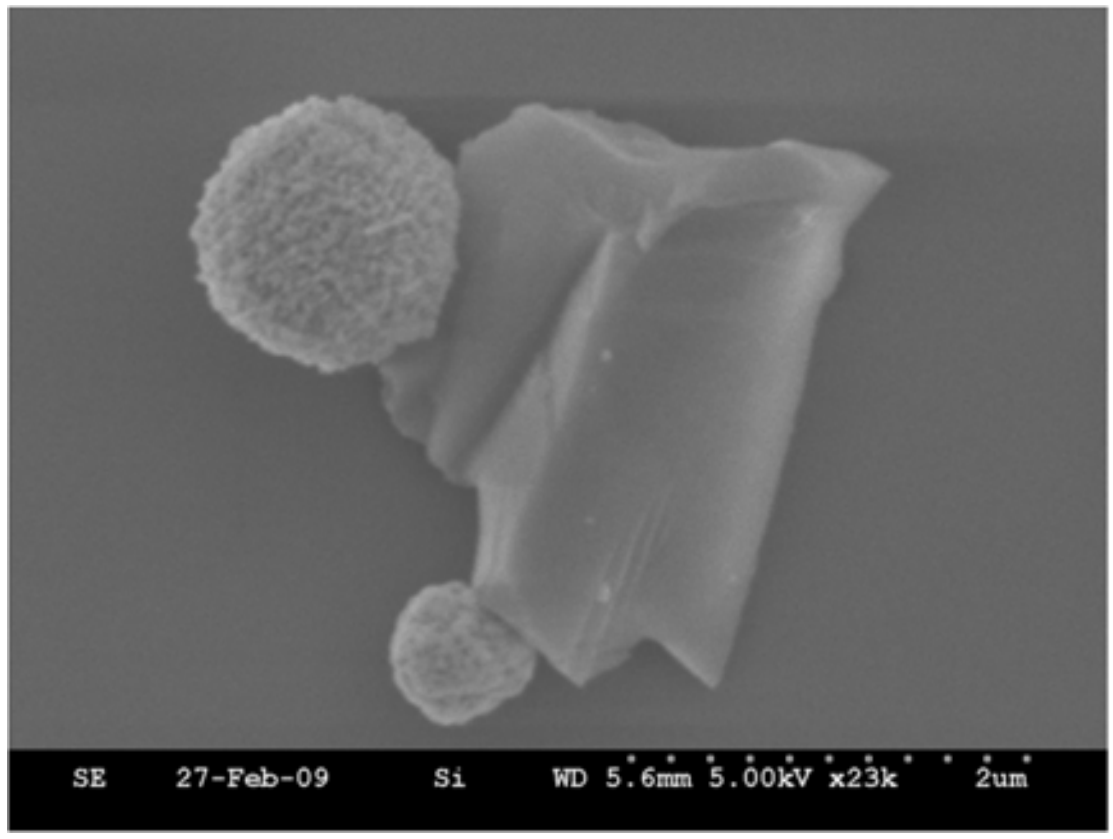

**Figure 9-2**. A gold-silicon-silver trimer formed by a modified version of the SQF technique.[217]

Once the doublets, trimers, or other desired aggregates are fabricated, another important problem arises: sorting. After the fabrication, the suspension consists of "unreacted" single particles or "singlets", the desired "doublets", and any higher order side-products that form. This is just like any chemical reaction with molecules. Sorting these particles is one of the primary challenges currently facing colloidal assembly operations, especially at a scalable level.

SQF provides a quick and easy method for fabricating assemblies. However, in order to achieve this simplicity, the method trades off specificity. A technique that is slightly more complicated to conduct, but has more specific capability in terms of placement of particles, is particle lithography.

---

[217] McDermott, Joseph J.; Chaturvedi, Neetu; Velegol, Darrell. "Functional Colloidal Trimers by Quenched Electrostatic Assembly." submitted to *Physical Chemistry Chemical Physics* (2010).

### *Particle Lithography*

In the early 2000s my lab group was studying charge nonuniformity on the surface of colloidal particles. At one point we decided we wanted to place known charges at particular locations on a particle surface, in order to test some of our ideas about charge nonuniformity. We pondered how to do this for some time, until developing the particle lithography method for placing nonuniform chemistry on individual colloidal particles. After we had succeeded in producing such charge, we realized that the technique had other useful possibilities.

The concept behind particle lithography is shown in Figure 9-3, and described in the figure caption. Essentially, we are using a surface to mask one part of a colloidal particle, while the rest of it is coated. The figure shows a flat surface being used, but in fact the "flat surface" could simply be a particle large enough to act as a flat surface, even say a 20 μm particle. Thus, the technique is scalable to larger quantities, since a beaker or other vessel can be filled with 20 or 50 μm diameter silica beads, which will act as the lithographing surface, and an entire 3-dimensional volume can be used to fabricate patterns onto particles.[218]

Once a surface is patterned with a certain type of chemistry, that region or those regions of the particle act as "bonding regions on a colloidal atom", and a "colloidal molecule"[219] can be formed. Figure 9-4 shows an example of a patterned particle, and of "colloidal water" that results from the final fabrication operation.

---

[218] Velegol, Darrell; Feick, Jason D.; Yake, Allison M.; Snyder, Charles E. "Particle lithography method and ordered structures prepared thereby." US 7771787 B2. Priority date 20014 Aug 26, publication date 2010 Aug 10.

[219] A. van Blaaderen, "Colloidal molecules and beyond", *Science* **301**, 470-471 (2003).

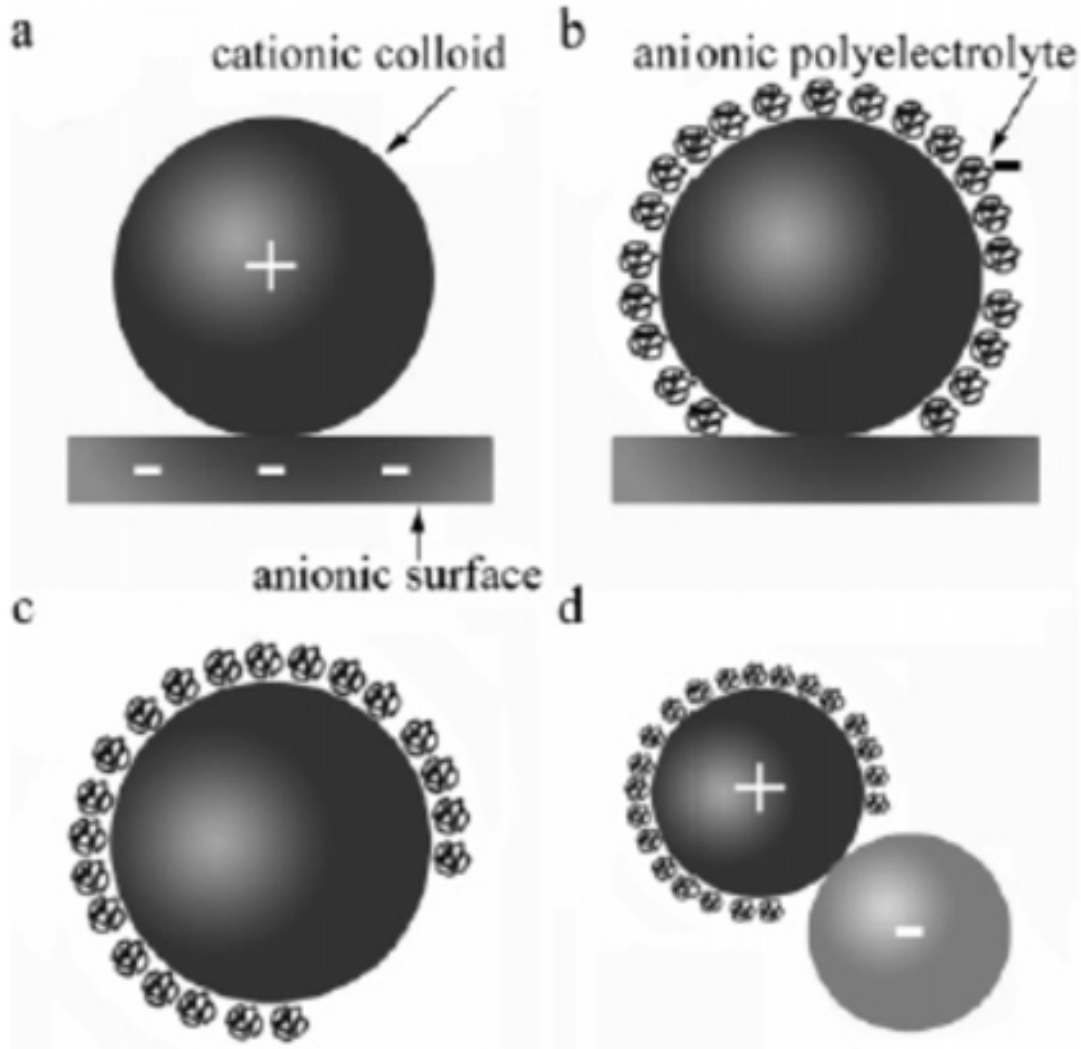

**Figure 9-3**. Schematic of the particle lithography method. (a) Amidine particles (positively charged) adhere to a negatively charged glass slide in water. (b) Negatively charged PSS polyelectrolyte (radius of gyration roughly 10 nm) is introduced, which covers the amidine particles except in the "lithographed" region near the plate where the PSS cannot access. (c) The particles are sonicated off the glass slide, exposing the nanoscale positively charged region. (d) Negative particles (e.g., silica, sulfated latex) are introduced, which adhere selectively to the positively charged region on the amidine PSL particles. (Figure taken from Ref 220).

The diameter of the nearly-circular patch(es) on the particles can be controlled. Using simple geometry, one can find the diameter of the excluded patch region on the particles to be

$$d = 4\sqrt{aR} \tag{9-1}$$

[220] Snyder, Charles E.; Yake, Allison M.; Feick, Jason D.; Velegol, Darrell. "Nanoscale Functionalization and Site-Specific Assembly of Colloids by Particle Lithography." *Langmuir*, **21**, 4813-4815 (2005).

where $a$ is the radius of the core sphere and $R$ is the radius of the coating particle or polyelectrolyte.

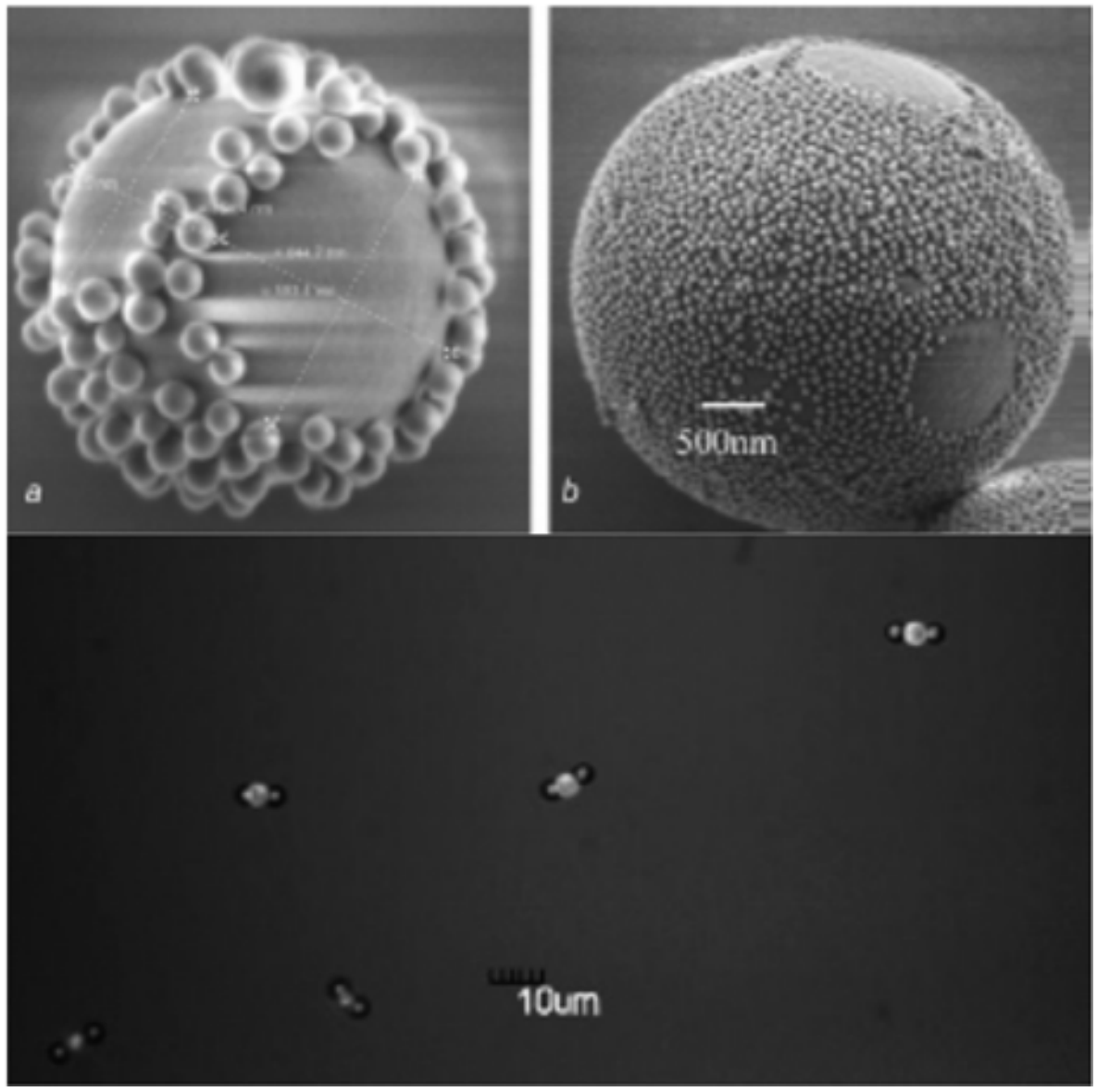

**Figure 9-4**. Patterned particles and resulting "colloidal water". The particles have been patterned using the particle lithography technique. (Upper left) Field emission scanning electron micrograph (FESEM) of a 1 μm colloid coated with 84 nm particles except at two patches. (Upper right) FESEM of a 4 μm diameter colloid coated with 60 nm particles except at two patches. (Lower) A confocal microscopy image merged with a DIC image of colloidal water assemblies. Only the center particle in each assembly (4 μm melamine formaldehyde, specific gravity 1.50) shows bright in fluorescence; the two outer colloids are polystyrene (SG = 1.055). The colloidal water assemblies are resting on a density gradient, but when flat, the bond angle turns out to be 90 degrees. (Figure taken from Ref 197).

The particle lithography technique has been extended from having local variations in chemistry, to local changes in geometry. We have used a particle flattening technique to produce doublets and chains of colloidal particles, which we call "polloidal chains" (Figure 9-5).

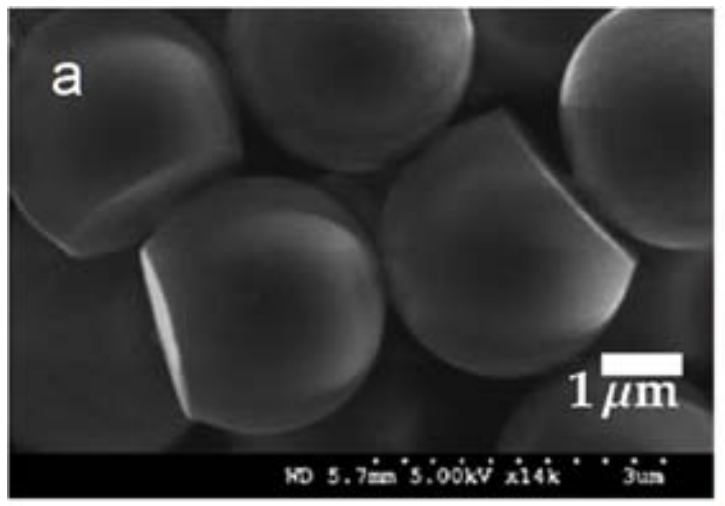

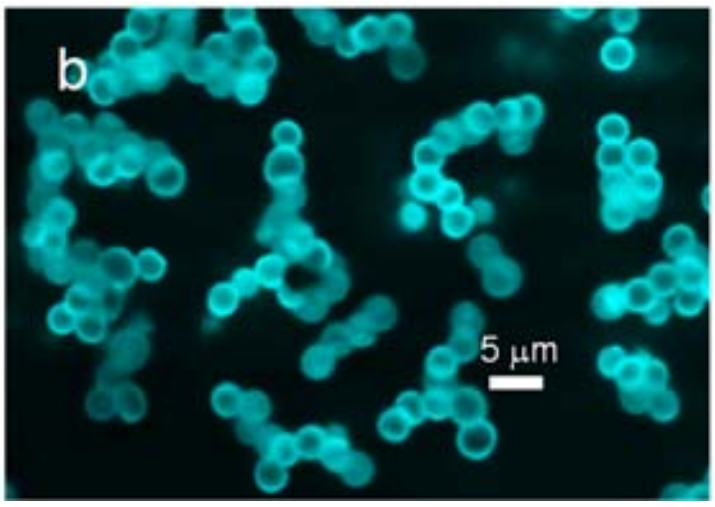

**Figure 9-5**. Flattened particles and poloidal chains. a) Polymer particles are flattened by raising the temperature above the glass transition temperature. b) The resulting flat surfaces produces stronger colloidal forces, which can result in doublets or longer chains of particles, which are flexible and stable like polymers.

### *Hydrodynamic Chromatography & Field Flow Fractionation*

Separating colloidal particles from their suspending fluid is a common operation. At the commercial scale, the milk industry separates an enormous quantity of fat droplets from milk using large disc-stack centrifuges, and a similar technique is used in the oil-sands industry to separate water and other solids from the bitumen.

But separating one type of particle from another type – what I call "sorting" – is a different type of operation. In fact, this operation is often the bottleneck to scaling up the fabrication of colloidal assemblies.

Two related techniques for sorting particles are hydrodynamic chromatography (HDC) and field flow fractionation (FFF).[221] Say we are using HDC to sort two particles, one with a diameter of 500 nm and the other with a diameter of 1000 nm. The particles travel down capillary tube, perhaps 100 μm in diameter, down a length of a few meters. During that time, the particles sample by Brownian motion the full cross section of the tube, and they do so many times. But since the small particle can get a bit closer to the wall than the large particle, the small particles spend a bit more time in the zero-fluid-speed region near the wall, due to the no-slip condition. Conversely, the larger particles spend a bit more time near the center of the tube, where the fluid is faster. As a result, the larger particle exits the tube first.

[221] Giddings, J. Calvin, *Unified Separation Science*, Wiley (New York) 1991. Giddings invented FFF in 1976. See Giddings et al, "Flow-field-flow fractionation: a versatile new separation method." *Science*, **193**, 1244-1245 (1976).

A related technique to HDC is FFF, where instead of the particles sampling the cross-section of the tube by Brownian motion, the particles are pulled toward a wall by an applied field, perhaps gravity, electrical, or magnetic. The particle that is pulled toward the wall therefore spends more time at the wall, and exits later, similar to HDC. The challenge with both HDC and FFF is that they are both analytical techniques, and are not well-suited to sorting a large quantity of particles.

### *Density Gradient Sedimentation*

An obvious technique for sorting a larger quantity of particles is to use sedimentation or centrifugation. When we first fabricated doublets in our lab, my naive thought was to simply put the particles in the centrifuge, spin them down for a few minutes, and collect the doublets at the bottom of the tube, while the singlets would not have made it as far down the tube.

Of course, the biologists have known for decades that this simple-minded idea fails, and in fact the entire suspension of particles mixes (Figure 9-6). In order to stabilize the suspension, they learned that they needed to add a “density gradient material” to the suspension, in order to make the fluid at the bottom of the container to have a higher density than the fluid at the top. Typical materials for such a density gradient are Ficoll[222] or sucrose.[223] Some researchers have used density gradient sedimentation with outstanding success, at least for obtaining single images.[201] However, in our experience, density gradient sedimentation is limited in its ability to scale up. We have found that at volume fractions of 0.001 or even 0.0001, the sedimenting material, which begins to settle in readily recognizable bands, begins to go unstable and mix. Tiny tendrils are the harbinger of an instability. This research is sometimes called a double-diffusive instability, and we have related it to the classic Rayleigh-Benard instability of heat transfer (Figure 9-7).[224,225]

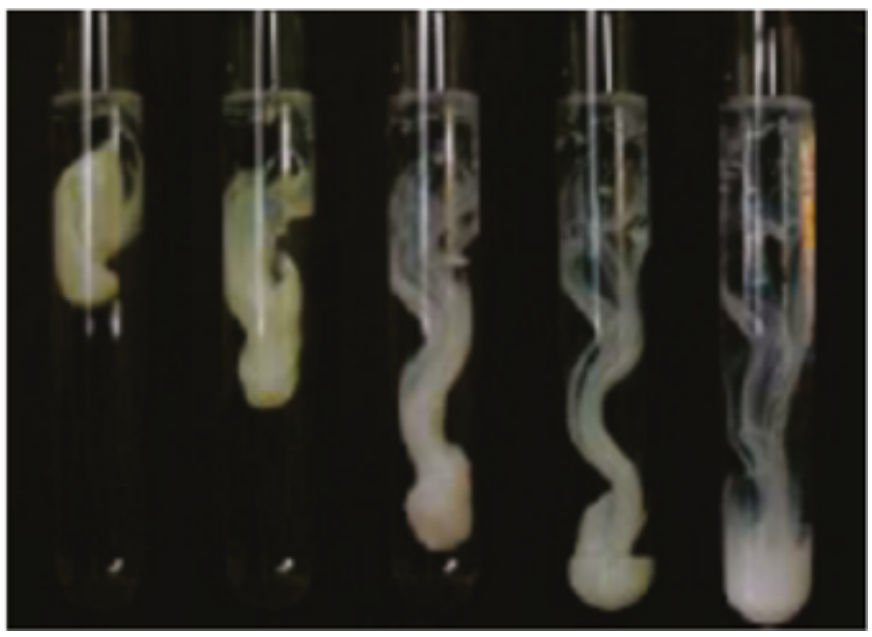

**Figure 9-6**. Mixing a droplet of suspension into water. Adding a 100 μL droplet of 0.01 volume fraction suspension containing 200 nm polystyrene particles in water to 5 mL of water. The suspension (particles plus water) has a greater density than the pure water and simply drops through the water convectively. (Figure from Ref 225).

---

[222] Ficoll is pronounced FI-kol, with a long I and a long O.

[223] Price, C. A. Centrifugation in Density Gradients; Academic Press: New York, 1982. See p 115, Table 5-1 for a list of materials.

[224] Velegol, Darrell; Shori, Shailesh; Snyder, Charles E. "Rayleigh-Bénard Instability in Sedimentation." *Industrial & Engineering Chemistry Research*, **48**, 2414-2421 (2009).

[225] Jerri, Huda A.; Sheehan, William P.; Snyder, Charles E.; Velegol, Darrell. "Prolonging Density Gradient Stability," *Langmuir*, **26**, 4725-4731 (2010).

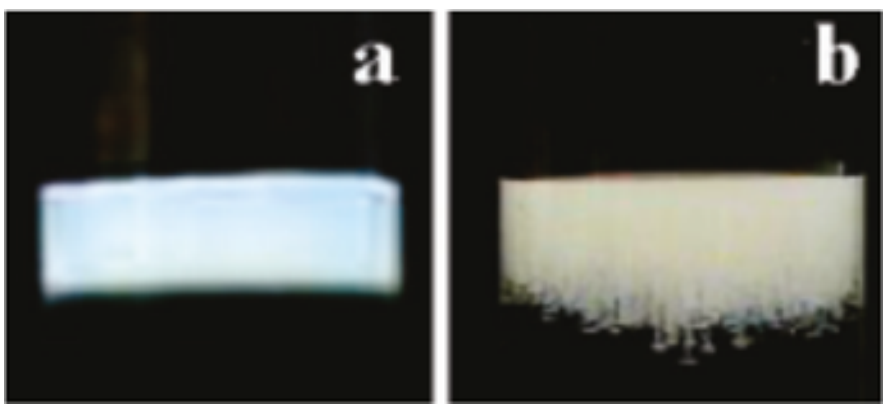

**Figure 9-7**. Instability at the top of a density gradient. The suspension has a volume fraction of 0.01, and it is supported above a sucrose density gradient of 305 kg/m$^4$. (a) At a time t = 0, the interface is flat. (b) At t = 15 min, tendril-like structures have formed. The suspension mixes very rapidly throughout the tube after this.

Another difficulty with density gradient sedimentation is the use of the density gradient material. In sensitive and serial assembly operations, one usually requires a clean particle surface for a subsequent step. Even "purified" sucrose samples contain a small amount of surface active agent, and since particles are excellent at "cleaning the solution" – in general, an undesirable trait for assembly! – they take up that surface active agent, which then contaminates the surface. We spent months working to sort particles using density gradient sedimentation, prior to realizing this fact.

### *Effective density sedimentation*

So far we have offered nothing but bad news regarding scalable sorting of two types of colloidal particles. In Fall 2008, my graduate student César was running experiments sorting particles, all of which were failing. At some point we decided to use particles of different colors, blue and green, so that we could identify quickly when a suspension sorted, rather than having to go to the microscope each time. One day he made too many samples, and put them aside for the night. The next day, just as I was getting ready to teach my class, he rushed in and yelled, "Darrell, look at this!" I looked at the tube of particles, which appeared to be perfectly sorted in layers of blue and green. We have now tried the experiments many times, with a similar result (Figure 9-8): When particles are allowed to settle to a dense suspension, up to the point where their EDLs overlap, the resulting sediment does not seem to depend on the dynamcis of how the particles reached the bottom, but the sediment follows the predictions obtained by minimizing the free energy of the system. That is, the most dense sediment goes to the bottom, while the less dense sediment goes to the top.[226]

Effective density results from two aspects, which will only be mentioned here. One part of the effective density results since individual particles have EDLs around them, and other particles cannot usually approach to within a few Debye lengths. Thus, larger particles will have more "particle material" per volume, since the thickness of the EDL does not depend upon the particle, but upon the ionic strength of the solution. Another factor is the packing fraction of mixtures. Monodisperse spheres can pack randomly to 64% volume fraction – and in this case, the spheres *plus their EDLs* seem to pack this tightly. This is true whether the particles are large or small. But *mixtures* of spheres can pack a bit more tightly. In Figure 9-8d, it is interesting that the "red" sediment at the bottom is actually a mixture of spheres, although it consists of roughly 80% red particles, while the top green layer consists of 100% green particles. This is because the mixture of spheres has the highest effective density. This technique has been extended to separate mixtures of multiple particles (Figure 9-8c).

### *Summary*

Many techniques exist for particle assembly. Two powerful techniques discussed in this chapter are the stimulate-quench-fuse (SQF) technique and the particle lithography technique. After fabricating assemblies, a sorting operation must usually be done, and this is a key bottleneck to larger production at present.

[226] González Serrano, César; McDermott, Joseph J.; Velegol, Darrell. "Sediments of Soft Spheres Arranged by Effective Density." *Nature Materials*, **10**, 716-721 (2011). This paper was inspired by a simple experiment that Linda Angus described to me. She said, "When I add sugar first into a container and then the tea bags, the tea sits on top. When I add the tea bags first and then the sugar, it all mixes up." Genius.

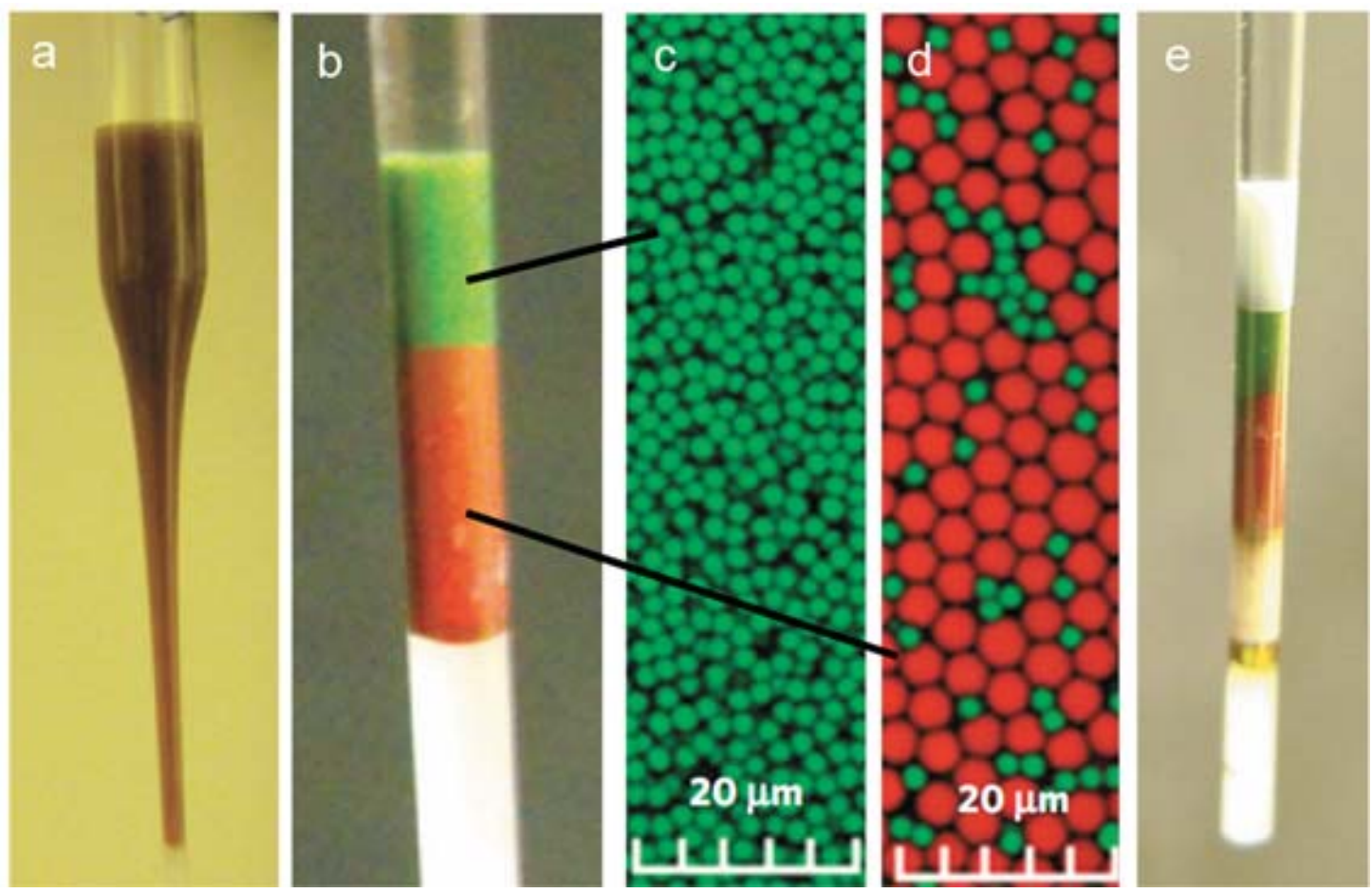

**Figure 9-8**. Sorting by effective density. a) A mixture of 2.0 μm green polystyrene and 2.0 μm red polystyrene in DI water. b) The particles spontaneously separate by density with time. c) The upper layer is 100% pure 2.0 μm green polystyrene, while d) the lower layer consists of a mixture, usually about 80% red. e) When 1.0 μm white polystyrene, 2.0 μm green polystyrene, 3.0 μm red polystyrene, 3.0 μm white silica, and 150 nm gold particles are mixed, they separate into 5 layers by gravity. Effective density sorting works only when the particles can close pack, usually in a random manner.

### *Symbols*

$a$ = sphere radius [=] m
$R$ = radius of coating particle or polyelectrolyte [=] m
$\zeta$ = zeta potential [=] V

### *Practice Problems*

1 Conceive of a particle assembly that has a functional use as a device.

2 Using the concepts in Chapter 9 – and others, if you need them – describe how you might build your colloidal device in #1. Detail the steps, describing how colloid science concepts have guided your thinking.

3 Use geometry to derive Equation 9-1 for the patch size on a lithographed particle, $d = 4\sqrt{aR}$ .

4 Use Equation 9-1 to find the patch size for a spherical particle with a diameter of 750 nm, and a coating particle of diameter 23 nm. *answer: patch diameter = 262 nm.*

# 10 Characterization

## *Classes of Experiment Measurements for Colloids*

Throughout this book, various measurements have been used and shown for colloidal particles. The measurements fall under several classes:

- visualization
- particle size
- zeta potential and mobility
- stability
- chemical composition
- interparticle forces

In this chapter the goal is to catalog a few of the most common measurement techniques, to compare and contrast the methods so that you know which is best for a given application, and to provide some sample data. I don't offer any deep insights into any of the methods.

## *Visualization*

Often the most powerful characterization that can be done is a visualization of the particles. Several methods are listed in Table 10-1. In our lab we use all these methods. We find brightfield microscopy the easiest to use, and confocal and FESEM to provide outstanding images in a fairly rapid manner.

**Table 10-1**. Methods for particle visualization.

| | |
|---|---|
| Brightfield microscopy | Quick, easy; allows real time measurements. |
| DIC microscopy | For particles with refractive index near to fluid. |
| Darkfield microscopy[227] | For small particles. |
| Fluorescence microscopy | Follow fluorescence particles in real time. |
| Confocal microscopy | Beautiful fluorescent images, almost real time. See Figure 10-1 for an image. |
| Atomic force microscopy | Map surface features with nm-scale precision |
| Scanning EM (SEM) | Images with ~30 nm precision; vacuum conditions needed. Goes by "SEM". |
| Transmission EM (TEM) | Images with ~2 nm precision; vacuum needed. Requires thin samples. |
| Field emission SEM (FESEM) | Images with ~2 nm precision. Gives topology like SEM. |

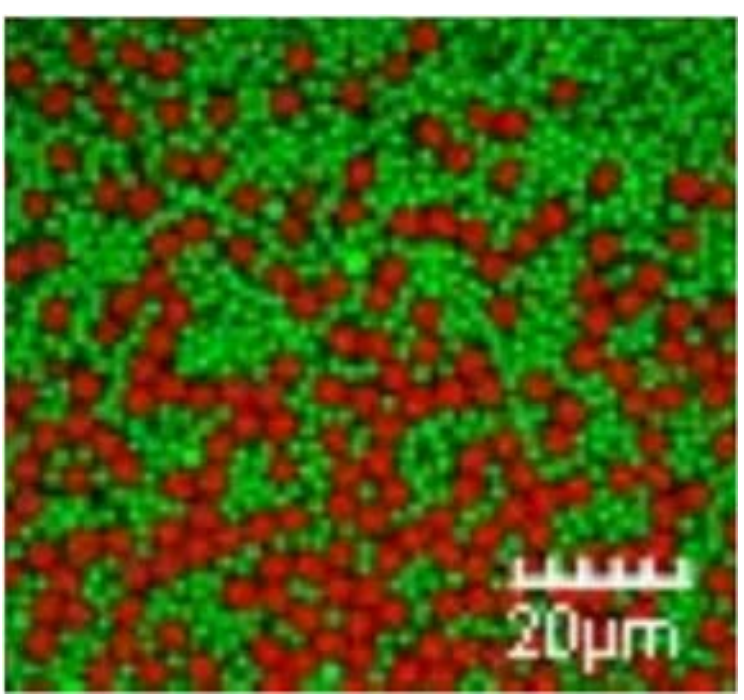

**Figure 10-1**. Confocal microscope image of 2 μm green fluorescent polystyrene particles and 4 μm red fluorescent polystyrene particles.

[227] Zsigmondy developed the "ultramicroscope", which uses darkfield microscopy, (http://en.wikipedia.org/wiki/Richard_Adolf_Zsigmondy), and won the 1925 Nobel Prize in Chemistry.

### *Particle size*

Methods are listed in Table 10-2. Some of these have been compared in the literature.[228] The sizing methods that we have most often used in our lab are electron microscopy, especially scanning electron microscopy (SEM) or field emmision SEM (FESEM), as well as dynamic light scattering.

**Table 10-2**. Methods available for particle sizing.

|  |  |
|---|---|
| Static light scattering | Size <40 nm. |
| Dynamic light scatteriing | Finds size from diffusion coefficients. |
| Electron microscopy | Particles must be dried. SEM and FESEM work well, but TEM required for <10 nm |
| Digital video microscopy[229] | Measures center-to-center distance for doublets; precise for particles >500 nm. |

### *Zeta potential*

Various methods (Table 10-3) work well for <100 mM aqueous solutions. Zeta potentials can be measured to about ±2-3 mV in my experience; I am always cautious in believing greater accuracy.

**Table 10-3**. Methods available for zeta potential.

|  |  |
|---|---|
| Microscope | Slow but sure method. The Rank Brothers instrument is seldom used (slow), but is reliable. Our lab has been working to develop this classic method for up to saturated salt conditions (e.g., ~5 M NaCl).[230] |
| ZetaPals | http://www.bic.com/ZetaPALS.html. Easy to use, reliable. Ionic strength <100 mM. |

---

[228] Davidson, John A.; Collins, Edward A. "Particle Size Analysis. Part IV: Comparative Methods fo Polyvinyl Chloride Latex." *J. Colloid Interface Sci.*, **40**, 437-447 (1972).

[229] Thwar, Prasanna K.; Velegol, Darrell. "Measuring particle diameter and particle-particle gap with nanometer precision using an optical microscope." *Industrial & Engineering Chemistry Research*, **40**, 3042 (2001). DVM is quick and easy to use.

[230] Garg, Astha; Cartier, Charles; Bishop, Kyle; Velegol, Darrell. "Finite particle zeta potentials at high ionic strength." to be submitted July 2016.

| | |
|---|---|
| Acoustosizer | www.colloidal-dynamics.com/acoustosizer.php. Measures size, zeta, other, even for concentrated |
| Colloid vibration potential | http://www.dispersion.com. Sound waves in, zeta potential out. Works well. Uses concentrated (1-50% volume fraction) particles. |

### *Stability*

See Table 10-4. We often simply observe our particles in a microscope, although uv-vis or static scattering also work well, and sometimes better depending on what you want to find.

**Table 10-4**. Methods available for assessing stability.

| | |
|---|---|
| Microscope | Slow but sure method. Take many frames of images and start counting. Slow. |
| Ultraviolent-visible | A type of static light scattering. Absorption or transmission changes as aggregation occurs. |
| Static light scattering | Scattering changes as aggregation occurs. |
| Dynamic light scattering | Particle size increases as aggregation occurs. |

### *Chemical composition*

For chemical composition, one type of measurement is a zeta potential measurement. As the surface material changes, so does the zeta potential usually. Another, more precise measurement is energy dispersive x-ray spectroscopy (EDS). EDS provides a fairly precise elemental map for particles. An example is shown in Figure 10-2.

### *Interparticle forces*

Interparticle forces for micron size or nano-size colloidal particles are very difficult to measure directly. Some methods are listed in Table 10-5. More commonly, one measures stability as a surrogate for measuring interparticle forces directly.

**Table 10-5**. Methods available for measuring interparticle forces.

Atomic force microscopy.[231] Sphere-plate of two spheres[232,233]. >10 pN. These are challenging measurements.

Total internal reflectance microscopy (TIRM).[234,235,236,237,238] Sphere-plate down to 0.01 pN. kT is the “ruler”. This is an outstanding technique, but you need to know what you are doing.

Differential electrophoresis.[239,240,241,242] Down to 0.1 pN. This is an easy technique for two particles with different zeta potentials.

---

[231] Ducker, William A.; Senden, Tim J.; Pashley, Richard M. “Direct measurement of colloidal forces using an atomic force microscope.” *Nature*, **353**, 239 (1991).
[232] Karman, Marilyn E.; Meagher, Laurence; Pashley, Richard M. “Surface Chemistry of Emulsion Polymerization.” *Langmuir*, **9**, 1220 (1993). AFM measurement between two spherical particles.
[233] Piech, Martin; Walz, John Y. “Direct measurement of depletion and structural forces in polydisperse, charged systems.” *J. Colloid Interface Sci.*, **253**, 117 (2002). Beautiful measurements of oscillatory depletion forces.
[234] Prieve, D.C.; Luo, F.; and Lanni, F. “Brownian Motion of a Hydrosol Particle in a Colloidal Force Field,” *Faraday Discuss. Chem. Soc.*, **83**, 297 (1987).
[235] Bevan, Micahel A.; Prieve, Dennis C. “Direct Measurement of Retarded van der Waals Attraction.” *Langmuir*, **15**, 7925 (1999).
[236] Sober, D.L.; Walz, J.Y. “Measruement of long range depletion energies between a colloidal particle and a flat surface in micellar solutions.” *Langmuir*, **11**, 2352 (1995).
[237] Prieve, Dennis C. “Measurement of colloidal forces with TIRM.” *Adv. Colloid Interface Sci.*, **82**, 93 (1999).
[238] Wu, H.J.; Pangburn, T.O.; Beckham, R.E.; Bevan, M.A. “Measurement and interpretation of particle-particle and particle-wall interactions in levitated colloidal ensembles.” *Langmuir*, **21**, 9879 (2005).
[239] Velegol, Darrell; Anderson, John L.; Garoff, Stephen. “Determining the Forces between Polystyrene Latex Spheres Using Differential Electrophoresis.” *Langmuir*, **12**, 4103-4110 (1996).
[240] Velegol, Darrell; Catana, Sebastian; Anderson, John L.; Garoff, Stephen. “Tangential Forces between Non-Touching Colloidal Particles.” *Physical Review Letters*, **83**, 1243-1246 (1999).
[241] Anderson, John L.; Velegol, Darrell; Garoff, Stephen. “Experimental Studies of the Forces between Colloidal Particles.” *Langmuir*, **16**, 3372-3384 (2000).
[242] Velegol, Darrell; Holtzer, Gretchen L.; Radović-Moreno, Aleksandar F.; Cuppett, Joshua D. “Force Measurements between Sub-100 nm Colloidal Particles.” *Langmuir*, **23**, 1275-1280 (2007).

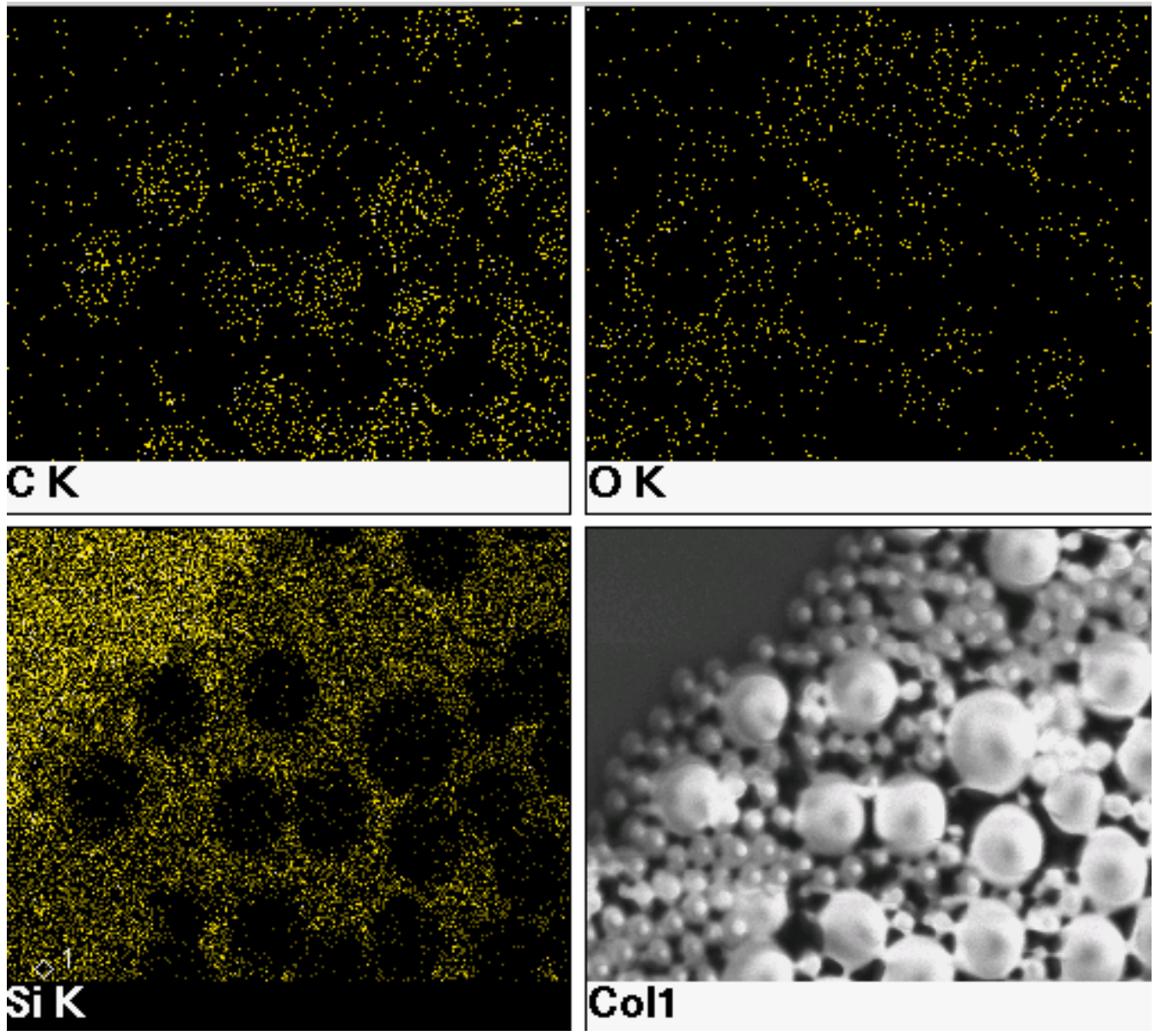

**Figure 10-2**. Energy dispersive spectroscopy. The lower right image shows a (not great) SEM of small silica ($SiO_2$) and larger polystyrene particles. Upper left: The larger polystyrene show up as having carbon in them. Upper right and lower left: The silica shows up a having silica and oxygen. Note that in the "Si" image, the upper left part shows a lot of silicon; this is because the particles are on a silicon wafer. Images taken by Neetu Chaturvedi and Maria Klimkiewicz.

### *Summary*

Many methods exist to characterize colloidal systems. Since many universities have the essential equipment for characterization, it makes sense to contact academic institutions for measurements you need to do only occasionally.

# *Index*